\documentclass{aa}
\usepackage{graphicx}
\usepackage{color}
\usepackage{txfonts}
\usepackage{hyperref}
\usepackage{lscape}
\usepackage{natbib}
\usepackage{amsmath}
\usepackage{amsfonts}
\usepackage{wasysym}
\usepackage{graphics}
\usepackage{times}
\usepackage{parskip}
\usepackage{pdflscape}
\usepackage{geometry}
\usepackage{marginnote}
\usepackage{multicol}
\usepackage{soul}
\usepackage{lmodern}
\usepackage{sidecap}
\usepackage{placeins}

\usepackage{natbibspacing}
    \renewcommand{\dbltopfraction}{0.9} % fit big float above 2-col. text
\newfont{\tlx}{cmssdc10 scaled 600}
\newfont{\mlx}{cmssdc10 scaled 770}
\newfont{\rlx}{cmssdc10 scaled 830}
\newfont{\nlx}{cmssdc10 scaled 900}
\newfont{\mfont}{cmssdc10 scaled 810}
\newfont{\rfont}{cmti12 scaled 840}
\newfont{\hvss}{cmssdc10 scaled 1540}
\definecolor{myblue1}{rgb}{0.0,0.604,0.831} 
\definecolor{myblue2}{rgb}{0.0,0.49,0.6745}
\definecolor{myblue3}{rgb}{0.0156,0.4078,0.9921}
\definecolor{myblue4}{rgb}{0.0,0.44,0.87}
\definecolor{myred1}{rgb}{0.529,0.019,0.017}
\definecolor{mycyan}{rgb}{0.63921569,0.0,0.48235294}
\definecolor{mygreen}{rgb}{0.3568,0.54902,0.2549}
\definecolor{applegreen}{rgb}{0.55, 0.71, 0.0}
\definecolor{cadmiumgreen}{rgb}{0.0, 0.42, 0.24}
\definecolor{lila}{rgb}{0.8,0.333,1.0}
\definecolor{reffig}{rgb}{0.0,0.6784,0.93725}

\newcommand{\brem}[1]{\textcolor{black}{\nlx #1}}
\newcommand{\irem}[1]{\textcolor{reffig}{\mlx #1}}

\newcommand{\imlabel}[1]{\textcolor{black}{\tlx #1}}

\newcommand{\PutLabel}[3]{\put(#1,#2){#3}}
\hypersetup{
    colorlinks=true,                            % false: boxed links; true: colored links
    linkcolor=cyan,                             % color of internal links (change box color with linkbordercolor)
    citecolor=myblue4,                          % color of links to bibliography
    filecolor=cyan,                             % color of file links
    urlcolor=cyan                               % color of external links
}
\def\starlight{\sc Starlight\rm}

\def\?{{\bf\color{red}?}}
\def\mbh{${\cal M}_{\bullet}$}

\def\reff{$R_{\rm eff}$}

\def\ha{H$\alpha$}
\def\hb{H$\beta$}

\def\msun{$\mathrm{M}_{\odot}$}
\def\zsun{$\mathrm{Z}_{\odot}$}

\def\D4000{$D_{4000}$}

\def\rr{$R^{\star}$}
\def\rbulge{$R_{\rm B}$}
\newcommand{\sbb}{mag/$\sq\arcsec$}

\def\mstar{${\cal M}_{\star}$}

\def\sstar{$\Sigma_{\star}$}

\def\ewha{EW(H$\alpha$)}

\def\mbulge{$M_{\mathrm{\star,B}}$}
\def\ml{\small ${\cal M/L}$\normalsize}
\def\tln2ha{$\log$([N\,{\sc ii}]${\scriptstyle 6584}$/H$\alpha$)}
\def\tlo3hb{$\log$([O\,{\sc iii}]${\scriptstyle 5007}$/H$\beta$)}
\newcommand{\tref}[1]{\textcolor{myblue4}{#1}}

\def\tmig{$\tau_{\rm m}$}
\def\pegase{{\sc P\'egase}}
\def\zet{$z$}
\newcommand\btab[5]{\begin{table*}[#1]\label{#3}{\parbox{#4}{\caption{#2}}\rule[-0.5ex]{0cm}{0.5ex} }
\begin{tabular*}{#4}{#5} \label{#3} }
\def\pano{\rule[0.0ex]{0cm}{2.5ex}}
\def\dmu{$\delta\mu\arcmin$\,}
\def\dmBD{$\delta\mu_{\rm BD}$}
\def\dcol{$\delta{\mathrm C}_{\rm BD}$}
\def\lya{Ly$\alpha$}
\def\z{\it z\rm}
\def\kc{{\it k}}
\def\ec{$\epsilon$}

\def\cmod{{\sc Cmod}}
\def\BD{B/D}
\def\BT{B/T}
\def\la{\lesssim}
\def\ga{\gtrsim}

\begin{document} 

% ==============================================================================================================
\title{Bulgeless disks, dark galaxies, inverted color gradients, and other expected phenomena at higher \zet}
\subtitle{The chromatic surface brightness modulation (\cmod) effect}
\titlerunning{CMOD: On the crucial importance of spatially resolved \kc\ corrections}
\authorrunning{Papaderos et al.}
% ===============================================================================================================
   \author{
          Polychronis Papaderos      % (Πολυχρὸνης Παπαδερὸς)
          \inst{\ref{IA-CAUP},\ref{IA-FCiencias},\ref{SU}}   
          \and          
          G\"oran \"Ostlin
          \inst{\ref{SU}}
          \and
          Iris Breda
          \inst{\ref{IAA-CSIC},\ref{UWien}}
          }
\institute{Instituto de Astrof\'{i}sica e Ci\^{e}ncias do Espaço - Centro de Astrof\'isica da Universidade do Porto, Rua das Estrelas, 4150-762 Porto, Portugal \label{IA-CAUP}
\and 
Instituto de Astrofísica e Ciências do Espaço, Universidade de Lisboa, OAL, Tapada da Ajuda, PT1349-018 Lisboa, Portugal \label{IA-FCiencias}
\and 
Department of Astronomy, Oskar Klein Centre; Stockholm University; SE-106 91 Stockholm, Sweden \label{SU}
\and
Instituto de Astrof\'isica de Andaluc\'ia (CSIC), Glorieta de la Astronom\'ia s/n, 18008 Granada, Spain \label{IAA-CSIC}
\and
 Department of Astrophysics, University of Vienna, T\"urkenschanzstr. 17, A-1180 Vienna, Austria \label{UWien}
\\
             \email{papaderos@astro.up.pt, ostlin@astro.su.se, iris.breda@univie.ac.at}
             }
\date{Received ??; accepted ??}
\abstract{The spectral energy distribution (SED) of galaxies varies both between galaxies and within them.
For instance, early-type spiral galaxies have a red bulge surrounded by a bluer star-forming disk with H{\sc ii} regions within.
When observing redshifted galaxies, a given photometric filter probes light at a bluer rest frame, and in relating the observed magnitudes 
to the rest frame of the filter, so-called \kc\ corrections are commonly applied to account for the relative dimming or brightening
in addition to the pure distance effect.
The amount of correction depends on the shape of the spectrum (SED), so different \kc\ corrections apply to galaxies of different spectral types.
This is, however, only part of the story, since any galaxy with a spatially non-homogeneous SED will experience a spatially varying 
relative dimming or brightening as a function of observed wavelength.
Also, the morphological appearance of galaxies will therefore change with redshift. For instance, an early spiral galaxy
observed in the $V$ band would show a prominent bulge at \zet=0, whereas, if at redshift \zet$\sim$1, the $V$ filter probes emission in
the rest-frame near-ultraviolet where the bulge is faint and the disk relatively brighter, thus the galaxy may appear as bulgeless.
One popular way of studying spatial variations in the stellar population and dust content of galaxies is the use of color maps. 
For star-forming galaxies that have an appreciable contribution from nebular emission (lines and continuum), an additional effect is that the
shifting of strong features in or out of filters will result in a non-monotonous color evolution with redshift.
Hence, unlike the effects of distance, cosmological surface brightness dimming, and gravitational lensing, which are all achromatic, 
the fact that most galaxies have a spatially varying SED leads to a chromatic surface brightness modulation (\cmod) with redshift.
While the \cmod\ effects are in principle easy to grasp, they affect multicolor imaging surveys and photometric properties derived 
from such surveys in a complex fashion.
Properties such as the bulge-to-disk ratio, S\'ersic exponent, light concentration, asymmetry index and effective radius, radial color gradients, and stellar mass determinations from SED fitting will depend on the redshift, the filters employed, and the rest-frame 2D SED patterns in a galaxy 
and will bias results inferred on galaxy evolution across cosmic time (e.g., the evolution of the mass-size, bulge-supermassive black hole,
and Tully-Fisher relation), and potentially also weak lensing, if these effects are not properly taken into account.
In this article we quantify the \cmod\ effects for idealized galaxies built from spectral synthesis models and from galaxies with observed integral field spectroscopy, and we show that they are significant and should be taken into account in studies of resolved galaxy properties and their evolution with redshift.}
% 5 {} token are mandatory
\maketitle
%
%________________________________________________________________
\section{Introduction \label{intro}}
% ::::::::::::::: Figure 1 ::::::::::::::::::::::::::::::::::::::::::::::::::::::::::::::
\begin{center}
\begin{figure*}
\begin{picture}(200,54)
\put(0,30){\includegraphics[height=2.3cm]{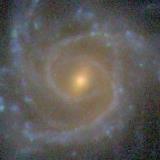}}
\put(0,5){\includegraphics[height=2.3cm]{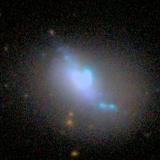}}
\put(30,0){\includegraphics[trim=0 140 0 0, clip, height=5.25cm]{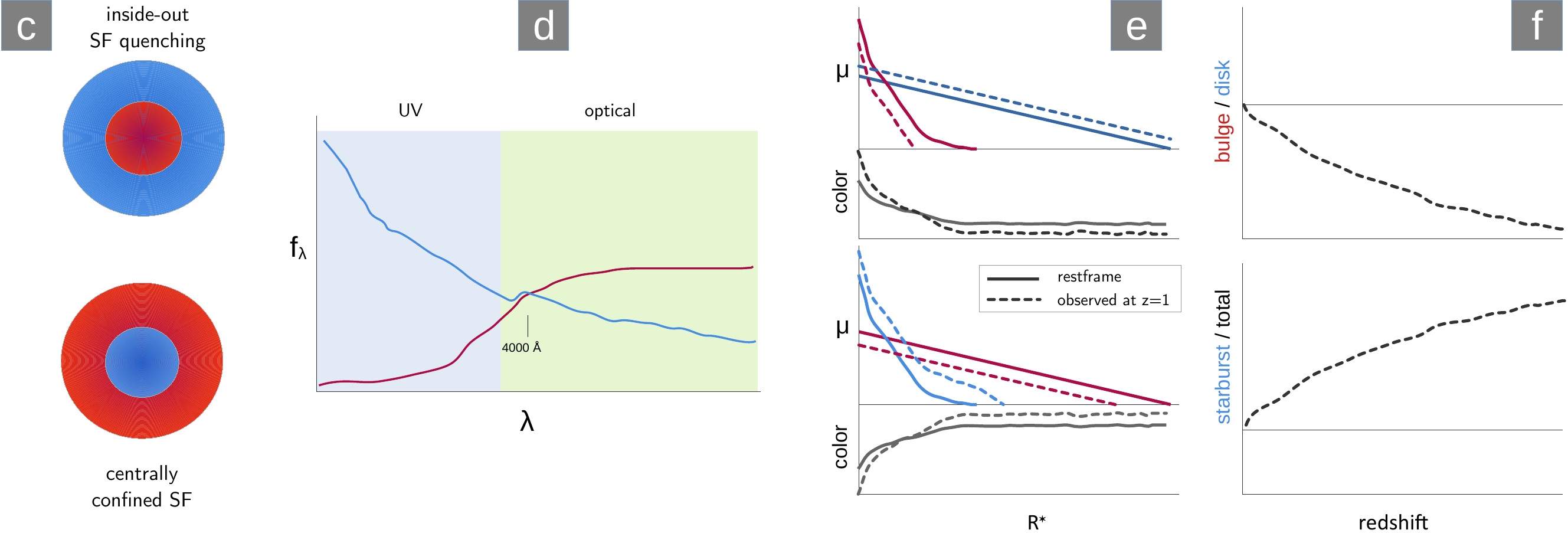}}
\PutLabel{21}{31}{\mfont \textcolor{white}{a}}
\PutLabel{21}{6}{\mfont \textcolor{white}{b}}
\end{picture}
\caption{Schematic illustration of the expected change of the structural properties of local cSED galaxies when shifted to $z\sim 1$. The two main types of cSED galaxies are shown with SDSS images in panels \brem{a} and \brem{b}: The case of a centrally SF-quenched spiral galaxy is exemplified by \object{NGC 309} \citep{BP23}, whereas \object{Haro 3} is a typical BCD galaxy \citep{LT86,Kunth88} with intense SF activity in the central part of an underlying old stellar host. These two main types of cSED galaxies are sketched in panel \brem{c} with the SF-quenched and star-forming zones of the galaxy depicted in red and blue, respectively.
Panel \brem{d} delineates the rest-frame optical-to-UV SED of these two zones, with the vertical line marking the 4000~\AA\ break.
Panel~\brem{e} attempts a qualitative description of the change in the $V$-band SBP $\mu_V$ (\sbb) and $V$-$I$ (mag) color profile when a local 
cSED galaxy is artificially redshifted to $z\sim 1$ such that the rest-frame NUV emission moves into the $V$ band. Cosmological surface brightness dimming is ignored for the sake of better visibility and because it is canceled out. Rest-frame and ObsF profiles are shown with solid and dashed curves, respectively.
In the case of a cSED galaxy such as \object{NGC 309}, the SF-quenched bulge experiences a dimming, whereas the opposite is the case for the surrounding star-forming disk.
This results in a systematically underestimated \BD\ ratio and in an amplified bulge-to-disk color contrast with increasing $z$, and thus also in a steepening of the radial color gradient within the bulge (upper diagram in panels \brem{e} and \brem{f}).
The opposite is the case in a cSED galaxy with a central starburst, such as \object{Haro 3}: in this case, the strong UV emission of the star-forming component leads to a surface brightness enhancement of the ObsF $V$-band emission simultaneously with a dimming of the surrounding older stellar host, and hence to an increasing starburst-to-total
luminosity ratio (lower diagram in panel \brem{f}). Similar yet reverse to the situation before, radial color gradients are amplified.
\label{sketch}} 
\end{figure*}
\end{center}      
% ::::::::::::::: Figure 1 ::::::::::::::::::::::::::::::::::::::::::::::::::::::::::::::
The last two decades saw an impressive amount of 2D bulge-disk decomposition studies aiming to elucidate the structural and morphological evolution of galaxies since \zet$\sim$3.
A remarkable characteristic of all of them is that they photometrically process high-\zet\ galaxies just like local ones, that is, with \kc\ corrections either entirely neglected or, in rare cases, applied adopting one single value for a galaxy as a whole (i.e., for both its red quiescent bulge and blue star-forming disk).

On this galaxy decomposition practice is founded a major part of our understanding of the evolution across cosmic time of galaxy ``size'' 
\citep[usually expressed by the effective radius, \reff; e.g.,][]{Trujillo04,Buitrago08},
morphology (approximated by, e.g., the asymmetry, light concentration, and S\'ersic exponent), and the bulge-to-disk (\BD) ratio.
The latter is a crucial observational constraint to other key topics, such as the formation mechanisms of bulges \citep[e.g.,][for a review]{KorKen04} and their co-evolution with the supermassive black hole (SMBH) they host \citep[e.g.,][]{KH13}, or the physical drivers and corresponding timescales for inside-out star formation (SF) quenching in spiral galaxies (hereafter, late-type galaxies; LTGs).
Likewise, the neglect of a \kc\ correction (be it spatially resolved or not) is a common feature of studies of color and color gradients in higher-\zet\ galaxies.

This article offers a critical view on this established practice. It reminds the reader of the warning voiced in \citet[][hereafter \tref{PO12}]{PO12}, that the application of a spatially uniform \kc\ correction to high-\zet\ galaxies with a spatially nonuniform spectral energy distribution (SED) leads to complex and serious observational biases that affect the characterization of their morphological and structural properties and stellar age patterns. Evidently, this is especially true when no \kc\ correction is applied at all. Extending our previous work, here we cursorily examine these biases and their implications for our understanding of the assembly history of galaxies. We argue that understanding and overcoming them is critically important for a meaningful study of galaxies at higher \zet\ ($\ga$0.1) and thus also a prerequisite for fully unfolding the potential of the \textit{James Webb} Space Telescope (JWST) and \textit{Euclid}.

Although the origin of the problem we discuss here is simple and generally acknowledged, its dimension and far-reaching implications have so far been largely overlooked. The starting point of our considerations is that the overwhelming majority of galaxies near and far are composite-SED (cSED) systems, that is, triaxial entities consisting of spatially and evolutionary distinct stellar populations with different star formation histories (SFHs), metallicities, and levels of nebular contamination and intrinsic extinction, and consequently also with different rest-frame SED evolutions across \zet.
In the local Universe, cSED galaxies can broadly be subdivided into two main morphological categories on the basis of the radial gradient of their specific star formation rate (sSFR), namely, centrally SF-quenched massive LTGs, with their typically red bulge and surrounding blue star-forming disk ($\nabla_{\rm sSFR}>0$ inside \reff), and star-forming dwarf galaxies, for example blue compact dwarfs (BCDs) and other extreme emission-line galaxies (EELGs), where SF typically takes place in the central part of a more extended, evolved stellar host ($\nabla_{\rm sSFR}<0$ inside their SF component), as sketched out in Fig.~\ref{sketch}\irem{a-d}.

The essence of the problem is illustrated in panels d-f of Fig.~\ref{sketch}: whereas the SED of the two main stellar populations in a cSED galaxy (for example, the bulge and the disk in an LTG) may not drastically differ from one another in the optical, they do so in the ultraviolet (UV), that is, in the rest-frame spectral window where a higher-\zet\ galaxy is observed.
Because \kc\ corrections for old, passively evolving stellar populations with a red rest-frame SED are far larger than for young stellar populations with a blue SED, it is apparent that applying one single \kc\ correction to a cSED galaxy (let alone no \kc\ correction at all) is a nonoptimal approach.

The next question is what implications the neglect of \kc\ corrections (spatially resolved or otherwise) has for the physical characterization of higher-\zet\ galaxies.
The sketch in Fig.~\ref{sketch}, although overly simplified, illustrates the principle effect and the reasoning behind the considerations in \tref{PO12}. 
We can consider the case of a massive LTG at, say, \zet$\sim$1 (upper panels) that consists of two circular-symmetric zones, the inner one representing its old SF-quenched bulge with its typical \citet{Sersic63} profile and the outer one its blue exponential star-forming disk. While in the rest-frame $V$ band the surface brightness, $\mu$, of the bulge is far brighter than that of the disk, the situation is different in the observer's frame (ObsF) $V$ band because of the faintness (copious emission) of the bulge (disk) in the near-ultraviolet (NUV).
As is apparent from panel \brem{e} (top), the expected effect is a differential dimming of the NUV-faint bulge relative to the NUV-bright disk, and vice versa. One consequence of this is the underestimation of the \BD\  ratio, or even the erroneous classification of a higher-\zet\ spiral galaxy as a virtually bulgeless disk.
Another implication is the amplification of the negative rest-frame NUV-$V$ color gradient into a steeper ObsF $V$-$J$ radial color gradient. These two effects together could be taken as evidence for a higher-\zet\ spiral containing a red, intrinsically faint bulge.

To the contrary, projection of a morphological analog of a local BCD (lower panels) to \zet$\sim$1 will lead to the opposite bias. In this case, the ObsF surface brightness of the centrally confined star-forming component will be elevated, whereas that of the surrounding old stellar host will be depressed. These two effects together will then act toward boosting the starburst-to-total luminosity ratio, eventually prompting the impression of a galaxy-wide starburst in a genuinely young galaxy. They will also lead to a compactification of the ObsF surface brightness profile (SBP) and thus to the shrinking of its \reff\ at a simultaneous increase of its light concentration index and S\'ersic exponent.
Reversely to the case of the centrally SF-quenched spiral galaxy, the positive rest-frame NUV-$V$ color gradient of the BCD will get amplified to an even steeper ObsF $V$-$J$ color gradient, reinforcing the youth interpretation.

Summarizing, the sketch in Fig.~\ref{sketch}, which merely combines simple and familiar facts, illustrates that the expanding Universe acts as a chromatic lens that
differentially modifies the morphology and color patterns of higher-\zet\ galaxies in a manner that depends on their 2D rest-frame SED. In the following, we refer to this
selective surface brightness depression (amplification) of old, passively evolving (young, star-forming) stellar populations relative to the systemic cosmological dimming as chromatic surface brightness modulation (\cmod).

This study attempts a concise yet quantitative analysis of the implications that the usual neglect of \cmod\ has on the characterization of higher-\zet\ spiral galaxies.
Its main goal is to invite the community to critically rethink established practices for structural decomposition and quantitative morphology studies of higher-\zet\ galaxies and motivate a joint exploration of new concepts that will allow \cmod\ to be overcome through unsupervised \kc\ correction tools that take the spatially varying SED of galaxies into account.
This endeavor is necessary because, as we show next, the neglect of \cmod\ introduces major biases in studies of the physical properties of high-\zet\ galaxies.
It is also especially timely since, if successful, it will allow JWST and \textit{Euclid} to fully realize their potential for elucidating the assembly history of galaxies since cosmic dawn.

In the following, we take two main approaches. The first uses evolutionary synthesis to reconstruct the ObsF properties of a mock two-zone (bulge+disk) spiral galaxy at $0 \leq z \leq 3$ (back to a cosmic age of 2 Gyr)\footnote{A standard cosmology with $H_0$=69.6, $\Omega_{\rm M}$=0.286, and $\Omega_{\Lambda}$=0.714 is assumed throughout.}.
The second uses spectral fitting of integral field spectroscopy (IFS) data for local galaxies to simulate the properties they would have if they were to be observed at a higher \zet.
More specifically, in Sect.~\ref{SED} we use synthetic SEDs for simple parametric SFHs and fixed ages to estimate the effect of \cmod\
on the bulge-to-disk surface brightness contrast and color contrast, both in the case of purely stellar emission (Sect.~\ref{sub:stSED}) and when
nebular emission is additionally taken into account (Sect.~\ref{sub:nebSED}). 
The effect of the latter is particularly relevant for near-infrared (NIR) studies of early evolutionary phases of spiral galaxies,  
when massive SF clumps emerging out of violent gas instabilities in the disk gradually migrate toward the galaxy's center.
Our discussion is supplemented by evolutionary consistent (EvCon) simulations (Sect.~\ref{sub:eSED}) that take the variation in the rest-frame SED across redshift into account.
Equipped with these simulations, we construct in Sect.~\ref{BD} synthetic galaxy SBPs, which are subsequently decomposed into a bulge and a disk in order to quantitatively
infer the variation in the ObsF versus rest-frame structural properties of a spiral galaxy as a function of its redshift and the photometric filter(s) it is studied with.
Among other quantities, we study the \BD\ ratio and the bulge-to-total (\BT) ratio, the S\'ersic exponent ($\eta$), the \reff, and various light concentration indices.

The insights obtained from evolutionary models are corroborated through simulations of higher-\zet\ galaxies that are based on population spectral synthesis models to spatially resolved IFS data (Sect.~\ref{simIFS}). These models show, consistently with our results in Sect.~\ref{SED}, that a distant analog (\zet$>$0.8) of a local bulge-dominated spiral will appear as a bulgeless disk in the optical, whereas the effect will be milder (or even reversed) in the NIR.
In Sect.~\ref{dis}, following a brief discussion of the conceptual limitations of this study (Sect.~\ref{dis:lim}), we comment on biases expected to be introduced by
\cmod\ in our understanding of the bulge assembly history (Sect.~\ref{dis:bulge-growth}) and the size evolution of galaxies (Sect.~\ref{dis:size}).
Further expected implications of \cmod\ include the systematic underestimation (by a factor of $\sim$2) of the stellar mass of higher-\zet\ galaxies 
from a fitting of their integral optical SED (Sect.~\ref{dis:sSFR}) and discrepancies between dynamical and photometric mass estimates with relevance to the Tully-Fisher (TF) relation
(Sect.~\ref{dis:TF}).

We also show that \cmod\ can globally change the morphology and color patterns of higher-\zet\ galaxies, thus greatly impacting quantitative morphology indicators and eventually justifying a closer examination of its possible effect on studies of weak lensing with \textit{Euclid} (Sect.~\ref{dis:morph}).
The complexity of interpreting ObsF color maps and color gradients of higher-\zet\ galaxies -- especially when they exhibit appreciable intrinsic stellar age gradients and strong, spatially inhomogeneous nebular emission or dust absorption -- is illustrated with an example of IFS-based simulations for the BCD \object{Haro 11} (Sect.~\ref{dis:CG}).
Finally, we argue that ``dark'' galaxies, that is, systems with an optically/NIR-invisible core surrounded by a patchy envelope, are a natural expectation from \cmod\
for protogalaxies in a short phase of vigorous SF, quick metal and dust enrichment, and eventually catastrophic cooling (Sect.~\ref{dis:DGs}).
Our main conclusions are summarized in Sect.~\ref{sum}.
Supplementing Sect.~\ref{SED}, the relation between the rest-frame \ha\ equivalent width (EW) and the enhancement of broadband magnitudes is discussed in Sect.~\ref{ap:nebSED}, and in
Sect.~\ref{ap:econsSED} we examine the variation in colors across \zet\ for EvCon models that include various SFH parameterizations.

% ::::::::::::::::::::::::::::::::::::::::::::::::::::::::::::::::::::::::::::::::::::::::::::::::::::::::::::::::::::::::::::
\section{Predictions based on evolutionary synthesis for a two-zone galaxy model \label{SED}}
\subsection{Simulations involving purely stellar SEDs \label{sub:stSED}} %\ccom{sub:stSED}
% ::::::::::::::::::::::::::::::::::::::::::::::::::::::::::::::::::::::::::::::::::::::::::::::::::::::::::::::::::::::::::::
Using the evolutionary synthesis code \pegase~2 \citep{FRV97} we computed the SED of stellar populations for seven SFH parameterizations
(Sect.~\ref{ap:SED}). These synthetic SEDs, covering an age between 0 and 13.7 Gyr, were then simulated in the range 0 $\leq$ \zet\ $\leq$ 3
(cf. Fig.~\ref{ap:synthSED}) taking into account bandpass shift and wavelength stretching, and convolved with filter transmission curves
to infer ObsF magnitudes in the Vega system. These values, referred to in the following as {reduced} surface brightness, $\mu\arcmin$,
are therefore luminosity distance independent and treat bulge and disk as point sources. This inconsistent usage of $\mu\arcmin$ is because cosmological dimming
affects bulge and disk equally, it is thus unimportant for the evolution of the bulge-to-disk surface brightness- and color contrast
as a function of \zet, which is the subject of this study. For the sake of simplicity, intrinsic extinction and attenuation by the intergalactic medium were ignored.

\begin{figure*}[h!]
\begin{picture}(200,125)
% l.h.s. (age: 13.7 Gyr)
\put(0,86){\includegraphics[clip, width=4.5cm]{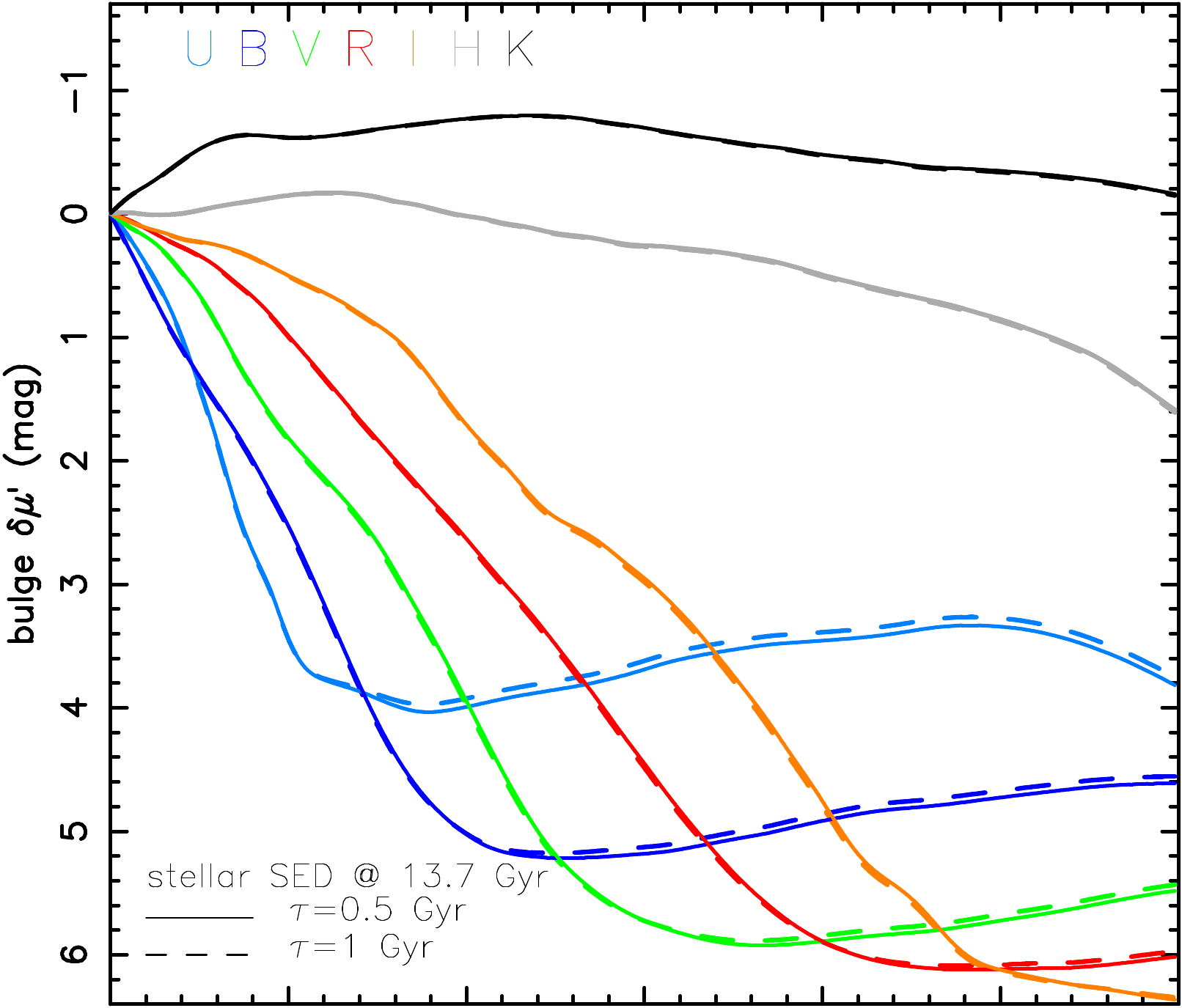}}
\put(0,45){\includegraphics[clip, width=4.5cm]{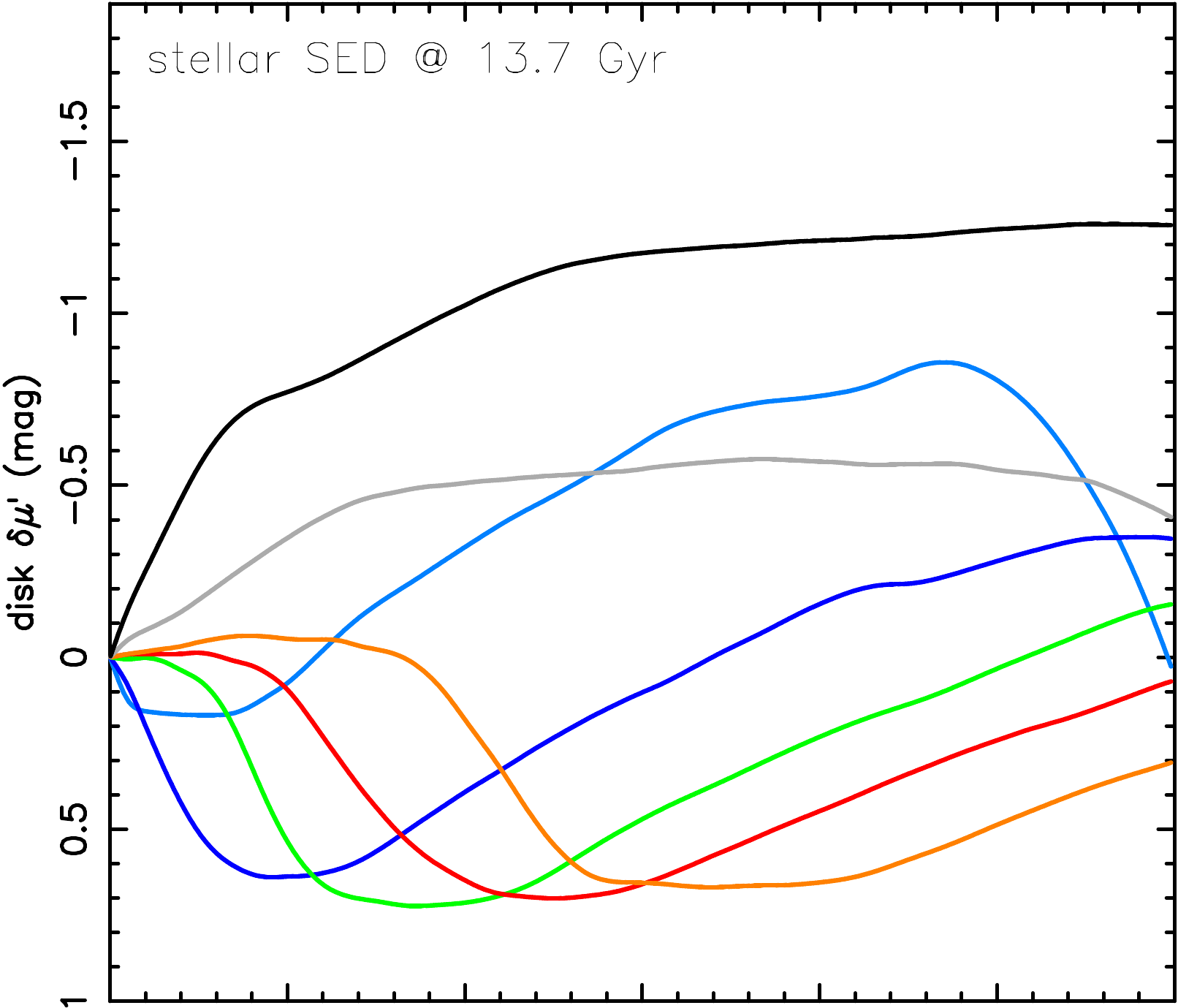}}
\put(0,0){\includegraphics[clip, width=4.52cm]{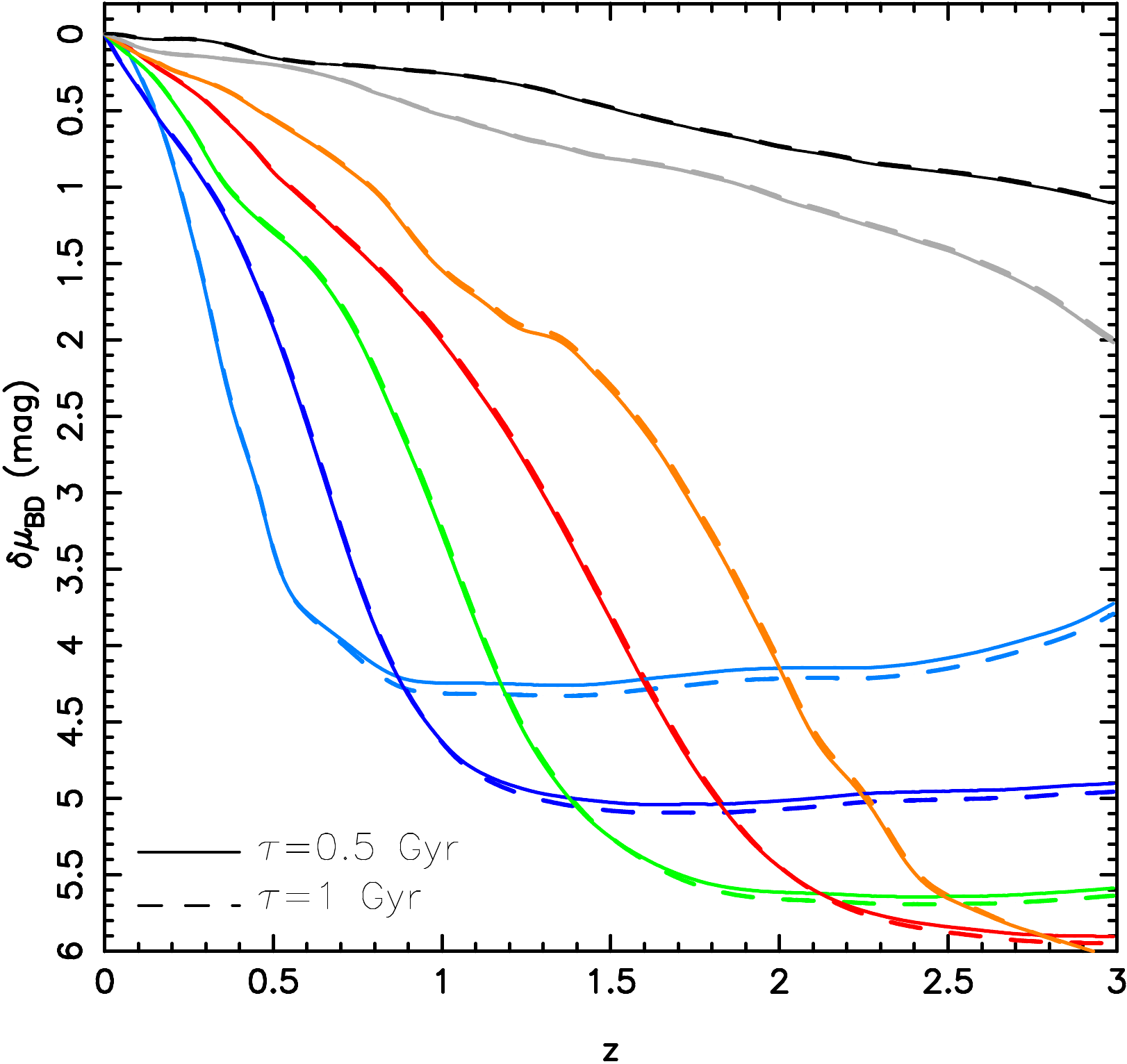}}
\put(46,86){\includegraphics[clip, width=4.5cm]{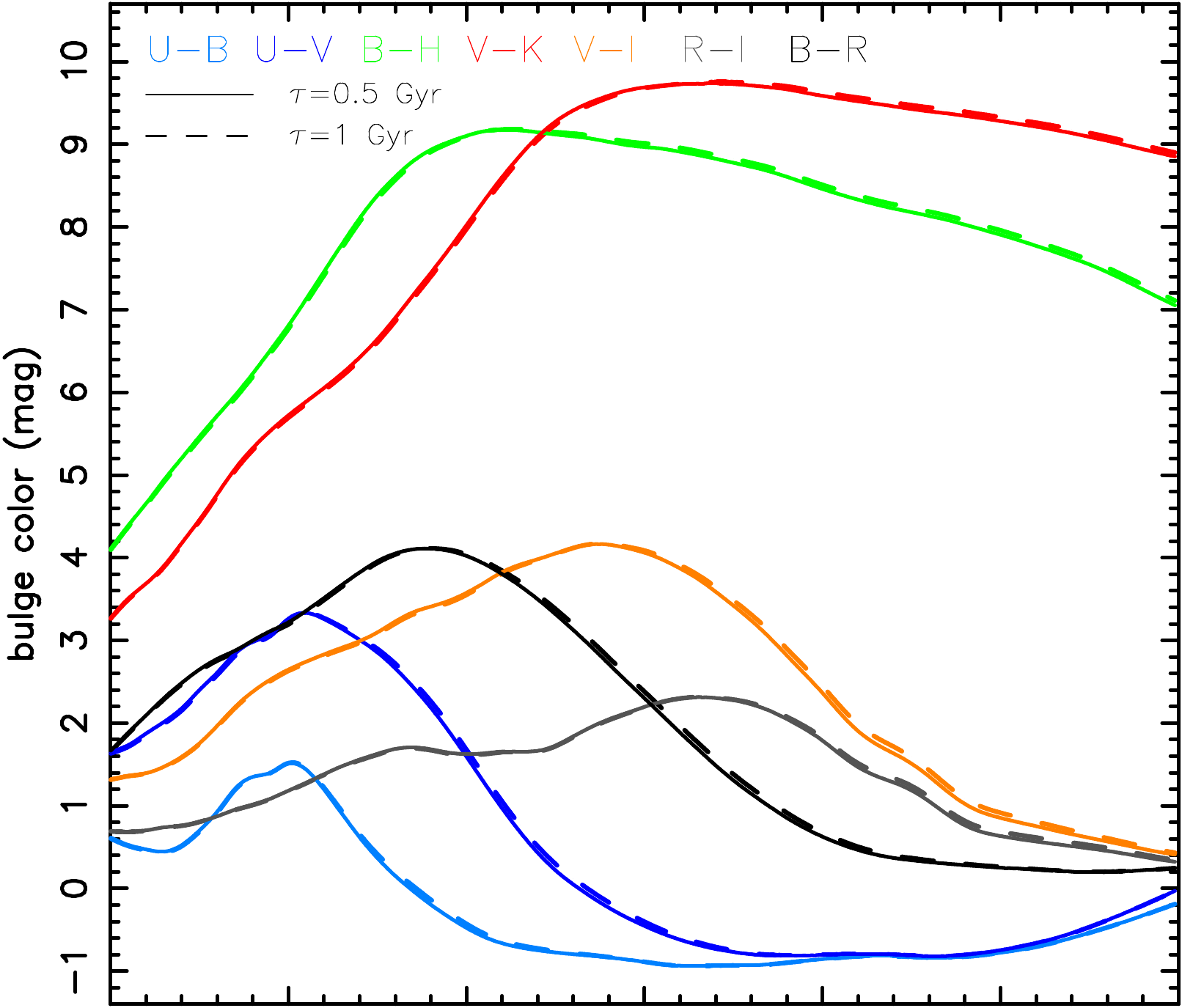}}
\put(46,45){\includegraphics[clip, width=4.5cm]{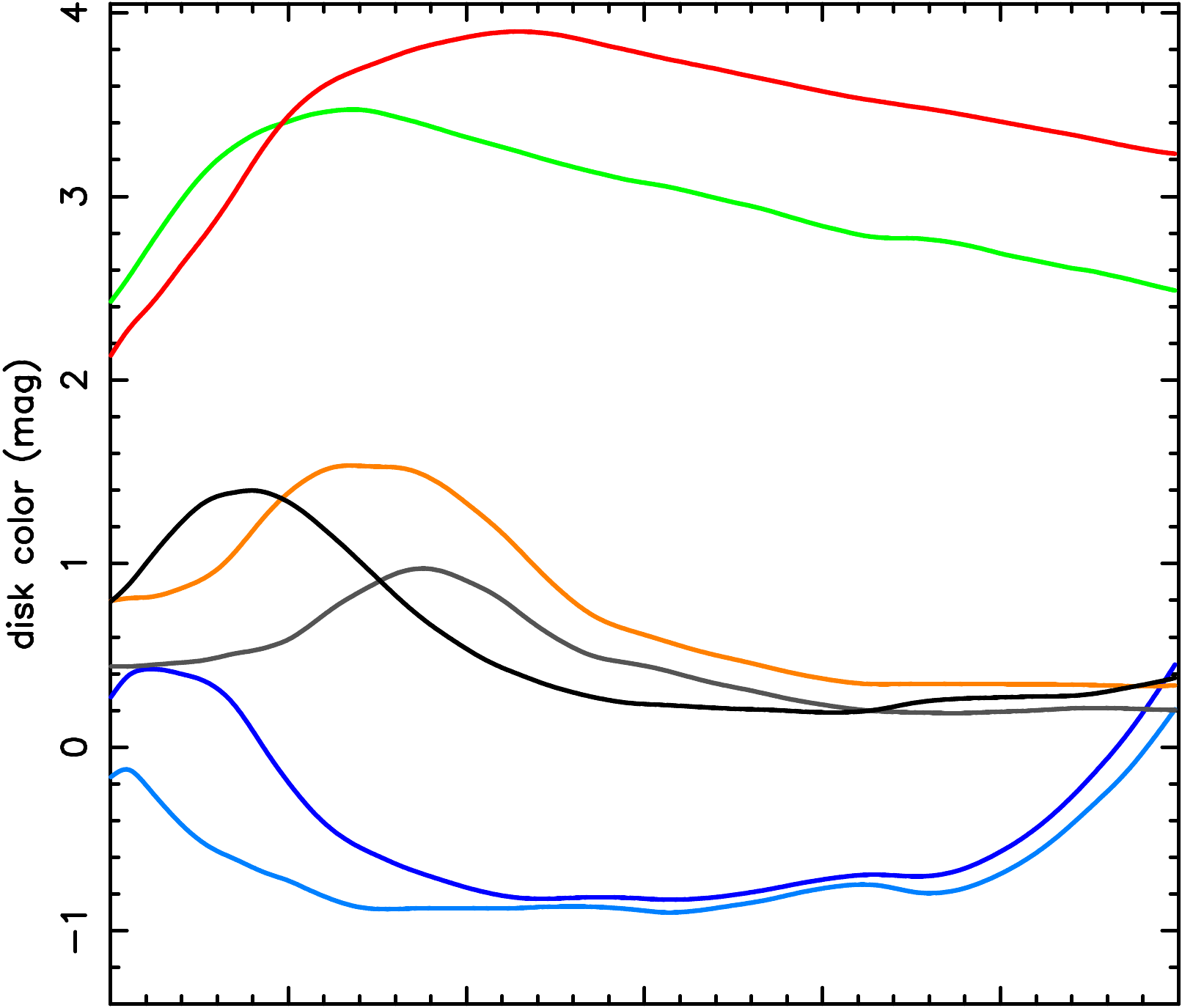}}
\put(46,0){\includegraphics[clip, width=4.52cm]{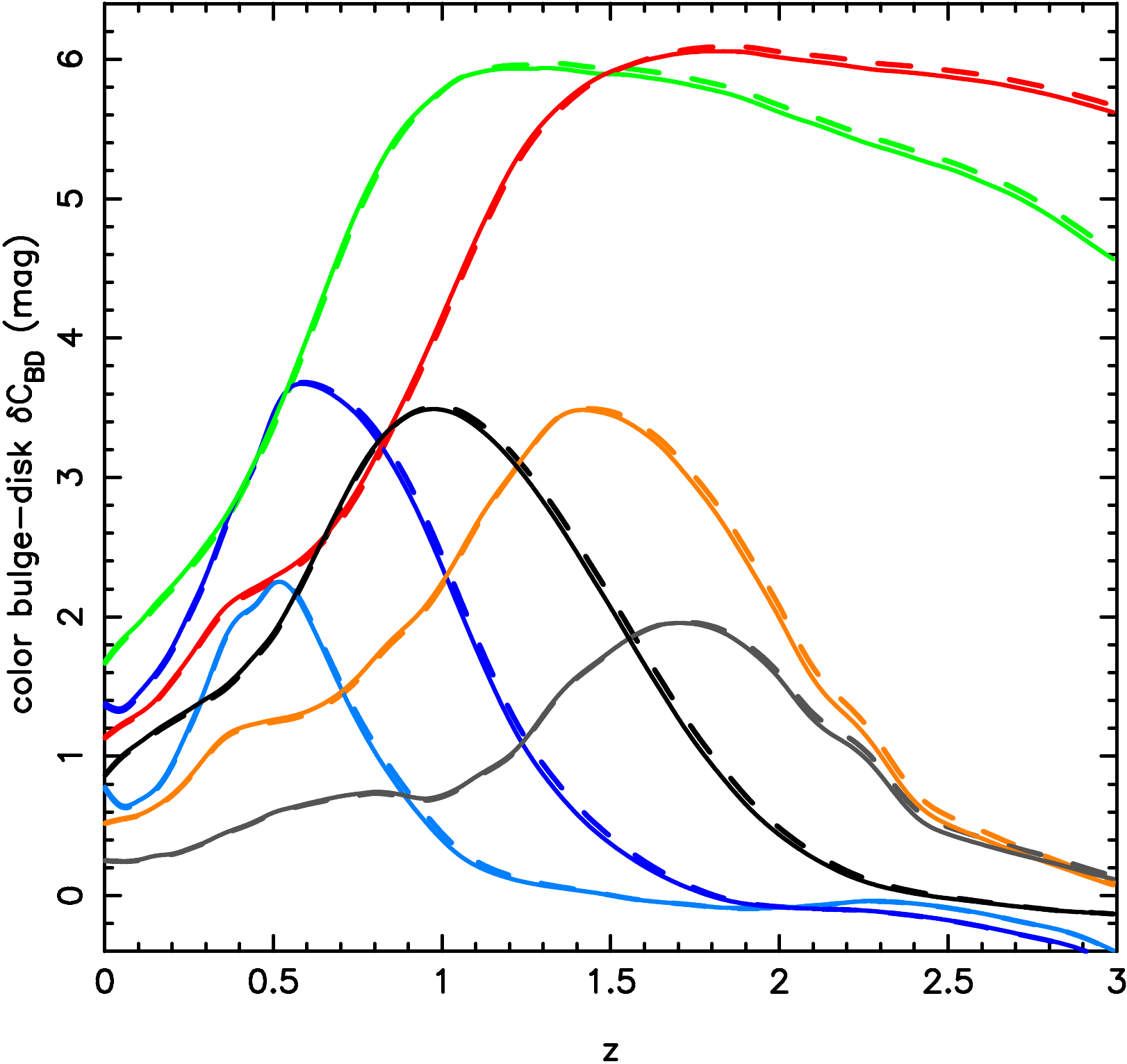}}
% r.h.s. (age: 4 Gyr)
\put(93,86){\includegraphics[clip, width=4.5cm]{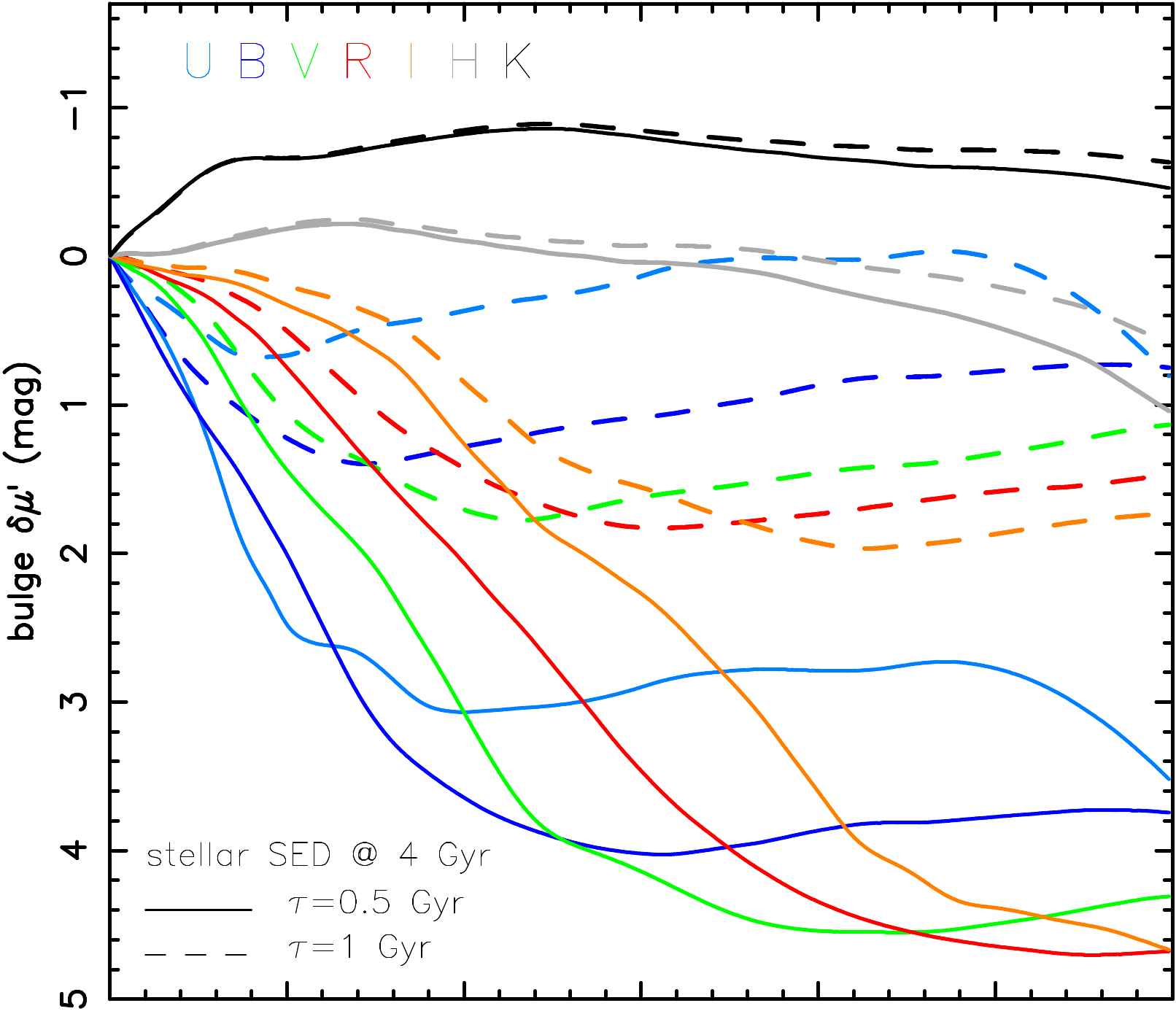}}
\put(93,45){\includegraphics[clip, width=4.5cm]{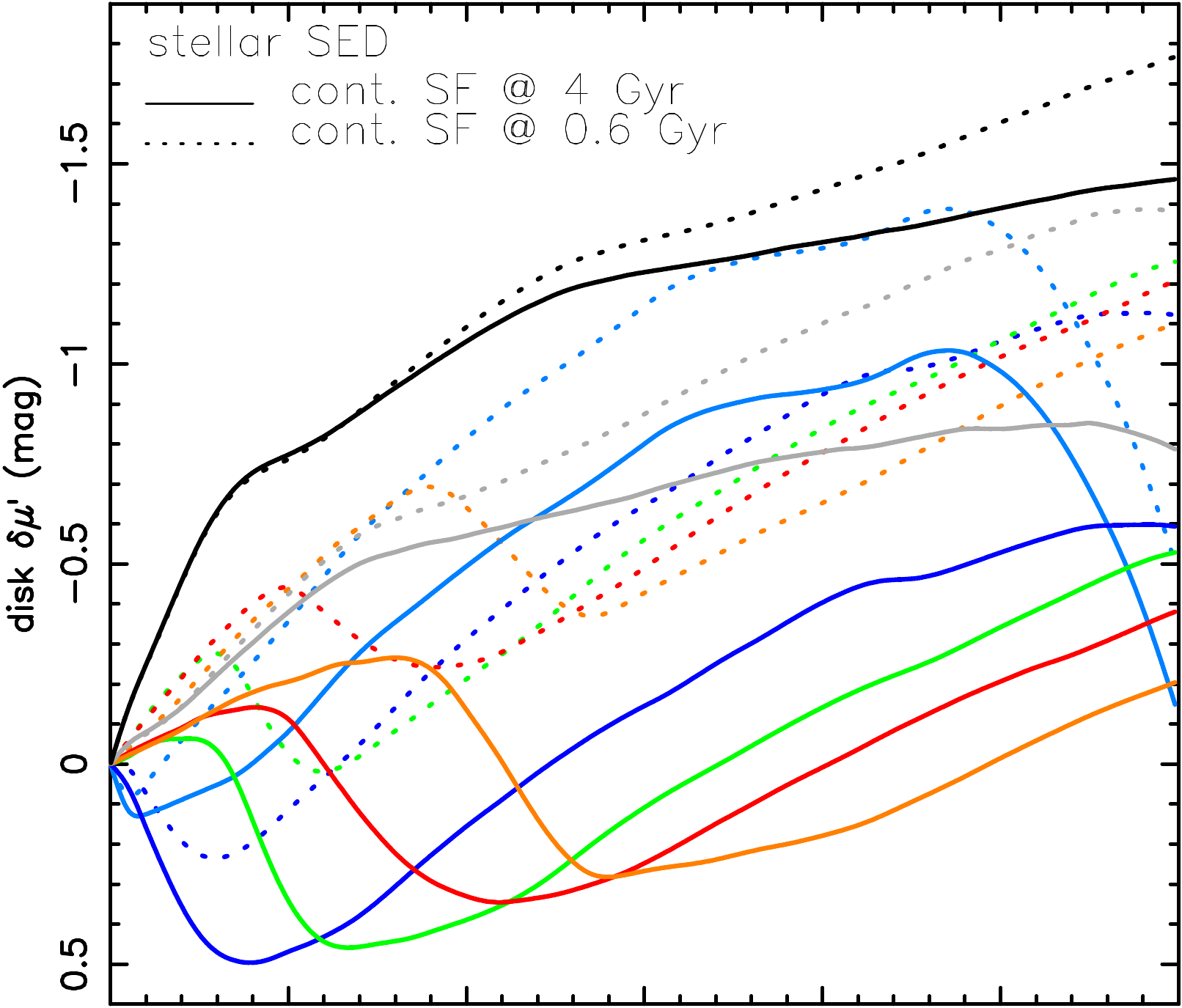}}
\put(93,0){\includegraphics[clip, width=4.52cm]{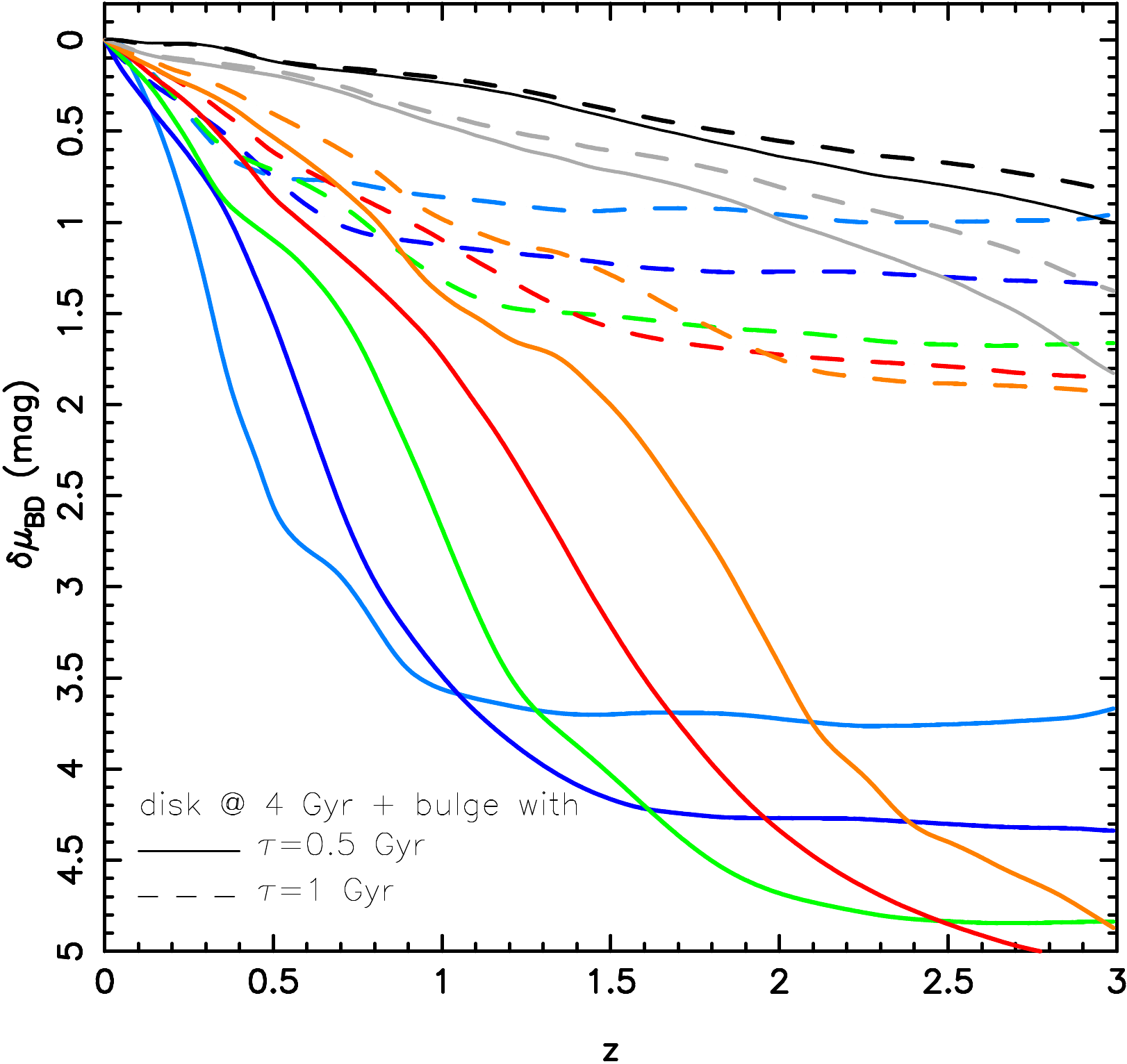}}
\put(139,86){\includegraphics[clip, width=4.5cm]{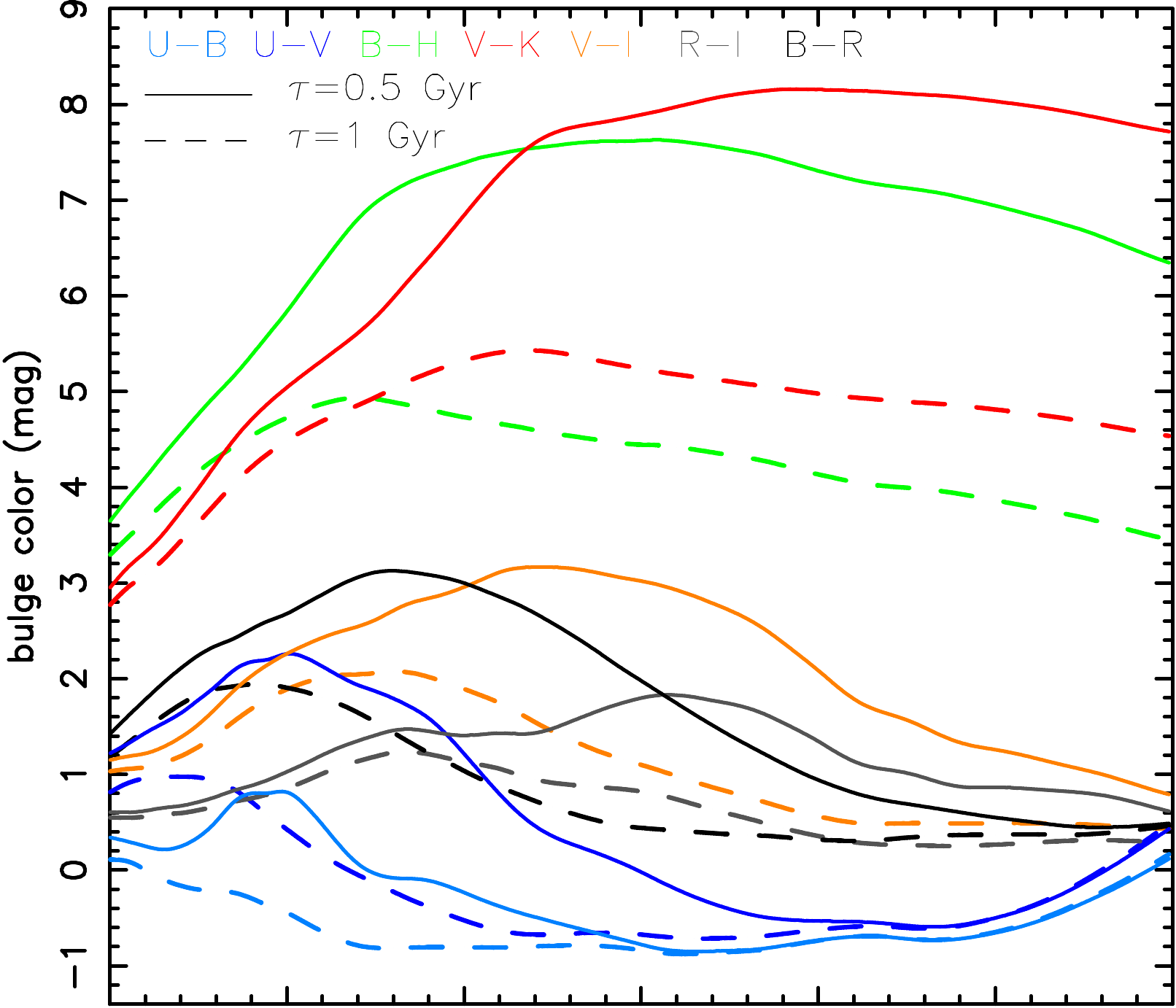}}
\put(139,45){\includegraphics[clip, width=4.5cm]{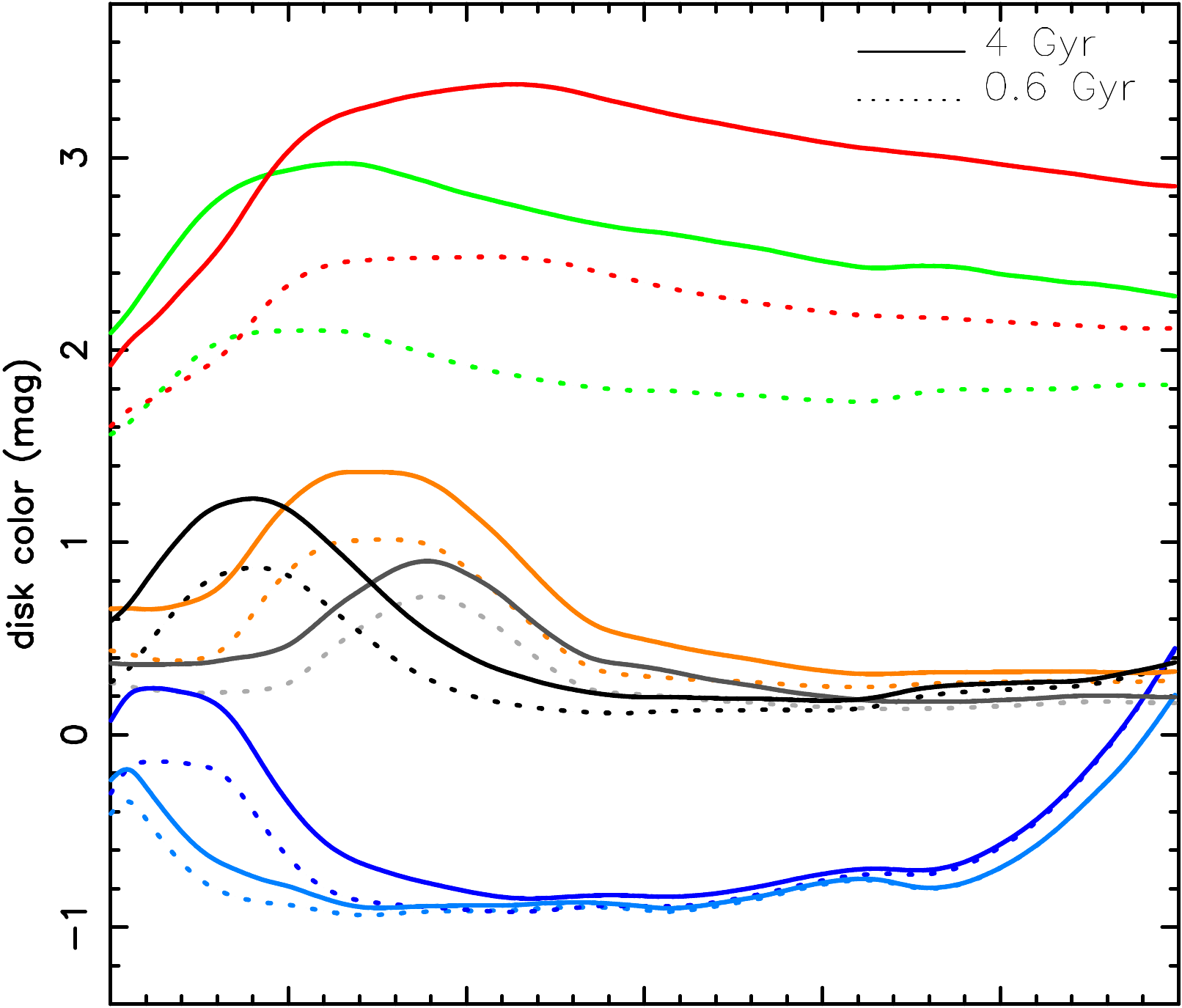}}
\put(139,0){\includegraphics[clip, width=4.52cm]{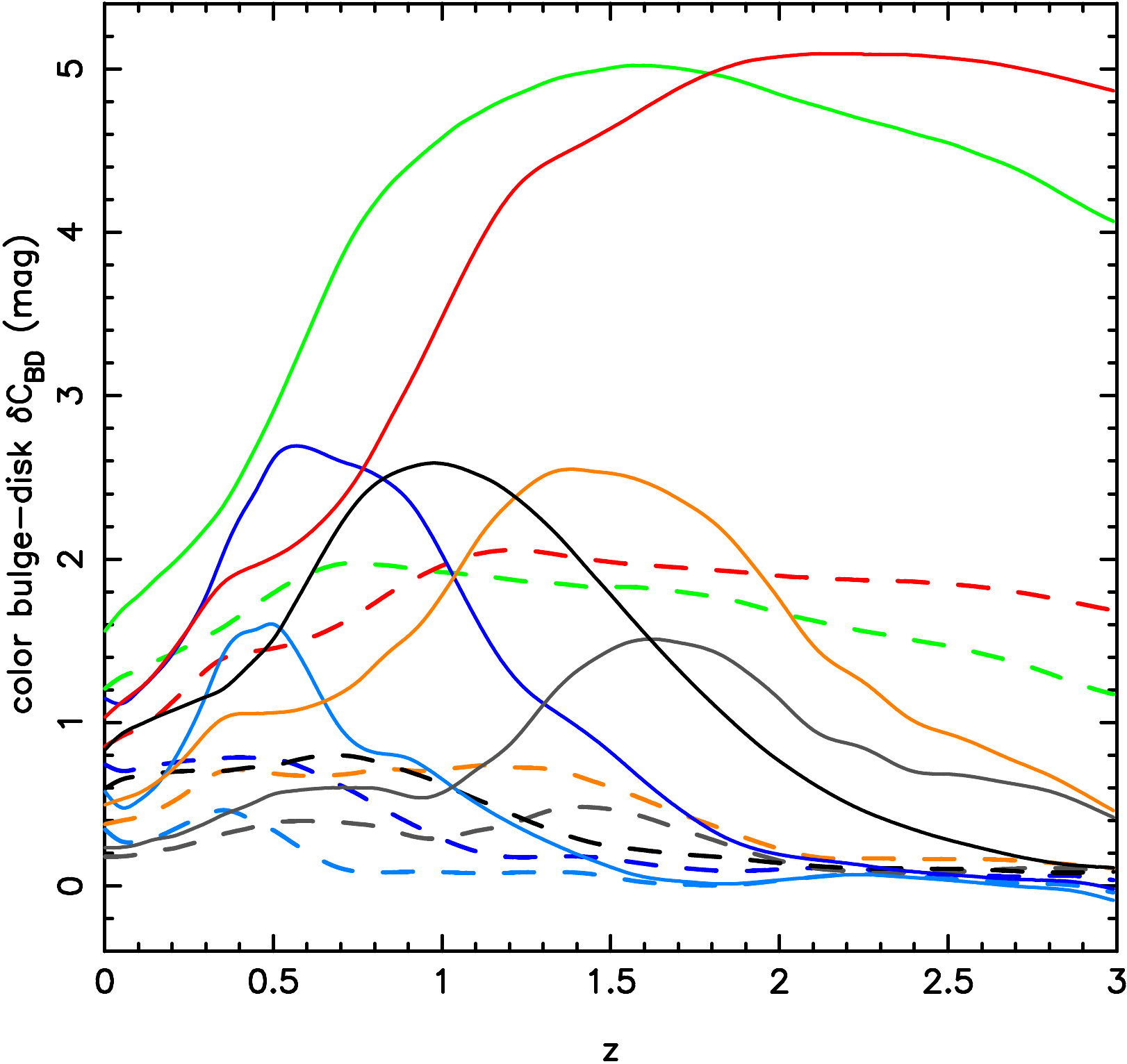}}

\PutLabel{40}{121}{\mlx \textcolor{black}{(a)}}
\PutLabel{86}{121}{\mlx \textcolor{black}{(d)}}
\PutLabel{133}{121}{\mlx \textcolor{black}{(g)}}
\PutLabel{180}{121}{\mlx \textcolor{black}{(j)}}

\PutLabel{40}{47}{\mlx \textcolor{black}{(b)}}
\PutLabel{86}{47}{\mlx \textcolor{black}{(e)}}
\PutLabel{133}{47}{\mlx \textcolor{black}{(h)}}
\PutLabel{180}{47}{\mlx \textcolor{black}{(k)}}

\PutLabel{40}{39}{\mlx \textcolor{black}{(c)}}
\PutLabel{52}{39}{\mlx \textcolor{black}{(f)}}
\PutLabel{133}{39}{\mlx \textcolor{black}{(i)}}
\PutLabel{145}{39}{\mlx \textcolor{black}{(l)}}
\end{picture}
\caption{Synthetic bulge+disk model for a spiral galaxy of age 13.7 Gyr (\brem{a}-\brem{f}) and 4 Gyr (\brem{g}-\brem{l}) simulated out to \zet=3.
The input SEDs to the simulation take only stellar emission into account and assume zero intrinsic extinction and attenuation by the intergalactic medium.
\brem{a)} Reduced surface brightness \dmu\!\!(\zet) for a bulge of solar metallicity forming with an exponentially decreasing SFR for an e-folding timescale, $\tau$, of 0.5 Gyr and 1 Gyr (solid and dashed curves, respectively). 
\brem{b)} \dmu\!\!(\zet) for a disk of \zsun/5 forming with a constant SFR.
\brem{c)} Bulge-to-disk surface brightness contrast, \dmBD\ (\dmu\!\!(bulge) -- \dmu\!\!(disk)).
\brem{d} and \brem{e)} ObsF color of the bulge and the disk. Values at \zet=0 correspond to the true (rest-frame) color.
\brem{f)} Bulge-to-disk color contrast, \dcol\ (the color of the bulge minus the color of the disk).
The layout of panels \brem{g}-\brem{l} is identical to that of panels \brem{a}-\brem{f} with the only difference being the addition 
of simulations for a 0.6 Gyr old stellar disk (dotted curves) in panels \brem{h} and \brem{k}.}
\label{sSED}
\end{figure*}
Figure~\ref{sSED} (panels \brem{a}-\brem{f}) shows simulations based on purely stellar SED templates for an age of 13.7 Gyr.
The variation in $\mu$\arcmin\ versus \zet\ is shown in $U$, $B$, $V$, Cousins $R$ and $I$, $J$, $H,$ and $K$, whereby negative (positive) 
values correspond to a $\mu$\arcmin\ enhancement (dimming) relative to the rest-frame value (modulo cosmological dimming, as pointed out above).
The bulge (panel \brem{a}) is approximated by a stellar population of solar metallicity (\zsun) forming with an exponentially decreasing 
star formation rate (SFR) for an e-folding time $\tau$ of 0.5 and 1 Gyr (models $\tau$0.5 and $\tau$1; solid and dashed curve, respectively). 
As for the disk (panel \brem{b}), we assume continuous SF at a constant SFR and a metallicity \zsun/5 (hereafter, the contSF model).

A central insight is that $\mu$\arcmin\ varies across \zet\ differently for the bulge and the disk.
At \zet$\sim$1, for example, its value for the bulge is nearly constant in the $H$ band
(\dmu$\approx$0 mag) and moderately brighter in $K$ (--0.76 mag), whereas its positive values in $B$, $V$ and $I$
(5, 4, and 1.7 mag, respectively, for both $\tau$ models) render an optical detection of the bulge difficult.
At \zet$\ga$2, \dmu\ increases to $>$5~mag even in the $I$ band, making an optical detection of the bulge practically impossible.
To the contrary, the disk shows at this \zet\ a significant brightening in the NIR (by --0.5 mag and --1.2 mag in $H$ and $K$, respectively),
whereas in the optical, following an initial dimming by 0.2--0.7 mag that becomes maximal at a \zet\ that increases with increasing filter central wavelength
$\lambda_{\rm c}$, it gradually returns to \dmu$\simeq$0~mag at \zet$\sim$0.6, 1.7, and 2.6 in $U$, $B,$ and $V$, respectively.
At \zet$\sim$3 the bulge appears in terms of $\mu\arcmin$ as bright as at \zet=0 in the $K$ band, whereas,
the disk recovers its rest-frame $\mu$\arcmin\ in the visual and becoming by $>$1~mag brighter in the $K$ band.
This divergent variation in \dmu\!(\zet) for the bulge and the disk does not qualitatively change when 4~Gyr old SEDs are used as input to simulations 
(panels \brem{g}\&\brem{h}) or when adopting a significantly younger SED of 0.6 Gyr for the disk (dotted curve in panel \brem{h}).

Panel \brem{c} shows the variation in the bulge-to-disk surface brightness contrast \dmBD=\dmu\!(bulge)-\dmu\!(disk) relative to its rest-frame value 
(set to 0 mag). If the SED of the bulge and disk were to be identical, that is, if a galaxy were a mono-component system with a spatially invariant SED, 
then \dmBD\ would be zero for all redshifts and filters.
The observed increase of \dmBD\ to $\ga$5 $V$ mag at \zet=1.5 and $>$1 $H$ mag at \zet=2, implies that, even with the naive assumption
of no spectrophotometric evolution (i.e., when the rest-frame SED of bulge and disk are kept fixed across \zet), the SBP of a distant spiral galaxy
differs from its true (rest-frame) shape both in the optical and NIR.
The main cause for this is the gradual disappearance of the NUV-faint bulge at a simultaneous constancy (or enhancement in the NIR) of the NUV-bright disk.

With other words, an implication of \cmod\ is the non-homologous change of the SBP of higher-\zet\ galaxies, with the bulge-to-disk surface brightness contrast \dmBD\ providing a metrics for this effect.
As we discuss in further detail in Sect.~\ref{BD}, one consequence of the \cmod-induced violation of homology is the systematic underestimation of the \BD\ 
and \BT\ ratios of higher-\zet\ spiral galaxies.
Also apparent from panel \brem{c} is that this bias depends on the rest-frame SED of the bulge and the disk (i.e., their relative SFH and intrinsic extinction),
the redshift of a galaxy, and the photometric filters considered.

Panels \brem{d} and \brem{e} display the ObsF color of the bulge and the disk, and the bulge-to-disk color contrast, \dcol\ (color in the bulge minus color in the disk, i.e., an inverse proxy to the radial color gradient; cf. \tref{PO12}) is shown in panel \brem{f}. It can be appreciated (also from panels \brem{j}-\brem{k}) that \cmod\ leads to a strong discrepancy between ObsF and rest-frame color at essentially any \zet ($>$0). For example, the disk attains its maximal $V$--$I$ color of 1.5 mag (0.7 mag redder than its rest-frame value) at \zet$\sim$0.7, whereas for the bulge this happens at \zet$\sim$1.4, where the observed color (4.2 mag) exceeds the rest-frame color by $\sim$2.8 mag.

The differential variation in ObsF colors with \zet\ propagates into \dcol, and its characteristic reversion toward bluer values after it becomes maximal
at 0.5$\la z \la$1.8, depending on the color considered.
For example, the \dcol\ in $B$--$H$ and $V$--$K$ exceed the rest-frame value (1.7 and 1.2 mag, respectively) by $\ga$4~mag, reaching
$\sim$6~mag at 1$\la$\zet$\la$1.8, while this trend is reversed at a higher \zet. Likewise, an inversion of \dcol\ can be read off panel \brem{f} at \zet$\sim$0.5 for $U$--$B$, \zet$\sim$0.6 for $U$--$V$, \zet$\sim$1 for $B$--$R$, \zet$\sim$1.45 for $V$--$I$ and \zet$\sim$1.7 for $R$--$I$.

Summarizing, interpreting colors and color gradients for higher-\zet\ galaxies is a nontrivial task that, if attempted without taking \cmod\ into account, can readily lead to contradictory (and in all cases erroneous) conclusions about the nature and evolutionary status of such systems.
For example, the high $V$-$I$ bulge-to-disk color contrast (3.5 mag) of an LTG at \zet=1.5 might prompt the interpretation that it
contains a dust-obscured bulge, whereas the nearly zero color contrast of the same galaxy at \zet$\sim$3 would suggest, consistently 
with its high \dmBD\ ($>$5 mag), that it is a bulgeless disk with a spatially uniform age and sSFR.
In this context it is worth noting that even a low amount of spatially inhomogeneous dust obscuration, leading to the depression of the rest-frame UV SED, adds another level of complexity to such potential biases. Finally, the confinement of intense nebular emission to the disk periphery in a centrally SF-quenched LTG further aggravates the problem as it can result to the reversion of radial color gradients in multiple narrow \zet\ intervals (cf. \tref{PO12} and Sect. 2.2).
% ::::::::::::::::::::::::::::::::::::::::::::::::::::::::::::::::::::::::::::::::::::::
\subsection{Simulations that include nebular emission \label{sub:nebSED}}
% ::::::::::::::::::::::::::::::::::::::::::::::::::::::::::::::::::::::::::::::::::::::
Nebular emission is an integral part of the SED of star-forming galaxies near and far. Its influence on broadband magnitudes and colors has been examined in several previous studies \citep[e.g.,][]{Huchra77,NussbaumerSchmutz84,Krueger95,FRV97,Lei99,Papovich01,Zackrisson01,Zackrisson08,SdB09,Reines10,CC16,GP17,Byler17,Inayoshi22}
and observationally documented in local BCDs and other EELGs at higher \zet\
\citep[e.g.,][]{Izotov97,P98,P02,Ostlin03,Cardamone09,Reines10,Izotov11,Amorin2012,Atek11,Atek22,Mobasher15,Breda22,LC22,Boyett22,Llerena23}.
Emission-line EWs in many of these high-sSFR ($\sim 10^{-8}$ yr$^{-1}$) systems, especially in extremely metal-poor BCDs with 12+log(O/H)$\la$7.6, exceed 
$10^3$ \AA\ \citep[e.g.,][]{Guseva04}, which implies a significant, if not dominant, contribution of nebular emission to broadband luminosities.

\begin{center}
\begin{figure}
\begin{picture}(86,26)
\put(7,0){\includegraphics[trim=0 190 0 0, clip, width=8cm]{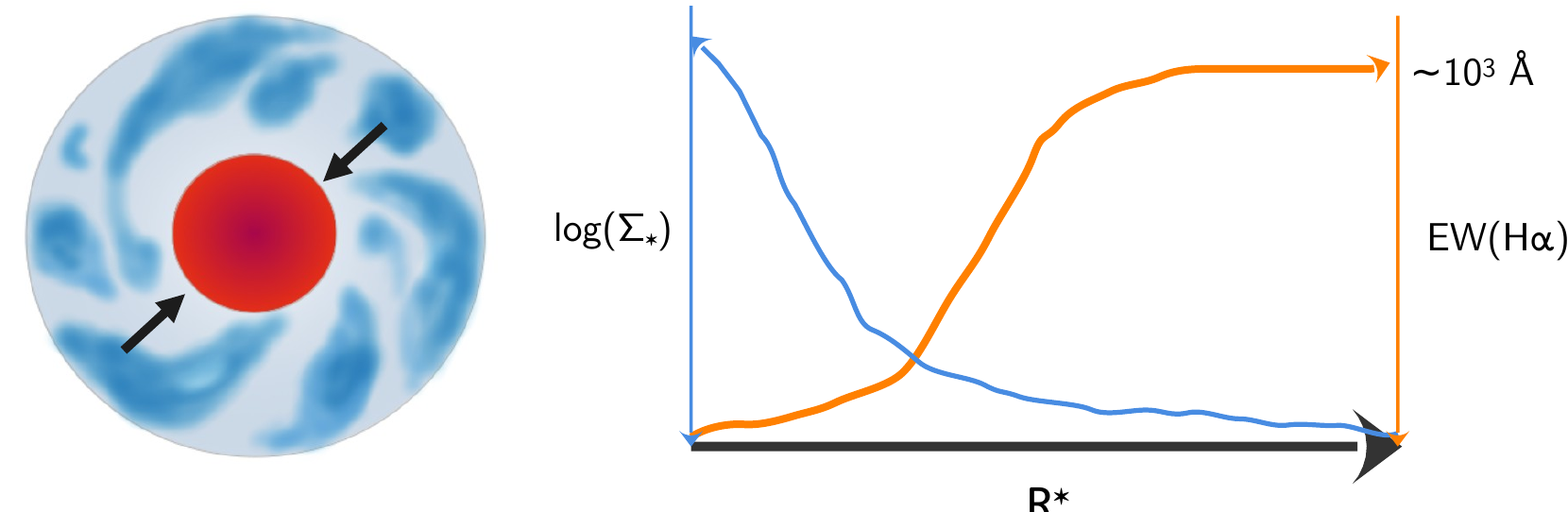}}
\end{picture}
\caption{Schematic illustration of a high-\zet\ spiral galaxy in the early stages of its evolution. Massive SF clumps (blue) forming in the disk 
migrate inward and coalesce, promoting the growth of a high-stellar surface density $\Sigma_{\star}$ protobulge \citep[e.g.,][]{Noguchi99,Bournaud07,Mandelker14}. 
The rest-frame \ewha\ of SF clumps, initially on the order of $10^3$ \AA\ (cf. Fig.~\ref{ew1} and discussion in \tref{P22}), decreases as they move inward, both because of their aging over the theoretically estimated migration timescale (\tmig=0.6-1 Gyr) and because of the increasing line-of-sight dilution by the underlying stellar background. This leads to a radial anticorrelation between \ewha\ and stellar surface density, similar to the case of the local BCD \object{I Zw 18} \citep[][and \tref{PO12}]{P02}, as delineated in the right-hand panel.}
\label{ew:LTG} 
\end{figure}
\end{center} 

\begin{figure}
\begin{picture}(200,125)
\put(-1,86){\includegraphics[clip, width=4.5cm]{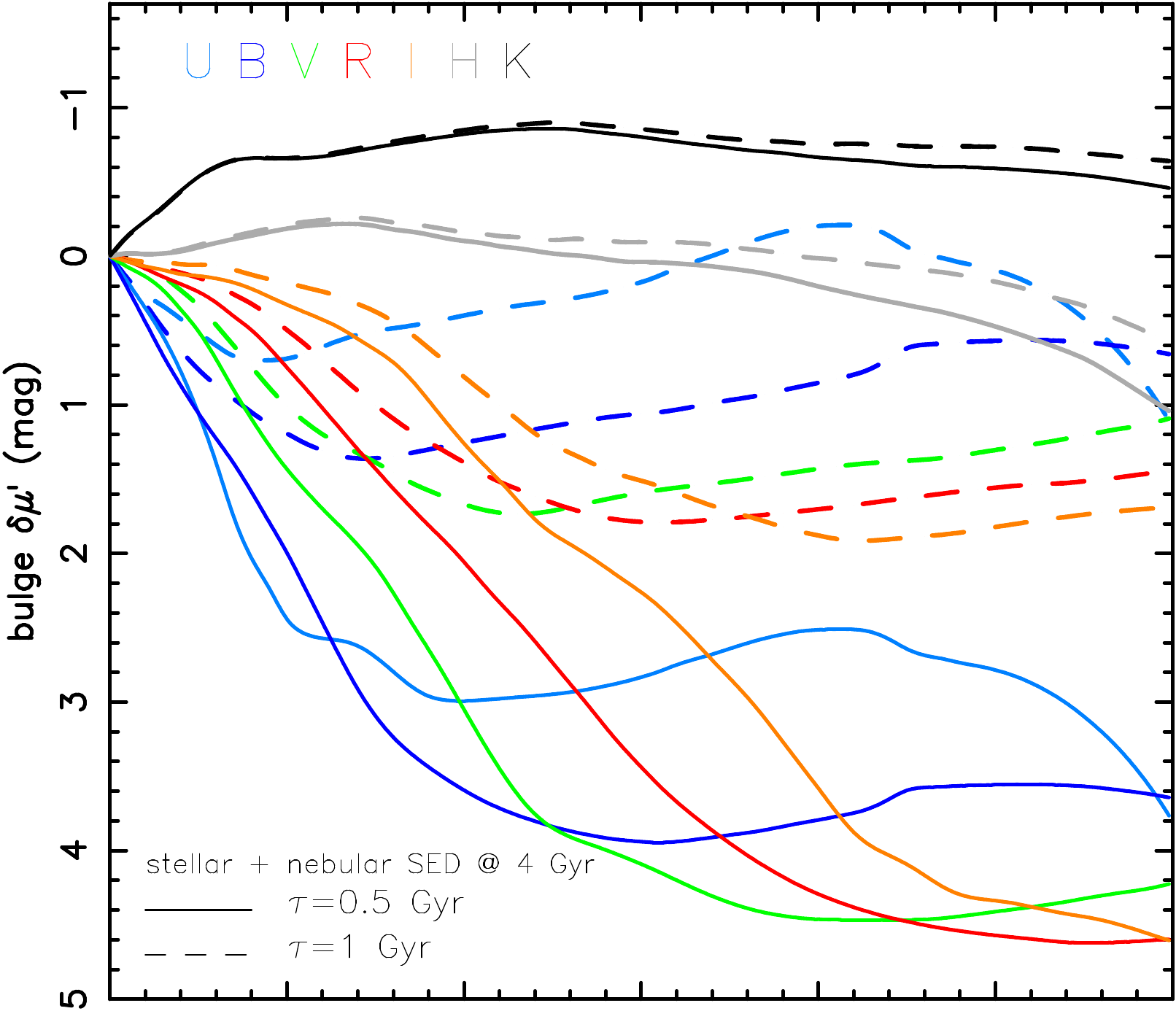}}
\put(-1,45){\includegraphics[clip, width=4.5cm]{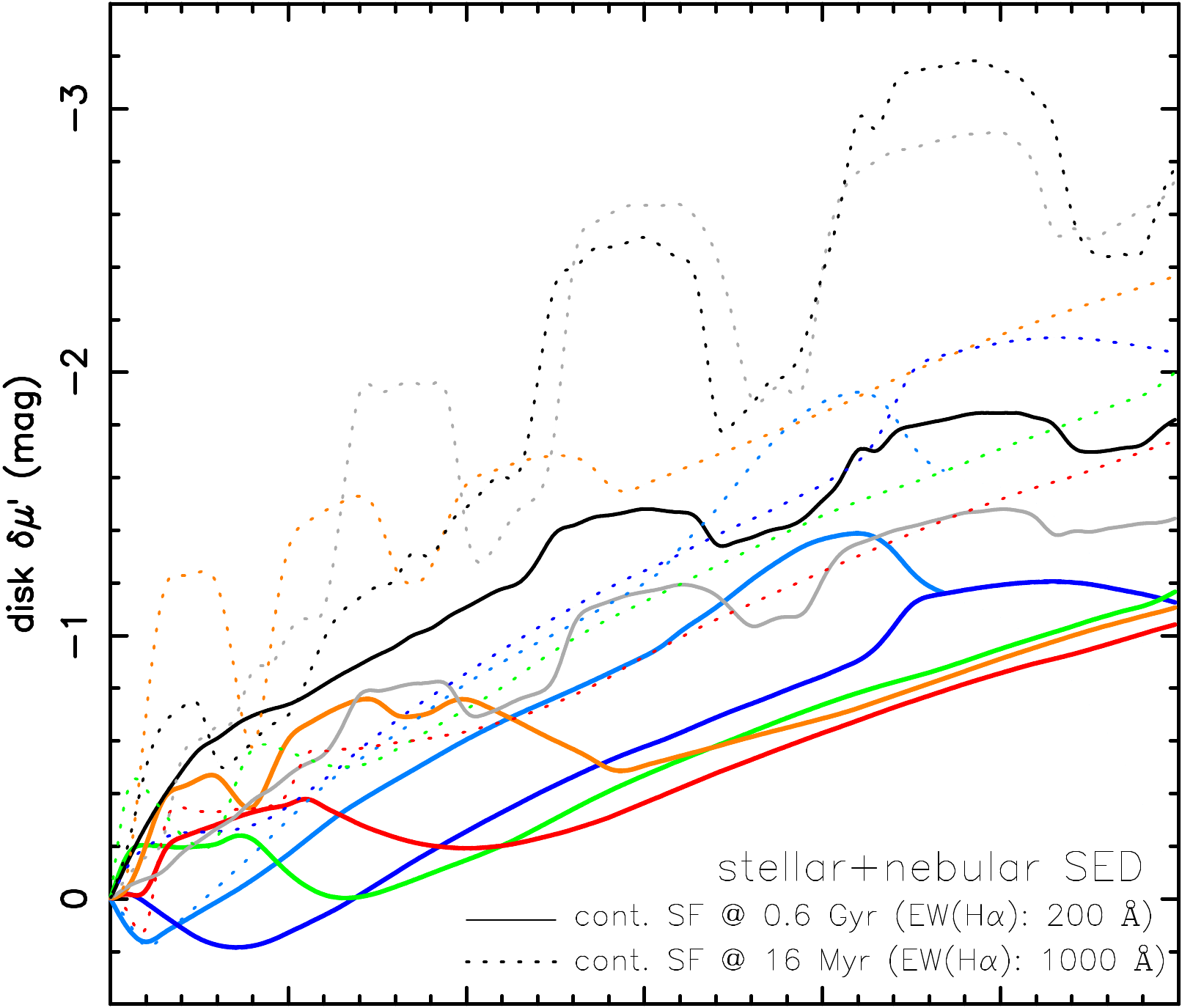}}
\put(-1,0){\includegraphics[clip, width=4.52cm]{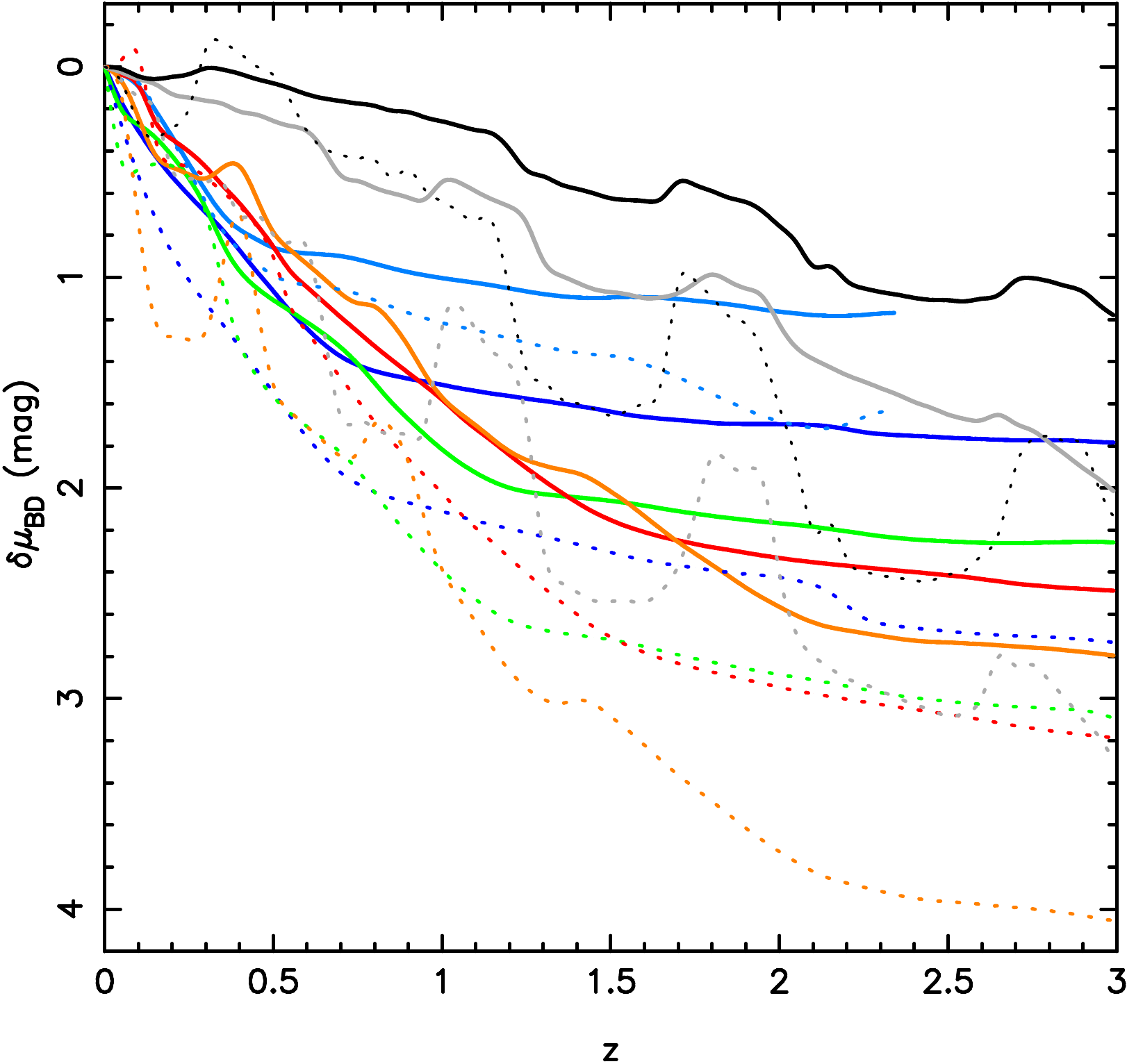}}
\put(44.5,86){\includegraphics[clip, width=4.5cm]{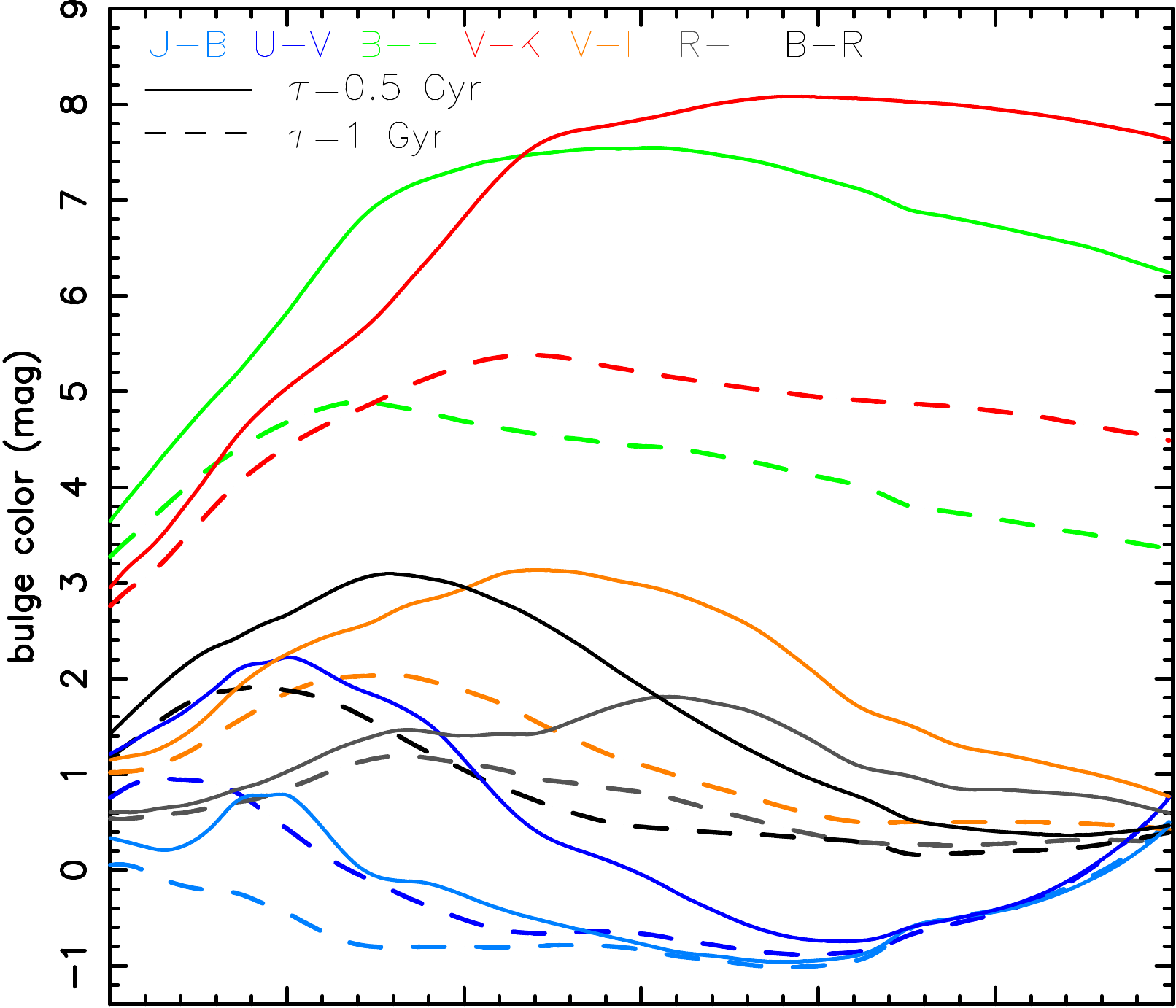}}
\put(44.5,45){\includegraphics[clip, width=4.5cm]{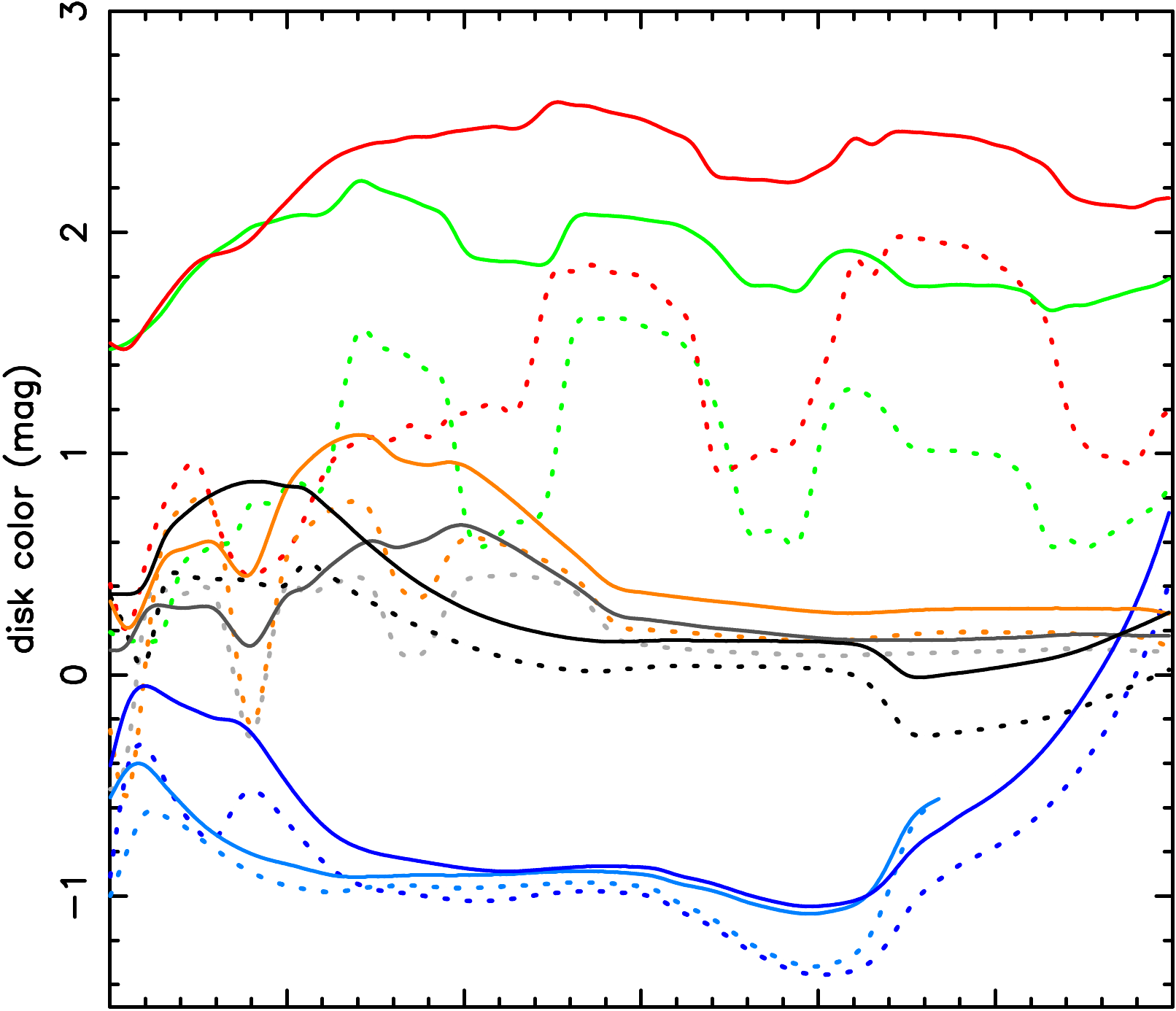}}
\put(44.5,0){\includegraphics[clip, width=4.52cm]{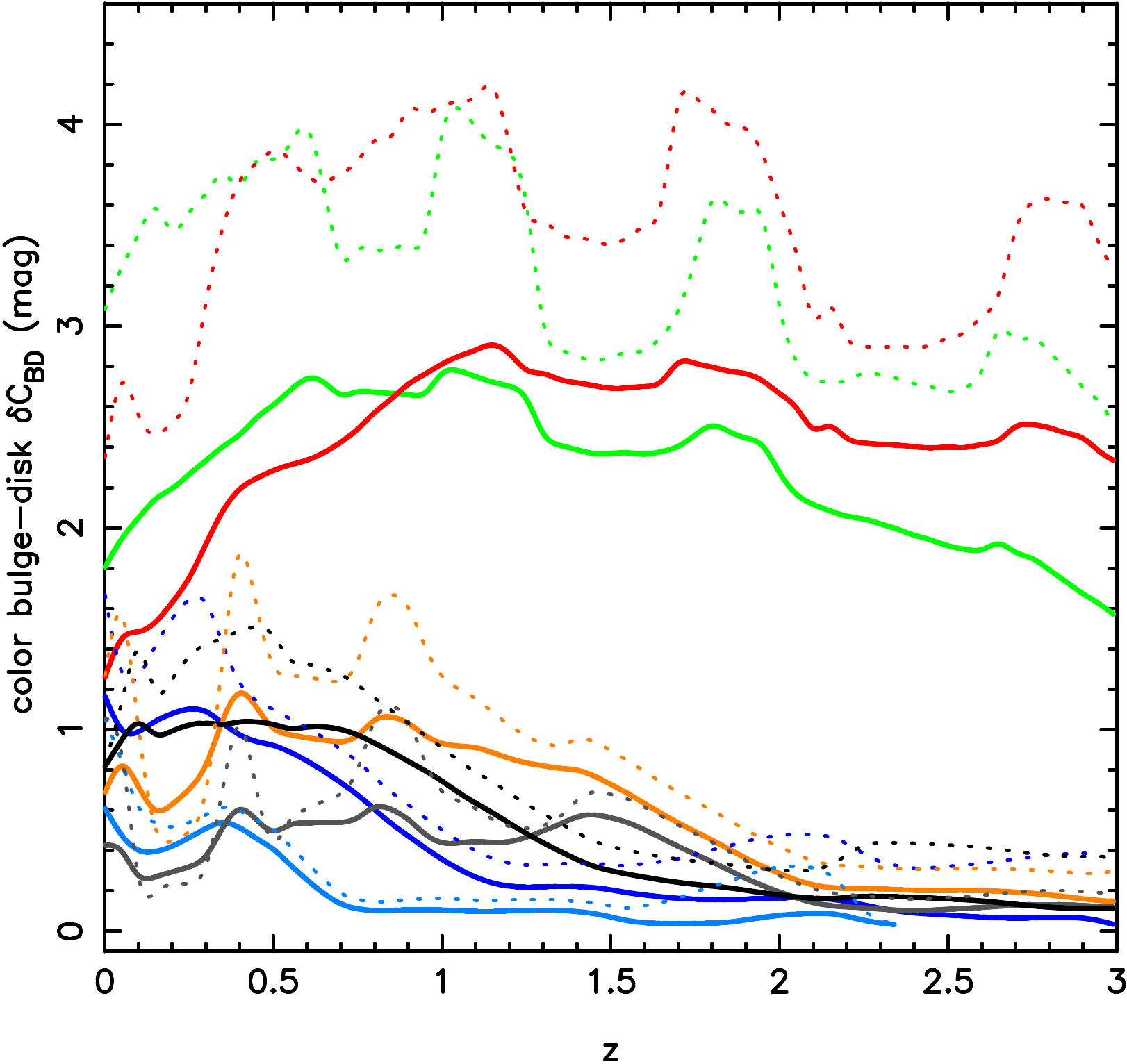}}
\PutLabel{39.5}{121}{\mlx \textcolor{black}{(a)}}
\PutLabel{4.5}{79.5}{\mlx \textcolor{black}{(b)}}
\PutLabel{39.5}{39}{\mlx \textcolor{black}{(c)}}
\PutLabel{85.5}{121}{\mlx \textcolor{black}{(d)}}
\PutLabel{85.5}{47}{\mlx \textcolor{black}{(e)}}
\PutLabel{85.5}{39}{\mlx \textcolor{black}{(f)}}
% %%%%%%%%%%%%%%%%%%%%%%%%%%%%%%%%%%%%%%%%%%%%%%%%%%%%%%%%%%%%%%%%%%%%%%%%%%
%\PutLabel{10}{126}{\imlabel{gSED4Gyr}}
% %%%%%%%%%%%%%%%%%%%%%%%%%%%%%%%%%%%%%%%%%%%%%%%%%%%%%%%%%%%%%%%%%%%%%%%%%%
\end{picture}
\caption{Variation in \dmu\ vs. \z\ for simulations that include nebular (line and continuum) emission. As in Fig.~\ref{sSED}\irem{g}, the bulge (panel \brem{a}) is approximated by a 4~Gyr old SED for a stellar population of solar metallicity that forms with an exponentially decreasing SFR with a $\tau$ of 0.5 and 1 Gyr (solid and dashed curve, respectively). As for the disk, we assume SEDs for a continuous SF scenario, one representing the case of moderately strong nebular emission (\ewha$\sim$200 \AA) in a 0.6~Gyr old stellar population of \zsun/5 (solid curves) and the second one of very intense nebular emission (\ewha$\sim 10^3$ \AA) excited by a 16~Myr old stellar population (dashed curve).
These two cases are meant to approximate, respectively, intermediate and very early stages of the evolution of massive star-forming clumps emerging from VDIs \citep[cf. the discussion in][]{P22} in high-\zet\ spiral galaxies (cf. Fig.~\ref{ew:LTG}). Whereas nebular emission has virtually no effect on the aging bulge, it strongly enhances the surface brightness of the disk at various broad redshift intervals where strong emission lines fall within various filter transmission curves.
At \zet$>$0.6 this effect is particularly strong in the NIR.
Given the uncertain \lya\ escape fraction, predictions in the $U$ and $B$ band should be considered only up to the redshift where the \lya\ line (1216 \AA) enters the blue edge of these filters (\zet$\sim$1.51 and $\sim$2.04, respectively). The layout is the same as in Fig.~\ref{sSED}.}
\label{gSED4Gyr}
\end{figure}

In local galaxy disks, nebular emission is on average weak (\ewha$\la$60 \AA) and thus implies a negligible enhancement ($\la$0.1 $r$ mag)
of their surface brightness \citep[e.g.,][hereafter \tref{P22}]{P22}\footnote{Further notes on this subject, and an approximation of the enhancement of broadband magnitudes as a function of rest-frame \ewha, can be found in Sect.~\ref{ap:nebSED}.}. However, the situation is most certainly different at high \zet.

The current consensus is that an important, if not the dominant, driver of early bulge growth is the coalescence of massive ($10^{8-9}$ \msun)
star-forming clumps that emerge out of violent disk instabilities \citep[VDIs;][]{Noguchi99,Bournaud07,Elm08,Mandelker14,Mandelker17}
and migrate to the center of a protodisk within a timescale \tmig\ of 0.6-1 Gyr. A further contribution to the stellar mass growth in a protobulge may be provided by
an initial short phase of ``wet compaction'' and intense SF that soon is succeeded by starburst-driven evacuation of cold gas and inside-out SF quenching \citep{DekelBurkert14}. Observational support for this picture comes from \textit{Hubble} Space Telescope (HST) imaging studies revealing that a SF-quenched core surrounded by SF clumps in the disk is typical for spiral galaxies at \zet$\ga$1 \citep[e.g.,][see also \tref{Me\v{s}tri\'{c} et al. 2022; Liu et al. 2023} for related studies at higher \zet]{Wuyts12,Zanella15,Wang22}.

A natural consequence of the aging of SF clumps as they migrate inwardly are negative (positive) stellar age (emission-line EW) gradients \citep[][hereafter \tref{BP18}; see also \tref{Papaderos et al. 2022}]{BP18}.
Spiral galaxies in early stages of their evolution should therefore be morphologically reminiscent of the local BCD \object{I Zw 18}, as discussed in \tref{PO12}, showing a  spatial segregation between a dense, more evolved stellar core (the protobulge) and a high-EW disk envelope (cf. Fig.~\ref{ew:LTG}).
The theoretically expected rest-frame \ewha\ of young SF clumps in the disk of $\ga 10^3$ \AA\ (Sect.~\ref{ap:nebSED}), boosted in the ObsF by a factor (1+\zet), is expected to further rise because of the low stellar metallicity of massive SF clumps at those redshifts, which implies a by an up to $\ga$2 higher specific Lyman continuum (LyC) production rate \citep[e.g.,][]{WeilbacherUF01,AFvA03}
compared to local galactic disks with 0.2$\la Z/Z_{\odot}\la$0.7 (cf., e.g., Fig.~7 in \tref{BP18}). 
Indeed, for a spatially constant sSFR, a negative radial stellar metallicity gradient implies a positive \ewha\ gradient.

Nebular emission selectively enhances the surface brightness of the high-EW disk (while having little effect on the quiescent bulge)
in distinct \zet-intervals as different strong emission lines are captured within different filters depending on the redshift of a galaxy (\tref{P22}). This differential amplification of the disk relative to the bulge, which manifests itself in ``jumps'' in the color of the disk and its color contrast relative to the bulge \citep[see Fig.~15 in \tref{PO12} for an analysis of this effect out to \zet=1.2 and the recent work by][which documents its importance for galaxies out to \zet$\sim$18]{Inayoshi22}, can systematically affect bulge-disk decomposition studies of high-\zet\ spiral galaxies.
This is because over-subtraction of the blue, nebular emission-enhanced disk from the red, SF-quenched bulge entails an underestimation (overestimation)
of the luminosity fraction (color) of the latter, potentially prompting the conclusion that bulge growth proceeds in major ``waves,'' eventually associated with discrete epochs of strong intrinsic obscuration (\tref{P22}).

Figure~\ref{gSED4Gyr} shows how this differential enhancement of the disk due to nebular emission influences the ObsF bulge-to-disk surface brightness and color contrast
of a young high-\zet\ spiral galaxy. The bulge is approximated by the same 4~Gyr old SED for models $\tau$0.5 and $\tau$1 (Fig.~\ref{sSED}) with the only difference being that nebular emission is taken into account. The photometric effect of massive SF clumps in the disk is simulated by adopting a stellar+nebular SED
for a continuously forming stellar population with an age of 16 Myr and 600 Myr. These two models yield, respectively, a rest-frame \ewha\ of $10^3$ \AA\ and $\sim$200 \AA\ and thereby approach the level of nebular contamination at the time of the formation of clumps in the disk and after $\sim$\tmig\ (i.e., when they are about to reach the bulge).
Clearly, the model is simplified, as it assumes that SF clumps survive feedback for hundreds of megayears, and ignores chemical self-enrichment and EW dilution by the underlying stellar background. Nevertheless, it is sufficient for a quantitative inference on the core-to-envelope surface brightness- and color contrast in a proto-LTG.

\begin{figure}
\begin{picture}(200,90)
\put(2,50){\includegraphics[clip, width=8.6cm]{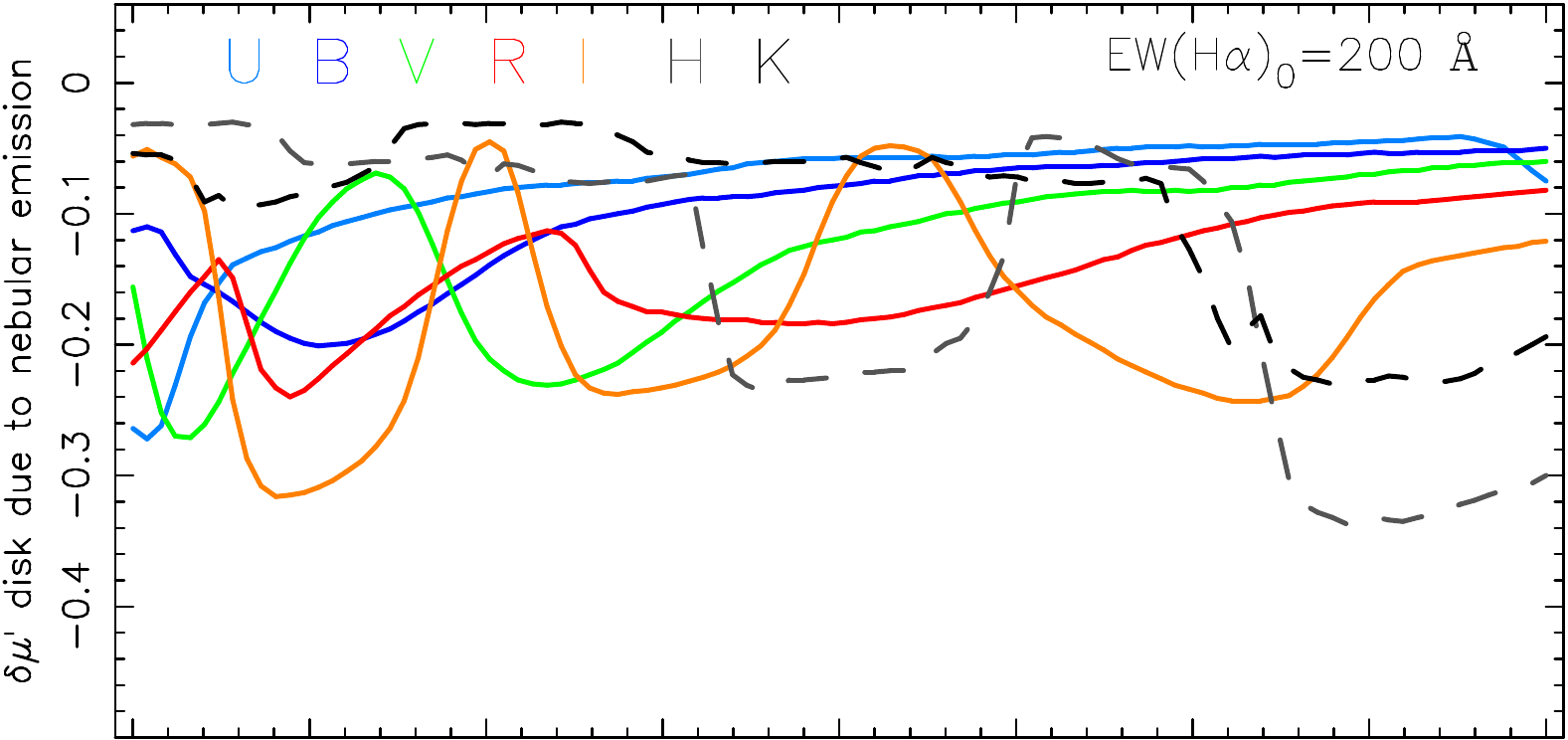}}
\put(2,0){\includegraphics[clip, width=8.622cm]{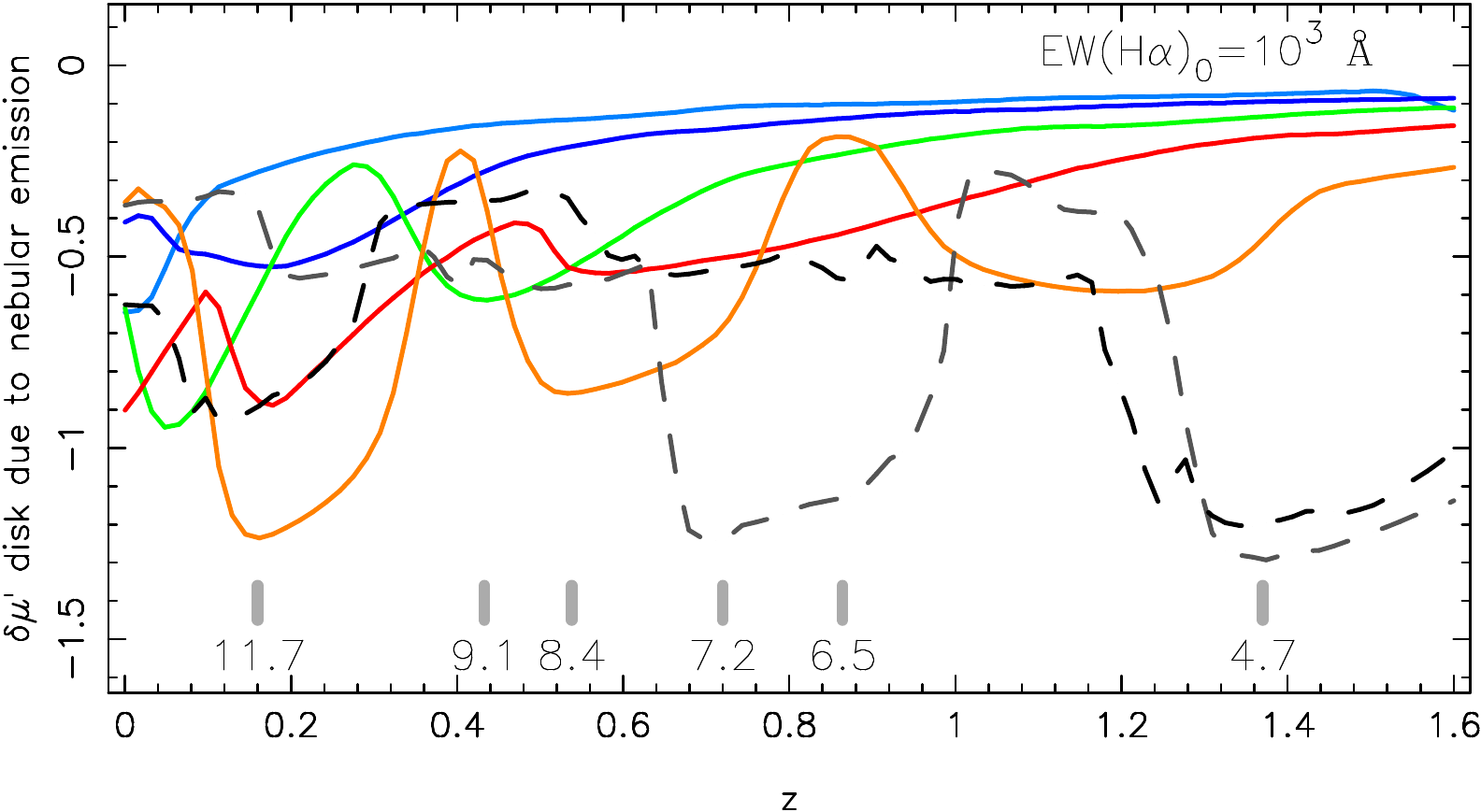}}
\end{picture}
\caption{Enhancement (in mag) of a purely stellar disk by nebular (line and continuum) emission vs. \zet. The case of a moderately evolved (0.6 Gyr),
continuously growing disk with a rest-frame \ewha$\sim$200 \AA\ is shown in the upper panel, whereas a very young disk (16 Myr) with an \ewha$\sim 10^3$ \AA\
is shown in the lower panel. Vertical bars mark the age of Universe (in Gyr) at some \zet\ intervals where nebular lines enhance the surface brightness of the disk most.
An insight from this figure is that NIR photometry is more affected by nebular emission than optical photometry for \zet$>$0.6.}
\label{zmag} 
\end{figure}

The photometric effect of nebular emission is better visible from the zoom-in in Fig.~\ref{zmag} where we show for the two cases considered the difference between predictions based on SEDs including nebular (line and continuum) emission and purely stellar SEDs. Vertical gray lines mark the cosmic age corresponding to some of the most pronounced discontinuities in the disk's surface brightness.
As is apparent both from Fig.~\ref{zmag} and \ref{gSED4Gyr}, when adopting an SED with 200~\AA\ ($10^3$ \AA), the $I$-band disk surface brightness increases by --0.3 (--1.2) mag at \zet$\sim$0.15, --0.23 (--0.4) mag at \zet$\sim$0.5 and 0.24 (--0.6) mag at \zet$\sim$1.2 because of, respectively, the \ha+[N{\sc ii}] blend, [O{\sc iii}]5007,4959 lines and the [O{\sc ii}] doublet entering the filter transmission curve.
Likewise, the $H$-band surface brightness is enhanced by --0.24 (--1.2) mag at \zet$\sim$0.7 and --0.34 (--1.3) mag at \zet$\sim$1.4, mainly because of Paschen-series lines and the \ha+[N{\sc ii}] doublet, while the $K$-band surface brightness in enhanced by --0.23 (--1.2) mag at \zet$\sim$1.4 and --0.23 ($<$ --1) mag at \zet$\sim$2.4.
Even though these jumps are less pronounced when a 0.6~Gyr old SED is assumed (upper panel), they remain well discernible at a level of $\sim$--0.3 mag, especially in the $I$, $H,$ and $K$ band.
To the contrary, the $U$ and $B$ filters are comparatively immune to nebular contamination for 0.4$\la$\zet$\la$1.6 because the near- and far-UV spectral range they encompass is free of strong emission lines (despite a significant contribution by the nebular continuum), and boosted only at z$\ga$1.5, when the Ly$\alpha$ (assuming that it escapes from a galaxy) enters the $U$-band transmission curve.
Evidently, similar considerations apply to HST (and JWST) filters. For example, the HST filters F814W ($\lambda_{\rm c}$=7880 \AA), F125W ($\lambda_{\rm c}$=12490 \AA), F160W ($\lambda_{\rm c}$=15430 \AA) encompass the \ha\ line at, respectively, \zet=0.2, 0.9, 1.35, whereas this is the case for the [O{\sc iii}]5007 at \zet=0.57, 1.49, 2.08.

Summarizing, nebular emission has a considerable photometric impact over broad redshift intervals that essentially replicate the filter transmission curves in \zet.
This makes an empirical a posteriori correction of their effect on bulge-disk decomposition studies a demanding, perhaps even intractable, endeavor.
Another insight from Figs.~\ref{gSED4Gyr} and \ref{zmag} is that the NIR photometry of star-forming galaxies at \zet$>$0.6 is far more affected by nebular emission than optical photometry. Whereas in a local starburst galaxy with a burst parameter (fraction by mass of stars formed in the starburst) of 0.1 more than half of the $K$-band luminosity can originate from the nebular continuum \citep[e.g.,][]{Krueger95}, the main contaminant of $H$ and $K$ magnitudes at higher-\zet\ are optical emission lines recorded in the NIR. Converting $K$ luminosities at those redshifts into stellar mass \mstar\ via an assumed stellar mass-to-light ratio (\ml) can thus lead to an overestimated \mstar\ by a factor of a few ($\sim$3 at \zet=1.4 for the model in the lower panel of Fig.~\ref{zmag}).

Finally, as is apparent from Fig.~\ref{gSED4Gyr}\irem{f}, nebular emission can in certain \zet\ intervals change the bulge-to-disk color contrast by up to 1 mag (at \zet$\sim$2 for $V$-$K$, for example), further contributing to the inversion of radial color gradients in distant spiral galaxies (cf. Sect.~\ref{sub:stSED}). This is especially true for hybrid optical-NIR color indices.

% :::::::::::::::::::::::::::::::::::::::::::::::::::::::::::::::::::::::::::::::::::::::::::::::::::::::::::::::::::
\subsection{Evolutionary consistent simulations \label{sub:eSED}}
% :::::::::::::::::::::::::::::::::::::::::::::::::::::::::::::::::::::::::::::::::::::::::::::::::::::::::::::::::::
Our previous considerations are based on SEDs with a single-age (13.7 Gyr and 4 Gyr) and thus explicitly ignore the evolution of the rest-frame SED across \zet\ \citep[see, however,][for \kc\ correction models that include an evolutionary correction term, \ec]{Poggianti97,Contardo98,BFvA05,Kotulla09}. EvCon predictions are evidently only possible on the basis of an assumed SFH in the different radial zones of a galaxy, and can hardly take into account its dynamical mass assembly history (e.g., multiple dry and wet minor mergers, creation of massive SF clumps in the disk periphery and their subsequent migration to the galaxy center), which is an important limitation (cf. Sect.~\ref{dis:lim}).

With these cautionary notes in mind, we supplement the analysis in Sect.~\ref{sub:stSED} with EvCon simulations\footnote{Section~\ref{ap:econsSED} provides details on the methodology used and Fig.~\ref{fig:zconsSEDs2} gives an illustrative example of the substantial difference between EvCon simulations and those based on SED templates with a fixed age. Figures~\ref{ub:zcons}--\ref{hk:zcons} show the variation in the ObsF \dmu\ and \dmBD, as well as of various color indices built from the seven photometric filters considered ($U$,$B$,$V$,$R$,$I$,$H$,$K$) for EvCon simulations and an extended set of SFHs. This material is supplemented by a graphical representation of the difference between rest-frame and ObsF color vs. \zet\ (Fig.~\ref{dif:zcons}), that is, the \ec+\kc\ correction that allows a conversion of observed to rest-frame colors.} in which the age of the rest-frame SED at a given \zet\ corresponds to the cosmic age at that redshift.
We note that, from an empirical point of view, bulges in massive (log(\mstar/\msun)$\ga$10.7) LTGs complete the dominant phase of their build-up early on and experience a largely passive evolution over the past $\ga$9 Gyr (e.g., \tref{BP18}), which implies a strong depression of their rest-frame NUV emission, especially in the first 2 Gyr of their evolution ($\equiv$2-4$\tau$), and therefore significant \kc- and evolutionary (\ec) corrections. To the contrary, the nearly constant SFR in the disk translates to a mild evolution of the rest-frame SED and thus to a weaker dependence of \ec-corrections on \zet.

\begin{figure}
\begin{picture}(200,125)
\put(-1,86){\includegraphics[clip, width=4.5cm]{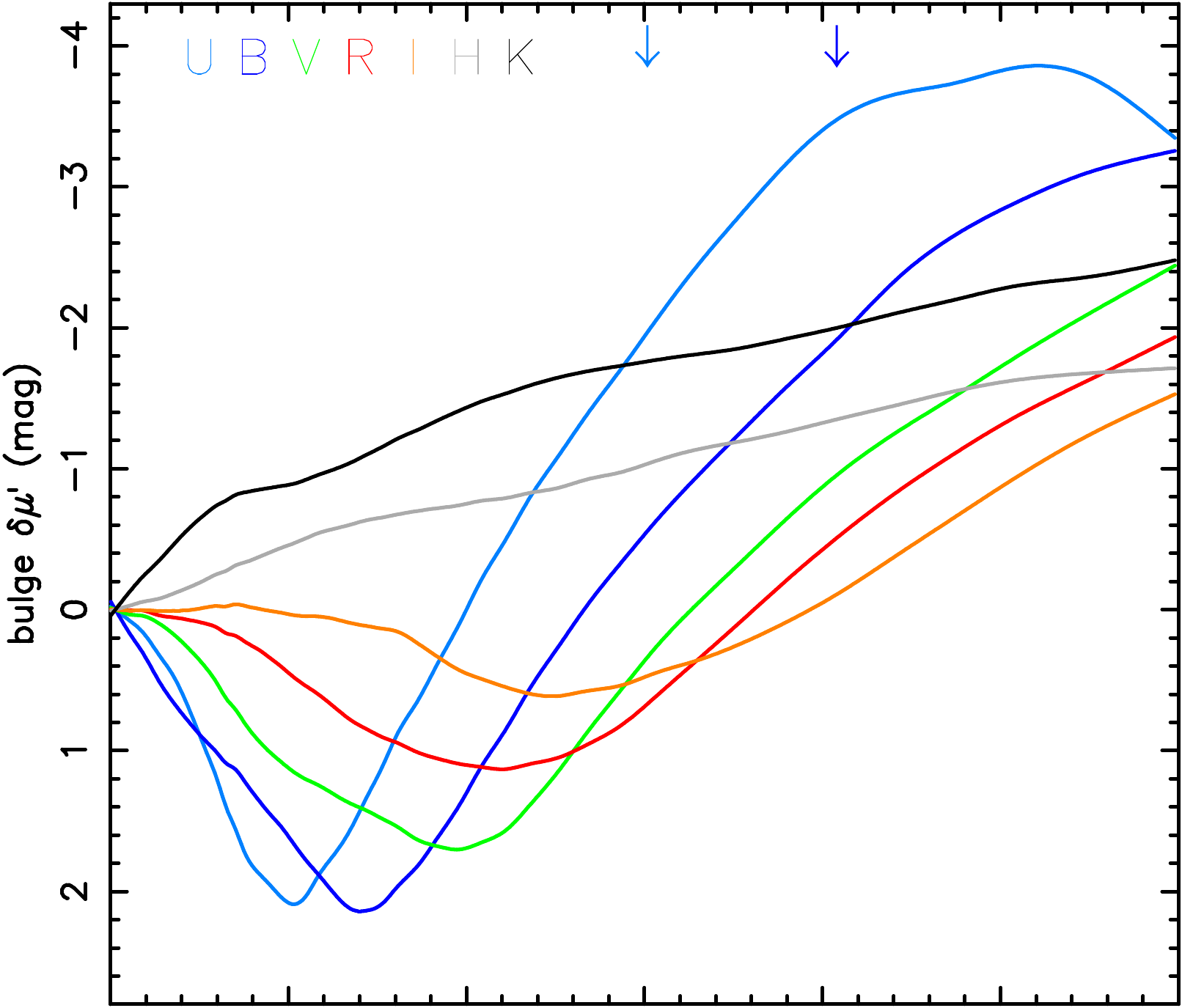}}
\put(-1,45){\includegraphics[clip, width=4.5cm]{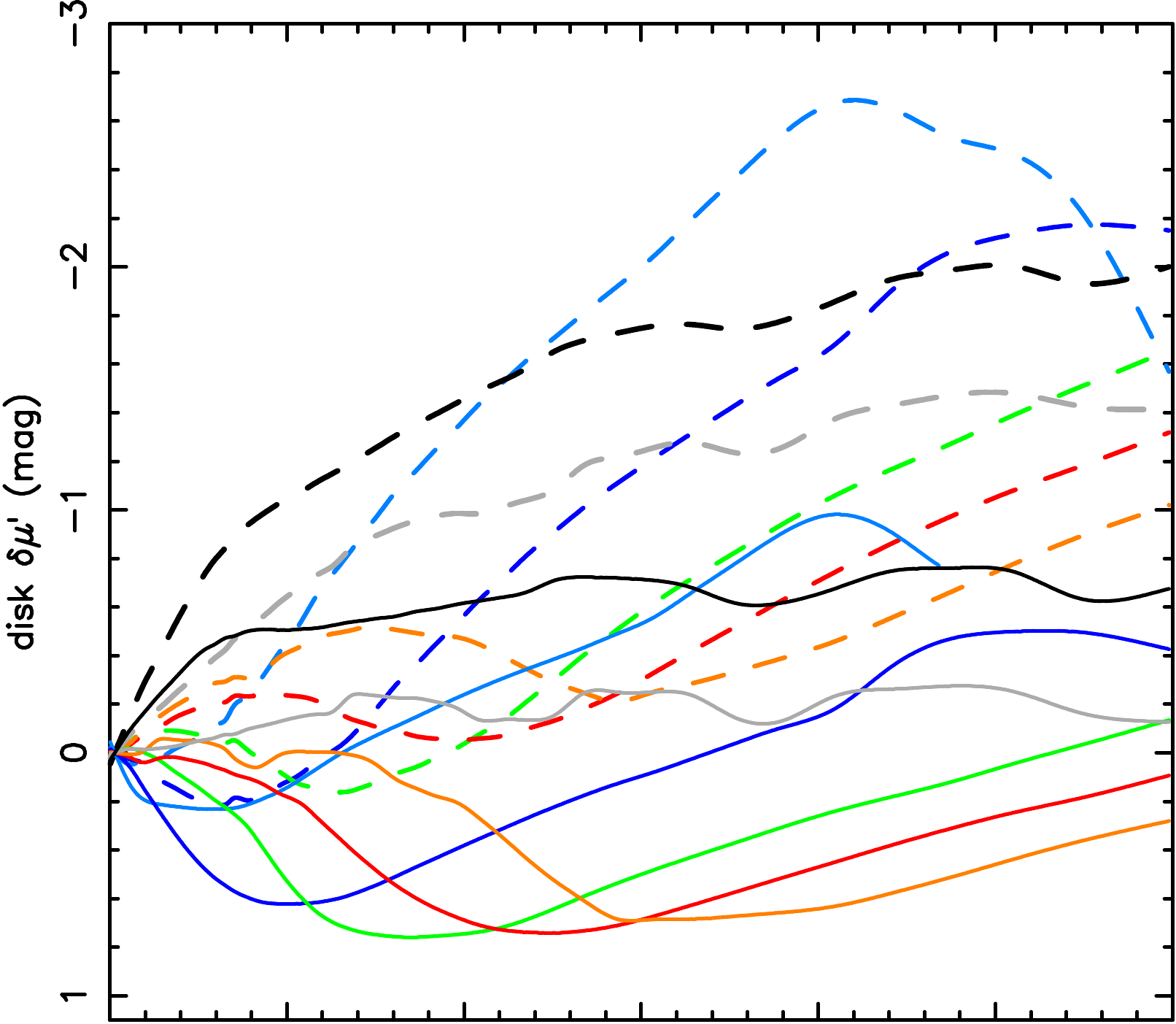}}
\put(-1,0){\includegraphics[clip, width=4.52cm]{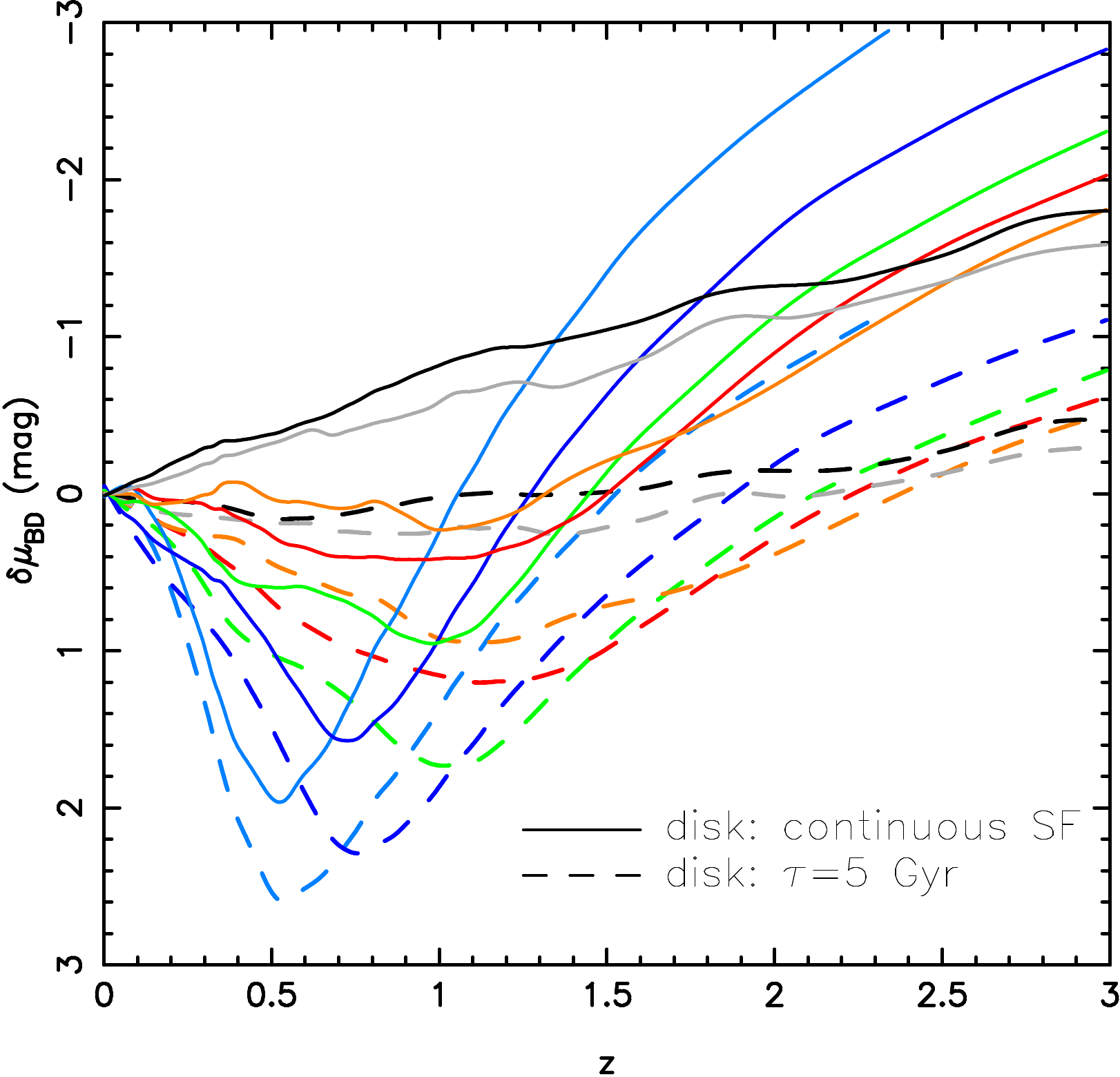}}
\put(44.5,86){\includegraphics[clip, width=4.5cm]{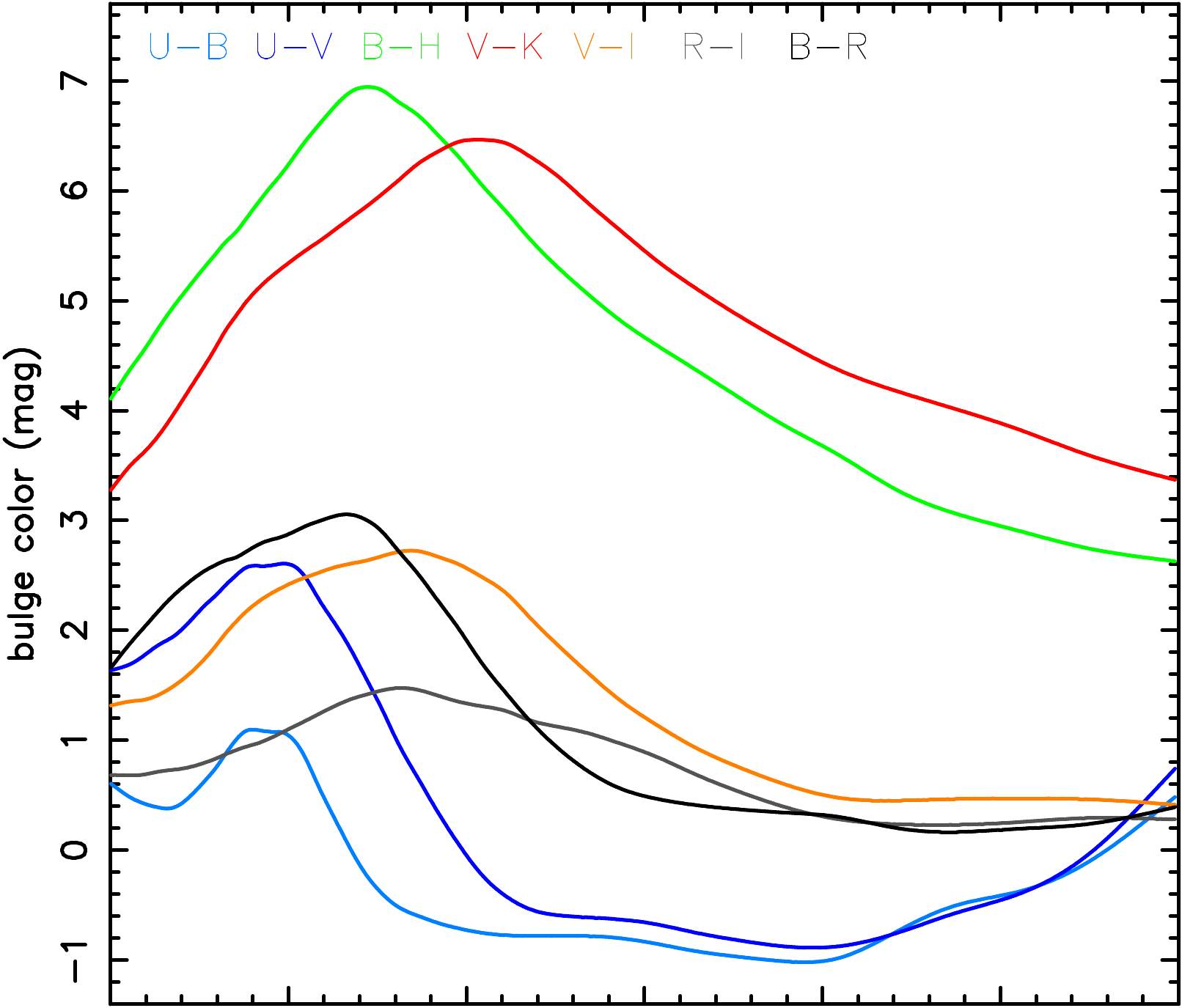}}
\put(44.5,45){\includegraphics[clip, width=4.5cm]{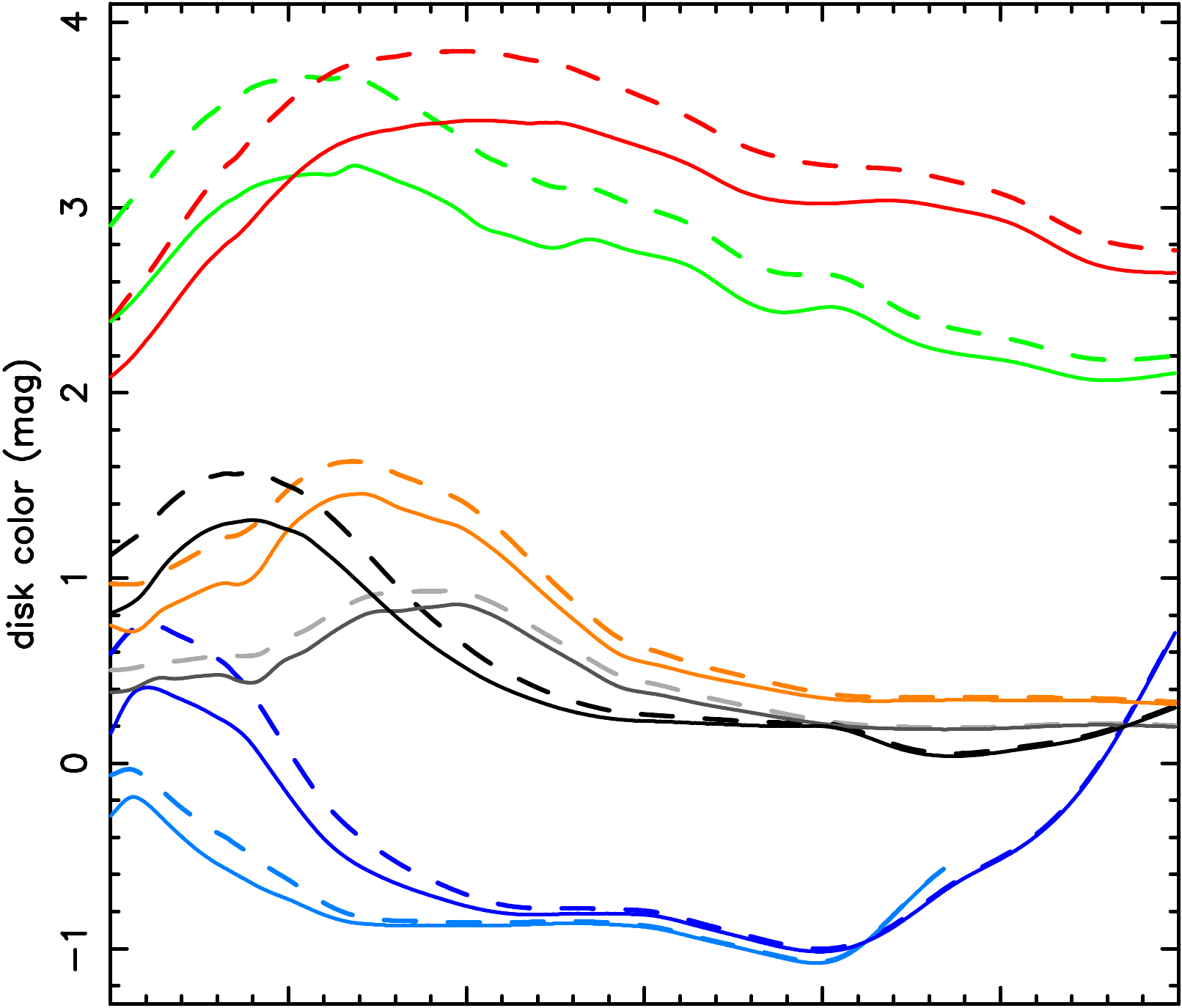}}
\put(44.5,0){\includegraphics[clip, width=4.52cm]{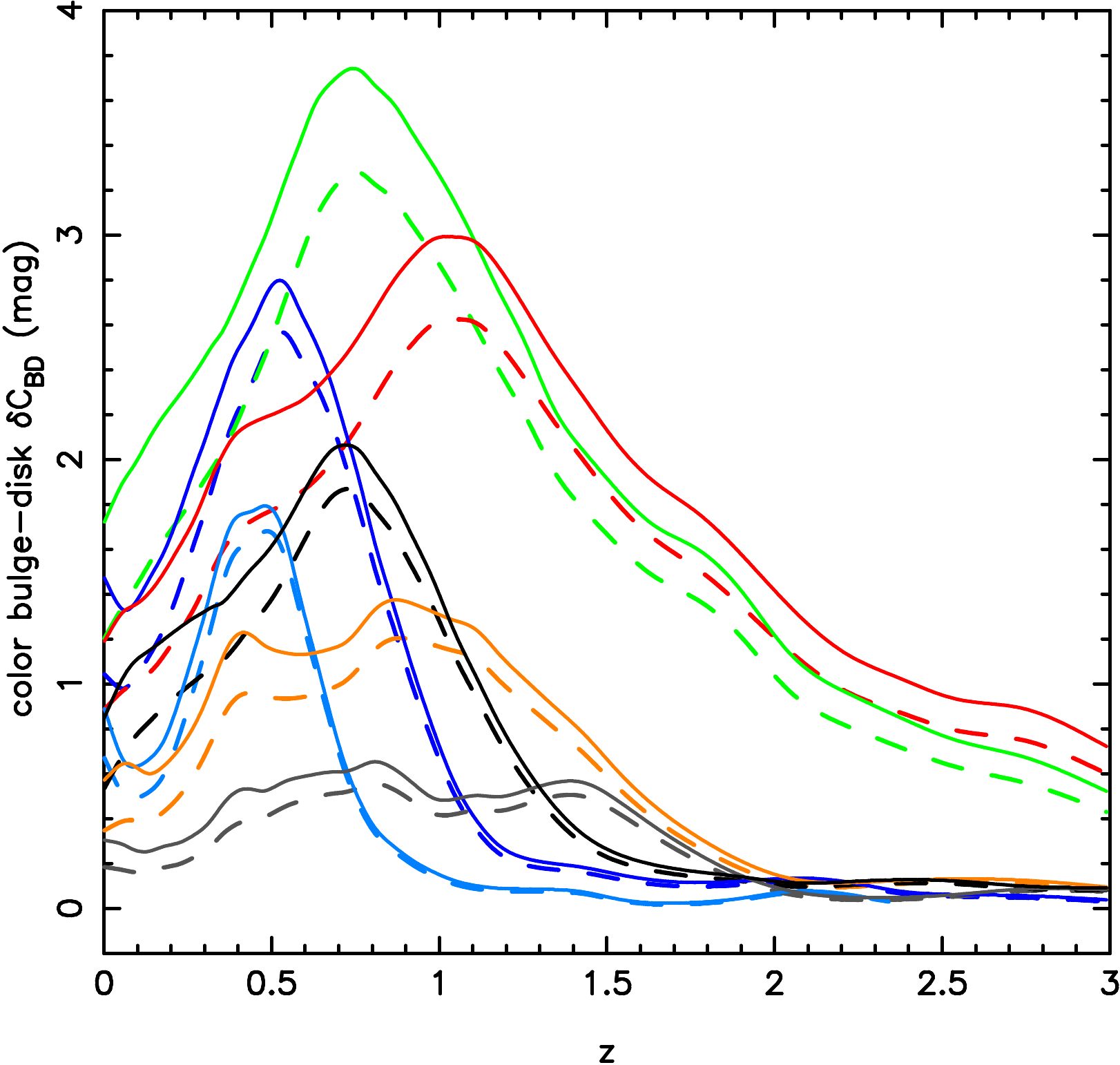}}
\PutLabel{39}{88}{\mlx \textcolor{black}{(a)}}
\PutLabel{39}{47}{\mlx \textcolor{black}{(b)}}
\PutLabel{5}{39}{\mlx \textcolor{black}{(c)}}
\PutLabel{85}{121}{\mlx \textcolor{black}{(d)}}
\PutLabel{85}{47}{\mlx \textcolor{black}{(e)}}
\PutLabel{85}{39}{\mlx \textcolor{black}{(f)}}
\end{picture}
\caption{EvCon simulation of a two-component galaxy that consists of a bulge (panel \brem{a}) of solar metallicity that started forming 13.7 Gyr ago with an exponentially decreasing SFR
($\tau$1 model) and a disk (panel \brem{b}) of \zsun/5 forming at a constant SFR (solid curves) or an exponentially decreasing SFR with an e-folding time of 5 Gyr ($\tau$5 model; dashed curves). Nebular emission is taken into account.
A noticeable difference from simulations that involve SED templates of a fixed age (Fig.~\ref{sSED}) is that the bulge gets brighter in the NIR with increasing \zet, reaching at \zet=3 a \dmu\ of --1.7 mag in $H$ and --2.4 mag in $K$. Another salient feature is an initial dimming in $\mu$\arcmin\ of up to $\sim$2 mag for optical filters at 0.5$\la z \la$1.5 that is followed by a reversion to negative values at higher \zet, with the brightening, \dmu, of the bulge in the optical becoming equal to that in the NIR at various \zet\ between 1.2 and 3, depending on filter. As for the disk, EvCon simulations do not drastically differ from those assuming a present-day SED
(panel \brem{b} in Fig.~\ref{sSED}) because of the mild evolution of the SED implied by continuous SF. 
The effect of the $\tau$5 SFH model for the disk is illustrated with dashed curves. The layout is similar to that of Fig.~\ref{sSED}. The arrows in panel \brem{a} mark the redshift at which the Ly$\alpha$ line enters the blue edge of the $U$ and $B$ filter.}
\label{eSED}
\end{figure}
The layout of Fig.~\ref{eSED} is similar to that of Fig.~\ref{sSED}, with panel \brem{a} referring to the bulge ($\tau$1 model with a constant stellar metallicity of \zsun) and panel \brem{b} to the disk. The latter is approximated by the contSF model (solid curves) and a model of exponentially decreasing SFR with an e-folding time of 5 Gyr ($\tau$5 model; dashed curves). Nebular emission is taken into account in both cases.

Some important differences relative to the single-age simulations (Fig.~\ref{sSED}) are apparent: In the NIR, the bulge gets brighter with increasing redshift, reaching a \dmu\ of --1.7 mag in $H$ and --2.4 mag in $K$ at \zet=3.
Another salient feature is that, in the optical, it initially becomes fainter for 0.5$\la z \la$1.5 and then shows a turnover at a higher \zet, where its high SFR for the assumed parametric SFH models translates to copious rest-frame UV emission and thus to a \dmu\ $<$ --3 mag in $U$ and $B$.

To the contrary, EvCon simulations for the disk (panel \brem{b}) do not substantially differ from those based on single-age SEDs (panels \brem{b} and \brem{h} in Fig.~\ref{sSED}).
More specifically, the contSF model implies in the optical a comparatively small variation in \dmu\ (between 0.8 and --1 mag), which is comparable to that in Fig.~\ref{sSED} (0.7 mag \dots\ --0.8 mag). Likewise, the disk shows throughout a brightening in the NIR, albeit with a lower amplitude (\dmu\ $\geq$ --0.8 $K$ mag compared to $\geq$ --1.5 $K$ mag for single-age simulations). The results are similar for the $\tau$5 model, which yields at \zet=3 a $\mu$\arcmin\ increase of up to --1.5 mag (--2 mag) in $H$ ($K$).

As for the variation in colors versus redshift, panel \brem{e} shows for the disk a reasonably good agreement with predictions based on single-age SEDs, whereas an important difference is visible for the bulge (panel \brem{d}) whose maximal optical-NIR colors at 0.5$\la$\zet$\la$1 are by $>$2 mag bluer than those in Fig.~\ref{sSED}.
The bulge-to-disk color contrast \dcol\ shows a qualitatively similar behavior as in Fig.~\ref{sSED}\irem{f}, with an initial rise of $\sim$2 (1) mag for optical-NIR (optical-optical) colors that is followed at \zet$>$1 by a gradual inversion to lower values than those in a present-day galaxy. For example, the bulge appears with respect to its ObsF $V$--$I$ ($B$--$R$) color by 1.4 (2.1) mag redder than the disk at \zet=0.9 (0.7) whereas at \zet$>$1.6 color gradients vanish in the optical and remain weakly traceable only when hybrid optical-NIR colors are used.

A combined inspection of \dmBD\ and \dcol\ therefore shows that within $0.4 \la z \la 1.2$ bulges appear throughout redder, and in terms of $\mu$\arcmin\ dimmer (brighter) in the optical (NIR) compared to \zet=0. 
It is important that this conclusion is supported both by EvCon simulations (Fig.~\ref{eSED}; see also simulations for an extended set of SFHs in Fig.~\ref{ub:zcons}--\ref{dif:zcons}) and those based on single-age SEDs (Fig.~\ref{sSED}).
% :::::::::::::::::::::::::::::::::::::::::::::::::::::::::::::::::::::::::::::::::::::::::::::::::::::::::::::::::::
\section{Simulations of galaxy bulge-disk decomposition across redshift \label{BD}}
\begin{center}
\begin{figure}
\begin{picture}(86,70)
\put(5,0){\includegraphics[clip, width=8cm]{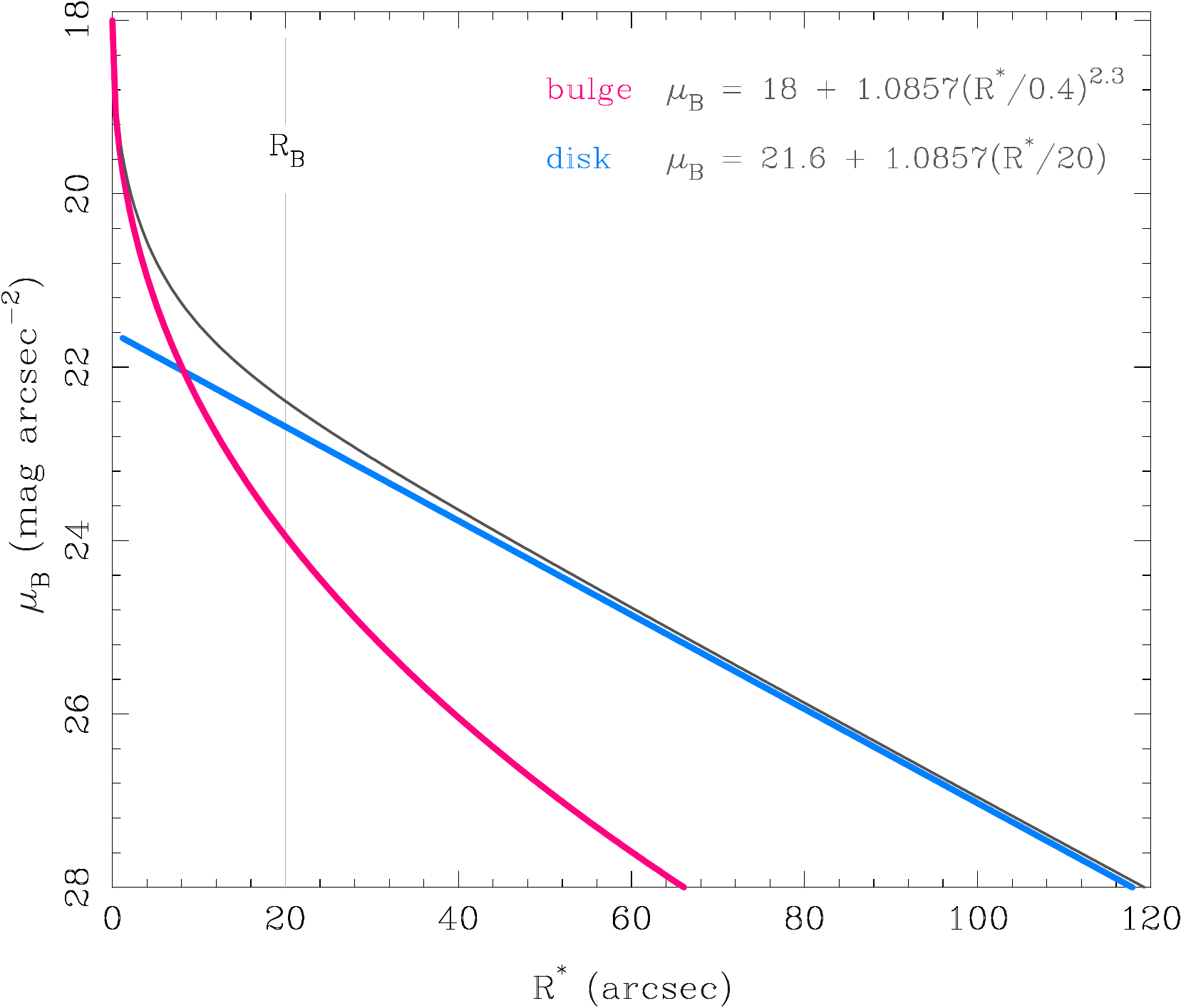}}
\end{picture}
\caption{Synthetic $B$-band SBP for a spiral galaxy that consists of a bulge and a disk (see labels on the upper right)
with structural properties typical of high-mass LTGs in the local Universe.}
\label{fig:sb3} 
\end{figure}
\end{center}

% :::::::::::::::::::::::::::::::::::::::::::::::::::::::::::::::::::::::::::::::::::::::::::::::::::::::::::::::::::
Equipped with estimates on \dmu($z$) we study next how the SBP of a cSED spiral galaxy varies across redshift. To this end, the \dmu(\zet) previously computed for various SFHs and filters (Sect.~\ref{sub:stSED} and Sect.~\ref{sub:eSED}) was added to the disk and bulge of a reference synthetic galaxy model
in order to simulate its total SBP and radial color profile as a function of \zet\ (cf. Sect.~\ref{ap:BD} for details). We adopt the same galaxy model as in \tref{P22} (cf. their Fig.~8), which roughly matches the structure of a present-day LTG. It is composed of a bulge with a central surface brightness of 18 $B$~\sbb\ and a S\'ersic exponent $\eta=2.3$, and an exponential disk with a  central surface brightness of 21.6 $B$~\sbb\ and a scale length $\alpha=20$\arcsec. The bulge radius \rbulge\ at \z=0, defined at an isophotal level of 24 \sbb, is equal to one $\alpha$ (20\arcsec). This model (Fig.~\ref{fig:sb3}) yields a B/D ratio of 0.35 and a \BT\ ratio of 0.26, in the range of typical values for local high-mass (log(\mstar/\msun)$\sim$11) LTGs \citep[e.g.,][]{Men17}.
For simplicity, we adopt the usual postulate that the exponential profile of the disk outside the bulge is preserved all the way to its center (i.e., beneath the bulge), which is equivalent to assuming a spatially invariant sSFR and zero radial color gradients throughout the disk \citep[see, however,][and \tref{P22}]{Breda20b}.

Synthetic (bulge+disk) SBPs in eight filters ($UBVRIJHK$) were constructed based both on single-age (Sect.~\ref{sub:stSED})
and EvCon simulations (Sect.~\ref{sub:eSED}) for the seven SFHs in Fig.~\ref{fig:SFHs}.
These SBPs were then decomposed into a bulge and a disk by fitting an exponential for \rr$\geq$70\arcsec\ (3.5$\alpha$)
and a S\'ersic model to the residual central luminosity excess. These fits (61 for each setup) allowed us to examine the variation within 0$\leq$\zet$\leq$3
of various commonly used photometric quantities, such as \reff\ and the \citet{Petrosian76} radius\footnote{$R_{\rm Petrosian}$ is defined as the photometric radius \rr\ (\arcsec) at which the Petrosian function $\mathrm{I(R^{\star})/\langle I(R^{\star})\rangle}$ drops to 1/3, where I(\rr) denotes the intensity
(erg s$^{-1}$ cm$^{-2}$ sr$^{-1}$).}
$R_{\rm Petrosian}$. These radii (marked with the superscript \arcmin\ and referred to the following as reduced radii) were kept uncorrected for the dependence of angular distance on \zet\ in order to isolate the effect coming purely from \cmod. Additionally, we determined for each SBP the S\'ersic exponent $\eta$,
the light concentration index log(R$_{80}$/R$_{20}$) involving the radii enclosing 80\% and 20\% of the total luminosity in a given band,
and the \citet{Trujillo01} concentration parameter (cf. Fig.~\ref{fig:BDdec2} for various combinations of the SFH of the bulge and the disk).

\begin{figure}
\begin{picture}(86,200)
\put(0.76,176){\includegraphics[clip, width=8.54cm]{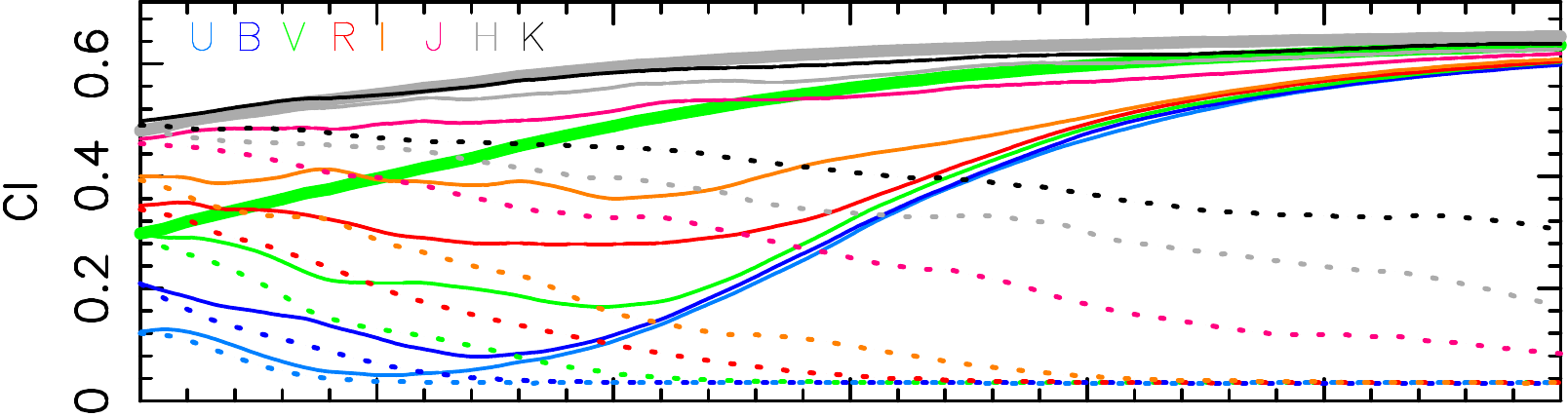}}
\put(0,151){\includegraphics[clip, width=8.6cm]{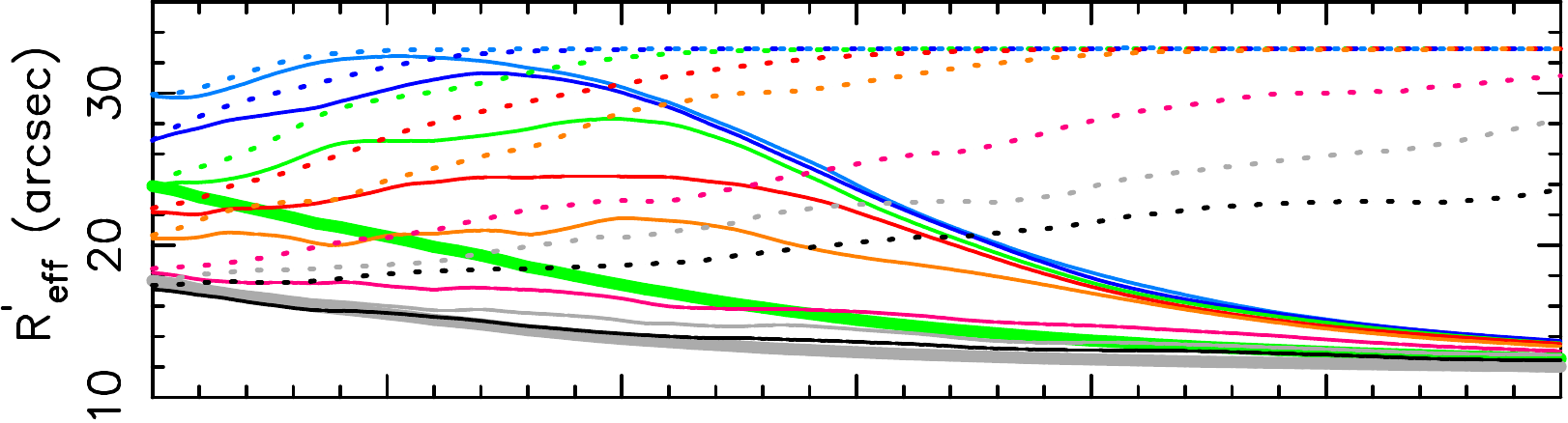}}
\put(0,126){\includegraphics[clip, width=8.6cm]{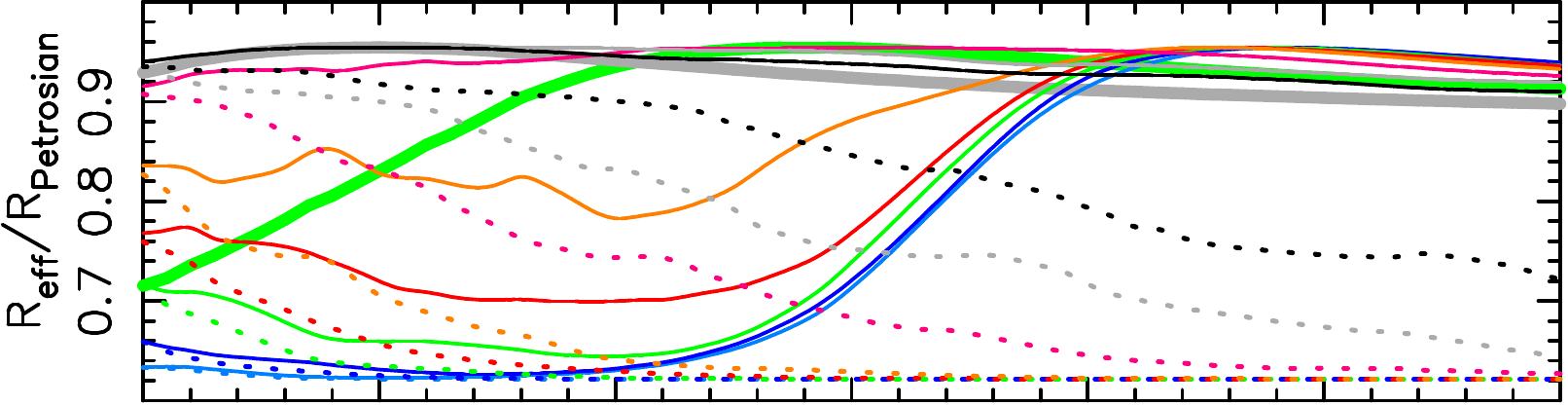}}
\put(0,101){\includegraphics[clip, width=8.6cm]{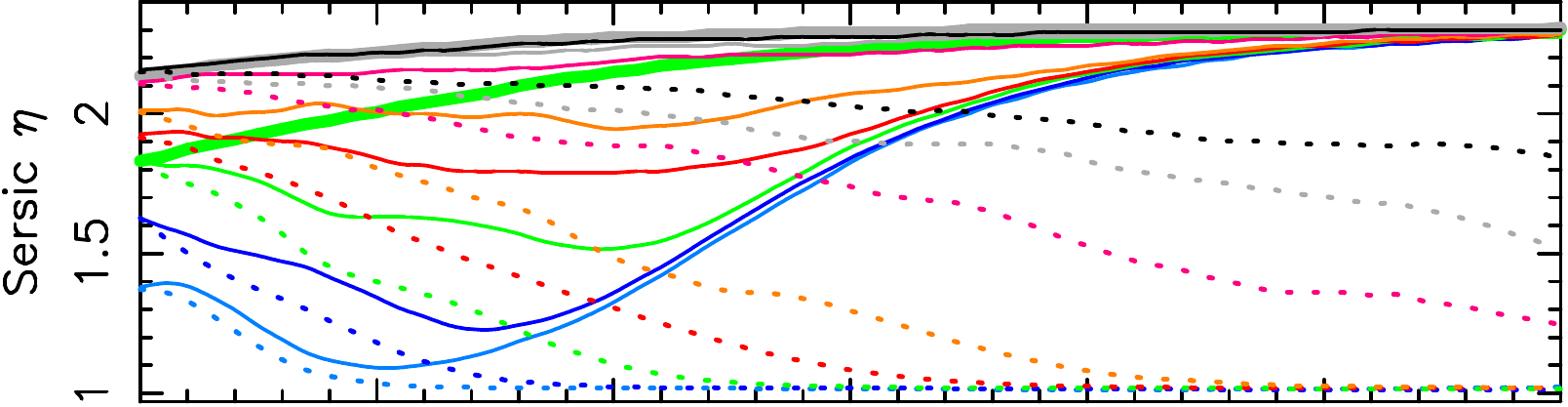}}
\put(0,52.5){\includegraphics[clip, width=8.6cm]{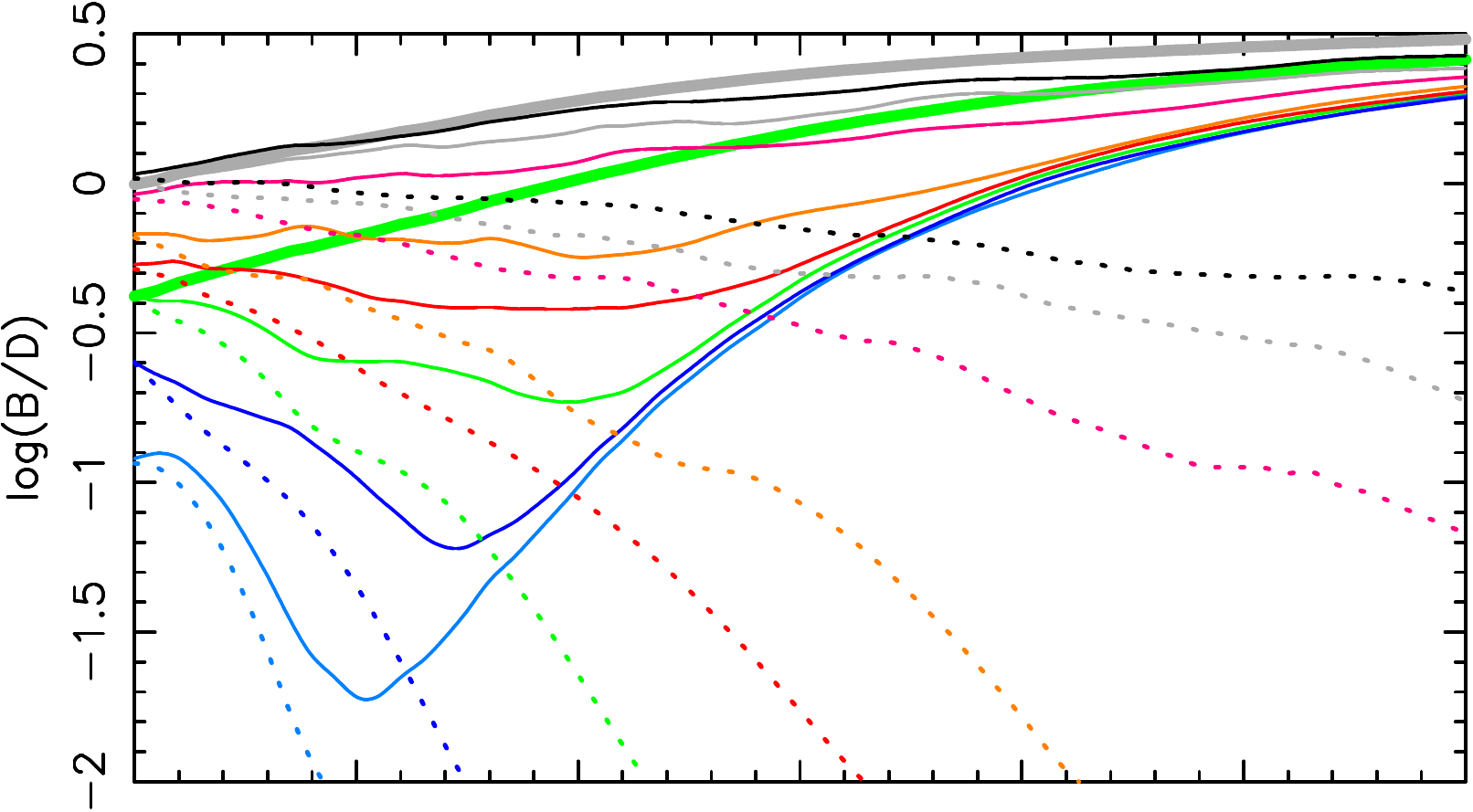}}
\put(0,0){\includegraphics[clip, width=8.63cm]{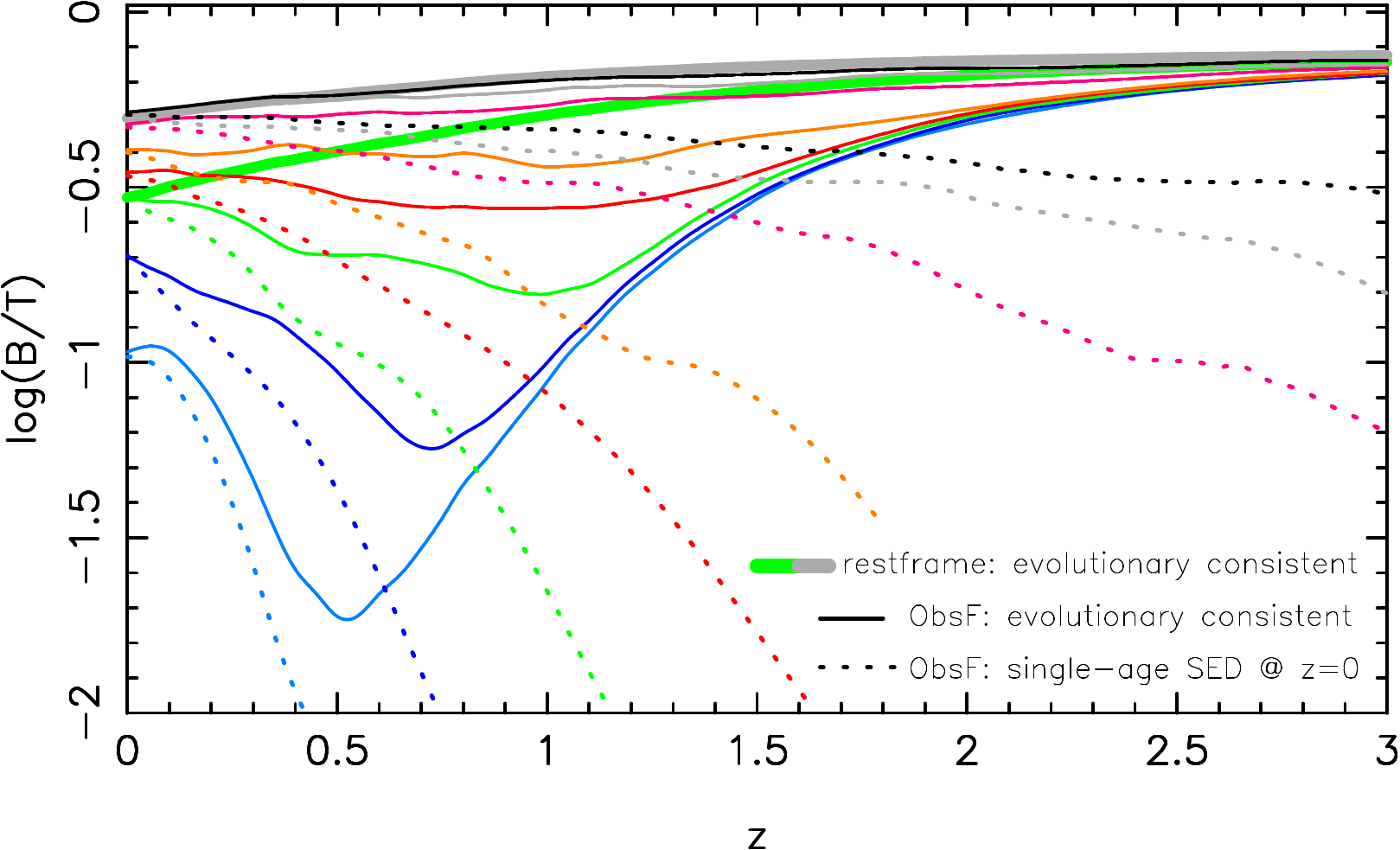}}
\PutLabel{81}{190}{\mlx \textcolor{black}{(a)}}
\PutLabel{81}{157}{\mlx \textcolor{black}{(b)}}
\PutLabel{81}{129}{\mlx \textcolor{black}{(c)}}
\PutLabel{81}{103}{\mlx \textcolor{black}{(d)}}
\PutLabel{81}{57}{\mlx \textcolor{black}{(e)}}
\PutLabel{81}{21}{\mlx \textcolor{black}{(f)}}
\end{picture}
\caption{Photometric quantities obtained by simulating the cSED galaxy model in Fig.~\ref{fig:sb3}
in the redshift interval $0\leq z \leq 3$. Our synthetic galaxy consists of a bulge and a disk that have been forming
for 13.7 Gyr according to, respectively, the $\tau$1 and contSF SFH model.
Single-age simulations based on the SED of a 13.7 Gyr old galaxy at \zet=0 that is projected out to \zet=3 are
shown with dotted curves, whereas solid curves refer to EvCon simulations (Sect.~\ref{sub:eSED}).
Thick curves show the true (rest-frame) values implied by EvCon models in $V$ (green) and $H$ (gray).
\brem{a)} Concentration index after Eq.~\ref{eq:CI} ranging between 0 (pure disk) and $>$0.54 (bulge-dominated galaxy).
\brem{b)} Reduced effective radius (i.e., with the dependence of angular distance on \zet\ not taken into account).
\brem{c)} Ratio between effective and Petrosian radius.
\brem{d)} S\'ersic exponent, $\eta$. \brem{e and f)} Logarithm of the \BD\ and \BT\ luminosity ratios.}
\label{fig:BDdec1}
\end{figure}

The prominence of the bulge was quantified through a concentration index,
\begin{equation}
CI = 1 - \frac{R_{\rm eff}}{1.7\alpha},
\label{eq:CI}
\end{equation}
that involves the ratio between the effective radius of the total SBP to that of the disk only.
A CI$\approx$0 corresponds to a nearly bulgeless galaxy, whereas higher values reflect an increasing \BD \ ratio.
For instance, from Fig.~8 of \citet{P06} one can read off that a $CI$ of 0.28, 0.54, and 0.68 corresponds to a \BT\ ratio of
0.25, 0.5, and 0.75, respectively\footnote{The fact that \reff\ is inversely related to the \BT\ ratio suggests caution should be taken when morphologically heterogeneous galaxy samples
at different \zet\ are inter-compared on the basis of \reff-normalized quantities. For example, the same (in terms of dex/kpc) metallicity gradient appears after normalization to \reff\ shallower in a bulge-dominated galaxy than in a disk-dominated one. This could introduce spurious correlations (or blur existing ones), as pointed out in \tref{BP18}.
Same considerations obviously apply to any other radially resolvable quantity (e.g., age, color, $D_{4000}$ index, Lick indices, \ewha).
The insight from Fig.~\ref{fig:BDdec1} that \reff\ for centrally SF-quenched high-\zet\ galaxies is overestimated in the optical because of the disappearance of the bulge
implies an amplification of \reff-normalized metallicity gradients, which in turn superficially suggests an enhanced level of chemical inhomogeneity.}.

Figure~\ref{fig:BDdec1} shows the main results from image decomposition. Dotted curves correspond to determinations one would obtain if using as reference a 13.7~Gyr old cSED galaxy (i.e., the same models as in Fig.~\ref{sSED} with the only difference being that nebular emission is taken into account)
whereas with solid curves we show results based on EvCon models (Fig.~\ref{eSED}).

It can be seen that the first set of simulations implies for all filters a nearly monotonous decrease in $CI$ with increasing \zet, which translates to a gradual disappearance of the bulge: the galaxy becomes ``bulgeless'' (a nearly pure disk with $CI\simeq0$) in all optical bands at a \zet\ that increases with increasing filter central wavelength $\lambda_{\rm c}$ (\zet$\sim$0.5 in the $U$ band and \zet$\sim$2 in the $I$ band), while the bulge remains detectable in the $H$ band at a CI$\approx$0.17 ($\sim$1/3 of its rest-frame value) even at \zet=3 (panel \brem{a}).
The disappearance of the bulge in the optical is accompanied by an inflation of the galaxy effective radius toward the value corresponding to a pure disk (1.7\,$\alpha \sim$ 34\arcsec; panel \brem{b}) and a decrease of the ratio \reff/$R_{\rm Petrosian}$ (panel \brem{c}), whereby optical filters show the strongest dependence on \zet.

As for the S\'ersic exponent $\eta$, it can be appreciated from panel \brem{d} that its value at \zet=0 depends on filter (e.g., $\sim$1.4 in $U$ and 2.1 in $K$) as a consequence of the decreasing prominence of the bulge in the blue spectral range. With increasing \zet, consistently with the evidence from panel \brem{a}, $\eta$ gradually decreases in the optical, prompting at \zet$\ga$1 the erroneous classification of the galaxy as a pure disk, whereas in the NIR this trend is milder, with $\eta$ decreasing from its original value of $\sim$2.1 to 1.5 (1.8) in $H$ ($K$) at \zet=3.
The \cmod-driven fading of the bulge with increasing (decreasing) \zet\ ($\lambda_{\rm c}$) is further reflected in the \BD\ and \BT\ ratios:
whereas at \zet=1 (2) the $H$-band \BT\ ratio is underestimated by merely 0.1 dex (0.23 dex) -- that is, by 20\% (60\%) -- the effect in the $I$ band amounts to 0.44 (1.4) dex (a factor of between $\sim$2.7 and $\sim$25).

Coming to EvCon simulations, the true value of the photometric quantities under study (i.e., their rest-frame value vs. \zet) is shown by the thick green (gray) curves for the $V$ ($H$) band. This set of simulations imply in the optical and out to \zet$\sim$0.5-1 (depending on filter) similar trends as the single-age simulations above, with the notable difference of a reversion of ObsF values toward a better agreement with rest-frame values at higher \zet.
However, as is apparent from determinations in the $V$ band, there is throughout a strong discrepancy between ObsF and rest-frame photometric quantities in all optical filters, with a near agreement between both appearing only at \zet$\sim$3.
An interesting (and possibly deceptive) coincidence is that, for certain redshift intervals, ObsF and present-day structural properties are nearly identical (e.g., at \zet$\sim$1.5 for the $V$ band), which at prima facie suggests that galaxies have not experienced evolution in their photometric structure over the past $\sim$9 Gyr.

The simulations in Fig.~\ref{fig:BDdec1} (see also other combinations of SFHs for the bulge and disk in Fig.~\ref{fig:BDdec2}) permit a quantitative assessment
of the systematic and substantial discrepancy between true (rest-frame) and observed structural properties of spiral galaxies at higher-\zet, and its
complex dependence on redshift and filters used.
Clearly, the two setups adopted for our simulations -- the first one based on single-age SEDs (dotted curves) thus assuming no evolution and the second one (EvCon; solid curves) approximating the galaxy assembly history by standard SFR parameterizations -- are oversimplified, and lack important evolutionary ingredients (e.g., intense and widespread nebular emission that, especially in the NIR, can influence the morphology and colors of distant galaxies, chemical evolution and spatially inhomogeneous dust extinction, bulge growth through inward migration and coalescence of SF clumps from the disk, or an empirical prescription for SF quenching; cf. Sect.~\ref{dis:lim}).
Nevertheless, we argue that they capture the essential implications of \cmod, and offer a conceptual framework for the refinement of simulations with more realistic assumptions.
% ::::::::::::::::::::::::::::::::::::::::::::::::::::::::::::::::::::::::::::::::::::::::::::::::::::::::::::::::::::::::::::
\section{Spatially resolved simulations based on integral field spectroscopy and population spectral synthesis \label{simIFS}}
% ::::::::::::::::::::::::::::::::::::::::::::::::::::::::::::::::::::::::::::::::::::::::::::::::::::::::::::::::::::::::::::
As a supplement to our previous considerations, we use IFS data to examine the effect of \cmod\ on the color and
reduced surface brightness of a cSED (bulge+disk) galaxy. This was done similar to Sect.~\ref{sub:stSED} with the main difference being that synthetic SEDs from \pegase\
were substituted by the spatially resolved panchromatic (91 \AA -- 160 $\mu$m) SED obtained from fitting the IFS data cube with \starlight\ \citep{Cid05}.
After combination of stellar and nebular emission, and by taking intrinsic extinction separately for both of these components into account (cf. Sect.~\ref{ap:simIFS} for details), the SED cube was used as input to simulations within 0$\leq$\zet$\leq$3.
\begin{figure}
\begin{picture}(86,60)
\put(7,0){\includegraphics[clip, width=7cm]{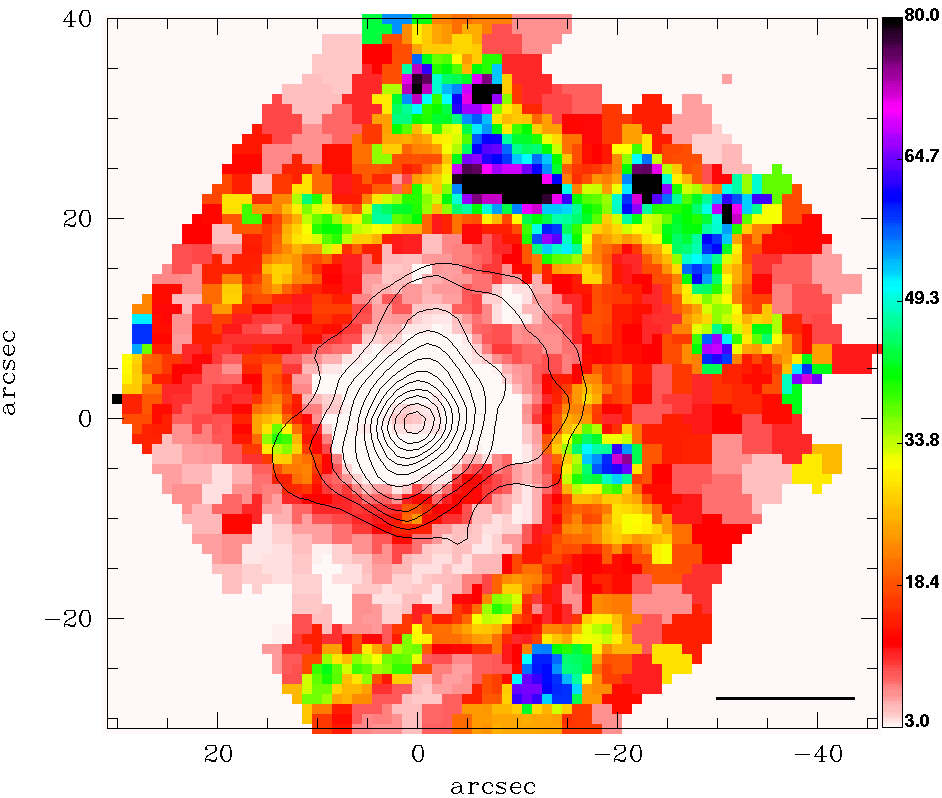}}
\end{picture}
\caption{\ewha\  map of the nearby spiral galaxy \object{NGC 0309}  (D=75.8 Mpc) displayed in the range between 3 and 80 \AA.
The \ewha\ map has been obtained from IFS data from the Calar Alto Legacy Integral Field Area (CALIFA) survey \citep{Sanchez12-DR1,GB15CALIFA} through their processing with the pipeline {\sc Porto3D} \citep{P13,GP16-ETGs}, as detailed in \tref{BP18}. Contours delineate the stellar surface density, $\Sigma_{\star}$, in \msun\,kpc$^{-2}$ on a logarithmic scale from 8.2 to 9.3 in increments of 0.1 dex. The horizontal bar corresponds to a projected linear scale of 5 kpc. The \ewha\ morphology of \object{NGC 0309} is typical of massive local LTGs with an old SF-quenched bulge immersed within an extended star-forming disk \citep[e.g.,][]{CatT17,Belfiore18,B19,Kalinova21}.}
\label{n309-ewha} 
\end{figure}

It should be noted that this exercise is not meant to reconstruct the rest-frame morphology and colors that a spiral galaxy used to have several Gyr ago\footnote{Even though this task is straightforward from the technical point of view, since the SFH, extinction and stellar velocity dispersion $\sigma_{\star}$ are encoded within the best-fitting population vectors (PVs) from \starlight, thereby permitting computation of the spatially resolved \ec-correction terms needed for reconstructing the past morphology of a galaxy, some skepticism is in place. This is not only because of the notorious age-metallicity-extinction degeneracy \citep[e.g.,][]{Worthey94,Guseva01} and the low time resolution of spectral synthesis models for old ages, but also because they obviously do not allow the inference of the initial location and mass fraction of ex situ formed stars in a given spatial cell. In other words, they are unable to account for the role of dynamical processes in the galaxy assembly history (e.g., multiple minor mergers, bulge formation through merging of inwardly migrating SF clumps from the disk; cf. Sect.~\ref{dis:lim}).}
but merely
\begin{figure}
\begin{picture}(86,107)
\put(0,72){\includegraphics[clip, height=3.6cm]{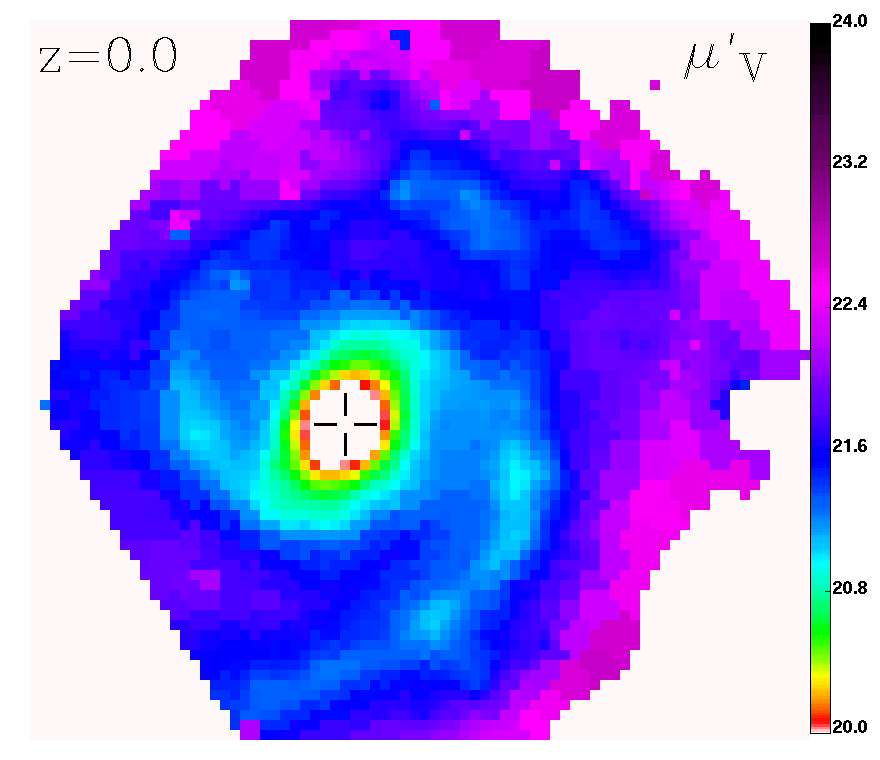}}
\put(47,72){\includegraphics[clip, height=3.6cm]{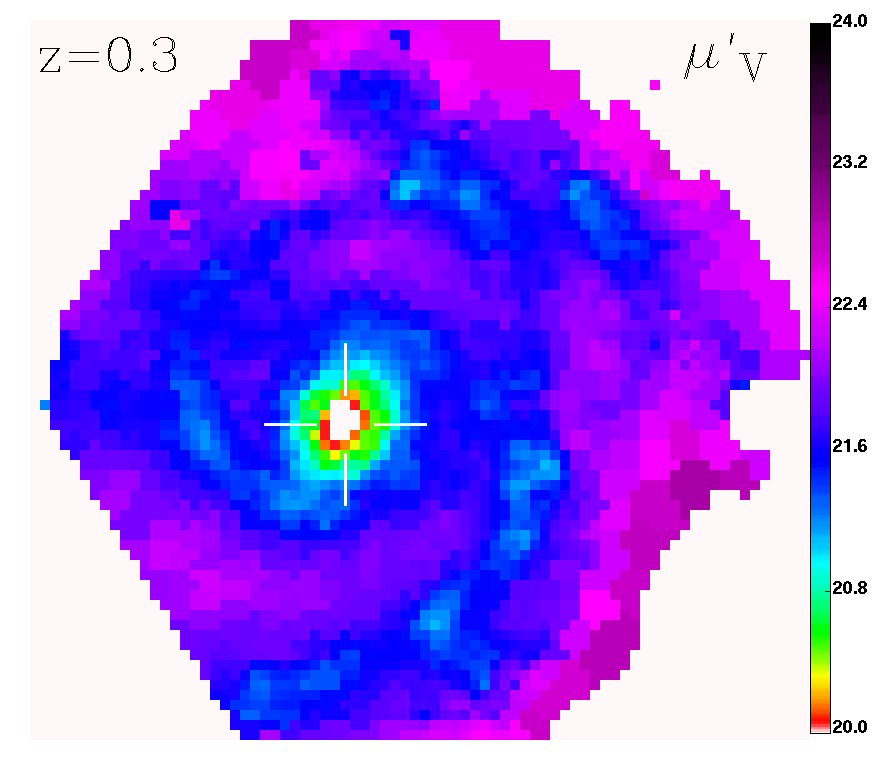}}
\put(0,36){\includegraphics[clip, height=3.6cm]{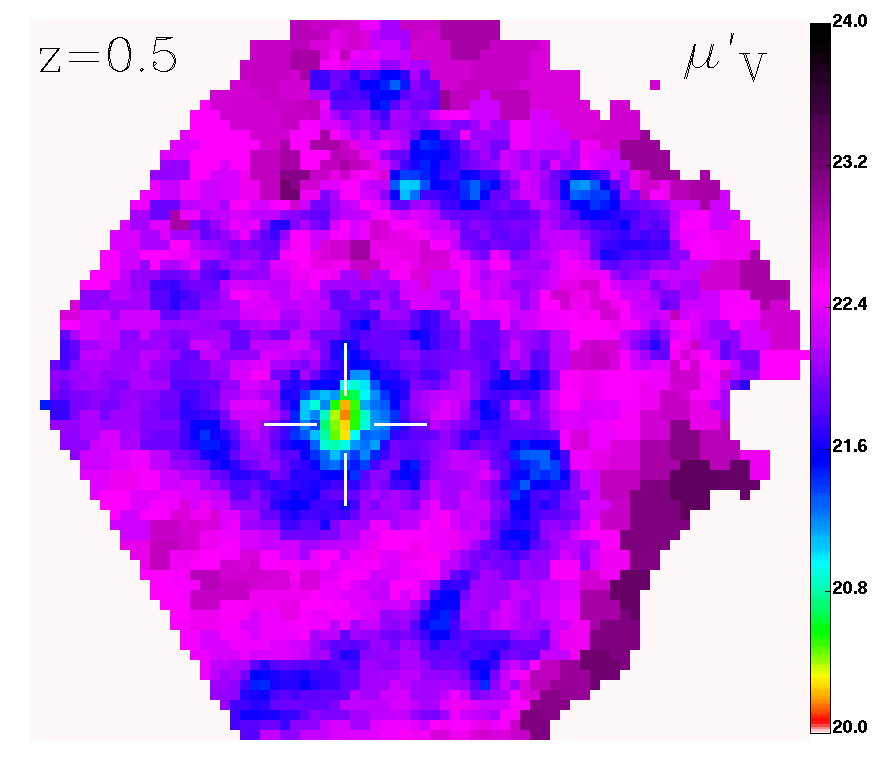}}
\put(47,36){\includegraphics[clip, height=3.6cm]{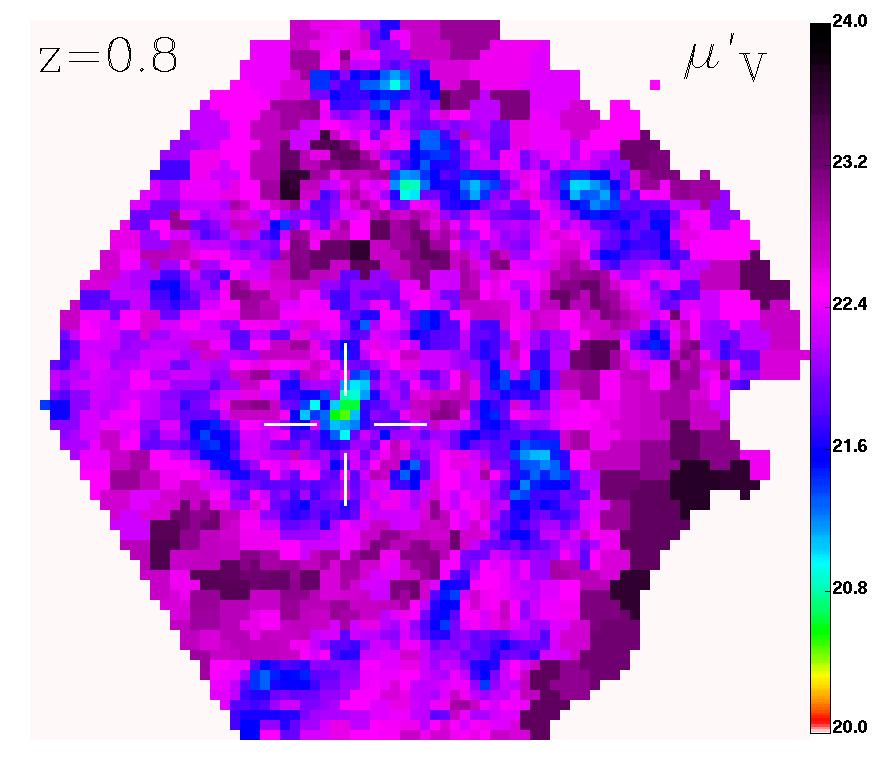}}
\put(0,0){\includegraphics[clip, height=3.6cm]{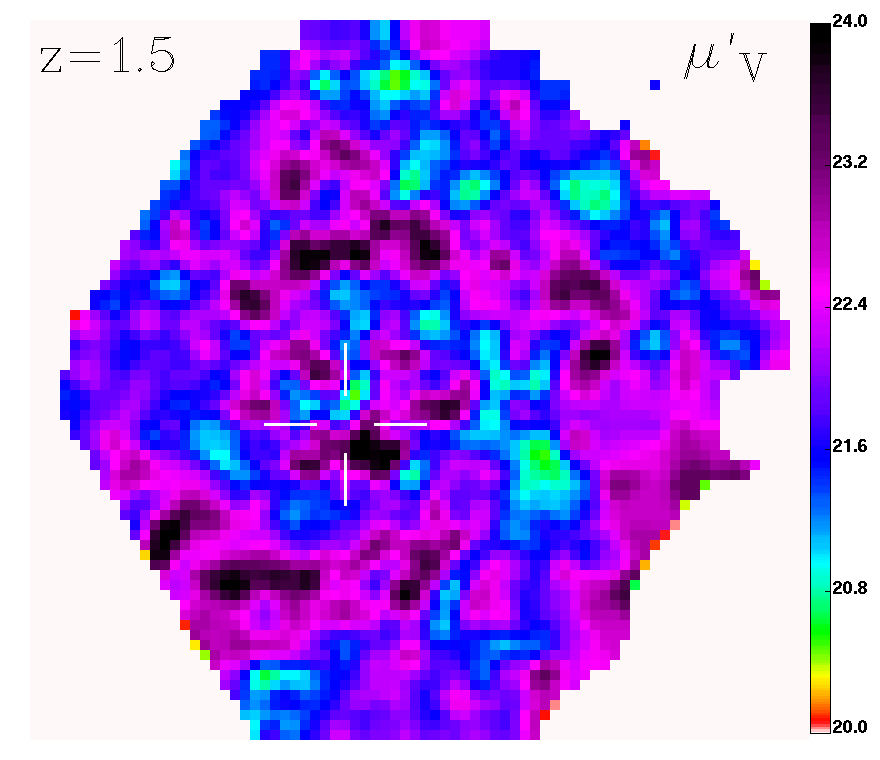}}
\put(47,0){\includegraphics[clip, height=3.6cm]{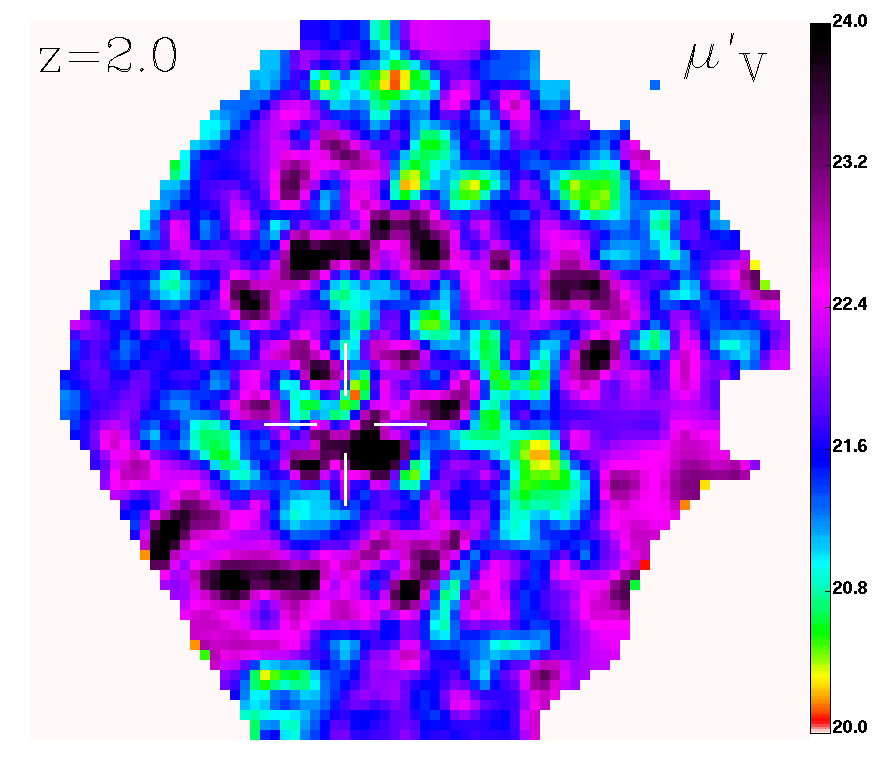}}
\end{picture}
\caption{Synthetic $V$-band images of \object{NGC 0309} computed on the basis of spectral modeling of
CALIFA IFS data. The maps show the reduced $V$-band surface brightness, $\mu\arcmin$, between 20 and 24 \sbb\ at \zet=0.3, 0.5, 0.8, 1.5, and 2.
It can be seen that the SF-quenched bulge becomes practically invisible for \zet$\geq$0.8 as a result of \cmod.}
\label{n309-zvsVmag} 
\end{figure}
\begin{figure*}
\begin{picture}(220,70)
\put(0,32){\includegraphics[clip, width=5.6cm]{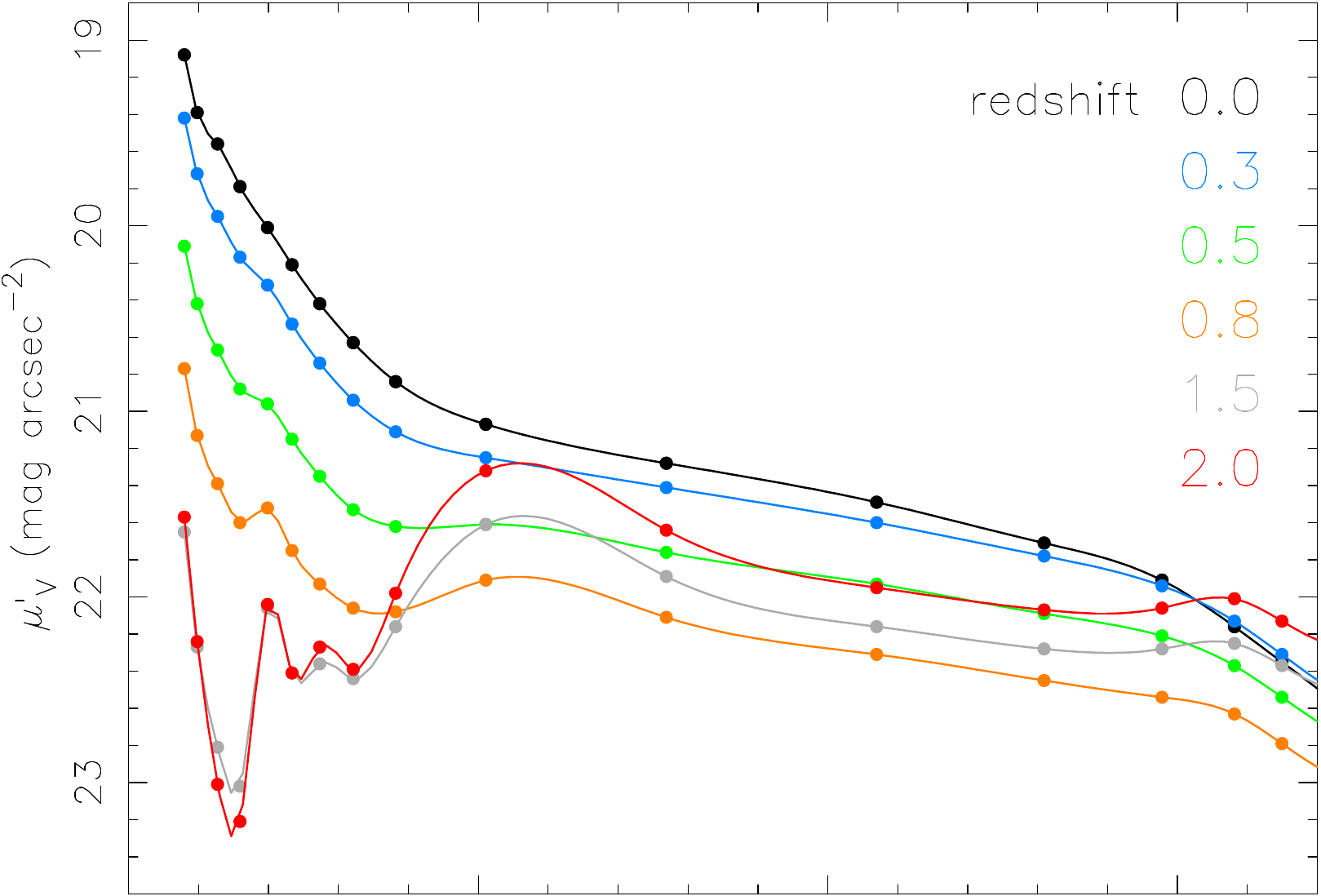}}
\put(0,0){\includegraphics[clip, width=5.62cm]{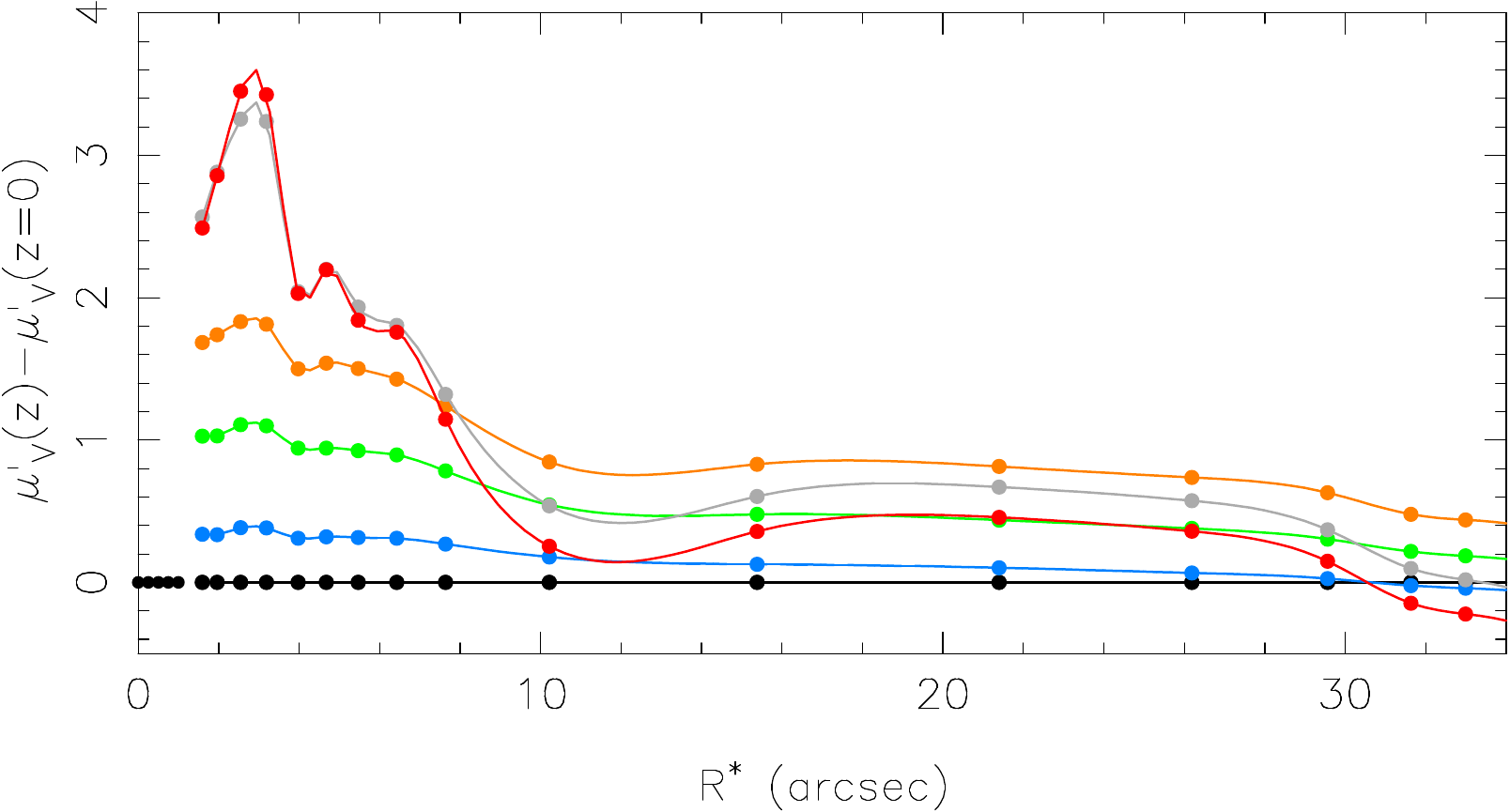}}
\put(64,32){\includegraphics[clip, width=5.6cm]{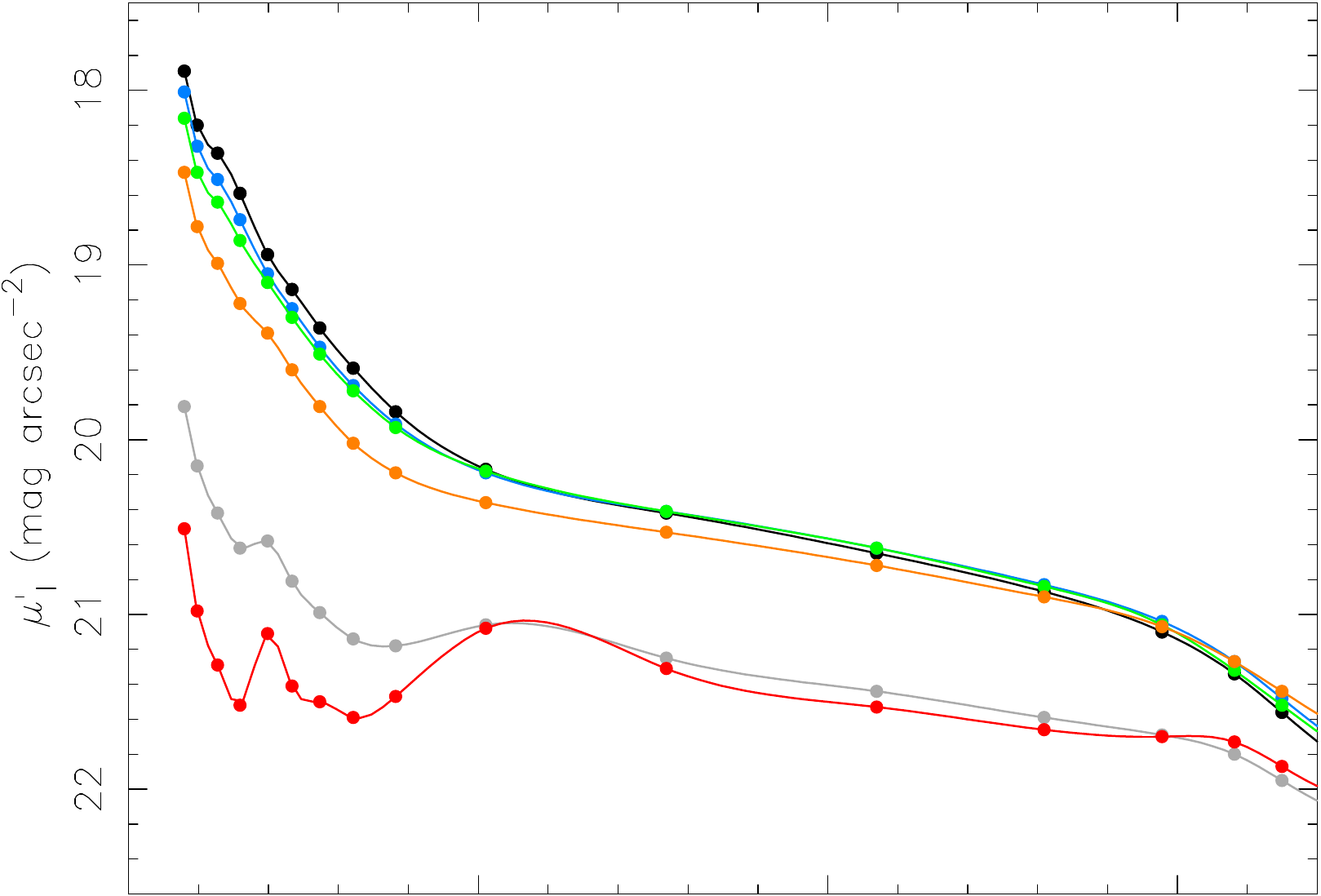}}
\put(64,0){\includegraphics[clip, width=5.62cm]{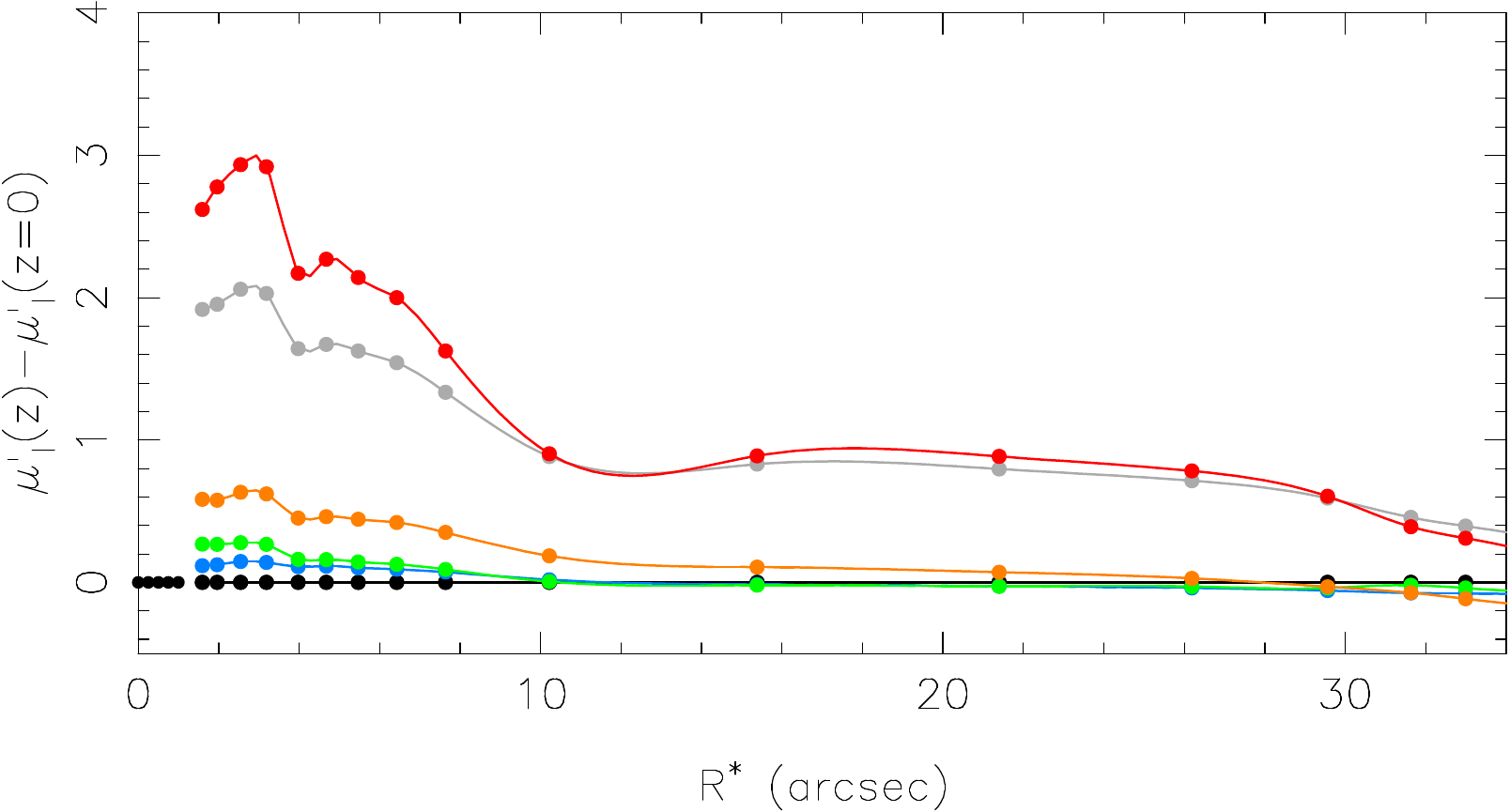}}
\put(128,32){\includegraphics[clip, width=5.6cm]{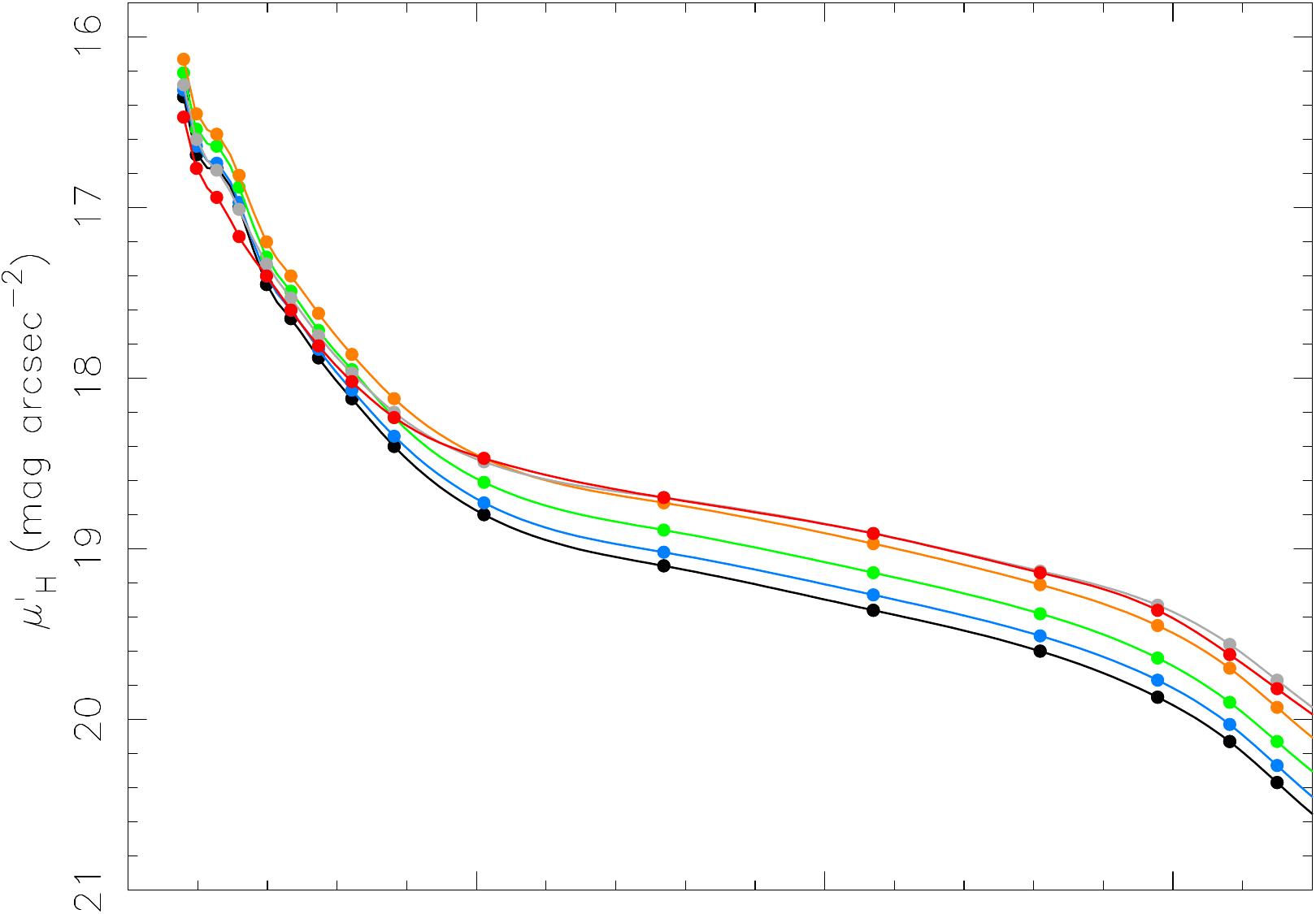}}
\put(128,0){\includegraphics[clip, width=5.62cm]{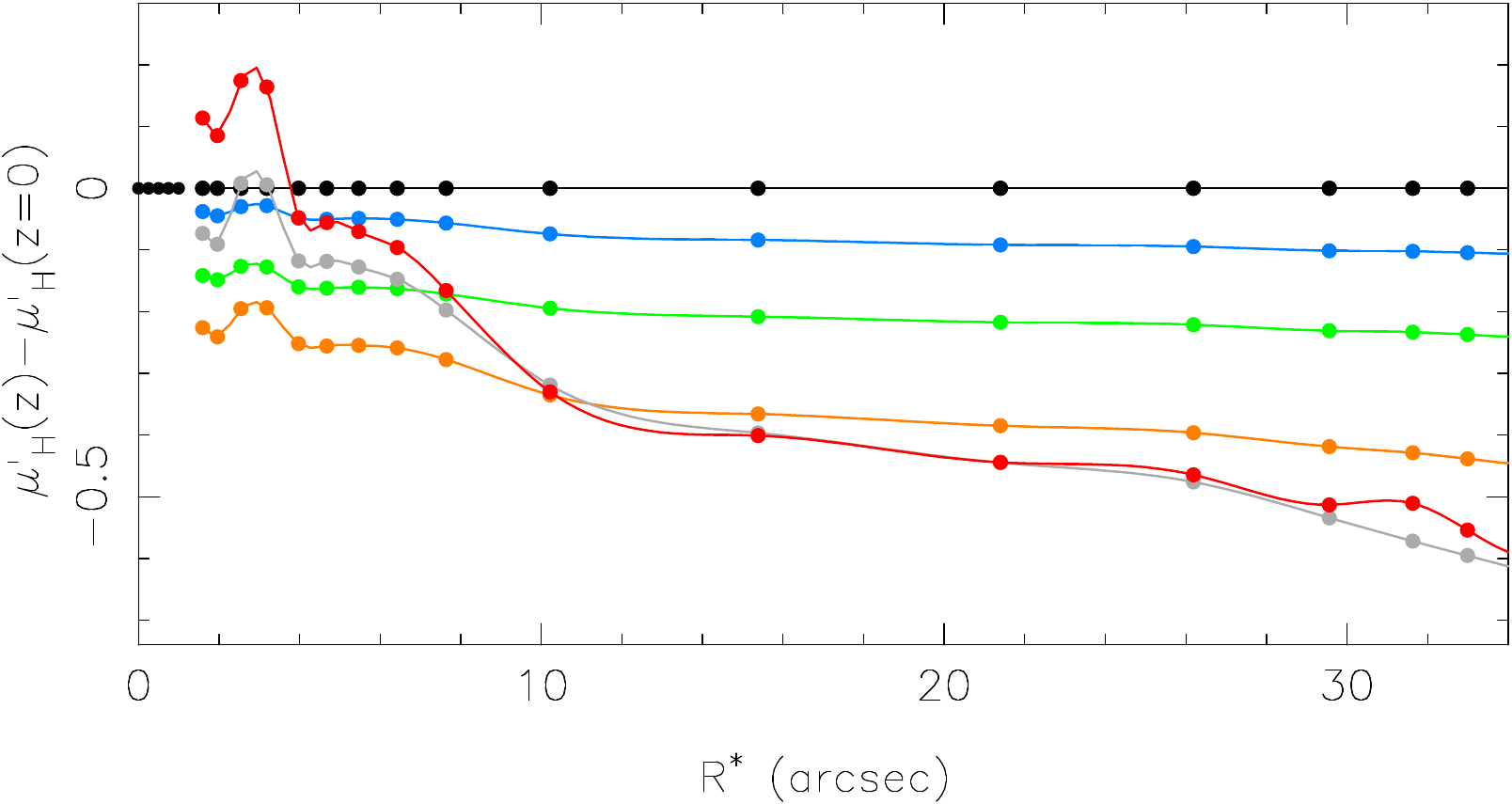}}
\PutLabel{30}{34}{\mlx \textcolor{black}{V\arcmin\ SBP}}
\PutLabel{30}{25}{\mlx \textcolor{black}{V\arcmin: ObsF-restframe}}
\PutLabel{98}{34}{\mlx \textcolor{black}{I\arcmin\ SBP}}
\PutLabel{98}{25}{\mlx \textcolor{black}{I\arcmin: ObsF-restframe}}
\PutLabel{161}{34.8}{\mlx \textcolor{black}{H\arcmin\ SBP}}
\PutLabel{161}{25}{\mlx \textcolor{black}{H\arcmin: ObsF-restframe}}
\end{picture}
\caption{SBPs of \object{NGC0309} in $V$\arcmin, $I$\arcmin\ and $H$\arcmin\ computed from simulated images
at \zet=0, 0.3, 0.5, 0.8, 1.5 and 2. The lower panels show the difference (in mag) between the ObsF and rest-frame $\mu$\arcmin.
Consistently with the evidence from Fig.~\ref{n309-zvsVmag}, the bulge (\rr$<$10\arcsec) fades to invisibility for \zet$>$0.8 in $V$\arcmin\ 
whereas the disk suffers only a modest dimming. The situation is different in the $H$\arcmin\ band where both bulge and disk get brighter with increasing \zet.}
\label{n309-zvsSBP}
\end{figure*}
to empirically explore the differential impact of \cmod\ on a present-day cSED galaxy, if it were to be observed at a higher redshift.
Even though this approach is evolutionary inconsistent and ignores
the dynamical assembly history of galaxies, it is not entirely unrealistic if the bulge has formed quasi-monolithically early on and the disk has maintained
a nearly constant SFR over the past few gigayears. This is because, in this case, the slope of the SED in both bulge and disk shows a modest evolution
back to \zet=1.6 (4 Gyr after the Big Bang).
For instance, one can infer from \pegase\ that the 1650\,\AA--$B$ color of a bulge forming according to the $\tau$05 model
(Sect.~\ref{sub:stSED}) increases by 5.5 mag within the first 4 Gyr, whereas by only 0.8 mag in the ensuing $\sim$10 Gyr of its evolution.
As for the disk (the contSF model), this color changes in the respective time intervals by 2.1 mag and 0.25 mag.
This comparatively mild evolution of the UV-through-optical SED of the bulge and disk offers a justification that the IFS-based simulations below
capture basic implications of \cmod.

Figure~\ref{n309-zvsVmag} shows synthetic $V$ images of the local spiral galaxy \object{NGC 0309} \citep[Figs.~\ref{sketch}\irem{a} and \ref{n309-ewha}; see also][for further details]{BP23} simulated for $0\leq z \leq 2$.
It can be seen that the SF-quenched bulge practically disappears for \zet$>$0.8 whereas the spiral features in the disk become after an initial dimming
at 0.3$\leq$\zet$\leq$0.8 prominent again for \zet$>$1.5, in agreement with predictions from Fig.~\ref{sSED}.
Even though the simulated images are increasingly noisy for \zet$\ga$0.8, as spectral modeling uncertainties in the optical are amplified in the NUV, it can be appreciated
that \cmod\ makes a higher-\zet\ analog of a local LTG appearing in the optical as a bulgeless disk.

The differential variation (fading in the optical and brightening in the NIR; Fig.~\ref{sSED}) of the ObsF surface brightness can be quantified from SBPs (Fig.~\ref{n309-zvsSBP}).
These were computed from simulated images with the irregular isophotal annuli technique by \citet[][see also \tref{Noeske et al. 2003,2007}]{P02}.
Upper panels show profiles in $V$, $I$ and $H$, and lower panels the difference $\mu\arcmin$(\zet)--$\mu\arcmin$(\zet=0) relative to the rest-frame surface brightness.
As already visual inspection of Fig.~\ref{n309-zvsVmag} indicates, the bulge (\rr$\leq$10\arcsec) suffers a dimming by $\sim$3 mag in the optical and shrinks
to a compact nuclear excess for \zet$\geq$0.8 (1.5) in $V$ ($I$).
The opposite happens in $H$, where the bulge appears brighter throughout, with only a minor dimming at \zet=2 in its very central part.
As for the disk (\rr$>$10\arcsec), after an initial dimming by $\sim$0.7 $V$ mag at 0.3$\la$\zet$\la$0.8, its $\mu$\arcmin\ gets brighter again at a higher redshift
where it approaches a mean $\mu\arcmin$(\zet)--$\mu\arcmin$(\zet=0) $<$0.5 mag. The situation is somehow reverse in $H$ where the disk gets brighter as \zet\ increases,
reaching a $\mu\arcmin$(\zet)--$\mu\arcmin$(\zet=0) $\simeq$ --0.5 mag at \zet$\geq$1.5.

\begin{figure*}
\begin{picture}(220,70)
\put(0,32){\includegraphics[clip, width=5.6cm]{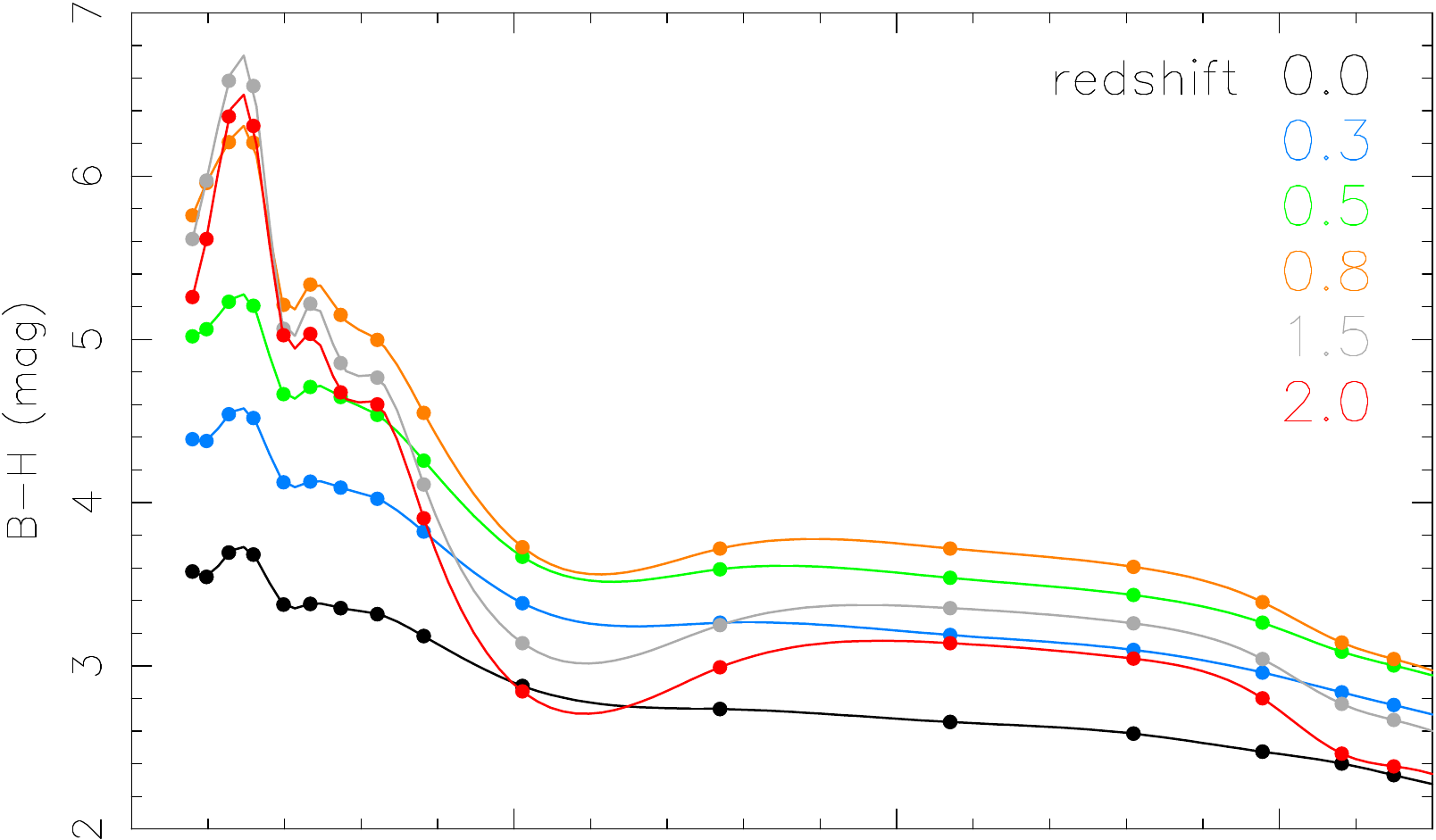}}
\put(0,0){\includegraphics[clip, width=5.62cm]{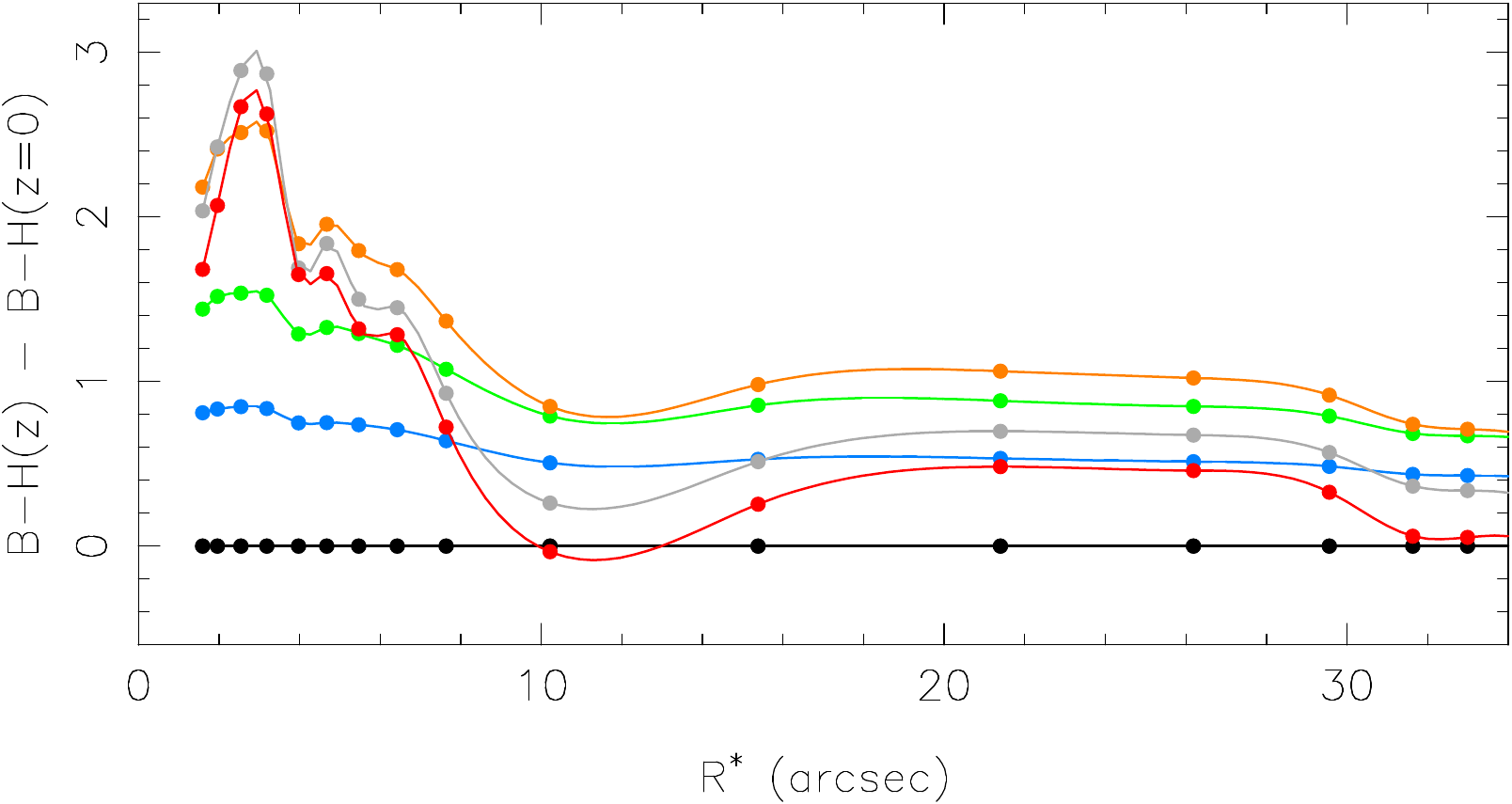}}
\put(64,32){\includegraphics[clip, width=5.6cm]{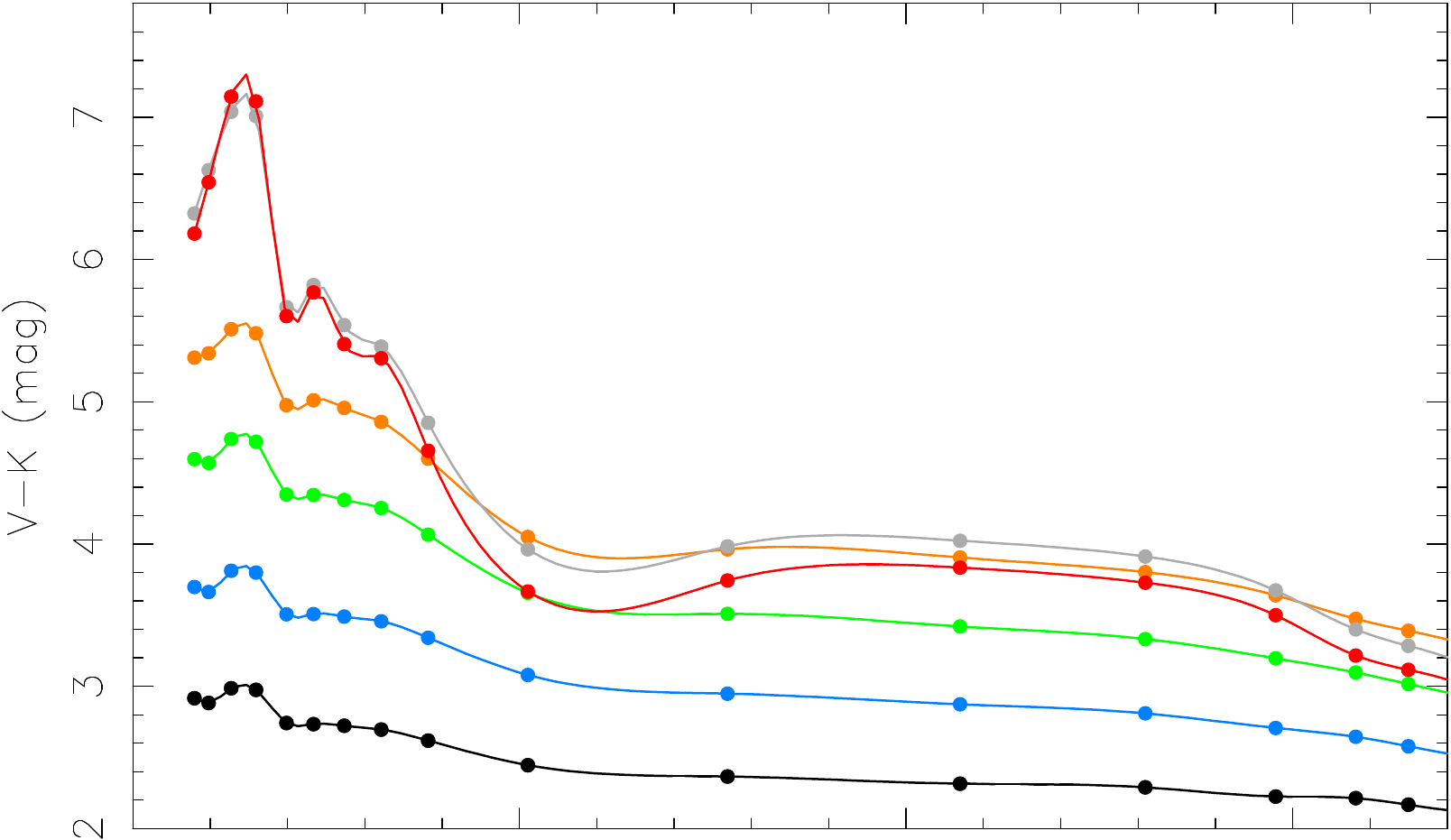}}
\put(64,0){\includegraphics[clip, width=5.62cm]{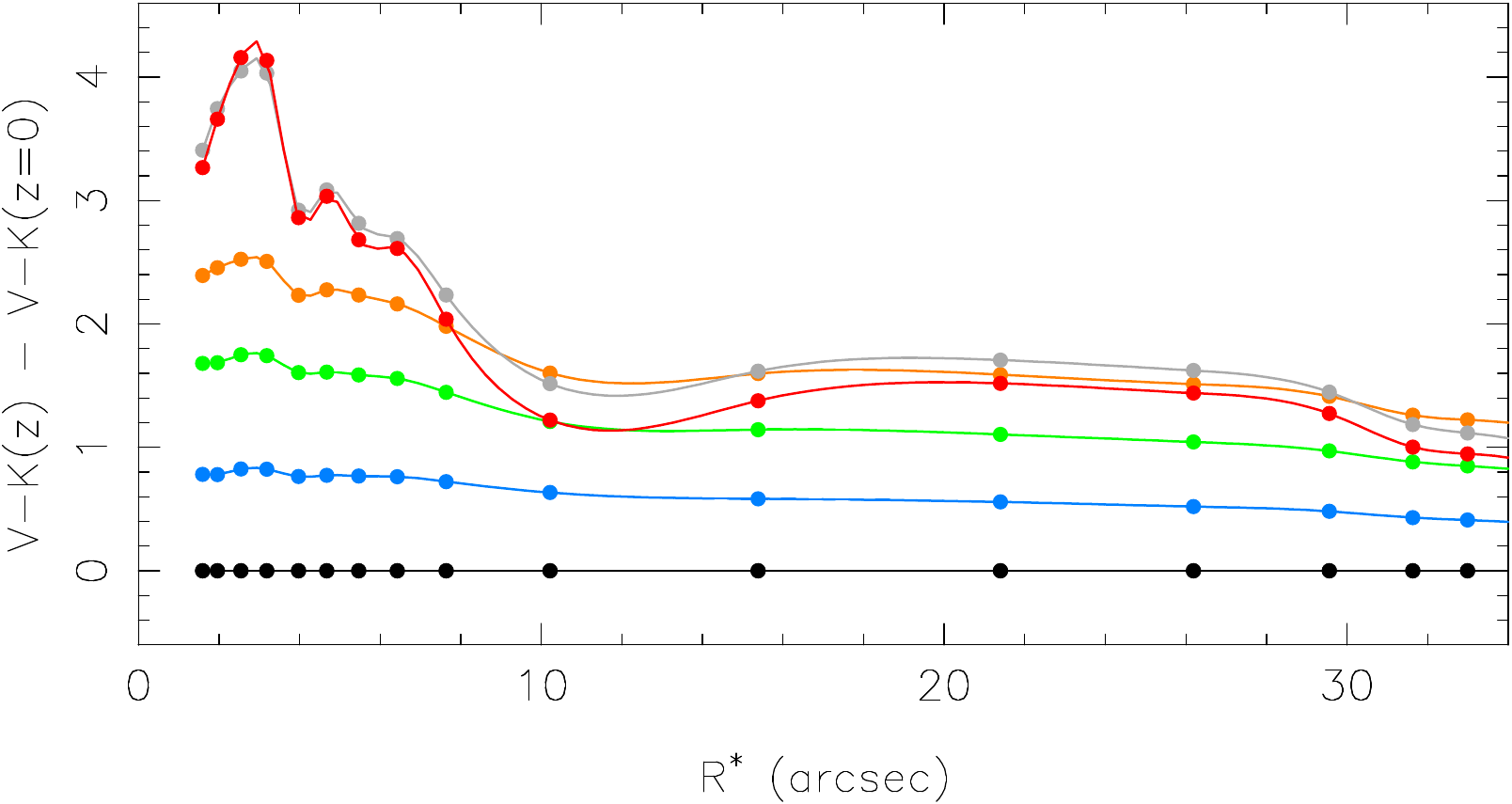}}
\put(128,32){\includegraphics[clip, width=5.6cm]{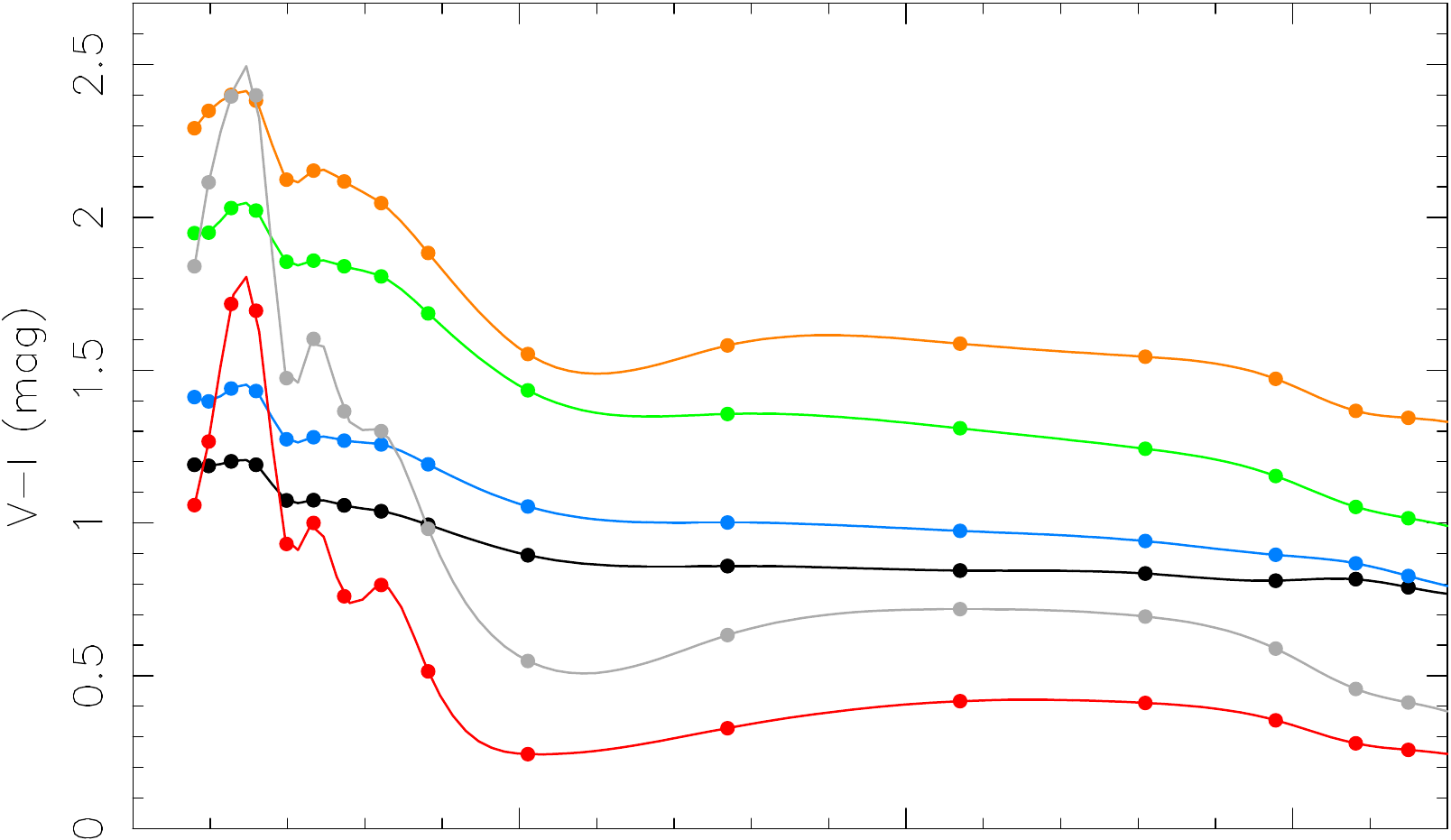}}
\put(128,0){\includegraphics[clip, width=5.62cm]{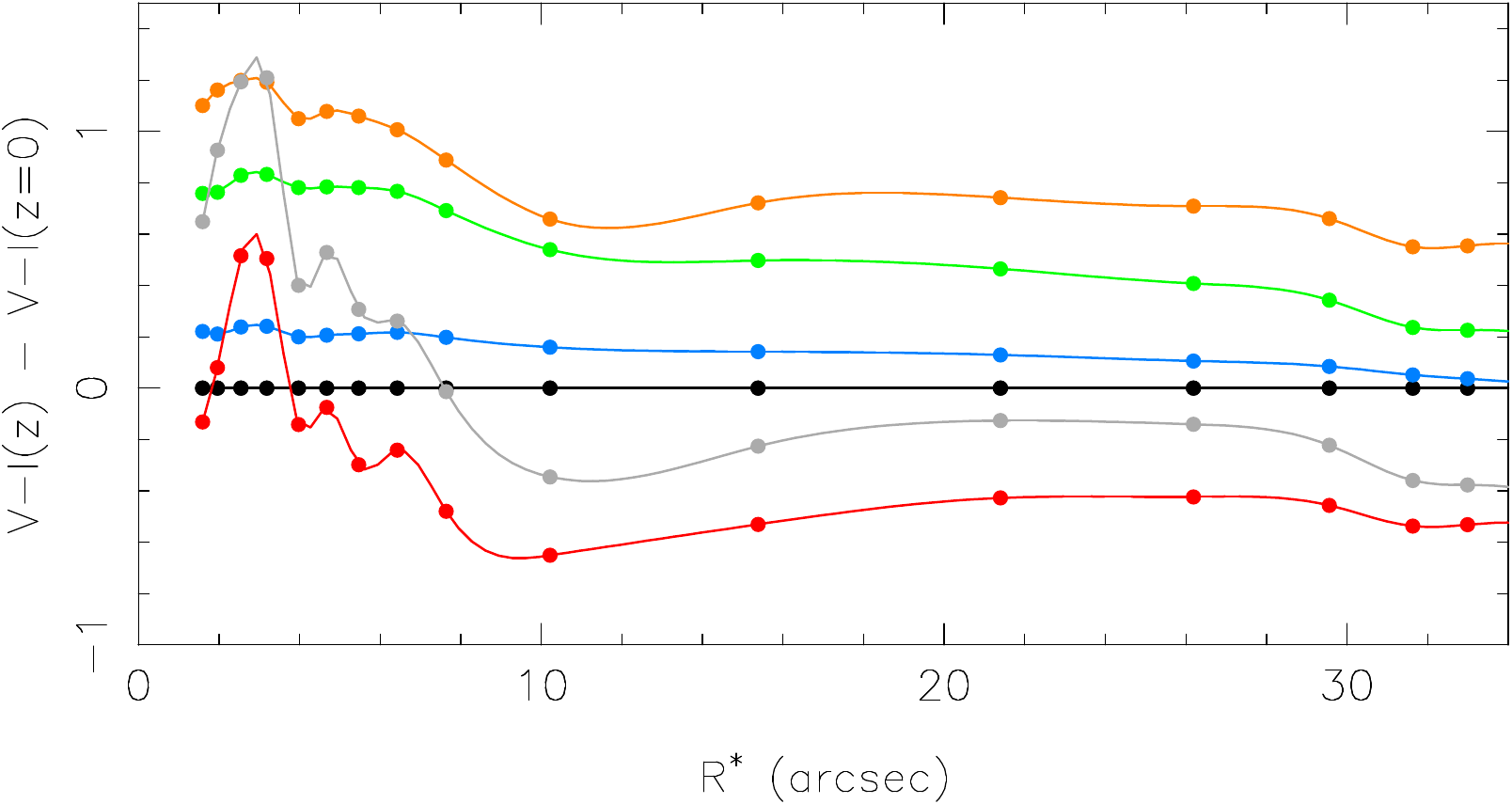}}
\PutLabel{30}{25.5}{\mlx \textcolor{black}{B-H: ObsF-restframe}}
\PutLabel{95.5}{25.5}{\mlx \textcolor{black}{V-K: ObsF-restframe}}
\PutLabel{161}{25.5}{\mlx \textcolor{black}{V-I: ObsF-restframe}}

\end{picture}
\caption{$B$-$H$, $V$-$K,$ and $V$-$I$ color profiles of \object{NGC 0309} at \zet=0, 0.3, 0.5, 0.8, 1.5, and 2.
The lower panels display the difference (in mag) relative to the true (rest-frame) color profiles of the galaxy. It can be seen that \cmod\
leads to a strong amplification of negative color gradients within the passively evolving bulge.}
\label{n309-zvscol}
\end{figure*}

It is also interesting to examine how \cmod\ impacts color gradients (and the bulge-to-disk color contrast \dcol).
Consistently with predictions from parametric SFHs (Figs.~\ref{sSED} and \ref{eSED}), radial color profiles from
simulated IFS data (Fig.~\ref{n309-zvscol}) present a complex dependence on redshift, with, for example, the ObsF $B$--$H$, $V$--$K,$ and $V$--$I$
color of the bulge appearing at \zet=1.5 more than 2, 3, and 0.5 mag redder than it is, respectively.
The $V$--$I$ and $B$--$H$ ($V$--$K$) color of the disk becomes reddest at \zet$\sim$0.8 ($\sim$1.5), with a reversion of this reddening trend at \zet$\geq$1.5
toward a $V$--$I$ color that is even bluer than the rest-frame value, in agreement with Fig.~\ref{sSED}\irem{e}.

Finally, an insight from Fig.~\ref{n309-zvscol} is that \cmod\ strongly amplifies negative color gradients within the bulge. For instance, the $V$--$K$ gradient
increases from --0.06 mag/\arcsec\ at \zet=0 to --0.12 mag/\arcsec\ at \zet=0.5, reaching --0.43 mag/\arcsec\ (a factor of $\sim$7 of its rest-frame value) at \zet=2.

% :::::::::::::::::::::::::::::::::::::::::::::::::::::::::::::::::::::::::::::::::::::::::::::::::::::::::::::::::::
\section{Discussion \label{dis}}
% :::::::::::::::::::::::::::::::::::::::::::::::::::::::::::::::::::::::::::::::::::::::::::::::::::::::::::::::::::
The \cmod\  effect is a simple consequence of the fact that galaxies inhabiting the expanding Universe
consist of spatially distinct structural components (e.g., bulge, bar, and disk) that differ from one another in their assembly history and SED.
Contrary to cosmological surface brightness dimming, which is achromatic and thus equally impacts all galaxy structural components regardless of their rest-frame SED,
\cmod\ is per se chromatic and thus inherently dependent on the time-evolving 2D SED of a galaxy, its redshift, and the photometric filters it is observed with.
Likewise, because gravitational lensing is in itself achromatic, it is not immune to \cmod.

The primary implication of \cmod\ is the differential amplification (suppression) of star-forming (passively evolving or dust enriched) spatial zones in a higher-\zet\ ($>$0.1) galaxy, thereby a drastic change in its morphology and 2D color patterns. It has a complex dependence on the relative SFH between the core (bulge) and the envelope (disk)
of a spiral galaxy and, hence, on its radius- and time-dependent rest-frame SED.
The neglect of this effect (i.e., of spatially resolved \kc\ corrections) can lead to a broad range of incoherent and even diametrically conflicting
conclusions on the nature and evolutionary status of a high-\zet\ galaxy.
A precise prediction on (and correction for) \cmod\ on the spatially resolved SED of a galaxy is a nontrivial task that requires knowledge
of its 2D SFH, and chemical, dust extinction and nebular emission patterns. The dynamical assembly history of the bulge via SF clump migration (cf. Sect.~\ref{dis:lim})
adds an extra level of complexity into this already demanding task.

It is important to reiterate that the effects of \cmod\ significantly depend on the spectral window in which a higher-\zet\ galaxy is observed:
simulations based on single-age SEDs (Fig.~\ref{sSED}) imply a virtual disappearance of the bulge for \zet$\sim$1 in the optical, whereas
NIR observations imply a modest dimming ($H$) or even enhancement ($K$).
EvCon simulations (Fig.~\ref{eSED}), on the other hand, indicate a monotonously increasing brightening of the bulge by up to $\sim$2 $K$ mag in the NIR,
whereas in the optical the situation is more complex and partly reverse, as the bulge first shows a strong dimming for \zet$<$1.5 that is followed by a brightening at a higher \zet.
Regardless of which type of simulation is used, the photometric effect of \cmod\ is far from negligible: at $\zet\sim$1 it changes
the $V$-band rest-frame bulge-to-disk surface brightness contrast \dmBD\ by $>$3 mag (a factor of $\sim$16) and reduces the bulge-to-disk luminosity ratio by 1.3 (0.8) dex
for single-age (EvCon) simulations (cf. Figs.~\ref{sSED}\irem{c} and \ref{eSED}\irem{c}).
Likewise, it has a strong effect on colors and the bulge-to-disk color contrast \dcol\ (up to 3 mag in $V$--$I$), even leading to the literal reversion
of radial color gradients at various redshifts.

The effect of of bandpass shift and wavelength stretching on ObsF colors has been examined in numerous studies that used SED
templates to estimate \kc\ corrections for higher-\zet\ galaxies
\citep[e.g.,][]{Humason56,OkeSandage68,Pence76,Coleman80,Kinney96,Mannucci01,Hogg02,Blanton03,Bernardi03,Bicker04,Barden08,Roche09,Chil10,Beare14,Salim18,Temple21,Fielder22},
some of those additionally providing evolutionary (\ec) correction terms \citep[e.g.,][]{Poggianti97,BFvA05} or
taking nebular emission into account \citep[e.g.,][]{BFvA05,BR07}. The luminosity fraction of the latter is known to be substantial in high-sSFR galaxies near and far,
especially in low-metallicity starburst galaxies \citep[e.g.,][]{Izotov97,P98,Guseva04,SdB09,Reines10,Atek11,Atek22,NO14,Maseda14,Micheva17,Kehrig18,PM21,Brinchmann22,Schaerer22,Matthee22}, and to vary with filter and redshift \citep[e.g.,][]{Zackrisson08,Inoue11,PO12,Inayoshi22}.

The substantial and extremely valuable work carried out so far in the context of \kc\ corrections refers to integral galaxy SEDs.
However, the integral SED of a distant composite galaxy is the luminosity-weighted sum of its evolutionary distinct constituent stellar populations and is therefore already ``preprocessed'' by \cmod\ prior to its acquisition (cf. Sect.~\ref{dis:sSFR}).
Thus, even if \kc\ corrections are applied to magnitudes determined from this integral SED, this
does not a posteriori rectify the \cmod\ bias but merely propagates it into the further analysis.

Remarkably, although the multicomponent nature of galaxies near and far is the starting assumption in all galaxy decomposition studies, the neglect
of spatially resolved \kc\ corrections in these studies is equivalent to the assumption that galaxies are invariably mono-component systems with a spatially invariant SED.
This seemingly subtle logical inconsistency is the root of complex, grave and highly interlinked biases in morphological, structural and evolutionary studies of higher-\zet\ galaxies, as pointed out in \citet{PO12}.
The dimension of the problem can better be realized when considering the entire chain of steps involved in decomposition studies of morphologically and physically heterogeneous galaxy samples covering a relevant range in redshift (0$\leq z \leq$3 in some cases), from the initial color-based selection and classification of galaxies, all the way to the extraction of their structural properties and color gradients. It is also important to reflect on how the latter quantities, including their byproducts (e.g., stellar masses \mstar\ estimated from luminosities via a \ml\ ratio) permeate into galaxy scaling relations and in our understanding of topics, such as the bulge versus SMBH relation or SF quenching.
In fact, it is difficult to name topics in contemporary extragalactic research for which \cmod\ is both directly and indirectly irrelevant,
and does not call for a critical inspection or even revision of previous work.

The goal of the following discussion is not to cover this immensely broad subject but merely to draw attention to the manifold implications of \cmod\
and invite the community to take part in a synergistic effort to better understand it and develop recipes that will hopefully allow its effects to be predicted and rectified.
This endeavor is especially timely in view of the prospects that JWST and \textit{Euclid} open to study early galaxy evolution and at the same time because of
the risk that the neglect of \cmod\ bears for jeopardizing these studies.

In the interest of brevity we limit the discussion to a few basic aspects, this mostly from the viewpoint of optical studies and simulations based on single-age SEDs.
For the same reason, spiral galaxies are reduced into just two components (bulge and disk), ignoring the bar and other structural
characteristics, such as thick, truncated or nuclear disks, circumnuclear SF rings \citep[e.g., ][among others]{BP94,ButaCombes96,Combes00,Courteau96,PT06,Men08,Men17,Comeron10,Comeron12,Laurikainen18,dLC19,Bittner20,Gadotti20,Barsanti21}, a possible central down-bending of the disk \citep{Breda20b}, extended stellar halos \citep{MD10,TB13,Duc15,Merritt16,Buitrago17,Trujillo21,Chamba22}, 
radial age and metallicity gradients, and biases that inside-out SF quenching entails for surface photometry studies \citep{P22}.
Likewise, high-\zet\ starburst galaxies are excluded from the discussion, except for a brief note on their color patterns.

Next, we discuss in Sect.~\ref{dis:size} some of the limitations of the evolutionary synthesis approach taken for modeling the SED of our synthetic two-component galaxy.
Since the advantages and limitations of evolutionary synthesis have been reviewed in numerous articles that followed the seminal works by
\citet{Tinsley68} and \citet{SMT78}, we here only briefly comment on one specific question, namely whether predictions based on standard SFH parameterizations
adequately describe the SED and colors of a stellar population that does not primarily grow through in situ SF but through superposition of ex situ formed stellar populations.
This is because of the importance of this dynamical process for the formation and growth of bulges (Sect.~\ref{sub:nebSED}).
The following sections provide a concise discussion of biases expected from the neglect of \cmod\ in determinations of the stellar mass growth of bulges (Sect.~\ref{dis:bulge-growth}), the size evolution (Sect.~\ref{dis:size}), stellar mass and specific SFR (Sect.~\ref{dis:sSFR}) and fundamental scaling relations (Sect.~\ref{dis:TF}) for higher-\zet\ galaxies.
Following a discussion of the effects of \cmod\ on morphology and color patterns (Sect.~\ref{dis:morph} and \ref{dis:CG}), it is pointed out
that the detection of optically/NIR dark galaxies at high \zet\ is a natural expectation from \cmod\ (Sect.~\ref{dis:DGs}).
% :::::::::::::::::::::::::::::::::::::::::::::::::::::::::::::::::::::::::::::::::::::::::::::::::::::::::::::::::::
\subsection{Migration-driven bulge growth versus parametric SFHs\label{dis:lim}} %\brem{dis:lim}
% :::::::::::::::::::::::::::::::::::::::::::::::::::::::::::::::::::::::::::::::::::::::::::::::::::::::::::::::::::
This study has adopted two main approaches, the first one based on evolutionary synthesis models for a bulge+disk galaxy model (Sect.~\ref{SED}) and the second on
spectral modeling of spatially resolved IFS data (Sect.~\ref{simIFS}). Both come with simplifying assumptions and limitations.

Simulations based on evolutionary synthesis, regardless of whether they use single-age SEDs (Sect.~\ref{sub:stSED}) or compute the evolution of bulge and disk in an EvCon manner (Sect.~\ref{sub:eSED}) are tied to simple parameterizations for the SFH. Following common practice \citep[e.g.,][]{Sandage86,GRV87,Poggianti97,Gavazzi02}, we assumed constant SF for the disk and an exponentially decreasing SFR with an e-folding time $\tau$ of 0.5 and 1 Gyr for the bulge.
Supplementary delayed-exponential SFR scenarios are discussed in Sect.~\ref{ap:pSFH}.

Whereas the assumption of constant SFR (or a prolonged SF with a $\tau > 3$ Gyr) is probably justifiable for intermediate-to-late stages of the evolution of the disk
\citep[e.g.,][and references therein]{Costantin21,BP23}, it is less clear whether exponential SFR models adequately approximate the spectrophotometric evolution
of the bulge. This is for three reasons at least. First, the presence of a variety of morphological substructures in the central part of spiral galaxies
document a more complex evolutionary history than what smooth parametric SFHs imply.
Secondly, simple $\tau$ models lack a prescription for SF quenching, a phenomenon that seems to commence in massive spirals once their bulge has reached a mass \mstar\ $\sim 10^{10}$ \msun\ and a mean stellar surface density \sstar\ $\sim$ $10^9$ \msun\,kpc$^{-2}$ \citep[e.g.,][]{Strateva01,Fang13,Mosleh17,BP18,WE19,Suess22}.
The timescales for SF quenching are poorly constrained, although indirect lines of evidence suggest a comparatively slow inside-out cessation of SF within $\sim$2 Gyr
\citep[e.g.,][\tref{BP18}]{Tacchella15}, which might reflect the superposition of different, non-mutually exclusive physical mechanisms \citep[cf. discussion in][]{Breda20a}.
Important in the context of this study is also that bulges follow a downsizing trend \citep[e.g.,][\tref{BP18}]{Ganda07} with spectral synthesis studies of IFS data
implying that those hosted by spiral galaxies with a present-day log(\mstar/\msun)$\ga$10.7 have assembled the bulk of their stellar mass earlier than 9 Gyr ago 
(\zet$\sim$1.3), whereas those in less massive systems are experiencing a more prolonged build-up \citep[e.g.,][cf. Fig.~\ref{fig:SFHs}]{BP23}

Thirdly, and perhaps most importantly, the current accepted view of bulge growth being primarily driven through inward migration and coalescence of massive SF clumps from the disk (Sect.~\ref{sub:nebSED}) implies for the bulge a substantially different spectrophotometric evolution than 
that expected from in situ SF according to standard SFH parameterizations.
Since these SF clumps are unlikely to survive SF feedback over the migration timescale \tmig\ of 0.6-1 Gyr, it is reasonable to assume that they arrive at the center of protogalaxies in a post-starburst state. In this case, the stellar mass filtering effect discussed in \citet[][see also \tref{PO12} and \tref{BP18}]{P02} becomes important: stellar ensembles forming in the outer disk according to a normal initial mass function (IMF) reach the galaxy's center depopulated from stars with a main-sequence lifetime shorter than \tmig\ (or, equivalently, a mass $>M_{\rm up}$).
With other words stellar migration imposes, in the specific context, a minimum age for stellar populations in the bulge of distant protogalaxies
and effectively mimics in situ SF with a top-truncated IMF.
From the observational point of view, the principle effect is a seemingly premature aging of the bulge because its SED remains reddish notwithstanding its continued high stellar mass growth rate. Using colors (whose time evolution typically refers to a standard, fully populated IMF up to $M_{\rm up}\sim 100$ \msun) to age-date the bulge then leads to an overestimation of age by a factor of $\ga$2 for $t\ga 0.7$\tmig.

\begin{figure}
\begin{picture}(220,84)
\put(0,42){\includegraphics[clip, height=4cm]{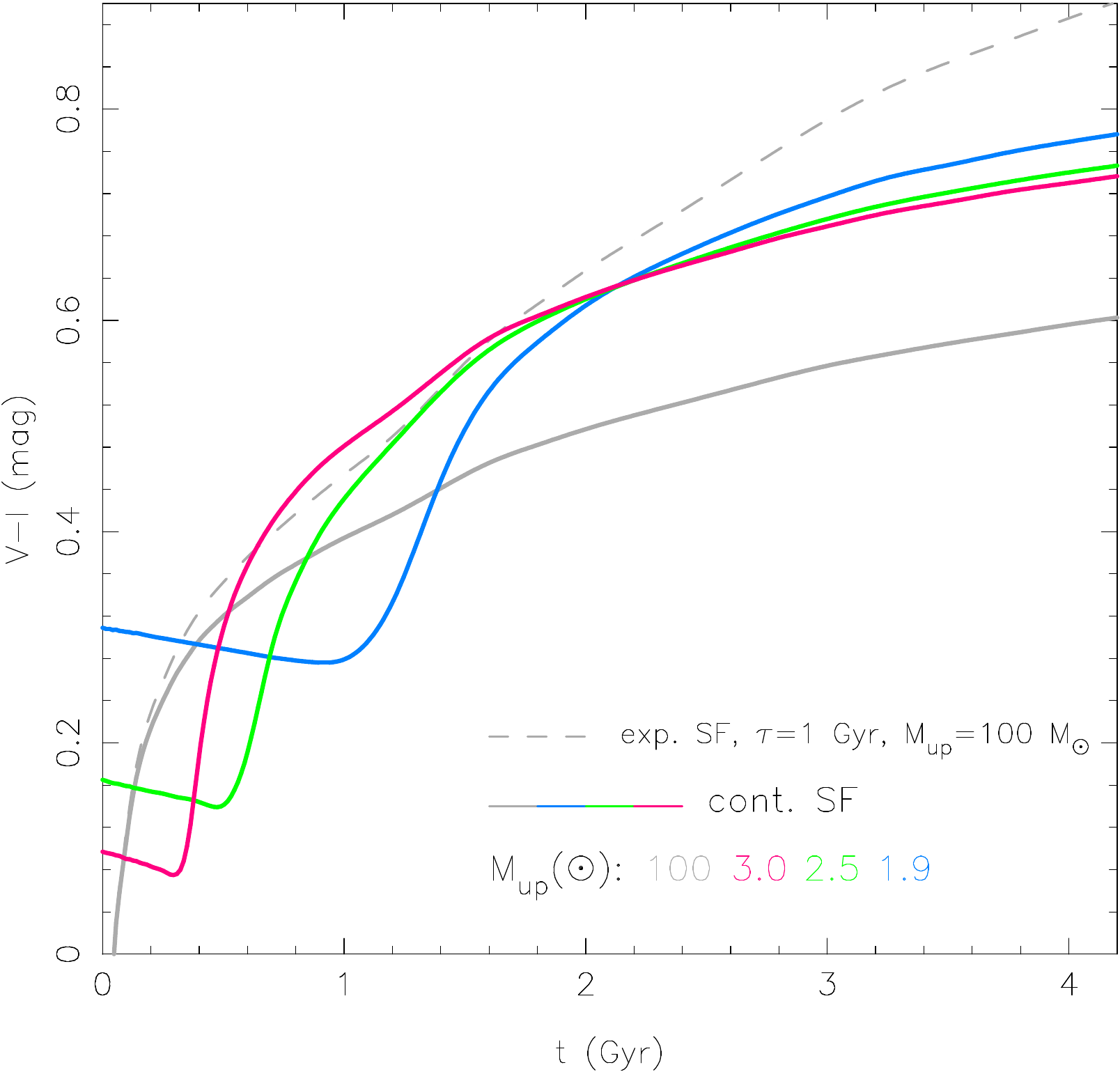}}
\put(48,42){\includegraphics[clip, height=4cm]{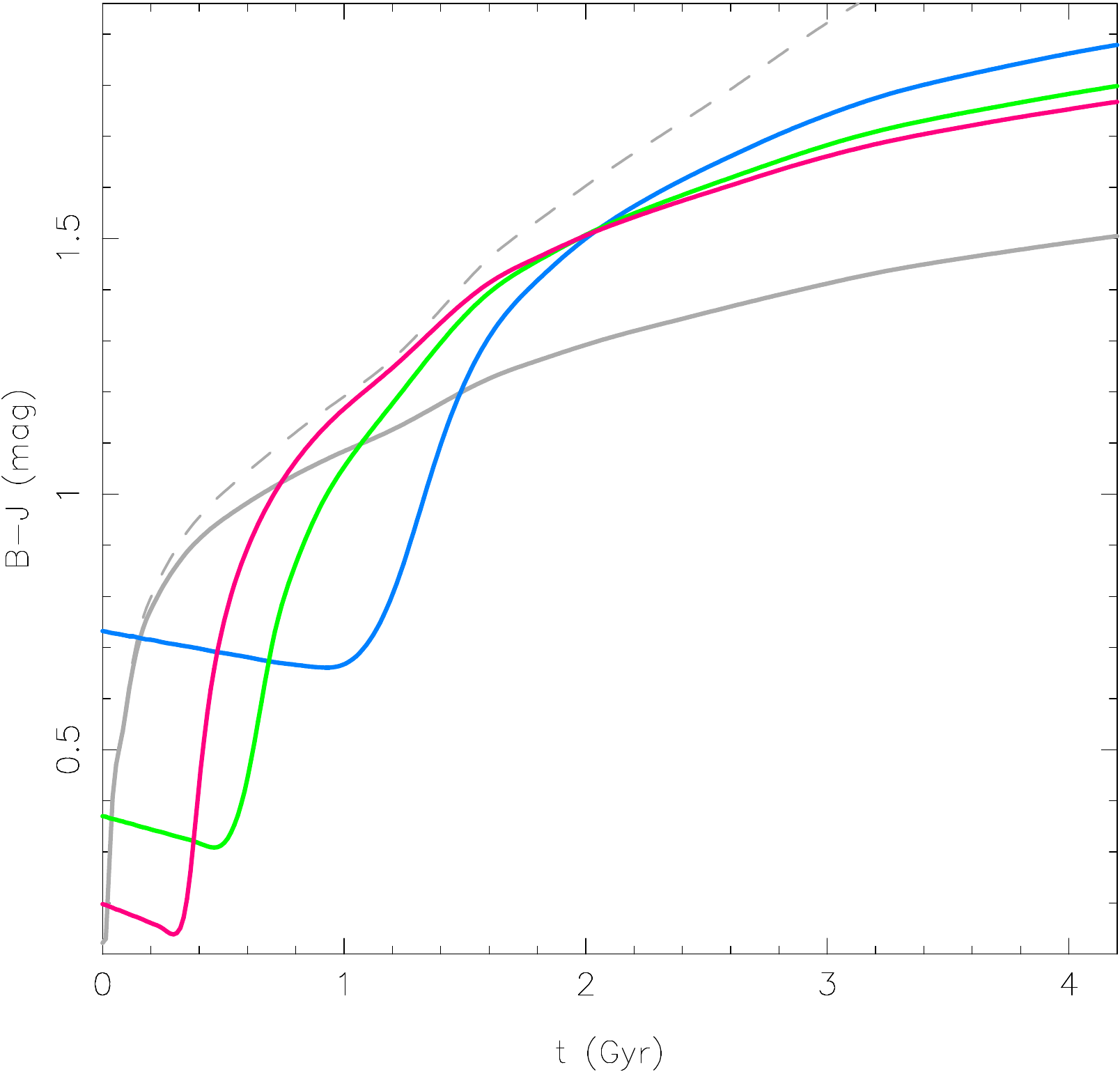}}
\put(0,0){\includegraphics[clip, height=4cm]{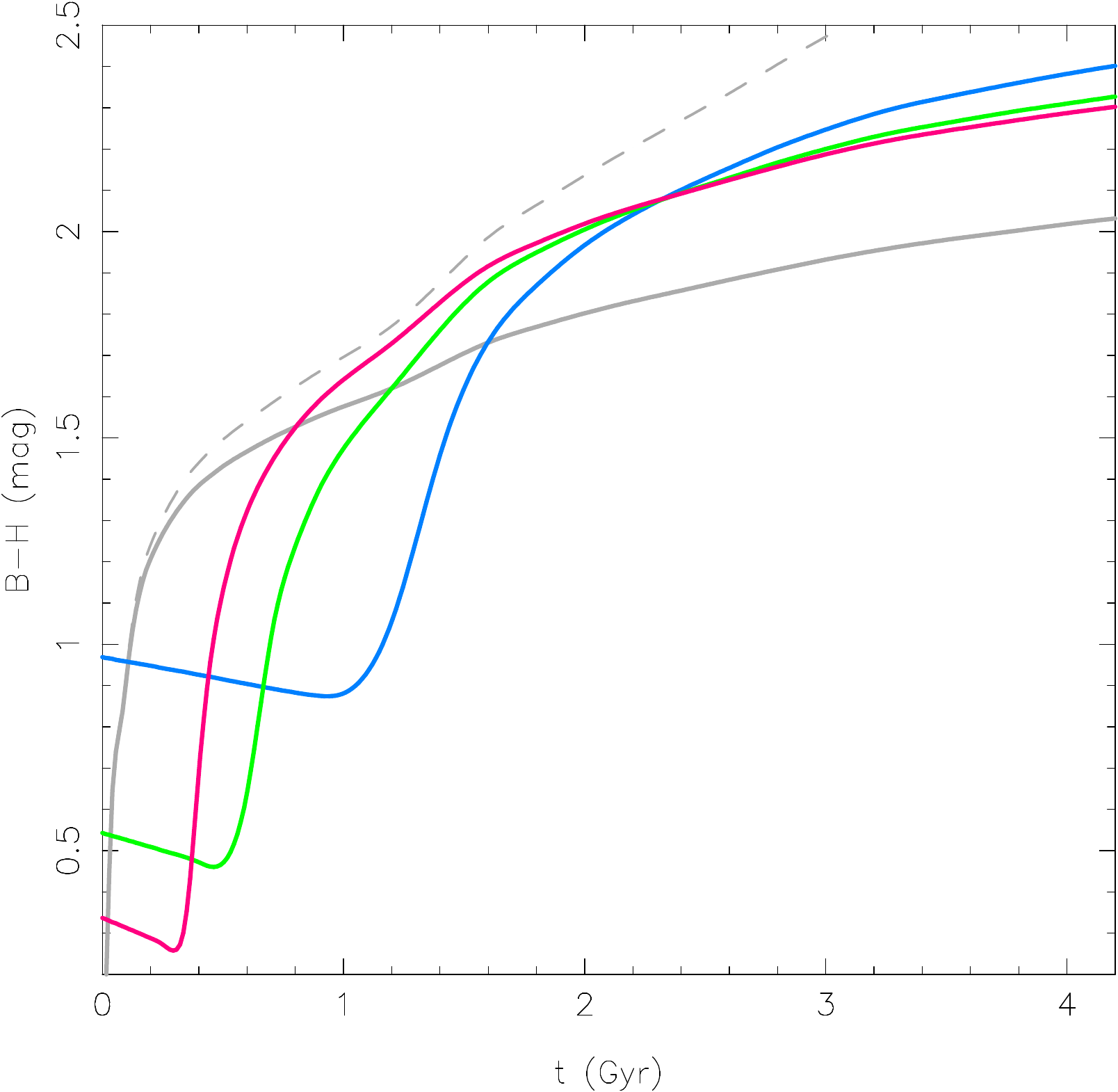}}
\put(48,0){\includegraphics[clip, height=4cm]{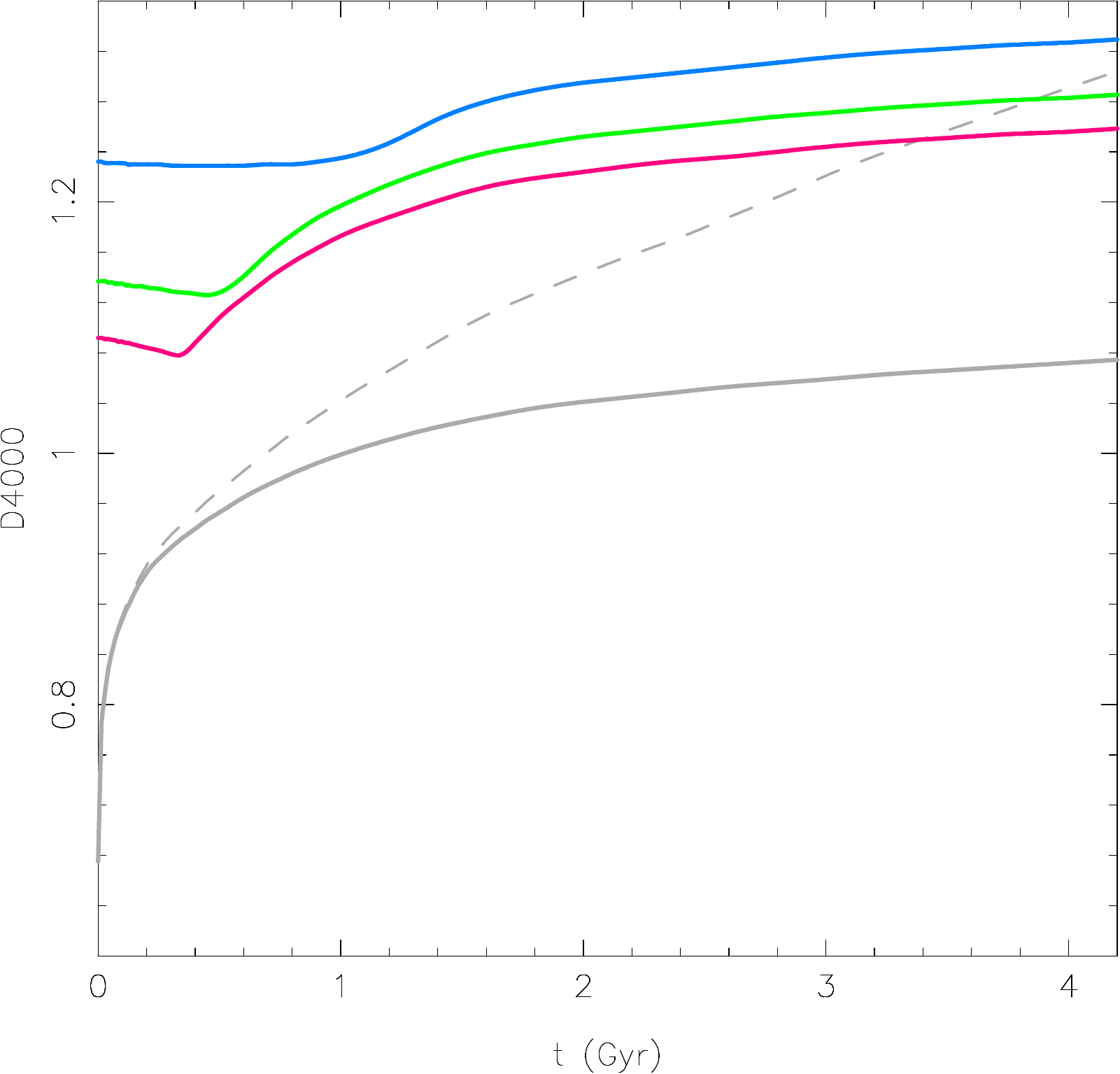}}
\PutLabel{6}{78}{\imlabel{a)}}
\PutLabel{86}{48}{\imlabel{b)}}
\PutLabel{37}{6}{\imlabel{c)}}
\PutLabel{86}{6}{\imlabel{d)}}
\end{picture}
\caption{Time evolution of the $V$--$I$, $B$--$J,$ and $B$--$H$ color and the D$_{4000}$ index for a stellar population of \zsun/5 forming at a constant SFR, according to a Salpeter IMF
that is truncated above $M_{\rm up}$=3, 2.5 and 1.9 \msun\ (red, green and blue, respectively), corresponding to a migration timescale \tmig\ of SF clumps from the disk of, respectively,
0.64, 1 and 2 Gyr. Predictions for a fully populated (0.1 -- 100 \msun) Salpeter IMF for a constant and exponentially decreasing ($\tau$1) SFR are shown with the gray solid and dashed curve, respectively.}
\label{migration}
\end{figure}

This is illustrated in Fig.~\ref{migration} where the time evolution of colors and the D$_{4000}$ index is shown for a Salpeter IMF that is truncated above
M$_{\rm up}$=3, 2.5 and 1.9 \msun, corresponding to a \tmig\ of 0.64, 1 and 2 Gyr. Solid and dashed gray curves correspond to predictions for an IMF between 0.1 -- 100 \msun\
for, respectively, the contSF and $\tau$1 model. It can be seen that a top-truncated IMF, if applying to in situ SF, initially keeps colors blue (for $t\la$\tmig/2), as young stars
with a mass $<M_{\rm up}$ that do not evolve into the red supergiant phase continue pilling up. For older ages, however, this IMF implies a steep color increase to
far redder values than those expected from a standard IMF.
For instance, it can be read off panel~\brem{a} that a $V$--$I$ color of 0.6 mag, which for top-truncated IMFs develops at 1.8 Gyr, corresponds to an age of 4 Gyr when a fully populated Salpeter
IMF is assumed.
Apparent is also (panel~\brem{d}) that the high D$_{4000}$ index for models involving a top-truncated IMF is only reproducible with the $\tau$1 model.
Its value of 1.2 at t=1 Gyr (\zet=5.7) for a $M_{\rm up}$=2.5~\msun\ translates for a standard IMF to an age of $\sim$3 Gyr (\zet=2.2), leaving no other option
for reconciling the high D$_{4000}$ index with the \zet-based youth of the galaxy than the interpretation of a dusty, monolithically formed bulge.

The cautionary notes on Fig.~\ref{migration} are meant to remind the reader that standard parametric SFHs are geared toward reproducing integral SEDs of local galaxies and can lead to substantial systematic errors when adopted in interpreting spectrophotometric observables on sub-galactic scales, especially when stellar mass growth is not solely driven by in situ SF: the SED of the bulge during the dominant phase of its assembly via SF-clump migration
could substantially differ from that assumed in Sect.~\ref{SED}. 
Nevertheless, since the effect of migration-driven bulge growth is the aging of the SED in a manner similar to
that predicted from the $\tau$1 model, we argue that the principle conclusions from Sect.~\ref{SED} remain valid.

Finally, a corollary from the considerations above is that reconstruction of the morphology that a present-day galaxy used to have several gigayears ago is barely achievable.
Even though the computation of spatially resolved evolutionary correction (\ec) terms is possible both with the evolutionary synthesis approach and from spectral modeling of IFS data (Sect.~\ref{simIFS}), the fact of migration-driven bulge growth hinters a backward simulation of galaxy morphology. Other dynamical phenomena (e.g., minor mergers and bar-driven stellar migration) pose further hurdles.
% :::::::::::::::::::::::::::::::::::::::::::::::::::::::::::::::::::::::::::::::::::::::::::::
\subsection{Bulge growth \label{dis:bulge-growth}}
% :::::::::::::::::::::::::::::::::::::::::::::::::::::::::::::::::::::::::::::::::::::::::::::
Determinations of mass, size and color of galaxy bulges across \zet\ are fundamental to key astrophysical subjects, such as the coevolution of SMBHs with stellar spheroids, the regulatory role of active galactic nuclei (AGN) on galaxy evolution, and the physical drivers and associated timescales for SF quenching.
Whereas possible via measurements of $\sigma_{\star}$, determinations of the bulge mass \mbulge\ are mostly done photometrically, through conversion of the net luminosity of the bulge, as inferred from bulge-disk decomposition, into mass via a \ml\ ratio. The latter is usually estimated from fitting the integral SED of a galaxy, which, as pointed out in Sect.~\ref{dis:sSFR}, is the (\cmod-biased) luminosity-weighted sum of the bulge and the disk.

\begin{figure}
\begin{picture}(86,37)
\put(0,0){\includegraphics[trim=0 70 0 0, clip, width=8.6cm]{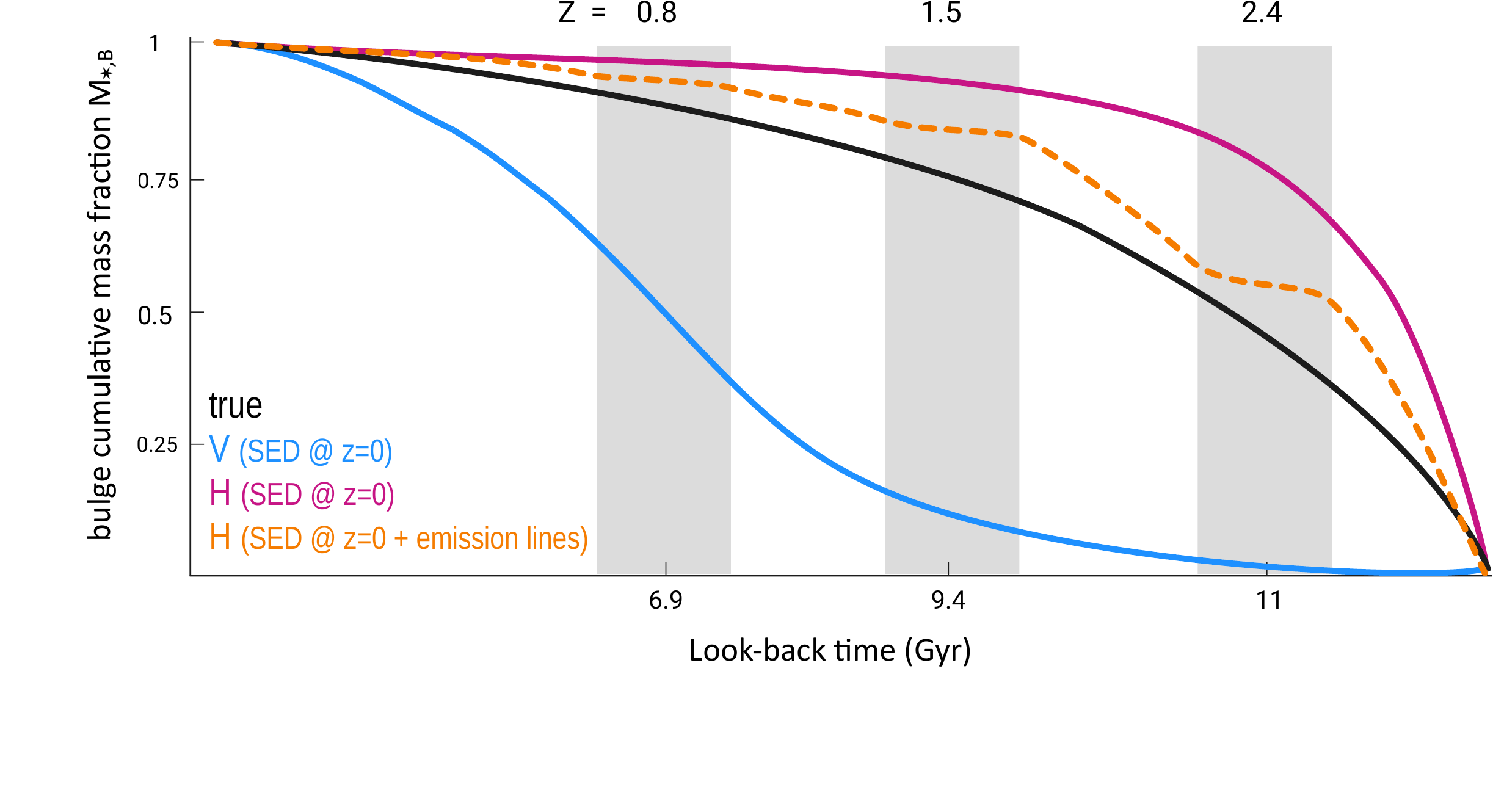}}
\end{picture}
\caption{Schematic view of the expected effect of \cmod\ on photometric determinations of the stellar mass of the bulge \mbulge.
The black curve shows the cumulative fraction of \mbulge\ as a function of lookback time, whereas solid blue and magenta curves the values expected
from image decomposition in the $V$ and $H$ band for purely stellar galaxy models (cf. Fig.~\ref{sSED}).
The orange-dashed curve qualitatively delineates the additional effect that massive SF clumps forming out of VDIs could have on \mbulge\ determinations: in redshift intervals where strong emission lines fall within the $H$ filter transmission curve (e.g., at \zet=1.5 in the case of \ha+[N{\sc ii}]; gray-shaded areas) the surface brightness in the disk periphery is differentially enhanced relative to that in the SF-quenched galaxy center. One consequence of this is the over-subtraction of the disk from the galaxy, leading to the under-estimation of the net luminosity (and \mbulge) of the bulge (\tref{P22}). This would translate to a flattening of the \mbulge\ growth curve in discrete \zet\ intervals, possibly prompting the conclusion that bulge formation proceeds in major waves (cf. Sect.~\ref{sub:nebSED} and Fig.~\ref{ew2}).}
\label{fig:bulge-growth}
\end{figure}

An insight from this study is that image decomposition in the optical fakes a delayed emergence and rapid growth of bulges since only $0.5\la z \la 1.5$ (depending on filter). This is sketched in Fig.~\ref{fig:bulge-growth}, where the cumulative fraction of \mbulge\ is shown by the black curve.
Photometry in the $V$ band will essentially miss the bulge for \zet$>$1, leading to the erroneous conclusion that galaxies initially evolve as disks and bulges come later.
The interesting coincidence that \zet$\sim$1 marks the epoch since the gradual decline in the cosmic SFR density \citep[][see also \tref{Madau \& Dickinson 2014} for a review]{Madau96,Lilly96}, could be taken as evidence for a causal link between the emergence of bulges (and the SMBHs inhabiting them) and the suppression of cosmic SF through negative AGN feedback.
Also, the occurrence of accretion-powered nuclear activity in seemingly bulgeless high-\zet\ disks could be seen as a tension with $\Lambda$ cold dark matter and prompt the
interpretation of super-Eddington accretion, or the need for tuning negative SF- and AGN feedback such as for simulations to suppress early bulge growth.
Finally, another consideration is that the seemingly delayed appearance of the bulge and the steeply rising \BD\ ratio since \zet$\sim$1 could be confused
with disk fading, leading to the conjecture that the latter process is responsible for a transition of spiral galaxies into S0s.

Largely opposite conclusions would emerge from bulge-disk decomposition studies in the NIR: because of their copious UV and optical emission during their early evolution, bulges are well detectable in $K$ at \zet$\geq$3, as their \dmu\ is $\sim$0~mag (and up to $<$--2 mag for EvCon models).
Depending on the \ml-ratio assumed, the conclusion one could draw is that massive mature bulges have been common 
$\sim$2 Gyr after the Big Bang, and their \mbulge\ has not significantly increased since then.

Another key point is that observational determinations of bulge growth via bulge-disk decomposition in the NIR can significantly depend on the importance of 
massive SF clump formation in the disk. The total stellar mass forming in these off-center clumps and the duration of this phenomenon, could vary from galaxy to galaxy.
If the latter process is irrelevant, the disk thus essentially follows a smooth SFH (contSF or $\tau$5) throughout its evolution, then image decomposition in the NIR allows a precise determination of rest-frame photometric properties of galaxies (e.g., \reff\ and \BD\ ratio) back to \zet=3, as explicitly demonstrated in Fig.~\ref{fig:BDdec1} through EvCon simulations. This is because, for these SFHs, nebular emission in the disk fades to an \ewha\ $\sim 100$ $\AA$ for $>$2~Gyr, becoming photometrically negligible at \zet$<$3.
The situation is somehow different if the SF clump migration scenario is valid, as is current consensus. Reminding the reader of the determinations in 
Sect.~\ref{sub:nebSED}, strong optical emission lines moving into the ObsF NIR will elevate the $H$ and $K$ surface brightness of the disk of higher-\zet\ spiral galaxies
by more than one mag, while having virtually no effect in their quiescent bulge. This differential enhancement of the disk will then lead to its over-subtraction 
from the bulge and thereby to the underestimation of the \BT\ ratio, superficially suggesting that bulge growth stagnates in certain redshift intervals 
(e.g., \zet=0.8, 1.5 and 2.4, corresponding to a lookback time of 6.9, 9.4 and 11 Gyr; cf. Fig.~\ref{zmag} and \tref{P22}). A conceivable conclusion from this might 
be that galaxy bulges assemble in major waves (Fig.~\ref{fig:bulge-growth}).
With other words, the advantage of NIR photometry as the method of choice for the study of relatively SF-quiescent galaxies near and far can turn into a disadvantage (unless \cmod\ is accounted for) when studying distant galaxies with strong and spatially inhomogeneous specific SFR.

Clearly, the effect of \cmod\ on determinations of bulge mass is worth a closer investigation, inter alia because of its relevance to the bulge-SMBH co-evolution, and the possible non-universality of the \mbh/\mbulge\ ratio \citep[e.g.,][\tref{BP18}]{HeckmanBest14}.

% :::::::::::::::::::::::::::::::::::::::::::::::::::::::::::::::::::::::::::::::::::::::::::::::::::::::::::::::::::::::::::::::::
\subsection{\cmod-induced non-homology and the size evolution of galaxies \label{dis:size}} %\brem{dis:size}
% :::::::::::::::::::::::::::::::::::::::::::::::::::::::::::::::::::::::::::::::::::::::::::::::::::::::::::::::::::::::::::::::::
A central implication of \cmod\ is the violation of homology as a result of the differential surface brightness dimming (amplification) of old, passively evolving (young star-forming) zones in a galaxy: the form of an SBP changes across \zet, with this non-homology effect being strongest for galaxies with substantial intrinsic radial gradients
in their rest-frame SED (i.e., in age, metallicity, dust content and nebular contamination) and weaker for more homogeneous systems, for example, early-type galaxies (ETGs).
In fact, only a perfectly homogeneous galaxy (i.e, the purely theoretical case of a mono-component system with a spatially uniform SED) has a \zet-invariant morphology and SBP.

This also means that any primary and secondary quantity inferrable from fitting an SBP and characterizing its shape (S\'ersic exponent $\eta$, \BD\ ratio, light concentration indices)
depends on \zet, even under the idealized (and obviously incorrect) assumption that a galaxy is non-evolving spectrophotometrically over time.
The bulge-to-disk surface brightness contrast \dmBD\ offers a metrics for this non-homology bias. As it can be read off Figs.~\ref{sSED} and \ref{eSED}, it reaches $>$4~mag in the optical and $>$1~mag in the NIR in the case of simulations based on single-age SEDs, whereas it varies in the range of $\ga \pm$2~mag for EvCon simulations.
Therefore, in both cases it is a substantial, if not the dominant, factor shaping the SBP and morphology of a higher-\zet\ galaxy.
The picture is further complicated by nebular emission that selectively enhances the surface brightness (in particular in the NIR) of star-forming zones in various broad \zet\ intervals (Sect.~\ref{gSED4Gyr}) as well as by attenuation by the intervening intergalactic medium. The latter somehow counter-acts \cmod\ since it impacts differentially the red bulge and blue disk, thereby reducing the bulge-to-disk color contrast \dcol. Even though the effect is on average small \citep[$\sim$0.2 $B$ mag at $z=1$ and about 0.5 mag at higher redshift; e.g.,][]{IK04}, this selective light attenuation might be higher along certain sightlines intersecting, for instance, metal- and dust-enriched foreground galaxy clusters, starburst-driven winds, or chemically pre-enriched gas halos associated with AGN host galaxies.

Quite importantly, a consequence of the \cmod-induced non-homology is that both the effective radius \reff\ and the Petrosian radius are not robust measures of size
for higher-\zet\ galaxies (Fig.~\ref{fig:BDdec1}\&\ref{fig:BDdec2}): although immune to cosmological surface brightness dimming, these commonly used quantities are robust against \cmod\ only for idealized mono-component galaxies. This entails a risk for spurious correlations (or the non-detection of existing ones) in size evolution studies. A related concern stems from the fact that \reff-normalized radial profiles are commonly used in studies of, for example, age and metallicity gradients and their evolution across cosmic time.

Several previous studies have demonstrated the capability of state-of-the-art 2D photometry tools to produce robust image decomposition solutions
\citep[e.g.,][and references therein]{Nedkova21,Bretonniere22,Merlin22}. Although reassuring from the technical point of view, this does not reverse the metamorphosis
of higher-\zet\ galaxies as a consequence of \cmod: if a galaxy is recorded at the ObsF as bulgeless, then no photometry code can recover
the existence of its prominent bulge in the rest frame.
While several primary and secondary implications of \cmod\ on surface photometry could be mentioned, we provide in the interest of brevity only one example: an appealing
approach taken recently in surface photometry is to force S\'ersic model parameters to smoothly vary along $\lambda$ according to an empirical functional form \citep{H13,Vika14}.
The latter function, which is calibrated for the local Universe, could, however, lead to systematic biases if imposed on high-\zet\ galaxies. The reason for this can be read off panel \brem{d} of Fig.~\ref{fig:BDdec1}: for single-age simulations, the local ratio of the S\'ersic exponent $\eta$ in $V$:$I$:$K$ is 1:1.1:1.2 whereas that of the same galaxy at \zet=1 is 1:1.35:1.9.

Summarizing, structural properties derived through bulge-disk decomposition of higher-\zet\ galaxies cannot be directly compared with those in local galaxies, in the quest of addressing, for example, the evolution versus \zet\ of the \BD\ ratio or the mass-size and mass-$\sigma_{\star}$ relation.
\cmod\ makes the transformation of galaxy structural properties from the ObsF into the rest frame both indispensable and critically important, and this task is nontrivial.

% ::::::::::::::::::::::::::::::::::::::::::::::::::::::::::::::::::::::::::::::::::::::::::::::::::::::::::::::::::::::::::
\subsection{Underestimation of the stellar mass through SED fitting and overestimation of the specific SFR \label{dis:sSFR}}
% ::::::::::::::::::::::::::::::::::::::::::::::::::::::::::::::::::::::::::::::::::::::::::::::::::::::::::::::::::::::::::
One consequence of \cmod\ is that the integrated optical SED of a star-forming high-\zet\ galaxy appears bluer than it is.
In the case of a spiral galaxy, for example, the SED is the sum of the fractional luminosity contribution of the SF-quenched bulge and that of the star-forming disk, which
implies a relation between its UV-through-NIR spectral slope and its \BD\ ratio.
The drastic optical dimming of the bulge as \zet\ increases, in conjunction with the comparatively mild effect of \cmod\ on the disk results at \zet$\sim$1 in a blue (disk-dominated) SED
(Sect.~\ref{sub:stSED}), which in turn forces any photometric SED fitting code to favor a much too low stellar \ml\ ratio, and thereby underestimate the total stellar mass \mstar.
This bias is obviously strongest for high-bulge-to-disk spirals that experience bulge assembly and ensuing SF quenching early on, whereas modest for those where bulge formation is prolonged, leading to a comparatively low bulge-to-disk age contrast.

This immediately entails a connection with galaxy downsizing.
Since these two spiral galaxy groups correspond to the antipodal ends of a galaxy mass sequence, with a present-day stellar mass
log(\mstar/\msun)$\ga$10.7 and $<$10, respectively, (e.g., \tref{BP18})
it follows that \cmod\ impacts \mstar\ determinations differentially, with the massive spiral galaxies affected most.
The implications of this are arguably complex and manyfold and permeate into various topics -- from galaxy downsizing and the slope and scatter of the SF main sequence \citep{Bri04,Noeske07}, all the way to our understanding of the assembly history of bulges (Sect.~\ref{dis:bulge-growth}).
One may invoke as a single example the mass-metallicity relation \citep{Lequeux79,Tremonti04}: since gas-phase metallicity determinations from monochromatic emission lines are unaffected by \cmod, contrary to determinations of \mstar, one may expect massive bulge-dominated galaxies to follow a steeper slope at high-\zet\ than in the local Universe\footnote{In reality, the situation is likely more complex, not only because of aperture effects but also because the most massive galaxies experience inside-out SF quenching first, and thus gas-phase metallicities inferred from their integral spectra may mostly reflect the lower-metallicity star-forming disk rather than the nearly emission-line-free bulge (cf. discussion in \tref{P22}).}, as sketched in Fig.~\ref{fig:MZTF}.

Recently, \citet{PA22} have applied various photometric SED fitting codes to simulated $griz$ images of local galaxies at $0.1 < z < 2$ that were computed based on spectral modeling of IFS data, in a similar manner as in Sect.~\ref{simIFS}. These authors found that these codes underestimate \mstar\ by up to 0.3 dex at \zet$\sim$1 and attributed this bias to the fact that the SED samples increasingly bluer wavelengths, thus giving weight to the younger stellar population that can outshine the older (higher-\ml\ ratio) one.
This empirically inferred bias is actually a natural outcome from \cmod, as shown in Sect.~\ref{SED} and Sect.~\ref{BD}, and primarily originates from the gradual disappearance
of the bulge from the optical SED.
In a present-day spiral galaxy with log(\mstar/\msun)$>$11 the bulge contributes $\geq 1/3$ of the total \mstar\
(e.g., \tref{BP18}, their Fig.~6c). This is arguably a lower limit for the bulge mass fraction in the galaxy progenitor at \zet$\sim$1, 
both because of the mass returned into
the ISM in the course of stellar evolution, and the continued growth (by a factor of $\sim$2) of the disk at a nearly constant SFR \citep[e.g.,][and references therein]{BP23}.
Therefore, a systematic underestimation through optical SED fitting of the stellar mass (by a factor of $\geq$2 at \zet$\sim$1) for seemingly bulgeless galaxies is consistent with the overall observational evidence.

Evidently, what applies to low-resolution photometric SEDs also applies to higher-resolution integral spectra: the selective suppression (amplification) of old (young) stellar populations poses a significant challenge to spectral synthesis studies of higher-\zet\ galaxies in general.
Strategies for overcoming this \mstar\ underestimation bias are crucially needed for a meaningful exploitation of spectroscopic surveys targeting the cosmic noon \citep[e.g., MOONS;][]{Cirasuolo20}.

Finally, the fact that, unlike \mstar\ estimates, SFRs from hydrogen Balmer-line luminosities are unaffected by \cmod\ entails a systematic overestimation
of the specific SFR out to \zet$\sim$1 (the redshift at which the bulge fades to invisibility in the optical).
This bias is further aggravated by a factor of $\ga$2 when SFRs in high-\zet\ metal-poor galaxies are estimated from their \ha\ luminosity assuming
solar metallicity for their ionizing stellar component, as is common practice (cf. Sect.~\ref{ap:nebSED}).
% :::::::::::::::::::::::::::::::::::::::::::::::::::::::::::::::::::::::::::::::::::::::::::::::::::::::::::::::::::::::::::::::::::::::::::::::::::::::
\subsection{Discrepancies between photometric and kinematical \mstar\ estimates, and the slope of the Tully-Fisher relation \label{dis:TF}} %\brem{dis:TF}
% :::::::::::::::::::::::::::::::::::::::::::::::::::::::::::::::::::::::::::::::::::::::::::::::::::::::::::::::::::::::::::::::::::::::::::::::::::::::
Since dynamical mass estimates via gas rotation curves or $\sigma_{\star}$ are robust against \cmod, contrary to \mstar\ estimates from SED fitting or an assumed \ml\ ratio,
discrepancies between dynamical and photometric masses may be expected to increase with \zet.
Specifically, a conceivable implication of \cmod\ is that massive spiral galaxies
at \zet$\sim$1 appear centrally dark-matter-dominated when studied in the optical but appear to have a lower dark matter (DM) content, or even be baryon dominated, when studied in the NIR.

This is particularly true for genuinely bulge-dominated systems that have experienced SF quenching in their centers early on
while continuing forming stars in their disk periphery. In this case, the filtering-out of the bulge in optical wavelengths causes
the dynamical-to-baryonic mass ratio to be overestimated within the bulge radius ($\sim$\reff), thereby implying a predominance of DM.
The situation is opposite in the NIR where, depending on the SFH assumed (cf. Figs.~\ref{sSED} and \ref{eSED}; cf. also Fig.~\ref{n309-zvsSBP})
the bulge can get by $>$1~$K$ mag brighter at \zet=1, possibly reconciling photometric with dynamical mass estimates, or even prompting the need for a bottom-light IMF.
Clearly, further work is needed for quantitatively assessing this issue.

\begin{figure}
\begin{picture}(86,32)
\put(42,0){\includegraphics[trim=0 75 0 0, clip, height=3cm]{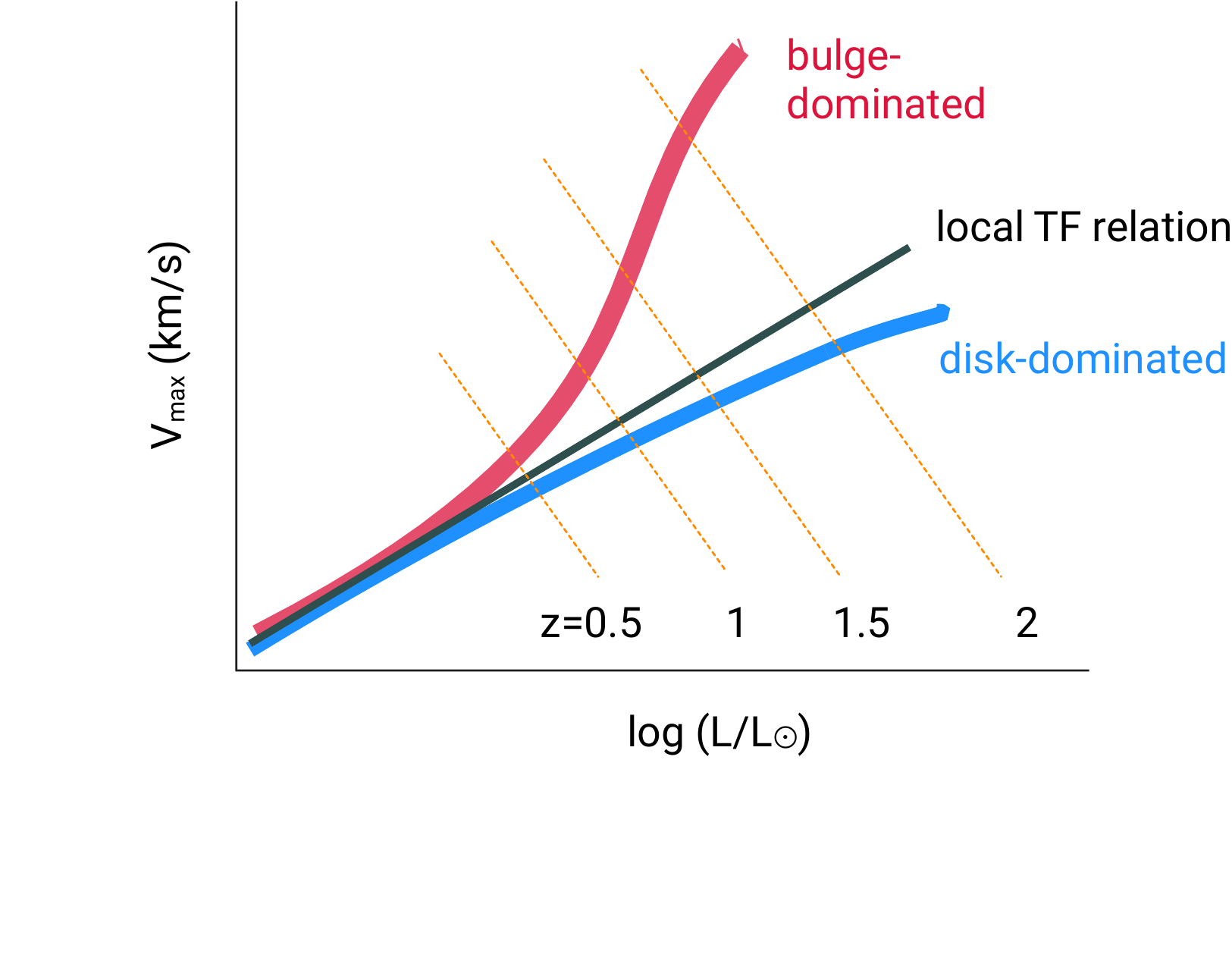}}
\put(0,0){\includegraphics[trim=0 75 0 0, clip, height=3cm]{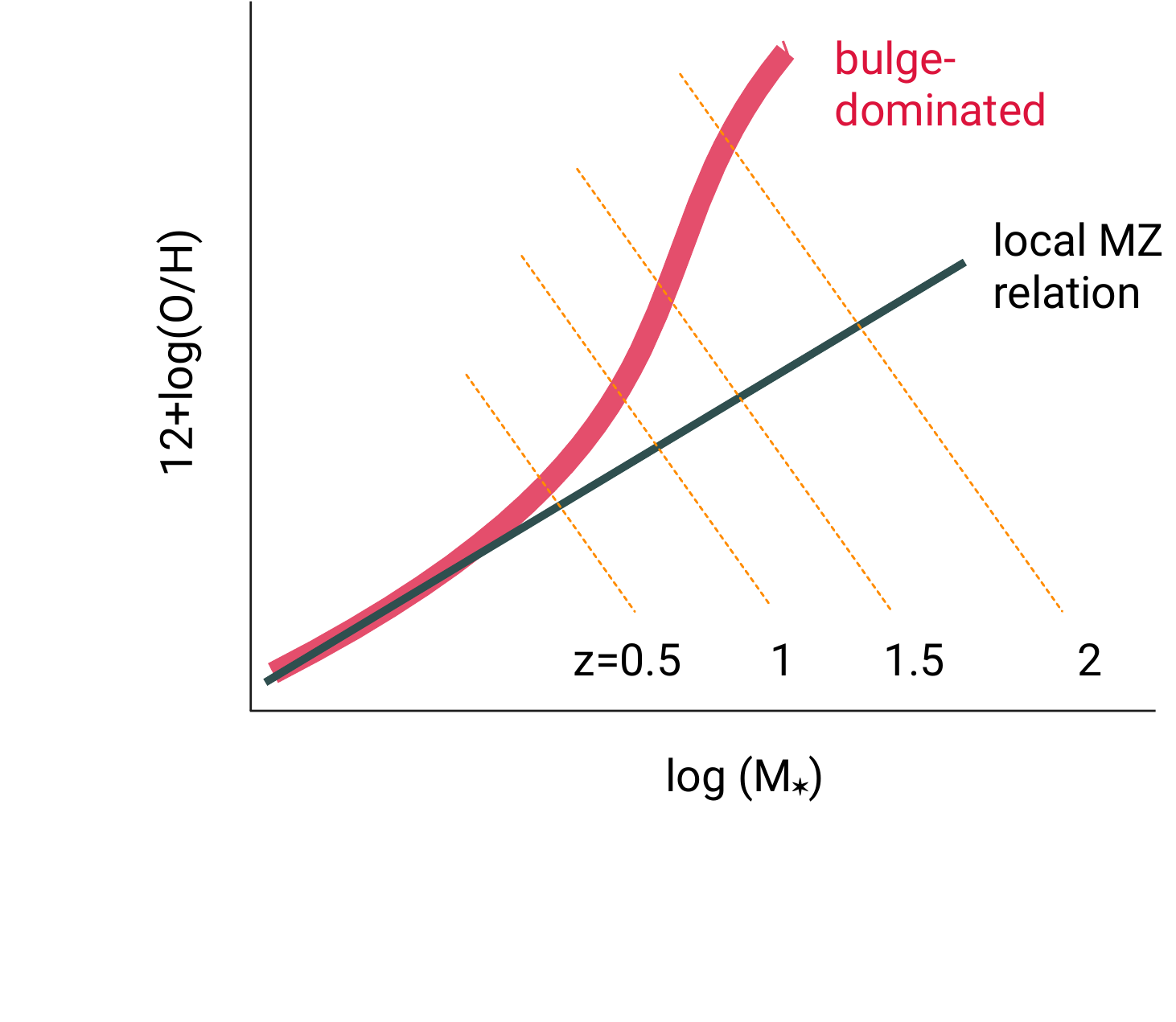}}
\end{picture}
\caption{Schematic view on the effect that underestimation of stellar mass \mstar\ through optical SED fitting is expected to have on determinations
of the mass-metallicity (MZ) and TF relation (left and right panel, respectively) for centrally SF-quenched bulge-dominated galaxies at higher-\zet.}
\label{fig:MZTF}
\end{figure}

The impact of \cmod\ on galaxy scaling relations is another topic of considerable interest.
One example is the TF relation \citep[][]{TF77}, which was initially established in the $B$ band and whose slope and scatter at higher \zet\ continue
stimulating numerous investigations \citep[e.g.,][and references therein]{Ziegler02,Boehm04,Puech08,Zaritsky14}.
By analogy to our previous considerations, a bulge-dominated galaxy (red curve in the right-hand panel of Fig.~\ref{fig:MZTF}) will appear at \zet$\sim$1 under-luminous for its circular velocity and deviate from the local TF relation (black line), whereas a disk-dominated one (blue curve) will likely better comply with the local TF.
This, of course, only applies if the slope of the TF relation is \zet-invariant, which is an issue that can only be conclusively addressed
after correction of photometric data for \cmod.

% :::::::::::::::::::::::::::::::::::::::::::::::::::::::::::::::::::::::::::::::::::::::::::::::::
\subsection{\cmod\ and the metamorphosis of high-\zet\ galaxies \label{dis:morph}} %\brem{dis:morph}
% :::::::::::::::::::::::::::::::::::::::::::::::::::::::::::::::::::::::::::::::::::::::::::::::::
While it is observationally established that the morphology of galaxies changes with redshift \citep[e.g.,][see \tref{Conselice 2014} for a review]{Driver95,Abraham96,vdBergh01,Windhorst02,Papovich03,Conselice05,Elmegreen07,Whitaker11,Buitrago13,HC15,Straughn06}
and becomes progressively smoother from the rest-frame UV to the rest-frame optical and NIR 
\citep[what is referred to as ``morphological \kc\ correction''; e.g.,][]{Bohlin91,Windhorst02}, 
it is less clear how it can be objectively quantified and compared with that of local galaxies.
Various approaches extending the popular CAS scheme \citep{Abraham96,Bershady00,Conselice03} and other quantitative morphology indicators \citep[e.g., the
Gini/M$_{20}$ coefficients;][]{Lotz04} are currently being tested \citep[e.g.,][]{HC20}, which clearly is an important and timely endeavor in view of JWST, and the synergy of the \textit{Euclid} Space Telescope with the \textit{Vera C. Rubin} Telescope \citep{Rhodes17,Guy22}.

The previously overlooked \cmod\ effect opens a new route to the understanding of the connection between morphology and underlying stellar surface density in higher-\zet\ galaxies, as it underscores the link between 2D rest-frame SED and ObsF photometric properties (e.g., the \BD\  ratio, color gradient, effective radius and light concentration; cf. Figs.~\ref{n309-zvsVmag}, \ref{fig:BDdec1} and \ref{fig:BDdec2}).
This study therefore adds to previous work that has explored that critical link with different methods, most notably in \citet{Papovich03,Papovich05}
and \citet{Conselice08,Conselice11}.

\begin{figure}
\begin{picture}(86,49)
\put(1,22){\includegraphics[clip, height=2.6cm]{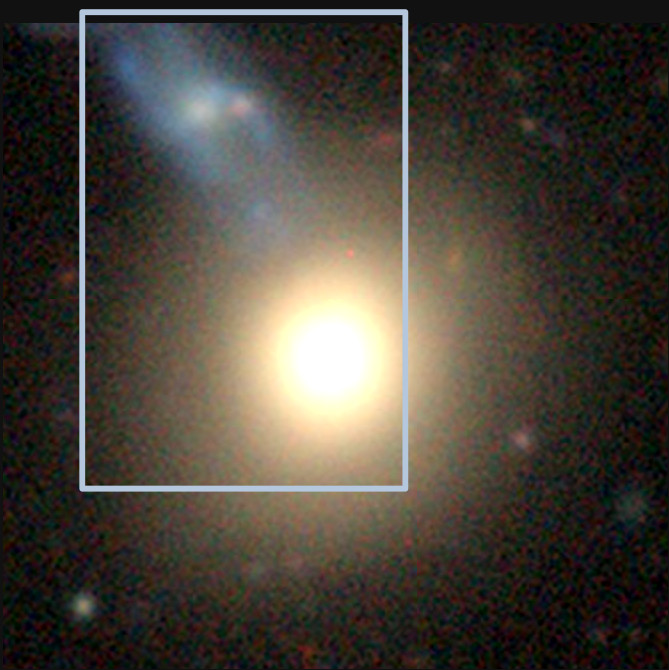}}
\put(30,22){\includegraphics[clip, height=2.6cm]{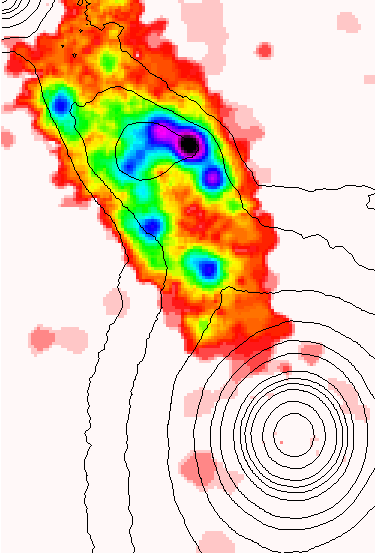}}
\put(51,22){\includegraphics[clip, height=2.6cm]{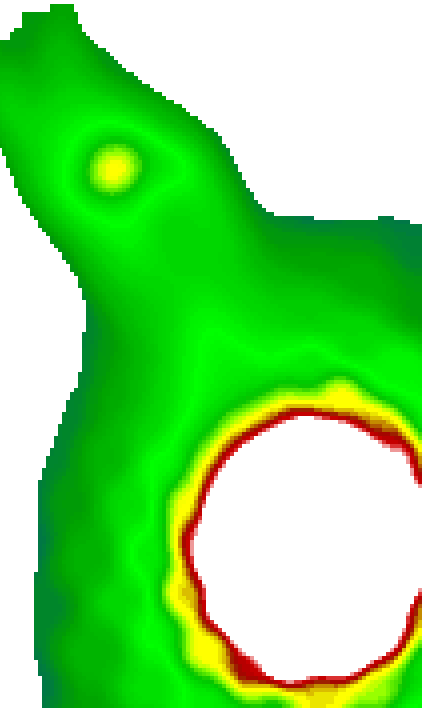}}
\put(72,22){\includegraphics[clip, height=2.6cm]{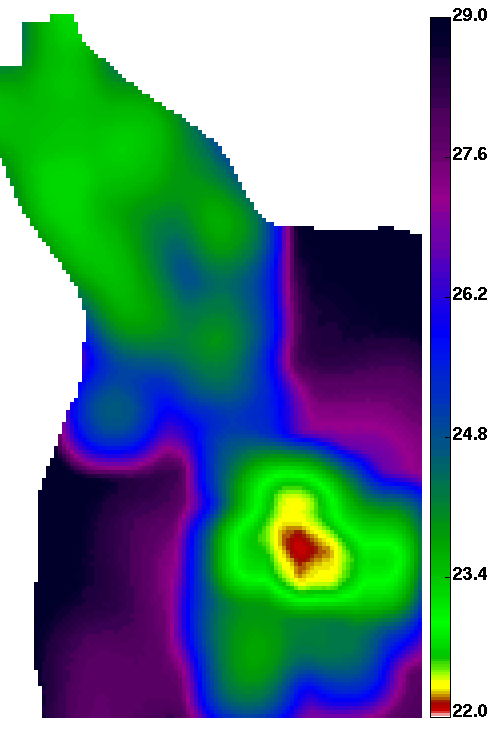}}
\put(5,0){\includegraphics[clip, height=2cm]{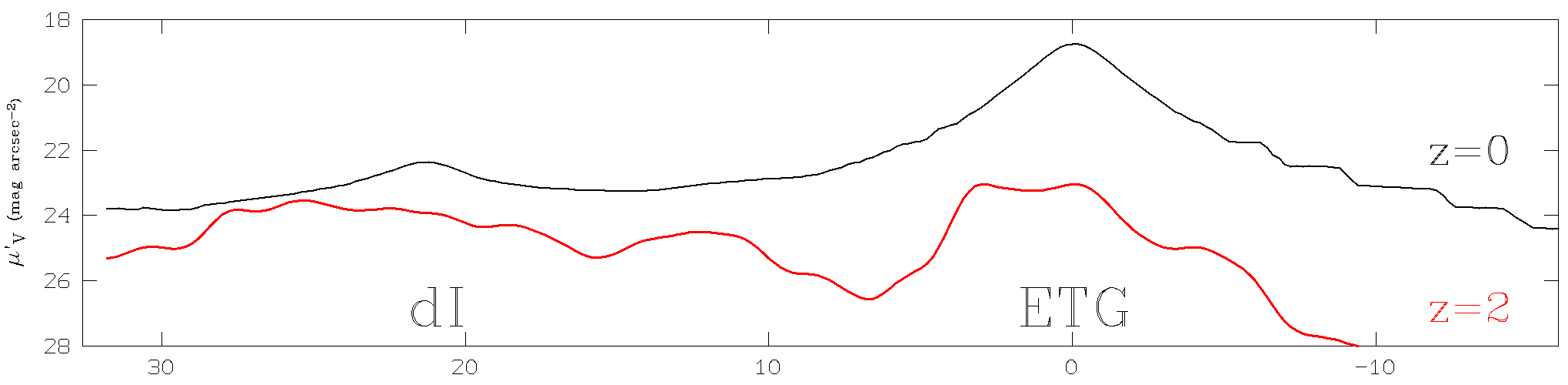}}
\PutLabel{24}{44}{\nlx \textcolor{white}{a}}
\PutLabel{45}{44}{\nlx \textcolor{black}{b}}
\PutLabel{64}{44}{\nlx \textcolor{black}{c}}
\PutLabel{84}{44}{\nlx \textcolor{black}{d}}
\PutLabel{11.6}{15}{\nlx \textcolor{black}{e}}
\end{picture}
\caption{Illustration of the impact of \cmod\ on the morphology of higher-\zet\ galaxies. \brem{a:} True-color composite image of \object{Mrk 1172} ($D$=172.2 Mpc) from the Dark Energy Camera Legacy Survey \citep[DECaLS;][]{Dey19}, combining exposures in the $g$, $r$ and $z$ filters.
\brem{b:} \ha\ map of the galaxy obtained with {\sc Porto3D} from archival MUSE data within the rectangular area in panel \brem{a}. The ETG is devoid of nebular emission, contrary to the faint dI NE of it \citep[cf.][]{Lassen21}, 
\brem{c\&d:} $V$\arcmin\ images simulated for \zet=0 and \zet=2 following the procedure in Sect.~\ref{simIFS} 
and displayed between 22 and 29 \sbb\ (the saturated white area at the center of the ETG reaches a $\mu_V\sim$18.6 \sbb), \brem{e:} $V$\arcmin\ surface brightness within a 3\arcsec-wide pseudo-slit crossing the center of the ETG and the dI at a position angle of 28$^{\circ}$. The reduced surface brightness of the ETG at \zet=2 (red curve) is by $\sim$4.4 mag fainter than its true (rest-frame) value, whereas the dI only suffers a modest dimming by 1.5 mag. Thus, although the rest-frame central $V$ surface brightness of the ETG is by 3.7 mag higher than that of the dI, both galaxies appear almost equally bright at \zet=2.}
\label{zima-Mrk1172}
\end{figure}

An illustration of the imprint of \cmod\ on the asymmetry index of higher-\zet\ galaxies is offered on the example of \object{Mrk 1172} (Fig.~\ref{zima-Mrk1172}). This system consists of an old ETG and a detached dwarf irregular (dI) where nebular emission reaches locally an \ewha$\sim$100 \AA, as processing of archival MUSE data  
with {\sc Porto3D} indicates \citep[see][for a detailed study of this system]{Lassen21}. Because the center of the dI is by $\sim$3.7 $V$ mag fainter than that of the ETG (22.4 and 18.7 \sbb, respectively) it has little influence on the global asymmetry index of \object{Mrk 1172}, which is low because of the dominant luminosity contribution of the circular-symmetric ETG. The situation is different when \object{Mrk 1172} is simulated at \zet=2, following the method in Sect.~\ref{simIFS}: despite its $\sim$2~dex lower mean stellar surface density, the dI now appears almost as bright as the ETG, which translates to a substantial increase in the asymmetry index of the ETG+dI system.

\begin{figure}
\begin{picture}(86,35)
\put(0,0){\includegraphics[clip, height=3.4cm]{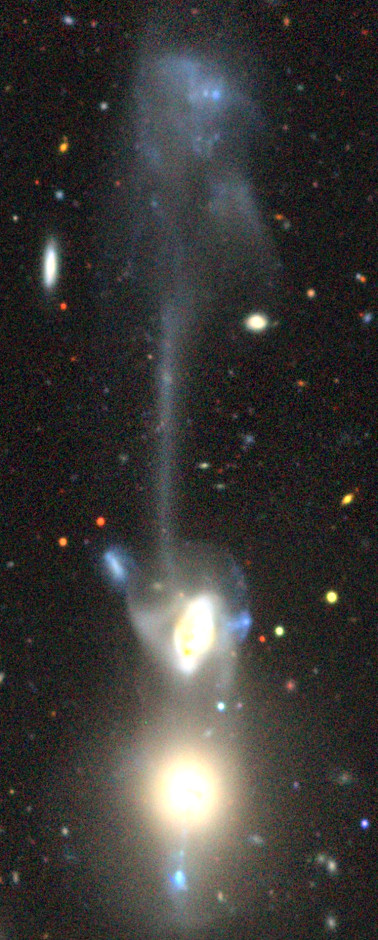}}
\put(16.5,0){\includegraphics[clip, height=3.4cm]{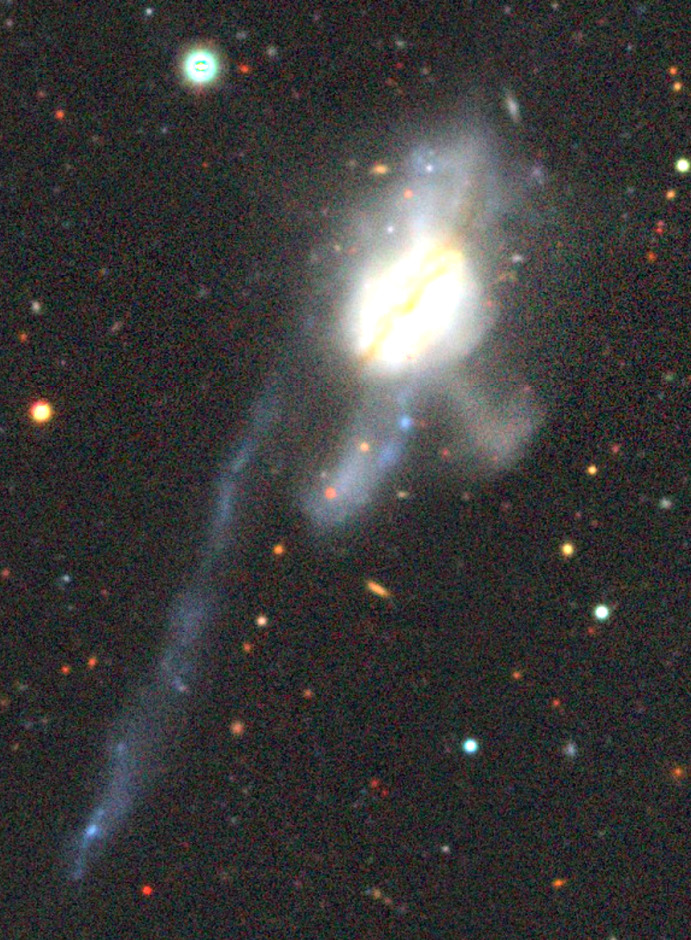}}
\put(44.5,0){\includegraphics[clip, height=3.4cm]{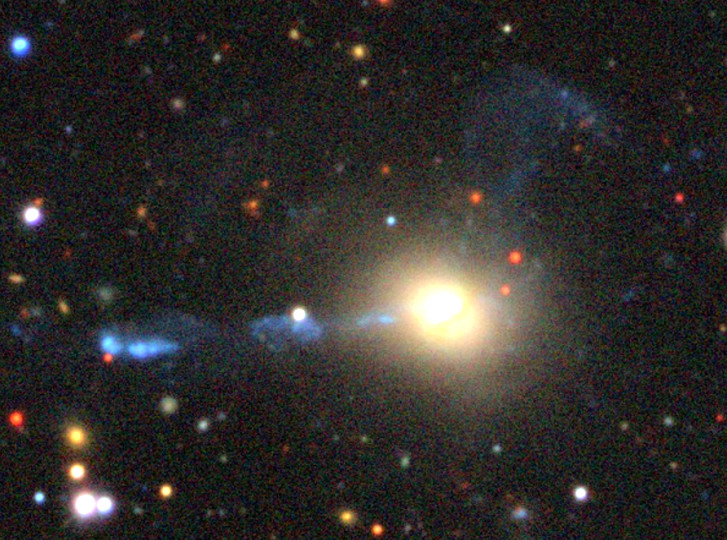}}
\end{picture}
\caption{True-color images from DECaLS of \object{Arp 105} (left), \object{UGC 2238} (middle) and \object{IC 1182} (right).}
\label{ig}
\end{figure}

Extending these considerations to other morphological configurations, such as interacting galaxy pairs consisting of a starburst and post-starburst source
\citep{Bernlohr93} or galaxy mergers, is straightforward. For example, tidal features in a high-\zet\ merger
would stand a good chance of being detected as long as they contain sites of ongoing SF whereas evade detection otherwise.
In a distant analog of the merger \object{Arp 105} \citep{DM94} one would probably detect the UV-bright tidally ejected
Magellanic irregular and southern BCD whereas missing the SF-devoid and UV-faint ETG (cf. Fig.~\ref{ig}).
Same applies to the blue tails of \object{UGC 2238} and \object{IC 1182}, and to several tidal dwarf galaxies \citep[e.g.,][]{Weilbacher02,Fensch16}.
Likewise, a ``jellyfish'' galaxy \citep[cf.][]{Poggianti19} such as \object{IC 3418} \citep{Hester10} will likely appear brighter in its UV-emitting ``tail''
than in its ram-pressure stripped ``head.''
The fact that \cmod\ selectively amplifies or suppresses sub-galactic features depending on their rest-frame SED obviously also influences the smoothness (or flocculency) of a galaxy in the CAS scheme \citep[see also][for a discussion of other sources of uncertainty in CAS determinations for high-\zet\ galaxies, primarily because of the non-detection of low-surface brightness features due to cosmological dimming]{Conselice11}, just like its fine-structure parameter \citep{Schweizer90}, tidal parameter \citep{Tal09}, or patchiness parameter \citep{Fetherolf23}.

Finally, what applies to sub-galactic scales is arguably also valid on scales of galaxy clusters, implying that \cmod\ amplifies the \citet{BO78,BO84} effect
and the negative gradient of the red-to-blue galaxy ratio as a function of cluster-centric radius \citep{FB94}.
It is thus relevant to the broader context of environmental studies of galaxies
and their preprocessing in infalling groups, the morphology-density relation \citep{Dressler80,PG84,Goto03},
and the relation of galaxy demographics with dynamical mass and virialization status of galaxy clusters, as reflected in their X-ray temperature and morphology \citep{MS97,MZ98}.
From the foregoing it is also apparent that an examination of the possible influence of \cmod\ on studies of cosmic 
shear with \textit{Euclid} and \textit{Rubin} is of considerable interest.

To summarize, \cmod\ poses a significant challenge, in addition to cosmological dimming, to established quantitative galaxy morphology concepts for higher-\zet\ galaxies that employ, for example, the \BD\ ratio or the CAS system.
It implies that these approaches are subject to systematic biases and thus incapable of warranting an objective characterization of the morphological evolution of galaxies across redshift.
At the same time, \cmod\ offers an ansatz for the development of a new generation of spectro-morphological indicators
within a novel conceptual framework for quantitative galaxy morphology.
% ::::::::::::::::::::::::::::::::::::::::::::::::::::::::::::::::::::::::::::::::::::::::::::::::::::::::::::::::::::::::
\subsection{Color gradients across redshift \label{dis:CG}} %\brem{dis:CG}
% ::::::::::::::::::::::::::::::::::::::::::::::::::::::::::::::::::::::::::::::::::::::::::::::::::::::::::::::::::::::::
% Haro 11
\begin{figure*}[h]
\begin{picture}(200,124)
\put(0,0){\includegraphics[clip, width=9.0cm]{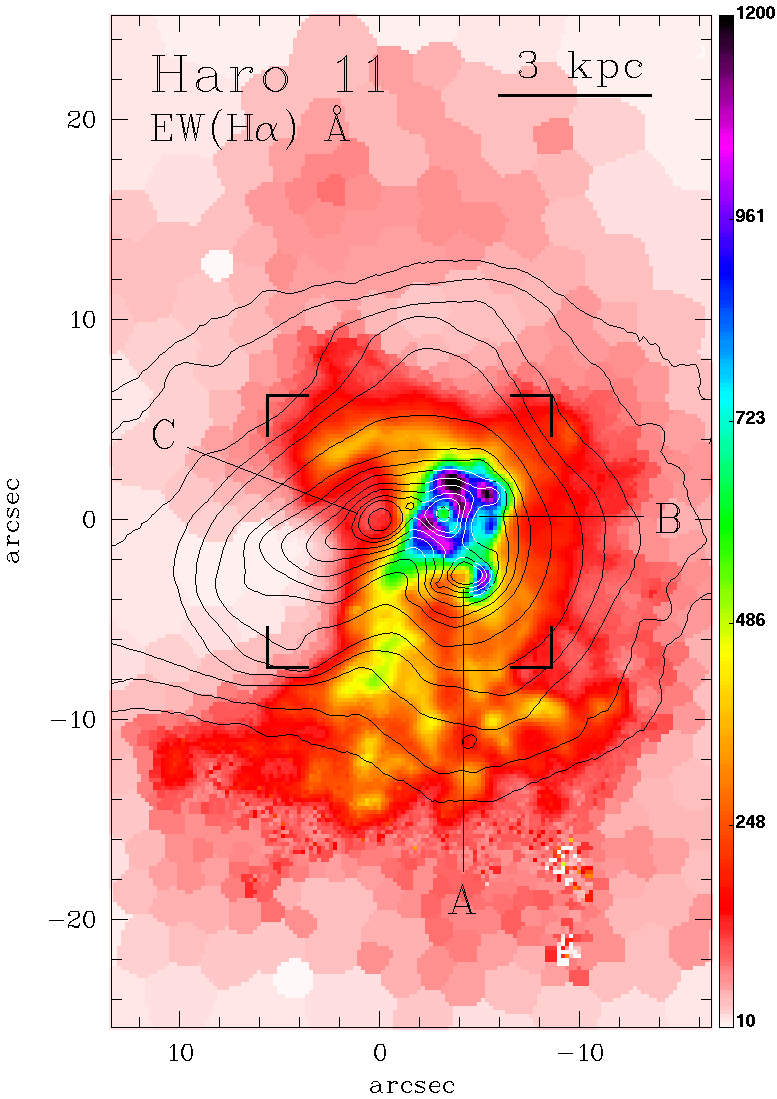}}
\put(92.6,92){\includegraphics[clip, height=3cm]{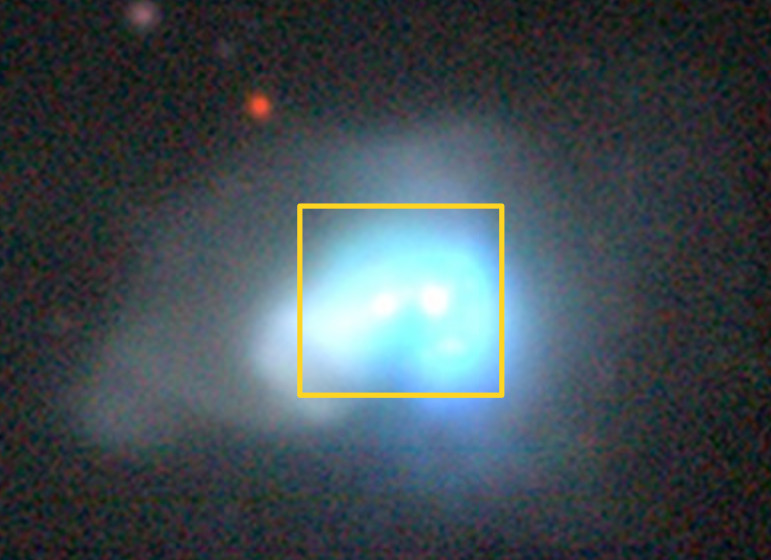}}
\put(92,60){\includegraphics[clip, height=3cm]{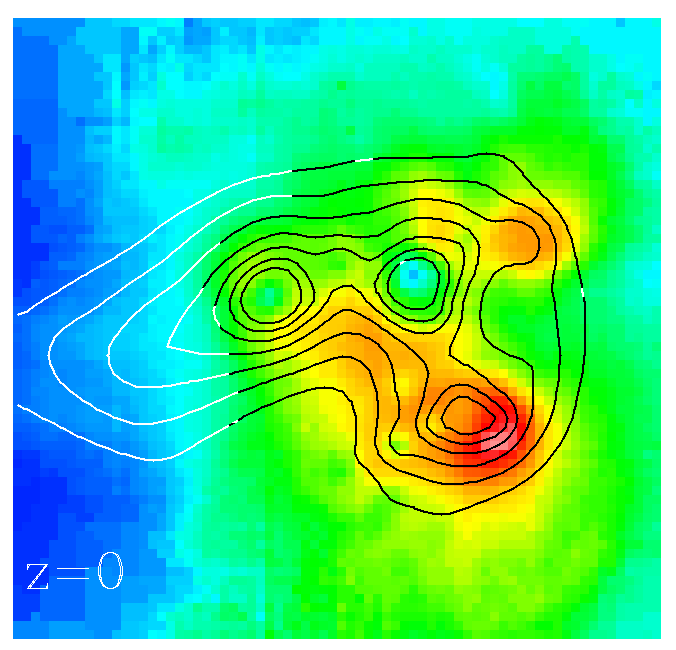}}
\put(124,60){\includegraphics[clip, height=3cm]{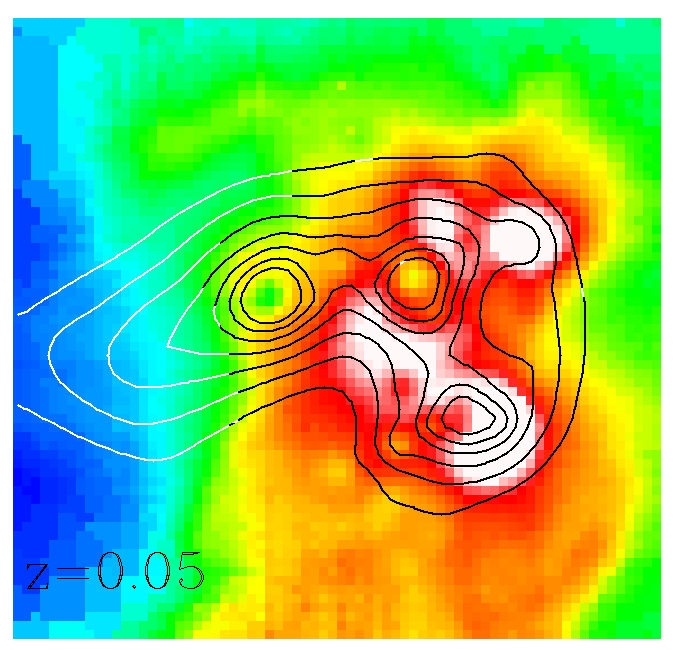}}
\put(156,60){\includegraphics[clip, height=3cm]{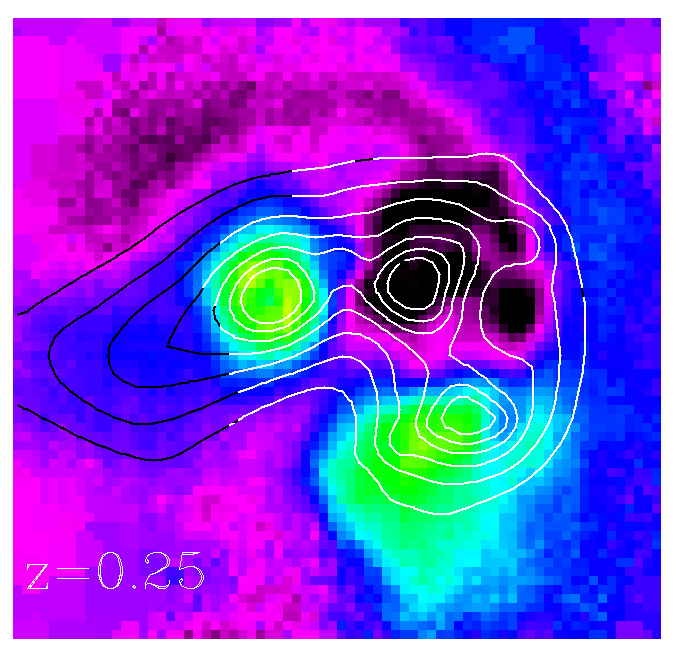}}
\put(92,32){\includegraphics[clip, height=3cm]{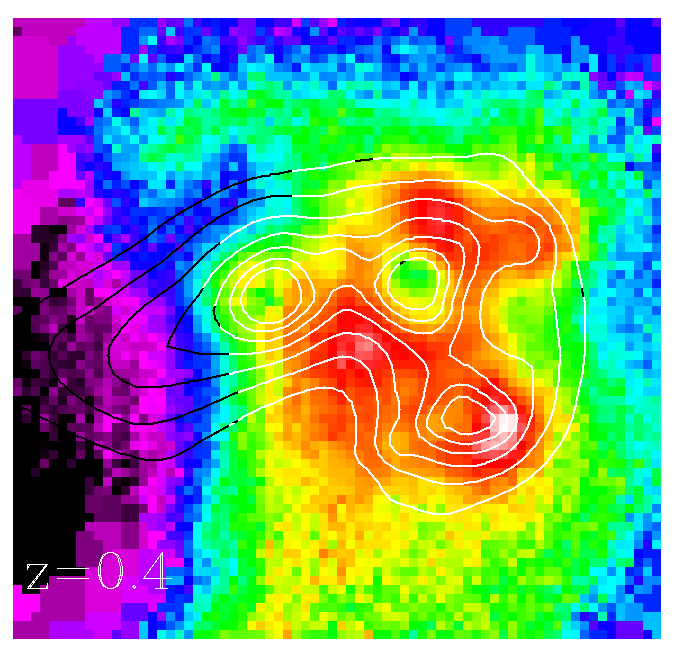}}
\put(124,32){\includegraphics[clip, height=3cm]{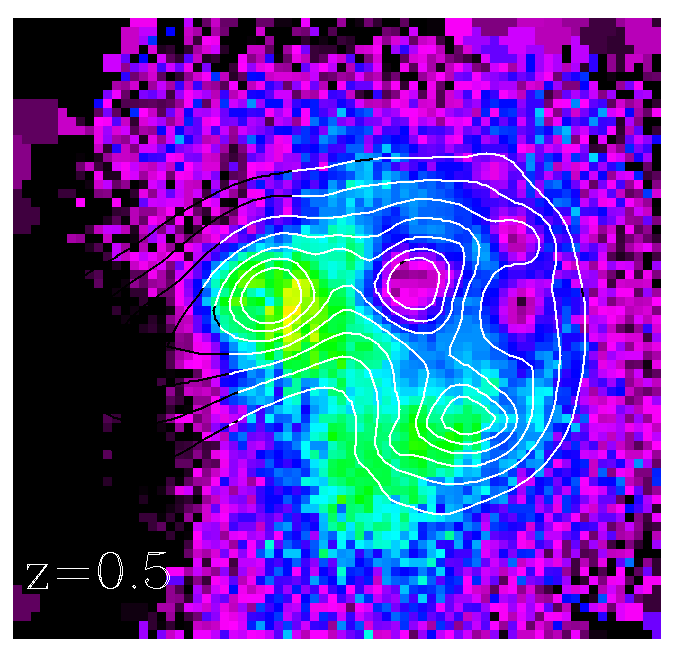}}
\put(156,32){\includegraphics[clip, height=3cm]{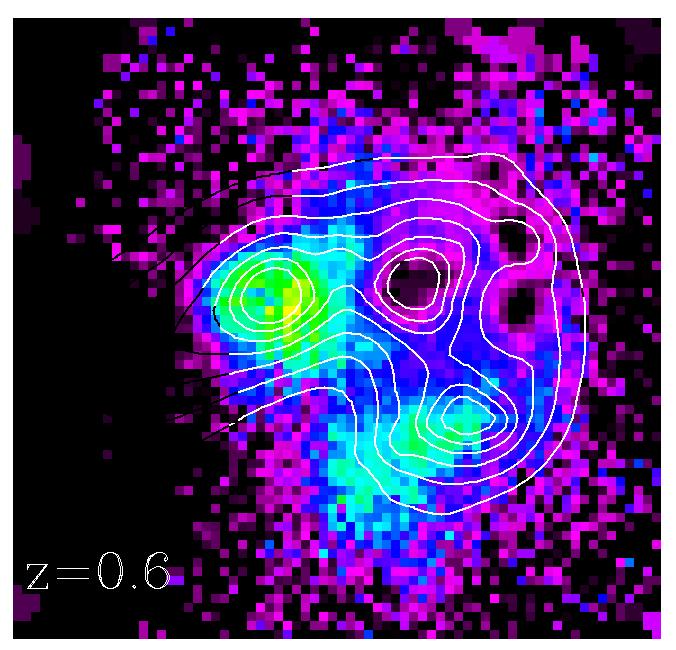}}
\put(92,0){\includegraphics[clip, height=3cm]{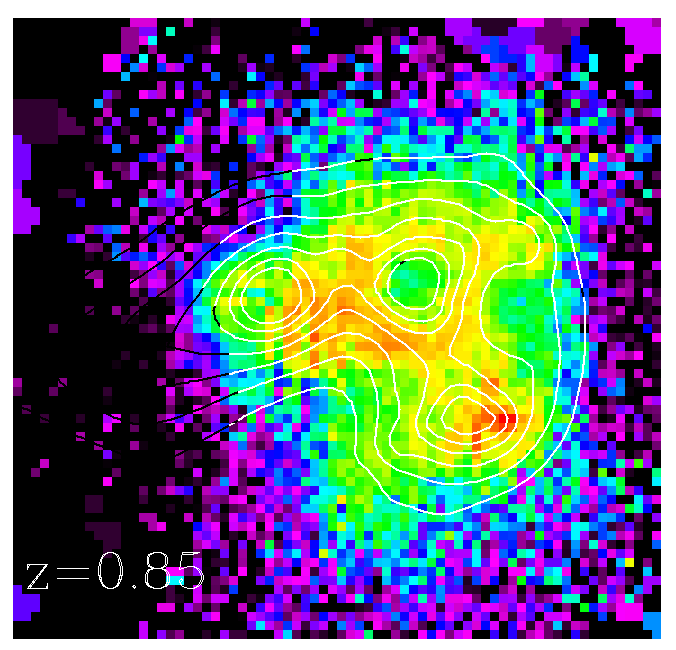}}
\put(124,0){\includegraphics[clip, height=3cm]{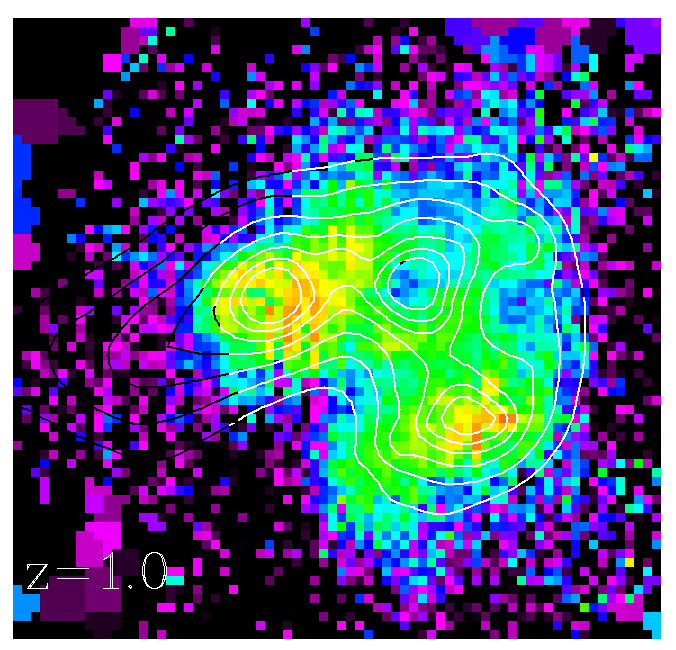}}
\put(156,0){\includegraphics[clip, height=3cm]{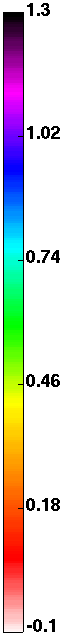}}
\end{picture}
\caption{Illustration of the impact of \cmod\ on color maps of higher-\zet\ starburst galaxies with their typically extremely intense and spatially inhomogeneous nebular emission.
\brem{left:} \ewha\ map of the nearby ($D$=82 Mpc) BCD galaxy \object{Haro 11} \citep{BO86}, computed from archival MUSE data with {\sc Porto3D} and displayed between 10 and 1200 \AA. The three dominant stellar knots (A-C) are indicated, and contours delineate the morphology of the emission-line-free continuum between 6390 \AA\ and 6490 \AA. It can be seen that nebular emission has its peak in the western half of the galaxy, where the \ewha\ locally exceeds 1200 \AA, in agreement with the morphological evidence from HST imaging \citep[e.g.,][]{Sirressi22}. 
\brem{top-right:} True-color image of \object{Haro 11} from DECaLS. The overlaid rectangle (14\farcs2$\times$13\farcs6) corresponds to the region marked on the \ewha\ map.
\brem{bottom-right:} Simulated $V$-$I$ color maps within the rectangular region marked in the \ewha\ image, computed from spectral synthesis models (cf. Sect.~\ref{simIFS}) for \zet=0, 0.05, 0.25, 0.4, 0.5, 0.6, 0.85 and 1.
The color maps are displayed in the range between --0.1 mag and 1.3 mag. Contours go from 17 to 20 $I$ \sbb\ (in the rest frame) in increments of 0.5 mag. It can be appreciated that the 2D color patterns of a starburst galaxy like \object{Haro 11} depend on redshift, as the rest-frame UV SED of its individual stellar populations varies with their age and metallicity,
and spatially inhomogeneous nebular contamination can be extremely important.}
\label{cp-Haro11}
\end{figure*}
As shown in Sects.~\ref{SED} and \ref{simIFS}, \cmod\ can enhance, erase, or even reverse rest-frame radial color gradients of distant galaxies,
depending on their redshift and 2D rest-frame SED, and the color index considered. Even though this effect has not been previously discussed, it is important to give credit to a few studies
that contain elements of our considerations and could perhaps have recognized it, would had they been further developed.
\citet[][see also \tref{Semboloni et al. 2013}]{Voigt12} point out the cSED nature of galaxies and address the issue that the SED varies across a galaxy, which in turn leads the point spread function (PSF) of a broadband exposure to slightly vary with position. Using a cSED bulge+disk galaxy model, similar to the one adopted here, these authors show that the bulge-to-disk color contrast (i.e., an analogous quantity to the radial color gradient) varies between \zet=0.6 and 1.2, and estimate
the effect that rest-frame color gradients have on determinations of the cosmic shear with \textit{Euclid} and other observing facilities.
Likewise, \citet{Er18} and \citet{Kamath20} point out that color gradients in galaxies result from their spatially varying SED, and
examine the effect of the (SED-dependent) PSF on measurements of galaxy shapes with \textit{Euclid}.

The effect of \cmod\ is particularly strong in galaxies with substantial intrinsic stellar age gradients and intense, spatially extended
nebular emission like BCDs \citep[cf. Fig.~15 in][]{PO12}.
This can be illustrated on the example of \object{Haro 11} \citep{BO86}, one of the most thoroughly investigated local starburst galaxies \citep{BO02,Menacho19,Menacho21} and an important laboratory to explore the escape of Ly$\alpha$ and LyC radiation \citep{Bergvall06,Hayes07,Leitet11,O21}.

The \ewha\ map of this system (Fig.~\ref{cp-Haro11}) reveals a complex nebular morphology \citep[see][for a detailed study of
gas kinematics and excitation conditions]{Menacho19} with a significant degree of spatial decoupling of nebular emission
from the ionizing background, as a typical feature of starburst galaxies \citep[e.g.,][]{P98}. The peak value of \ewha\ in the surroundings of
region B ($\sim$1200 \AA) translates via Fig.~\ref{ew2} to an enhancement by $>$1 (0.7) mag of the $R$ ($V$) surface brightness by nebular emission.
The right-hand panel of Fig.~\ref{cp-Haro11} shows $V$--$I$ color maps of the galaxy, simulated within $0 \leq z \leq 1$ with the method outlined in Sect.~\ref{simIFS},
with the addition that spectral modeling with \starlight\ has been carried out for the purely stellar component (the MUSE IFS data cube after bidimensional subtraction
of nebular emission), and that the gaseous (line+continuum) SED, after taking its extinction from the \ha/\hb\ ratio into account, was subsequently added to the synthetic UV-through-NIR stellar SED model.

It can be appreciated that even a small ($\sim$0.05) shift in \zet\ can appreciably alter color maps, as strong nebular lines move into different regions of filter transmission curves.
Another salient feature is a spatial anticorrelation of emission-line EWs with the stellar surface density and $V$--$I$ color,
a phenomenon that is typical for local starburst galaxies \citep{P02,Guseva04,PO12}.
For example, in the innermost part of region B, where the \ewha\ has a local minimum of $\sim$800 \AA, the $V$--$I$ color changes from $\sim$0.5 mag at \zet=0
to $\sim$1.3 mag at \zet=0.25, as a consequence of the fact that the [O{\sc iii}]$_{5007}$ line almost drops out of the $V$ filter while simultaneously the \ha\ line
moves close to the maximum of the $I$-band transmission curve. To the contrary, in component C, where nebular emission is comparatively weak (\ewha$\la$150 \AA),
a shift from \zet=0 to \zet=0.25 leads to only a minor color increase from 0.65 to 0.74 mag.
Likewise, the $V$--$I$ color of the older component C remains at \zet=0.6 moderately blue ($\sim$0.6 mag) whereas that of the starburst region B becomes as red as 1.2 mag,
superficially pointing to a passively evolving (or dusty) stellar population.

Evidently, the principal biases illustrated in Fig.~\ref{cp-Haro11} apply to any higher-\zet\ system having an intrinsic age and sSFR pattern 
(e.g., interacting galaxy pairs, spiral galaxies containing an old bar or a young circumnuclear star-forming ring), or spatially inhomogeneous nebular emission (e.g., directional outflows powered by an AGN) and dust obscuration.
For instance, one point to consider is that \cmod\ could lead to an overestimation of the observationally determined inclination of a distant spiral galaxy, which could then affect its position on the TF relation. This is because the combination of \cmod\ with dust attenuation along the line of sight 
will strongly dim the receding half of the disk, making its major-to-minor axial ratio appearing larger than it is. 

Important is also that in a higher-\zet\ spiral galaxy with a centrally decreasing (increasing) sSFR (color), \cmod\ will make 
negative color gradients appearing steeper, this way promoting the impression of an outwardly expanding ``quenching wave''.
Such enhanced color gradients, if uncorrected for \cmod, would then dilute signatures of fast (AGN-driven) SF quenching,
since \citep[if assuming a uniform stellar age at the initial phase of bulge formation, cf.][]{Breda20a}
their amplitude scales (scales inversely) with the duration (mean velocity) of the inside-out SF quenching process. 
Thus, a correction for \cmod\ is essential for an objective assessment of the relative importance of various mechanisms 
behind SF quenching in the cosmic noon, such as inhibition of inflow of cold gas from the cosmic web due to virial shocks in the galactic halo \citep{Dekel09},
ram-pressure stripping \citep[][see also, \tref{Peng et al. 2015}]{LarsonTinsleyCaldwell1980}, morphological quenching \citep{Martig09,Genzel14}, or negative AGN feedback \citep{Silk97,DMSH05,Croton06,Cattaneo09}.

In a nutshell, color maps of high-\zet\ galaxies can be deceptive: if taken at face value, without a prior correction for \cmod, they can lead
to severe misinterpretations on their nature, evolutionary status and physical drivers of their inside-out SF quenching.
% :::::::::::::::::::::::::::::::::::::::::::::::::::::::::::::::::::::::::::::::::::::::::::::::::::::::::
\subsection{Dark galaxies \label{dis:DGs}} %\brem{dis:DGs}
% :::::::::::::::::::::::::::::::::::::::::::::::::::::::::::::::::::::::::::::::::::::::::::::::::::::::::
\begin{figure}
\begin{picture}(86,60)
\put(10,0){\includegraphics[clip, height=6cm]{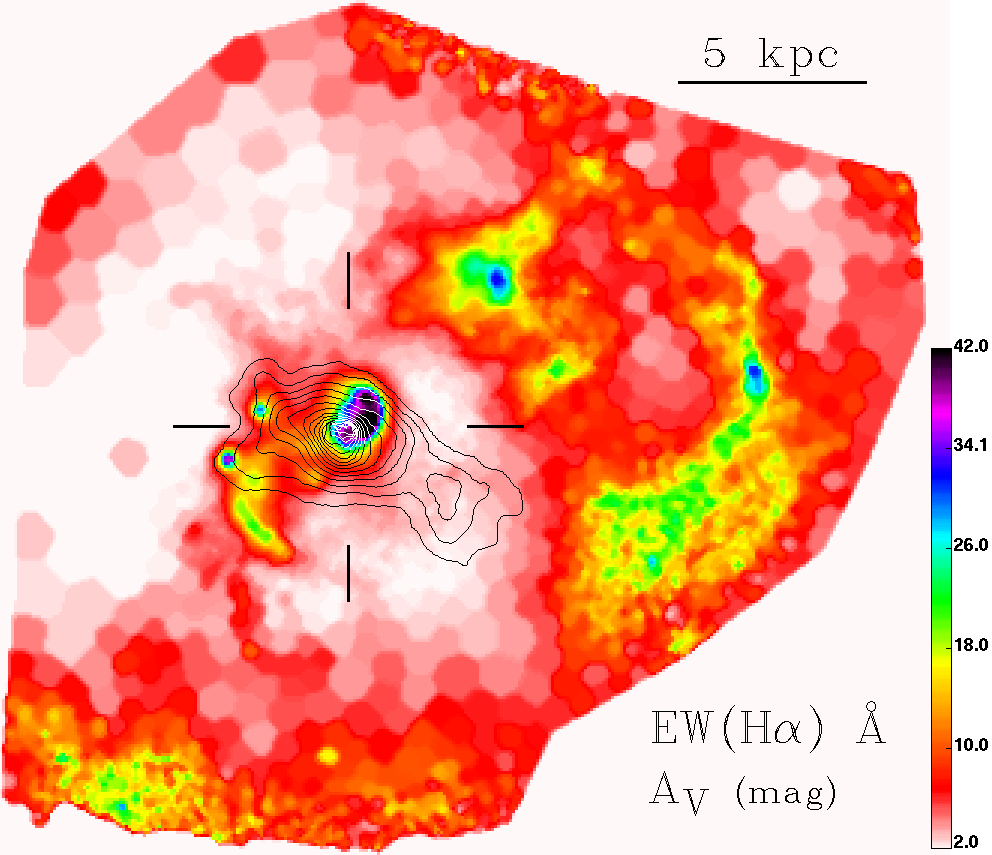}}
\end{picture}
\caption{\ewha\ map of the ULIRG \object{Arp 220} (D=83 Mpc) obtained from archival MUSE data. Contours delineate the $V$-band extinction in the stellar component
and go from 0.5 to 4 mag in steps of 0.25 mag. The cross marks the surface brightness maximum of the galaxy in the $H$ band.}
\label{ewha-Arp220}
\end{figure}
A question of considerable interest concerns the morphology of a centrally obscured protogalaxy that is rapidly assembling its stellar mass out of dense molecular gas in its nucleus.
Even though a qualitative answer can be read off Fig.~\ref{sketch}, it is worth briefly addressing this issue following the empirical approach in Sect.~\ref{simIFS}.
For this we use as reference the prototypical ultra-luminous infrared galaxy (ULIRG) \object{Arp 220} \citep{JosephWright85,Armus87}, in which a vigorous burst of SF is fed by a H$_2$ reservoir of $\sim 3\times 10^9$ \msun\ that is mixed with large amounts ($\sim 1.3\times 10^8$ \msun) of dust \citep{DA2020}.
As is apparent from the \ewha\ map in Fig.~\ref{ewha-Arp220}, nebular emission in this system is comparatively weak with only some gaseous shells 
in its periphery witnessing SF feedback \citep[see][for a detailed study with MUSE IFS data]{Perna20}.
The fact that the stellar $V$-band extinction inferred from spectral modeling reaches only $\sim$4 mag (contours) suggests that optical observations merely capture the
moderately obscured outer layers of the ULIRG and entirely miss its opaque interior where the $V$-band extinction is estimated from IR data to exceed 50 mag \citep{Sturm96}.
\begin{figure}
\begin{picture}(86,52)
\put(0,27){\includegraphics[clip, height=2.6cm]{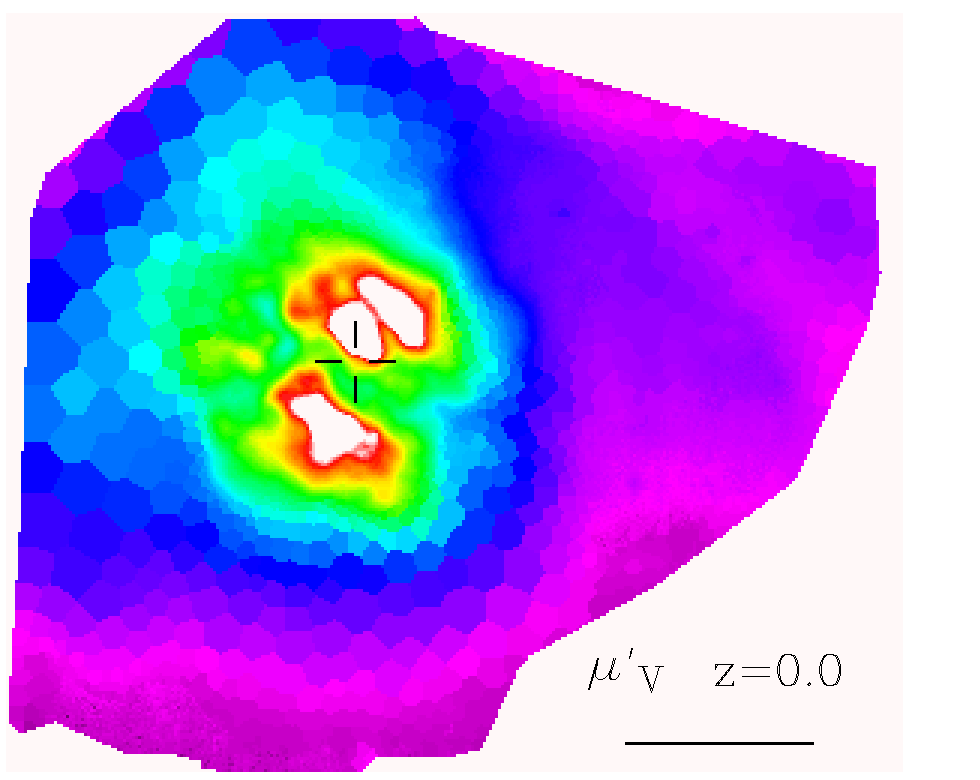}}
\put(29,27){\includegraphics[clip, height=2.6cm]{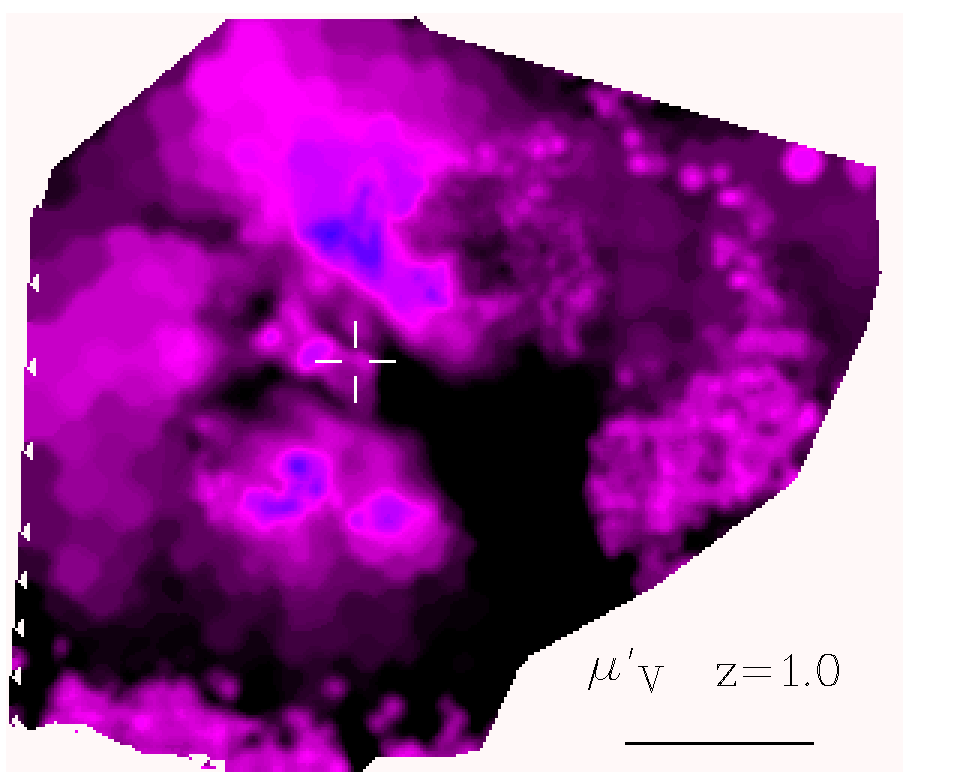}}
\put(58,27){\includegraphics[clip, height=2.6cm]{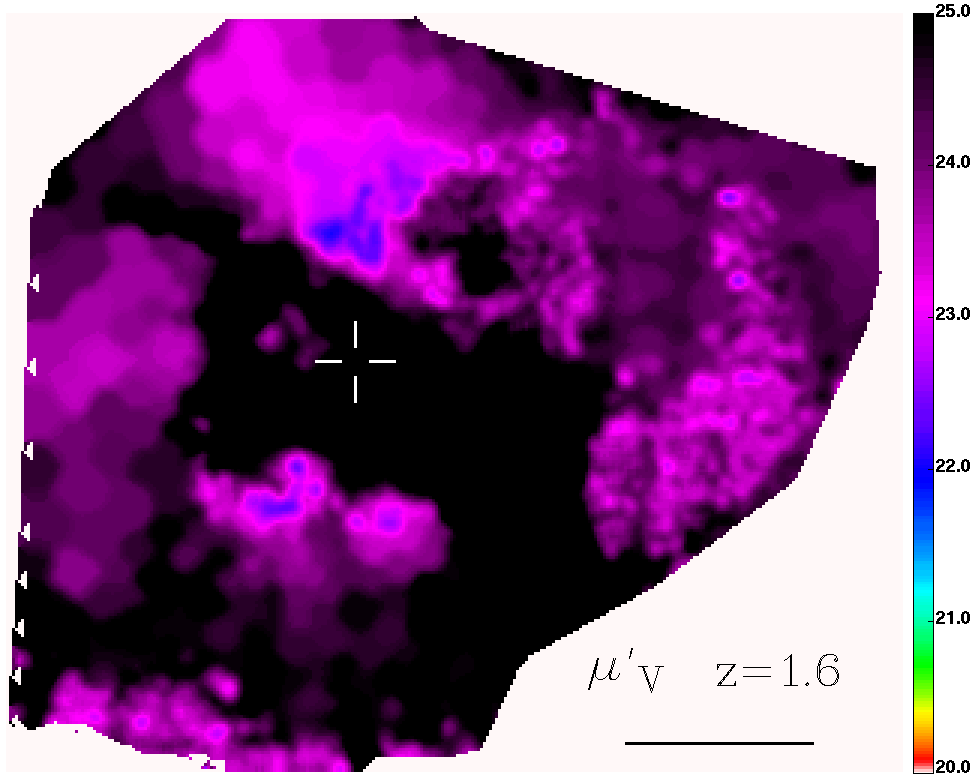}}
\put(0,0){\includegraphics[clip, height=2.6cm]{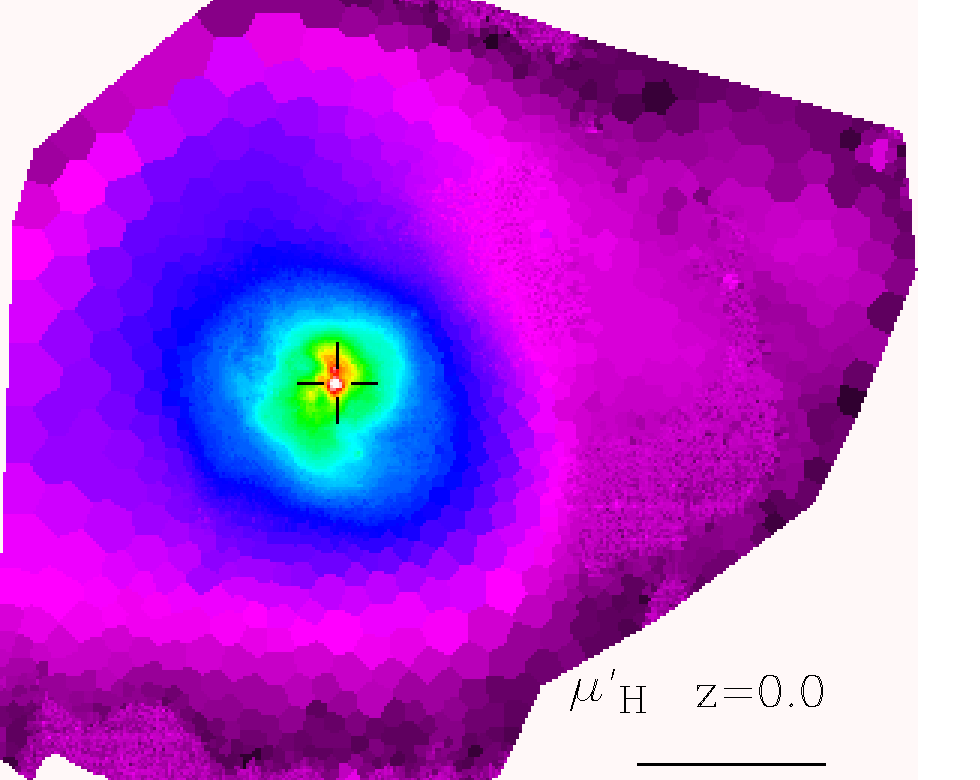}}
\put(29,0){\includegraphics[clip, height=2.6cm]{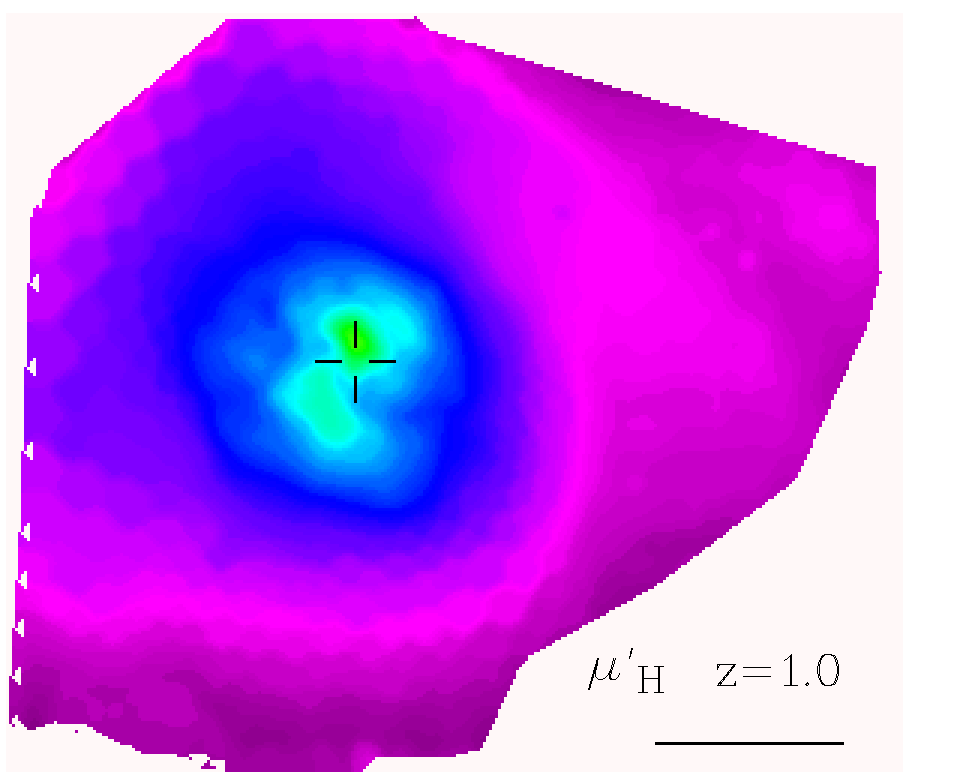}}
\put(58,0){\includegraphics[clip, height=2.6cm]{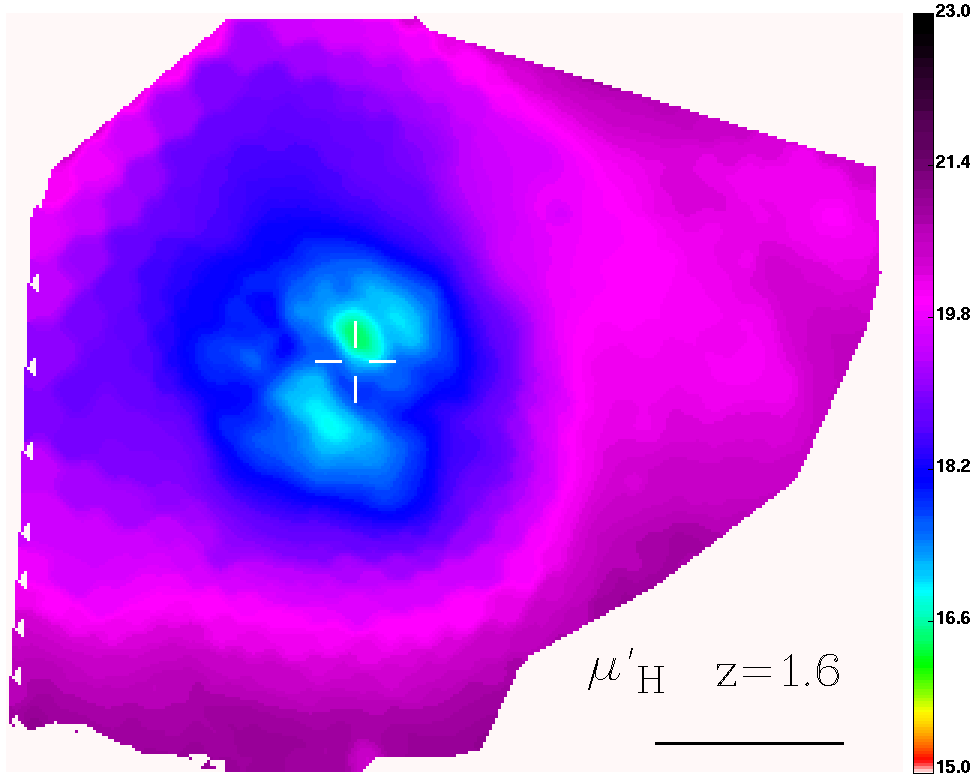}}
\end{picture}
\caption{Simulated images of \object{Arp 220} in $V$ and $H$ (upper and lower panels, respectively) at \zet=0, 1 and 1.6.
The cross marks the nucleus of the galaxy and the horizontal bar depicts a linear scale of 5 kpc.}
\label{zima-Arp220}
\end{figure}

\begin{figure}
\begin{picture}(86,140)
\put(0,96){\includegraphics[clip, width=8.6cm]{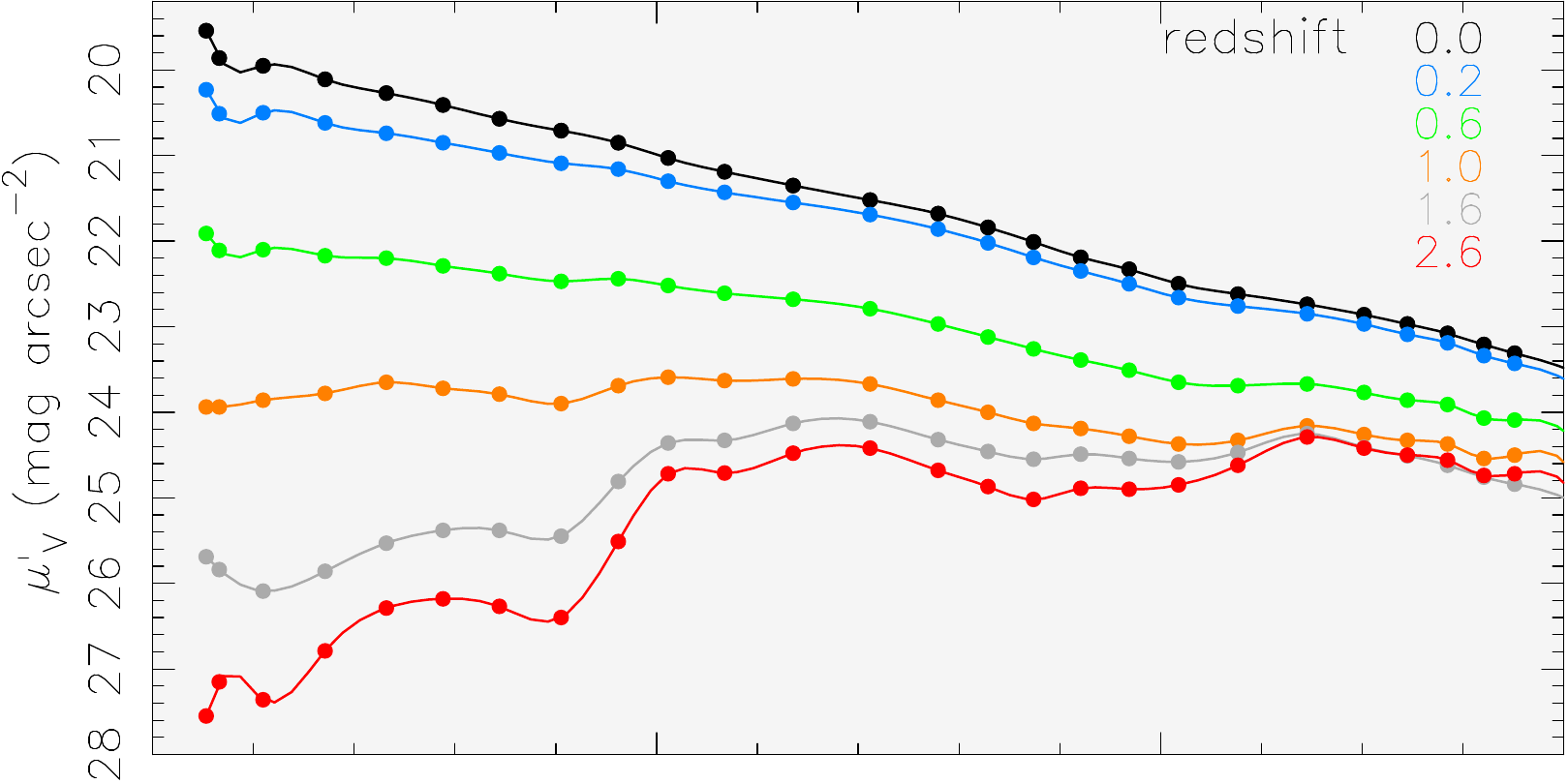}}
\put(0,53){\includegraphics[clip, width=8.6cm]{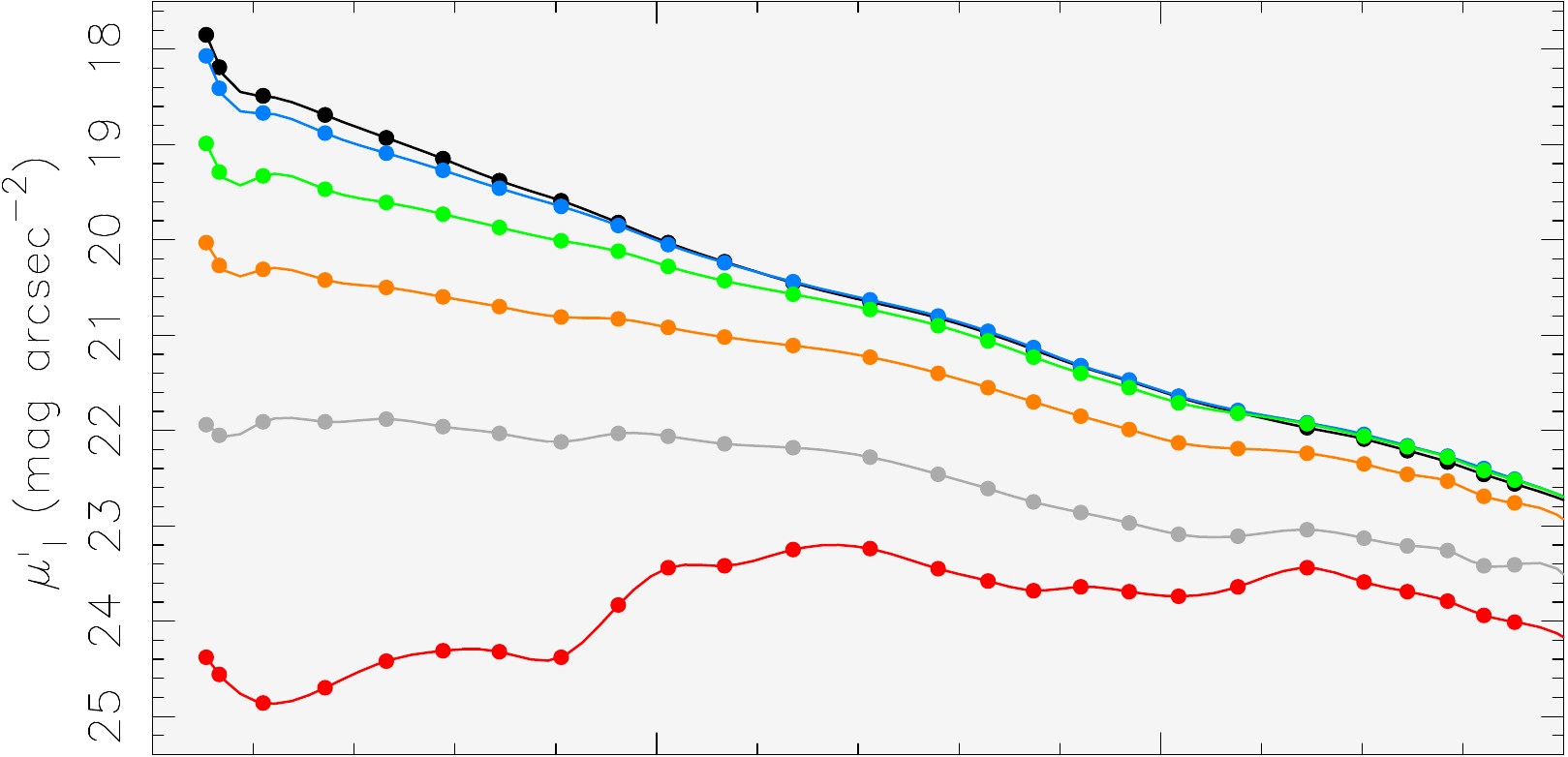}}
\put(0,0){\includegraphics[clip, width=8.6cm]{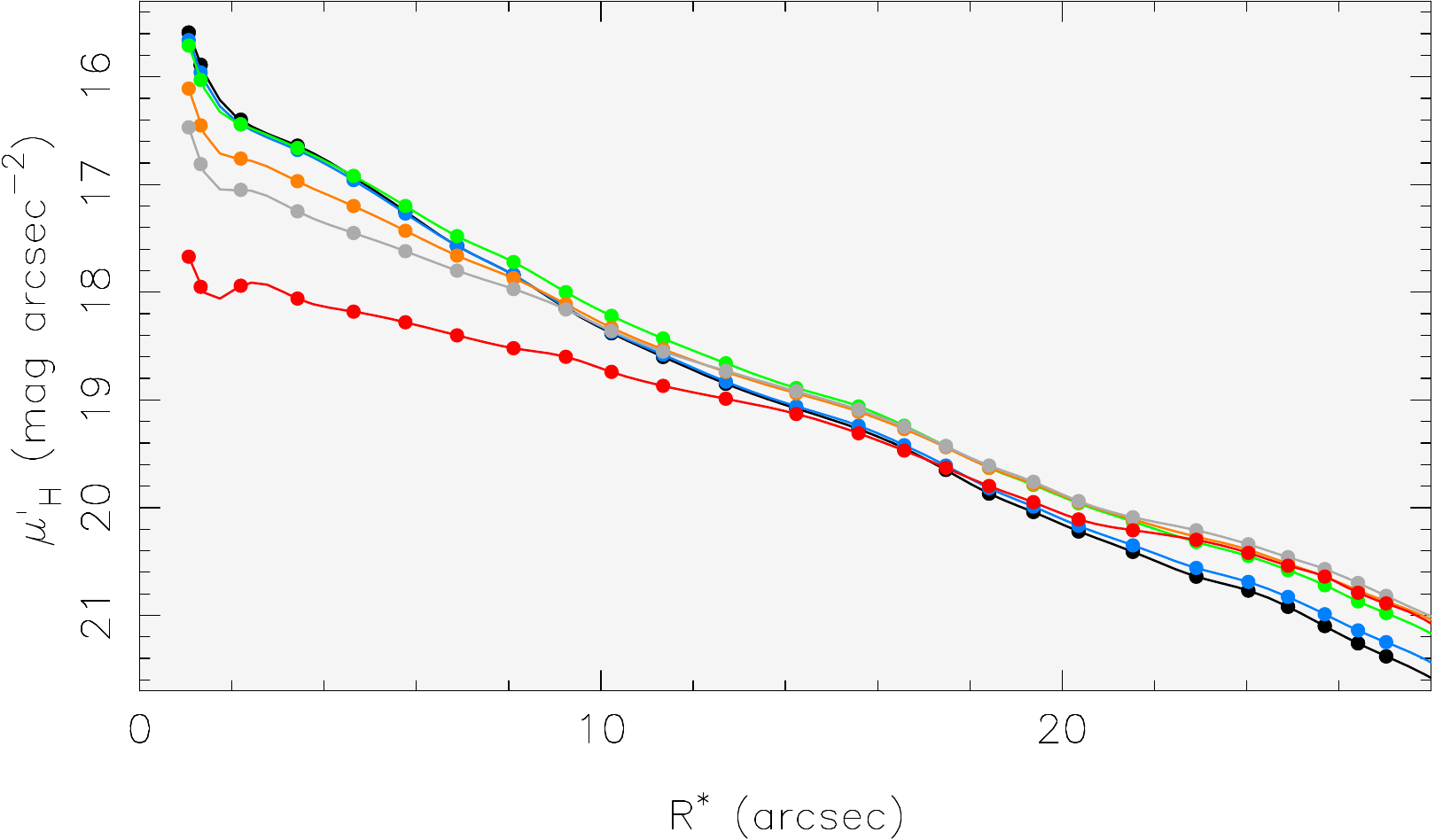}}
% %%%%%%%%%%%%%%%%%%%%%%%%%%%%%%%%%%%%%%%%%%%%%%%%%%%%%%%%%%%%%%%%%%%%%%%%%%
\PutLabel{72}{100}{\mlx \textcolor{black}{V\arcmin\ SBPs}}
\PutLabel{72}{88}{\mlx \textcolor{black}{I\arcmin\ SBPs}}
\PutLabel{72}{44}{\mlx \textcolor{black}{H\arcmin\ SBPs}}
% %%%%%%%%%%%%%%%%%%%%%%%%%%%%%%%%%%%%%%%%%%%%%%%%%%%%%%%%%%%%%%%%%%%%%%%%%%
\end{picture}
\caption{SBPs of \object{Arp 220} computed from simulated $V$, $I$, and $H$ images (from top to bottom) in the range $0\leq z\leq 2.6$.
The central part of the galaxy suffers with increasing redshift a strong dimming in the optical (by $\sim$6 $V$ mag at \zet=1.6) whereas the effect is
modest in $H$ ($\sim$0.7 mag at \zet=1.6 and $\sim$1.6 mag at \zet=2.6). It can be appreciated that, as the optical SED moves at a higher \zet\ into
the NIR, the $H$ surface brightness of the less attenuated periphery of the ULIRG gets slightly enhanced, in agreement with the evidence from Fig.~\ref{sSED}.}
\label{sbp-Arp220}
\end{figure}

Figure~\ref{zima-Arp220} shows simulated $V$ and $H$ images of \object{Arp 220} at \zet=0, 1 and 1.6. It can be seen from the left-hand images that the nucleus is well distinguishable in the $H$ band as a single compact source (cross), whereas in the visual it splits into two peaks as a result of dust obscuration.
Another conspicuous feature is the appreciable change of the $H$ morphology with redshift, with the single nucleus disappearing at \zet=1.6
and the center gradually approaching the $V$-band morphology of the ULIRG at \zet=0.
As for the optical morphology, it drastically changes with redshift, as the $V$ surface brightness drops by $>$3 mag at \zet=1, and by $\sim$6~mag at \zet=1.6,
rendering the entire central 25\arcsec$\times$10\arcsec\ of the galaxy invisible.

\begin{figure}
\begin{picture}(86,36)
\put(0,0){\includegraphics[trim=0 170 0 0, clip, width=8.6cm]{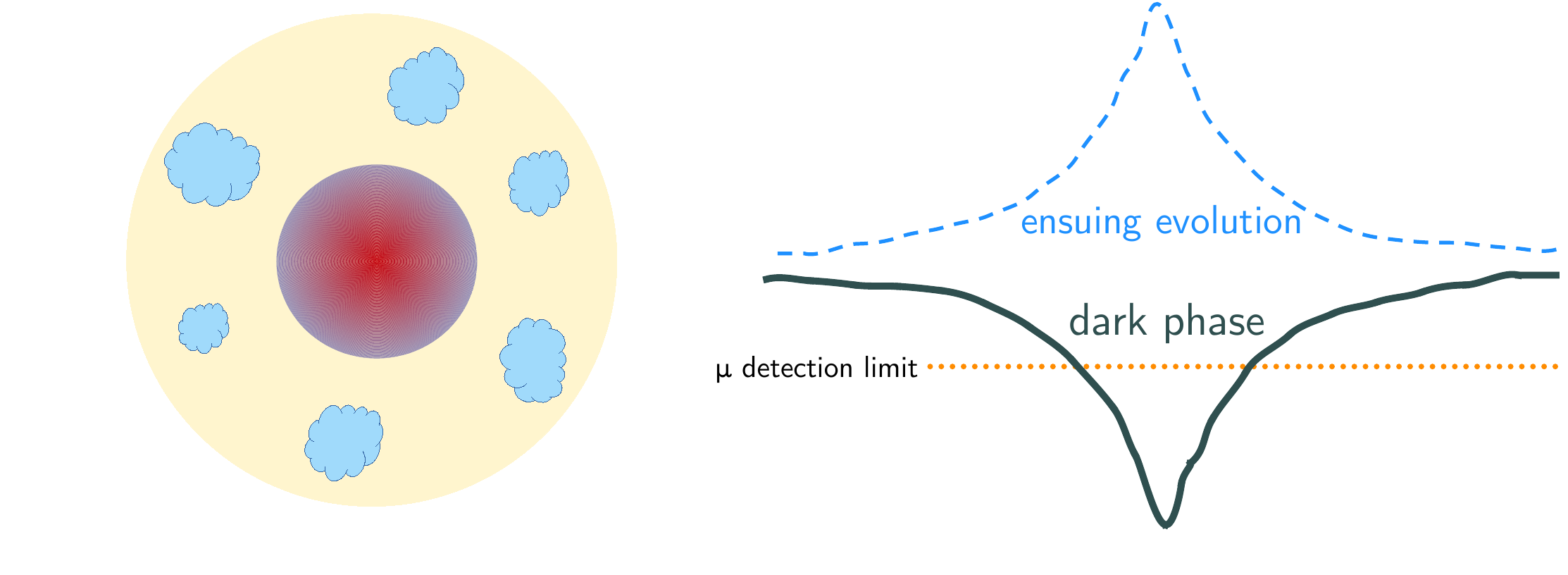}}
\end{picture}
\caption{Schematic view of a ``dark'' protogalaxy in the early Universe.
\brem{left:} Protogalactic unit experiencing rapid growth of its stellar mass through a dust- and molecular-gas enshrouded nuclear starburst (red) while gas continues accreting onto its optically thin periphery, where it feeds localized SF. \brem{right:} \cmod\ in combination with the intrinsically red SED of the nucleus results in a strong suppression of the ObsF optical and NIR surface brightness, thus preventing the detection of the central zone of the galaxy. The dark phase is terminated once the energy release from the starburst leads to a large-scale disruption and evacuation of absorbing gas and dust from the nuclear region of the galaxy.}
\label{fig:DG} 
\end{figure}

A distant analog of \object{Arp 220} will therefore appear optically dark, with only fragments of its less extincted periphery remaining
visible, while its core becoming increasingly difficult to detect at higher \zet\ even in the NIR (cf. lower panel of Fig.~\ref{sbp-Arp220}).
As long as the dense molecular nucleus withstands starburst-driven feedback, the galaxy will likely present a donut-like morphology
as a result of its intrinsically red SED and the \cmod\ effect.
During this presumably short yet energetic evolutionary phase, in which catastrophic cooling \citep{Silich03,TT05,TT07} might be important,
the hidden starburst and eventually also a seed-AGN, might only be traceable as an ``orphan'' X-ray source whose hard ($>$8 keV), 
inverse Compton-scattering amplified, radiation could penetrate the obscuring dust and gas.
Radio synchrotron and [C{\sc ii}]158$\mu$m emission might offer additional tracers of hidden starburst and accretion-powered nuclear activity.
The ensuing disruption of the shielding molecular layer by momentum-driven superwinds \citep[e.g.,][]{Fiore22} and eventually directional AGN outflows
will then turn the galaxy visible over its entire extent, as sketched in Fig.~\ref{fig:DG}, with fragments of molecular gas perhaps surviving only in its periphery,
much like the case of the local BCDs \object{NGC 1569} \citep{Taylor99} and \object{Mrk 86} \citep{GdP02}.

To summarize, dark galaxies, such as those that have been reported recently \citep[e.g.,][]{Marino18,Nelson22,PG22}, are naturally expected as a result from \cmod.
Deep searches of [C{\sc ii}]158$\mu$m emission with ALMA in conjunction with near- and mid-IR imaging with JWST could offer precious observational insights in this regard.

% =====================================================
\section{Summary and conclusions \label{sum}}
% =====================================================
The \cmod\ effect is a simple consequence of the fact that galaxies inhabiting the expanding Universe consist of spatially and evolutionary distinct structural components that differ from one another in terms of their time-evolving rest-frame SED. This SED obviously depends on the local SF and stellar mass assembly history,
stellar metallicity, and level of intrinsic extinction and nebular contamination.

Contrary to cosmological surface brightness dimming or gravitational lensing, which are achromatic and thus equally impact all galaxy structural components regardless of their rest-frame SED, \cmod\ is per se chromatic. Its essential implication is the differential dimming (amplification) of the surface brightness of red SF-quenched (blue star-forming) 
zones in a higher-\zet\ galaxy. A rectification of this effect, that is, the 2D reconstruction of the rest-frame characteristics of a galaxy from a set of observed images, is a nontrivial task that requires a spatially resolved \kc\ correction. Such a correction is evidently vastly different for an old (or dusty) UV-faint bulge and a young UV-bright star-forming disk. 
Applying a uniform \kc\ correction term to a galaxy as a whole, or entirely skipping this step, as is common practice in state-of-the-art bulge-to-disk decomposition studies 
of higher-\zet\ galaxies, is a problematic approach that deterministically leads to serious and highly interlinked biases in our understanding of galaxy evolution in the early cosmos, as pointed out in \citet{PO12}.

The essential goal of this study is to invite the community to critically rethink established practices for structural decomposition and quantitative morphology studies of high-\zet\ galaxies and motivate a synergistic effort toward a better understanding and overcoming of the \cmod. This endeavor appears especially timely since, if successful, it will allow JWST and \textit{Euclid} to fully realize their potential for elucidating galaxy evolution in the early Universe. 

This article attempts a concise yet quantitative analysis of the main implications that the usual neglect of \cmod\ has on the characterization of higher-\zet\ spiral galaxies. The methodology adopted is as follows:

\brem{i)} We simulated a spiral galaxy that consists of a bulge and disk forming with, respectively, an exponentially decreasing and a constant SFR. The SED for these two components, computed with the evolutionary synthesis code \pegase~2, was simulated for $0\leq z \leq 3$, taking the passband shift and wavelength stretching into account, and convolved with
($UBVRIJHK$) filter transmission curves to infer the variation in their {reduced} surface brightness (\dmu) as a function of \zet.
Hereby two approaches were adopted, the first one (Sect.~\ref{sub:stSED}) based on synthetic SEDs of fixed ages of 13.7 Gyr and 4 Gyr and the second (Sect.~\ref{sub:eSED}) carried out in an EvCon manner, that is, with the age of the SEDs at a given \zet\ being equal to the cosmic age at that redshift. From these two sets of simulations, we computed the variation across \zet\ in the bulge-to-disk surface brightness and color contrast (\dmBD\ and \dcol, respectively).

\brem{ii)} By approximating the SBP of the bulge (disk) via a S\'ersic (exponential) model, we constructed a synthetic reference galaxy model with properties typical of those for a massive present-day spiral. In turn, using the output from \brem{(i)}, we studied how the combined (bulge+disk) SBP of the synthetic galaxy varies with \zet. More specifically, we derived a set of commonly used photometric quantities, such as the effective radius (\reff), Petrosian radius, S\'ersic exponent ($\eta$), and various light concentration indices. Additionally, from bulge-disk decomposition we quantified the variation across \zet\ of the \BD\ and \BT\ ratios (Sect.~\ref{BD}).

Two extra components were added in our analysis to cursorily address the fact that standard parametric SFHs do not account for the now accepted picture of early bulge growth being primarily driven through inward migration and coalescence of massive SF clumps emerging out of VDIs. The first 
(\brem{iii}) shows that strong nebular line emission from these young SF clumps selectively elevates the surface brightness of the disk over several broad redshift intervals, with the effect being particularly relevant ($>$1 mag) in the NIR (Sect.~\ref{sub:nebSED}). As recently pointed out in \citet{P22}, this differential enhancement of the star-forming disk relative to the SF-quenched bulge in these discrete \zet\ intervals can lead to the over-subtraction of the former from the total SBP and the underestimation of the luminosity of the latter.
Spatially inhomogeneous nebular contamination therefore poses a significant challenge for bulge-disk decomposition studies of higher-\zet\ protogalaxies, especially in the NIR.

The second analysis task (\brem{iv}; Sect.~\ref{dis:lim}) addresses the color evolution of the bulge when its growth is driven by migration of ex situ formed stellar clumps, instead of in situ SF. We show that, from the photometric point of view, migration-driven bulge assembly mimics 
the effect of in situ SF according to a top-truncated IMF. The principle effect is an overestimation by a factor of $\sim$2 of the stellar age determined from rest-frame colors, which potentially forces the interpretation of mature dusty bulges already being present $\sim$2 Gyr after the Big Bang.

\brem{v)} As an empirical check of the results from \brem{(ii)}, we used spatially resolved IFS data in conjunction with population spectral synthesis models to simulate broadband images and color maps of local galaxies if they were to be observed at a higher \zet\ (Sect.~\ref{simIFS}). Hereby, nebular emission and intrinsic extinction in the stellar and nebular component are taken into account.

The main insights from this study are as follows:

\brem{vi)} \cmod\ in practice prevents the optical detection of old, quasi-monolithically built bulges at \zet$\sim$1, leading to the erroneous
classification of galaxies as virtually bulgeless disks. Specifically, our simulations imply that at those redshifts the $V$-band \BD\ ratio is underestimated by a factor of
between $\sim$6 and $\sim$20. The situation is different in the $H$ band, for which the effect is merely 20--30\%.
The false (\cmod-induced) impression of the delayed emergence and accelerated growth of bulges since only \zet$\la$1 is additionally documented through simulations 
based on IFS data.

\brem{vii)} An implication of \cmod\ is the non-homologous variation of the form of the SBP of a galaxy as a result of the differential dimming (amplification) 
of its old, passively evolving (young star-forming) zones. A consequence of this is that any primary and secondary photometric quantity inferrable from profile 
fitting and which characterizes the galaxy shape (e.g., S\'ersic exponent, \BD\ ratio, light concentration index, effective radius \reff) depends on \zet, even in the idealized case 
of a galaxy not spectrophotometrically evolving over time. Quite importantly, \reff\ is no robust (\zet-invariant) measure of galaxy size
and therefore should not be used for normalizing radially resolvable quantities (e.g., age, metallicity, or EW) when inter-comparing galaxy samples at different \zet.

\brem{viii)} The integrated SED of a distant spiral galaxy is the luminosity-weighted sum of its evolutionary distinct constituent stellar populations and is therefore already 
preprocessed by \cmod\ prior to its recording: the red SF-quenched bulge is filtered out due to its drastic dimming, making the optical SED of 
the galaxy appear bluer than it is. This, in turn, is expected to drive photometric SED fitting codes toward a much too low stellar \ml\ ratio, and 
thereby underestimate the total stellar mass, \mstar, of a higher-\zet\ galaxy by up to a factor of  $\sim$2.
Framed within the picture of galaxy downsizing, this bias is expected to be strongest for massive, high-bulge-to-disk\ spirals that experience early bulge formation 
(and vice versa) and thus to impact galaxy mass determinations differentially.

\brem{ix)} Contrary to \mstar\ determinations from SED fitting, physical quantities that are inferred from luminosities, EWs, or kinematics of 
monochromatic emission lines are immune to \cmod. 
An expected consequence of this is an increasing overestimation of the specific SFR of galaxies out to \zet$\sim$1 (or, equivalently, a decline in sSFR since \zet$\sim$1)
and an artificial steepening of the mass-metallicity relation for massive galaxies at high \zet. 
Further qualitative predictions from \cmod\ include that discrepancies between kinematical and photometric mass estimates will increase with \zet\ and galaxy mass, 
with galaxies at \zet$\sim$1 appearing DM (baryon) dominated in the optical (NIR), and that the TF relation will steepen for bulge-dominated galaxies above \zet$\sim$0.5. 

\brem{x)} \cmod\ can globally change the morphology of higher-\zet\ galaxies, especially when these contain appreciable spatial inhomogeneities in their
stellar age, metallicity, extinction, or nebular contamination. For this reason, it strongly limits the applicability of established quantitative galaxy morphology 
concepts, such as CAS, and makes an objective assessment of the morphological evolution of galaxies across \zet\ a nontrivial task.
At the same, \cmod\ offers an ansatz for the development of a new generation of spectro-morphological indicators toward a novel concept for quantitative galaxy morphology.

\brem{xi)}  \cmod\ can enhance, erase, or even reverse radial color gradients of distant galaxies, depending on their redshift and 2D rest-frame SED and the  
color indices specifically considered. Its neglect leads almost unavoidably to serious misinterpretations of the nature (e.g., age and extinction patterns)
of higher-\zet\ galaxies that directly affect, for example, our understanding of the physical drivers and corresponding timescales for inside-out SF quenching.
Selective attenuation by the intergalactic medium is highly unlikely to compensate for these biases.
A correction for \cmod\ is especially important for the study of color and stellar age patterns in EELGs at high \zet.

\brem{xii)} A natural expectation from \cmod\ is that massive protogalaxies in a brief energetic phase, during which they form their first generation of
stars (and eventually also a seed SMBH) in their dense molecular core, should be centrally invisible, even in the NIR. 
The existence of such dark galaxies might only be witnessed by an orphan radio and hard X-ray source surrounded by fragments
of visible matter that delineates their less obscured periphery.

In summary, it is virtually impossible to name a topic related to galaxy evolution across cosmic time for which \cmod\ does not call for a critical inspection or even revision of previous work. Understanding and overcoming the \cmod\ effects is crucial for a meaningful study of galaxies at higher \zet\ ($\ga$0.1), and thus also a prerequisite for fully unfolding the potential of JWST and \textit{Euclid}.

\begin{acknowledgements}
We thank the anonymous referee for an insightful and thorough report that improved the paper.
We thank Dres. Matthew Lehnert, Bodo Ziegler and Andrew Humphrey for valuable comments.
Polychronis Papaderos gratefully acknowledges support by the Wenner-Gren Foundation and the hospitality of the Astronomy Department at Stockholm University.
He also thanks Funda\c{c}\~{a}o para a Ci\^{e}ncia e a Tecnologia (FCT) for managing research funds graciously provided to Portugal by the EU.
This work was supported through FCT grants UID/FIS/04434/2019, UIDB/04434/2020, UIDP/04434/2020 and the project "Identifying the Earliest Supermassive Black Holes with ALMA (IdEaS with ALMA)" (PTDC/FIS-AST/29245/2017). 
Göran Östlin acknowledges support from the Swedish Research Council (VR) and the Swedish National Space Administration (SNSA).
Iris Breda acknowledges financial support from the State Agency for Research of the Spanish MCIU through the “Center of Excellence Severo Ochoa” award to the Instituto de Astrofísica de Andalucía (SEV-2017-0709). She also acknowledges HORIZON-TMA-MSCA-2021-PF-01 postdoctoral fellowship contract 101059532 (GALYKOS).
This study uses data provided by the Calar Alto Legacy Integral Field Area (CALIFA) survey (http://califa.caha.es), 
funded by the Spanish Ministry of Science under grant ICTS-2009-10, and the Centro Astron\'omico Hispano-Alem\'an.
It is based on observations collected at the Centro Astron\'omico Hispano Alem\'an (CAHA) at Calar Alto, operated jointly 
by the Max-Planck-Institut f\"ur Astronomie and the Instituto de Astrofísica de Andalucía (CSIC).
This research has made use of the Cosmology Calculator for the World Wide Web \citep{Wright06} and 
of NASA/IPAC Extragalactic Database (NED) which is operated by the Jet Propulsion Laboratory, 
California Institute of Technology, under contract with the National Aeronautics and Space Administration.
\end{acknowledgements}

% ::::::::::::::::::::

% :::::::::::::::::::::::::::::::::::::::::::::::::::::::::::::::::::::::::::::::::::::::
\begin{appendix}
\section{Supplementary notes on the computation and processing of synthetic SEDs \label{ap:SED}}
\subsection{SFH parameterizations \label{ap:pSFH}} %\ccom{(ap:pSFH)}
\begin{center}
\begin{figure}[h]
\begin{picture}(86,46)
\put(0,0){\includegraphics[width=8.6cm]{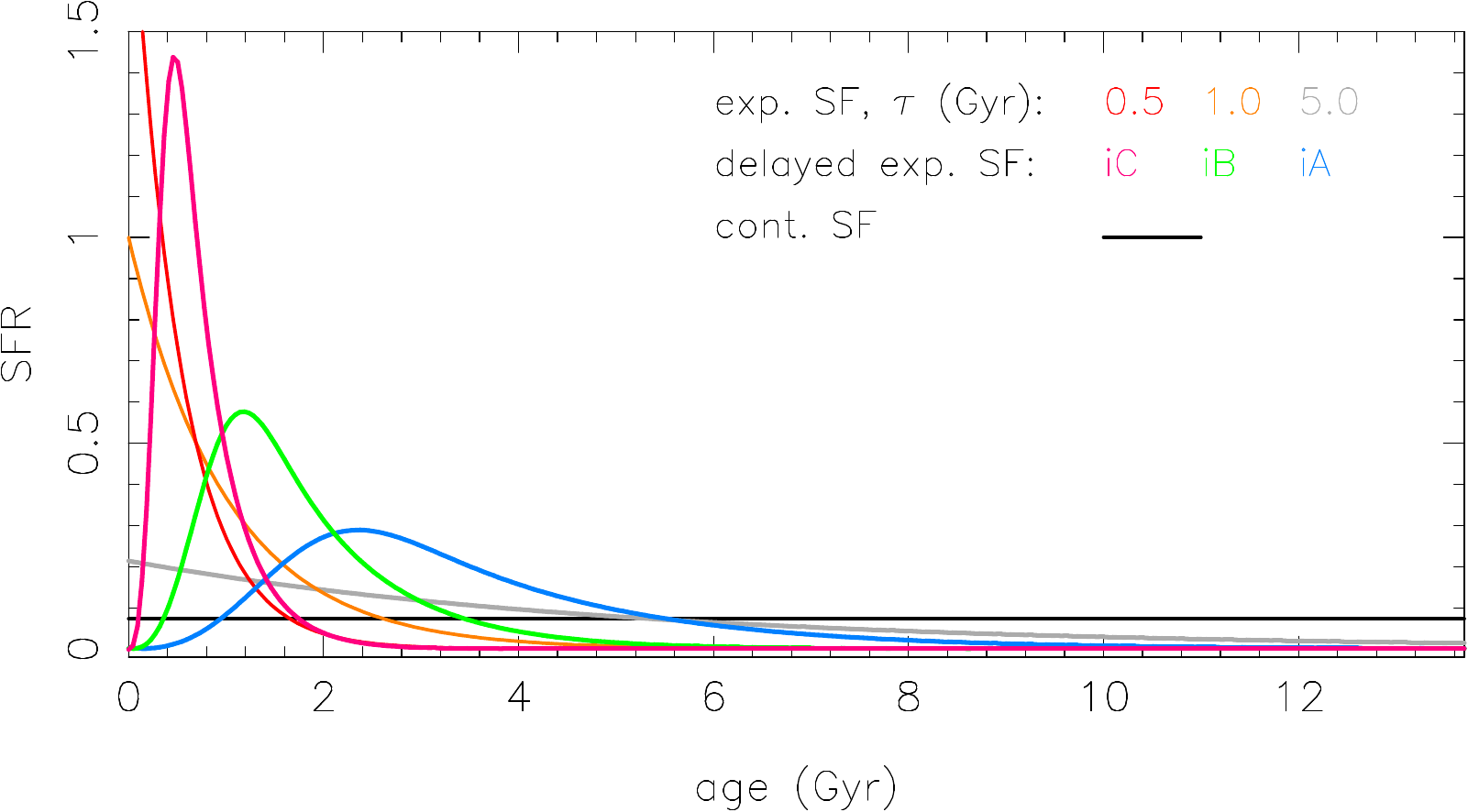}}
\end{picture}
\caption{SFR (in \msun/yr) vs. age (Gyr) for the seven parametric SFHs assumed. The total stellar mass produced amounts in all cases to $10^9$ \msun.}
\label{fig:SFHs}
\end{figure}
\end{center}

\begin{center}
\begin{figure*}[ht!]
\begin{picture}(200,154)
\put(3,104){\includegraphics[width=18.2cm]{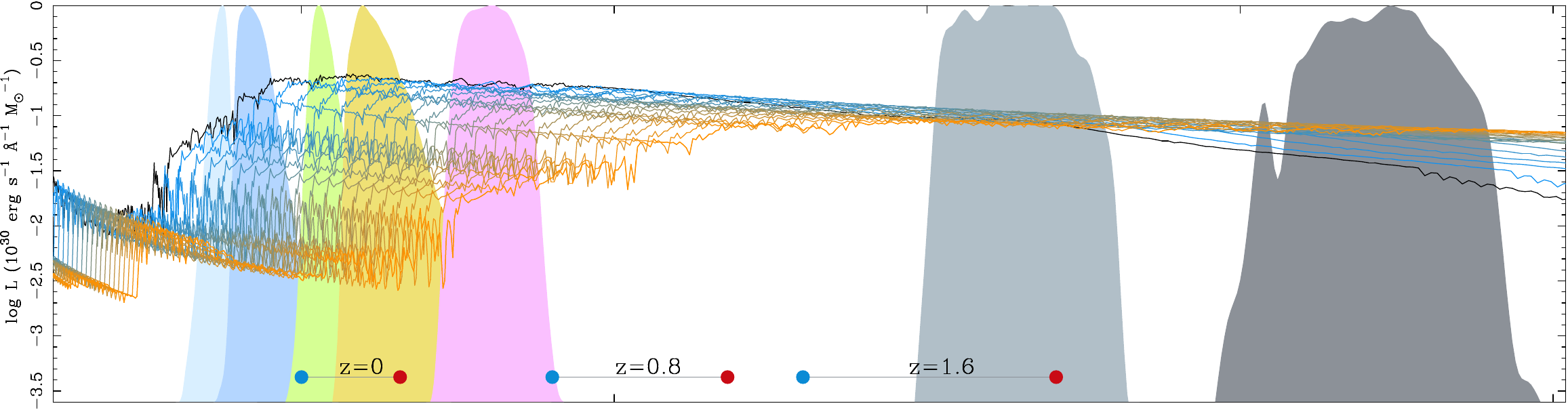}}  % bulge, purely stellar emission, age: 4 Gyr
\put(3,55.3){\includegraphics[width=18cm]{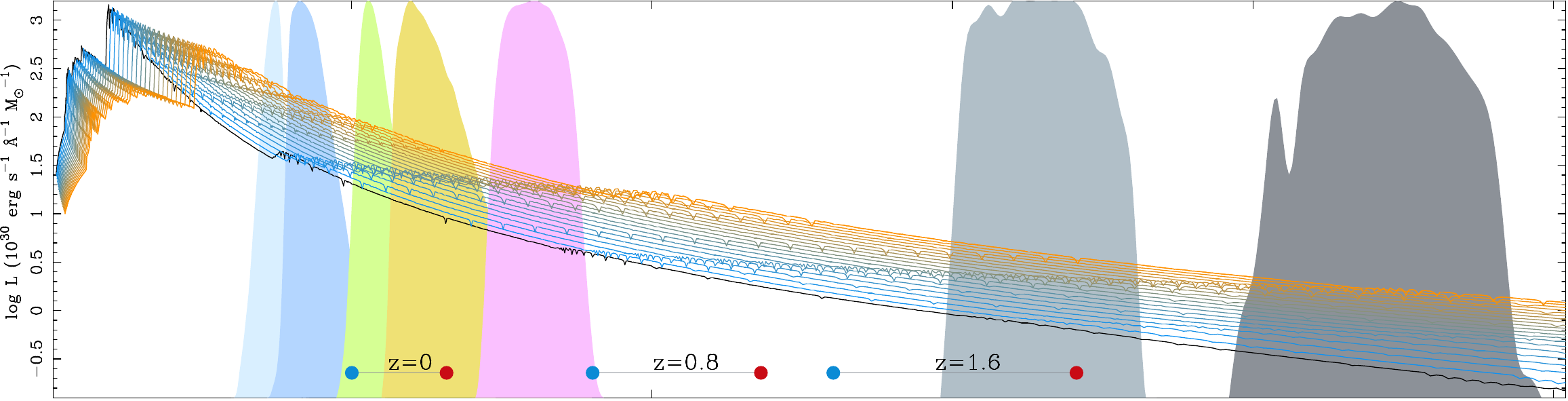}}
\put(3,0){\includegraphics[width=18.2cm]{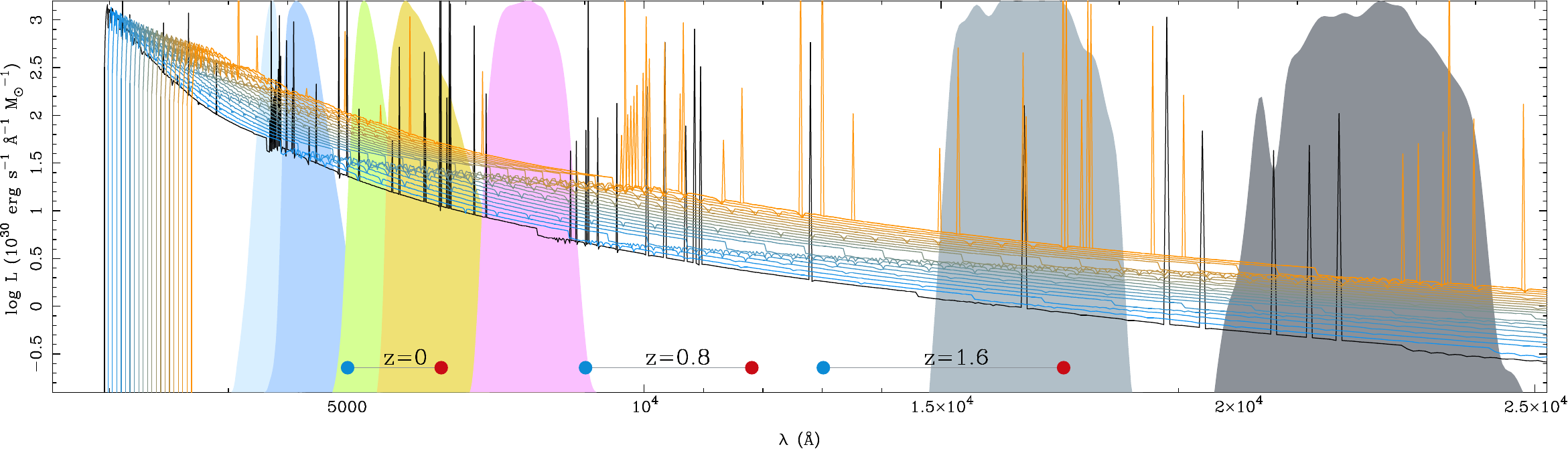}} % disk, 16 Myr, stellar+nebular emission
\PutLabel{75}{152}{\mlx \textcolor{black}{bulge (4 Gyr: stellar emission)}}
\PutLabel{75}{101.7}{\mlx \textcolor{black}{disk (16 Myr: stellar emission)}}
\PutLabel{75}{52.6}{\mlx \textcolor{black}{disk (16 Myr: stellar + nebular emission)}}
\PutLabel{27.4}{152}{\nlx \textcolor{black}{U}}
\PutLabel{31}{152}{\nlx \textcolor{black}{B}}
\PutLabel{39}{152}{\nlx \textcolor{black}{V}}
\PutLabel{44.7}{152}{\nlx \textcolor{black}{R}}
\PutLabel{59.4}{152}{\nlx \textcolor{black}{I}}
\PutLabel{120}{152}{\nlx \textcolor{black}{H}}
\PutLabel{164}{152}{\nlx \textcolor{black}{K}}
\PutLabel{177}{142}{\hvss \textcolor{black}{a}}
\PutLabel{177}{93}{\hvss \textcolor{black}{b}}
\PutLabel{177}{46}{\hvss \textcolor{black}{c}}
\end{picture}
\caption{Illustration of the effect of bandpass shift and wavelength stretching on the example of SEDs approximating the bulge (panel \brem{a})
and the disk (panels \brem{b} and \brem{c}) that are simulated from $z=0$ (black) to $z=1.6$ (dark orange).
Shaded areas show normalized transmission curves of the $U$, $B$, $V$, Cousins $R$ and $I$ and the NIR $H$ and $K$ filters. Horizontal bars mark the wavelength separation between the [O{\sc iii}]5007 and \ha\ emission lines (blue and red dot, respectively) at a redshift of 0, 0.8 and 1.6. The SEDs, computed with \pegase~2, refer to 1 \msun\ and a Salpeter IMF between 0.1 and 100 \msun.
\brem{a)} Stellar SED for a 4~Gyr old stellar population that forms with an exponentially decreasing SFR with an e-folding time of 0.5 Gyr ($\tau$0.5 model). It can be seen that, whereas the emission registered in the $K$ filter increases with increasing \zet, the ObsF emission in $U$ through $R$ decreases by more than 1 dex (cf. Fig.~\ref{sSED}\irem{g}).
\brem{b)} Stellar SED for a 16 Myr old stellar population of \zsun/5 forming at a constant SFR (contSF model).
\brem{c)} As in panel \brem{b}, however, with the contSF model including both stellar and nebular emission. The drop of the SED shortward of 912 \AA\ is because of the
reprocessing of the ionizing LyC radiation into nebular and dust emission \citep[see][for details]{FRV97}.
For the sake of better readability, emission lines are shown only for the SED at \zet=0 and \zet=1.6.}
\label{ap:synthSED}
\end{figure*}
\end{center}

% :::::::::::::::::::::::::::::::::::::::::::::::::::::::::::::::::::::::::::::::::::::::
Using the evolutionary synthesis code \pegase~2 \citep{FRV97} synthetic SEDs were computed for seven parameterizations for the SFH (cf. Fig.~\ref{fig:SFHs}), involving an exponentially decreasing SFR with an e-folding time $\tau$ of 0.5, 1 and 5 Gyr for solar metallicity \zsun\ (henceforth referred to as $\tau$0.5, $\tau$1, and $\tau$5, respectively), continuous SF at constant SFR for \zsun/5 (contSF), and three delayed-exponential SFR models, which are referred to as iA, iB, and iC. These three models are meant to simulate a downsizing scenario for galaxy bulges in low-, intermediate- and high-mass LTGs \citep[log(\mstar/\msun)$<$9.7, $\sim$10.5 and $>$10.7, respectively;][their Fig.~13; see also \tref{BP18}]{P22} for which the age of a bulge at its maximum SFR and the e-folding time for the ensuing exponential decline in the SFR scale inversely with the present-day stellar mass \mstar. In these delayed-exponential SFR models the dominant phase of bulge assembly occurs, respectively, 2.4, 1.2, and 0.47 Gyr after the Big Bang.
% SFHs
Synthetic SEDs (90 per SFH, covering an age between 0 and 13.7 Gyr, and a spectral range between 90 \AA\ and 160 $\mu$m)
for a Salpeter IMF between 0.1 and 100 \msun\ were computed in two versions, the first one involving purely stellar emission and the second  additionally nebular emission \citep[see][for details]{FRV97}. A zero intrinsic extinction and attenuation by the intergalactic medium have been assumed throughout.
This grid of synthetic SEDs was then simulated in the redshift range $0\leq$ \zet\ $\leq 3$ (cf. Fig.~\ref{ap:synthSED}) as $F(\lambda,z)=F_0(\lambda/(1+z))/(1+z),$ where $F(\lambda,z)$ (erg s$^{-1}$ cm$^{-2}$ \AA$^{-1}$) denotes the observed spectrum and $F_0$ that in the rest frame \citep[e.g.,][]{Boehm04}, and subsequently convolved with various filter transmission curves to compute ObsF magnitudes in the Vega system. These magnitudes, referred to in the following as {reduced} surface brightness $\mu\arcmin$, only take bandpass shift and wavelength stretching into account; that is to say, they are luminosity distance independent and treat the bulge and disk as point sources. Cosmological surface brightness dimming \citep{Tolman30,Tolman34} is not taken into account since it equally impacts the bulge and the disk and is thus unimportant for the evolution of the surface brightness contrast and the color contrast between the two. Leaving it out also makes Figs.~\ref{sSED}, \ref{gSED4Gyr}, and \ref{eSED} easier to interpret.

% FIGURE 1

An example of the decrease by $>$1 dex of the ObsF flux in the $U$, $B$, $V$, $R$ and $I$ filters for the passively evolving bulge of a higher-\zet\ galaxy is given in the upper panel of Fig.~\ref{ap:synthSED} where the rest-frame SED (black) of a 4~Gyr old stellar population forming according to the $\tau$0.5 model is simulated
at \zet=1.6 (dark orange). The increase in the ObsF flux in the optical for young higher-\zet\ populations (e.g., massive SF clumps forming out of VDIs) is illustrated in the middle panel where the purely stellar SED of a 16~Myr old stellar population (contSF) is shown. The combined nebular+stellar SED of this young stellar population (lower panel) illustrates the shift of various emission lines into the transmission curves of different filters as \zet\ increases.
\newpage
% :::::::::::::::::::::::::::::::::::::::::::::::::::::::::::::::::::::::::::::::::::::::::::::::::
\subsection{Evolutionary consistent simulations \label{ap:econsSED}} %\ccom{ap:econsSED}
% :::::::::::::::::::::::::::::::::::::::::::::::::::::::::::::::::::::::::::::::::::::::::::::::::
Estimates of ObsF colors of distant galaxies based on single-age SED templates have the disadvantage of not taking the spectrophotometric evolution of stellar populations into account and being dependent on the somehow subjective choice of SEDs. The EvCon approach \citep[e.g.,][]{Poggianti97,Contardo98,BFvA05}, in which the age of simulated SEDs
at a given \zet\ corresponds to the respective cosmic age provides an important supplement in this regard (cf. Sect.~\ref{sub:eSED}).

\begin{figure}
\begin{picture}(900,70)
\put(0,68){\includegraphics[angle=270, width=8.6cm]{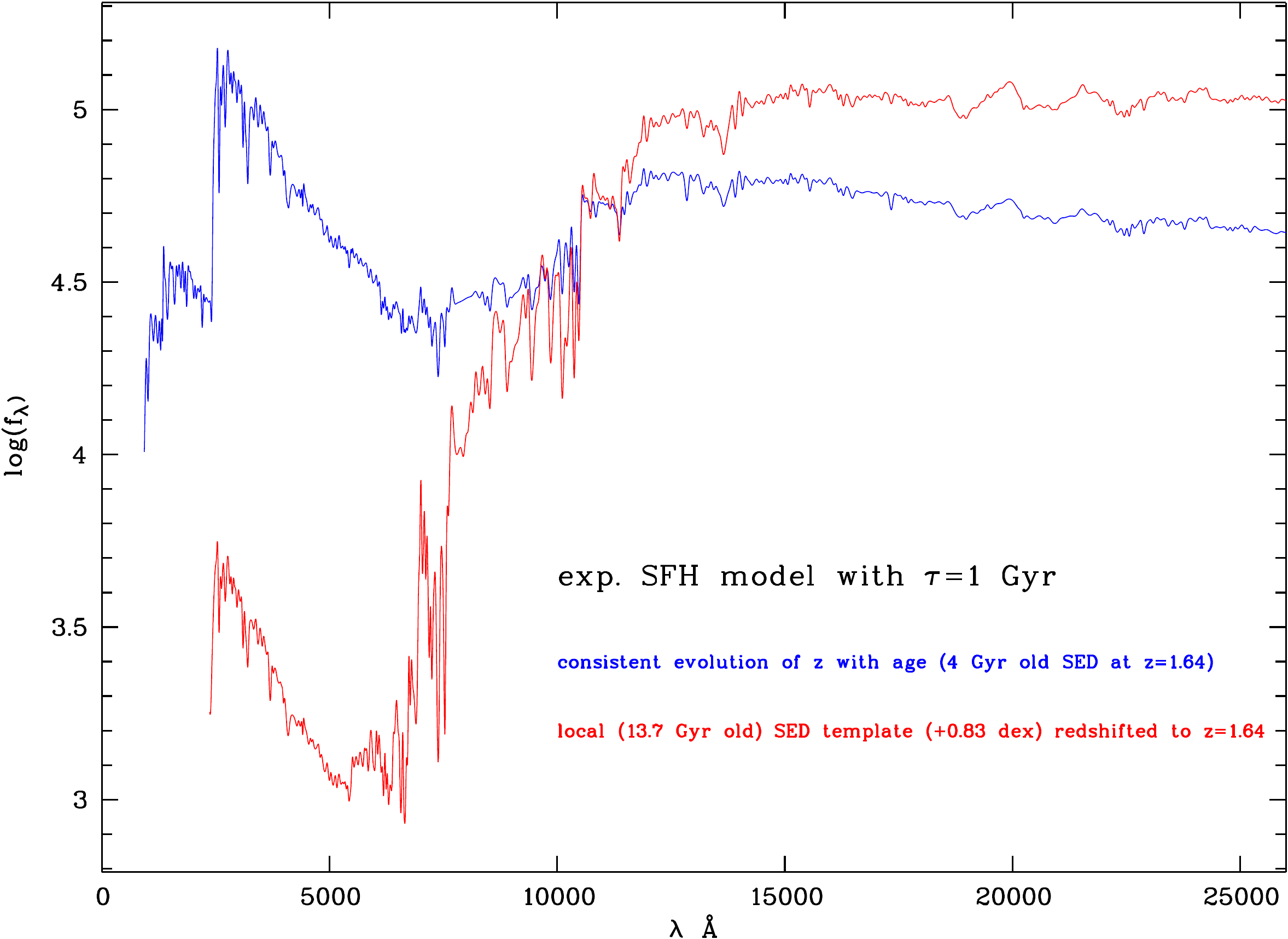}}
\end{picture}
\caption{Comparison of the SED of a 13.7 Gyr old bulge ($\tau$1 SFH model) that is simulated at \zet=1.64 (corresponding to a cosmic age of 4~Gyr; red)
with a 4~Gyr old SED for the same SFH and redshift (blue). To ease a comparison, the 13.7 Gyr old SED has been shifted by +0.83 mag to match the 4 Gyr old SED at the rest-frame wavelength of 4100 \AA. It can be seen that, whereas these SEDs do not strongly differ from one another in their slope for $\lambda > 1$$\mu$m, they do so in the ObsF optical
where the 4~Gyr old SED is by a factor of $\sim$5.8 brighter than the 13.7 Gyr old SED. This example illustrates the significant difference between \kc\ corrections computed in an EvCon manner and those based on SED templates with a fixed age.}
\label{fig:zconsSEDs2}
\end{figure}

Figure~\ref{fig:zconsSEDs2} gives an example of the difference of EvCon models compared to those based on single-age SEDs. It compares the SED of a 13.7 Gyr old bulge forming according to the $\tau$1 model when redshifted to \zet=1.64 (i.e., back to a cosmic age of 4 Gyr) with that of a 4~Gyr old bulge at the same redshift.
The first SED was vertically shifted by +0.83 dex to normalize it to the rest-frame flux of the second one at 4100 \AA\ to ease comparison: it can be appreciated that, although both SEDs have a similar slope for $\lambda>$1$\mu$m, they differ by more than 2 dex in the rest-frame NUV.
This illustrates that \kc\ corrections and rest-frame colors estimated from SED templates with a single-age can substantially differ from those computed in an EvCon manner, which is especially true for passively evolving stellar populations. On the other hand, as discussed in Sect.~\ref{sub:eSED}, differences between these two approaches
are comparatively small in the case of continuous SF at a constant SFR.

\begin{figure*}
\begin{picture}(200,132)
\put(0,114){\includegraphics[clip, width=18.96cm]{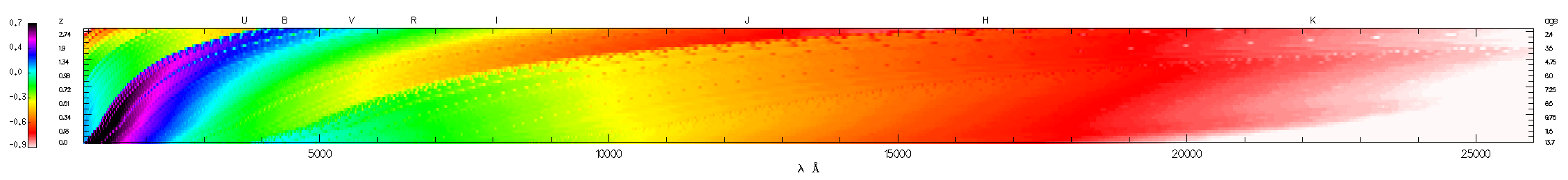}}
\put(0,92){\includegraphics[clip, width=18.6cm]{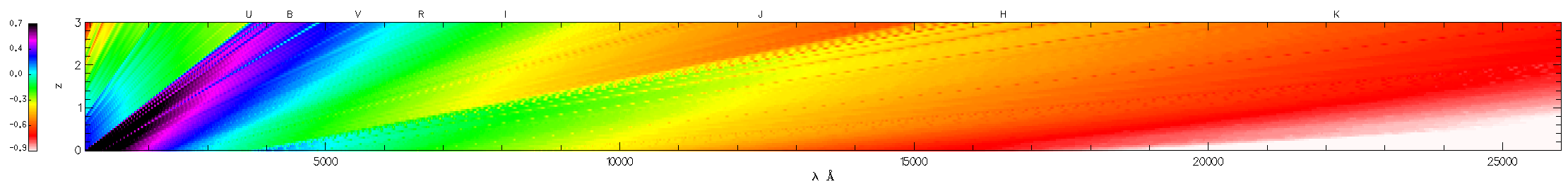}}
\put(0,70){\includegraphics[clip, width=18.6cm]{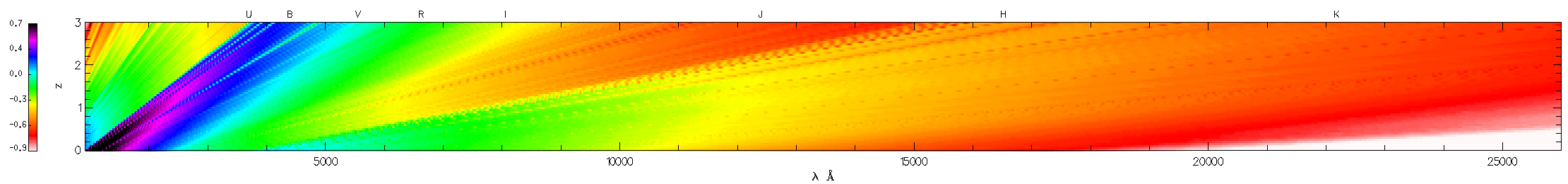}}
\put(0,44){\includegraphics[clip, width=18.96cm]{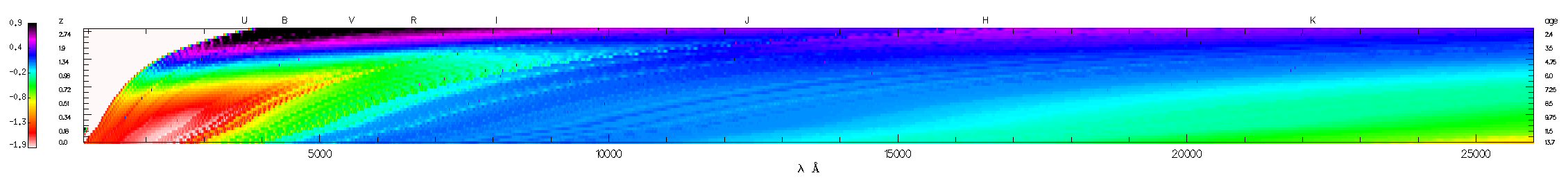}}
\put(0,22){\includegraphics[clip, width=18.6cm]{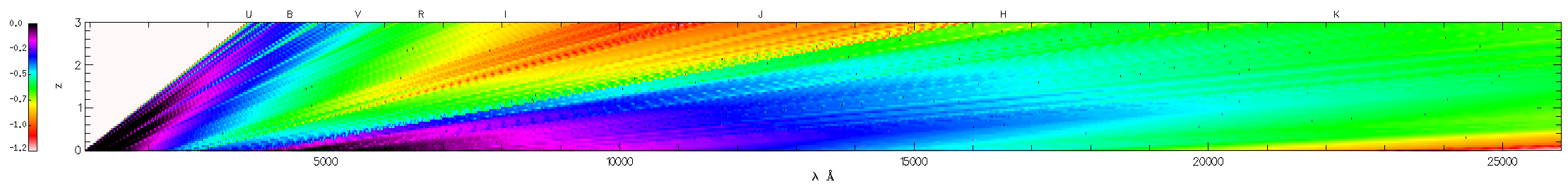}}
\put(0,0){\includegraphics[clip, width=18.6cm]{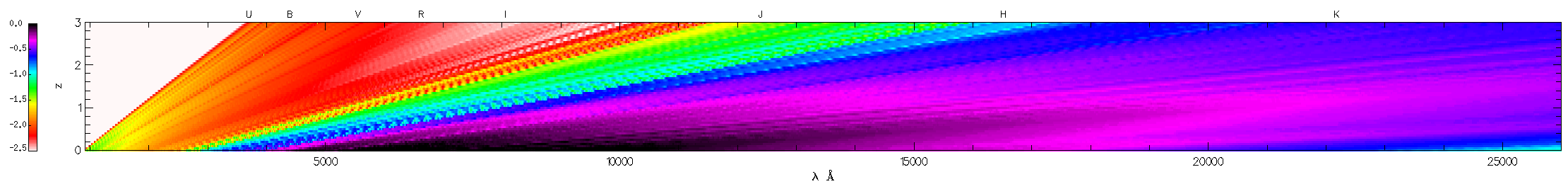}}
\PutLabel{181}{120}{\hvss \textcolor{black}{a}}
\PutLabel{181}{97}{\hvss \textcolor{black}{b}}
\PutLabel{181}{75}{\hvss \textcolor{black}{c}}
\PutLabel{12}{57}{\hvss \textcolor{black}{d}}
\PutLabel{12}{37}{\hvss \textcolor{black}{e}}
\PutLabel{12}{14}{\hvss \textcolor{black}{f}}
\end{picture}
\caption{2D illustration of the difference between EvCon simulations and those based on single-age SED templates.
All panels cover an observers-frame wavelength range from 912 \AA\ to 2.6 $\mu$m for a redshift 0$\leq$\zet$\leq$3 (from bottom to top).
The SEDs, displayed in a logarithmic representation, take only bandpass shift and wavelength stretching into account and are normalized
to the flux density at $\sim$5500 \AA\ of the SED at \zet=0 (bottom line in each panel) to ease a comparison.
Labels at the upper abscissa mark the central wavelength $\lambda_{\rm c}$ of the $U$, $B$, $V$, $R$, $I$, $J$, $H,$ and $K$ filters.
\brem{a)} EvCon simulation that uses a grid of (42) purely stellar SEDs covering an age between 2.2 Gyr (\zet=3; upper row) and 13.7 Gyr (\zet=0; lower row)
for the contSF model and a stellar metallicity of \zsun/5.
\brem{b)} Simulation based on a 4~Gyr old SED that is simulated out to \zet=3 in increments of 0.05.
\brem{c)} As in panel \brem{b} for a SED corresponding to an age of 13.7 Gyr.
Panels \brem{d}-\brem{f} follow the same layout yet refer to an exponentially decreasing SFR with an e-folding time $\tau=1$ Gyr that started
13.7 Gyr ago and additionally include nebular emission. The blank area at the upper-left is due to the reprocessing of the LyC radiation ($<$912 \AA) into nebular and dust emission.}
\label{fig:zconsSEDs1}
\end{figure*}

Another illustration is provided in Fig.~\ref{fig:zconsSEDs1} where we compare a single-age SED simulated between \zet=0 and \zet=3 (panels \brem{b}-\brem{c} and \brem{e}-\brem{f}; SEDs are arranged from bottom to top in order of increasing \zet) with a grid of SEDs computed in an EvCon manner and covering an age between 2.2 and 13.7 Gyr
for a redshift between 3 and 0 (panels \brem{a} and \brem{d}; left-hand labels inform about the age in Gyr).
The first three panels refer to purely stellar models, whereas the models in the three lower panels include both stellar and nebular emission.
Panels \brem{b} and \brem{c} show, respectively, simulations for a 4~Gyr and 13.7~Gyr old SED forming according to contSF, and panels \brem{d}-\brem{f} refer to the same ages
for the $\tau$1 model. As pointed out before, the blank area in the upper-left of panels \brem{d}--\brem{f} is due to the reprocessing of the ionizing LyC radiation
into nebular and dust emission.

\begin{center}
\begin{figure*}[h]
\begin{picture}(200,217)
\put(0,165){\includegraphics[clip, width=8.6cm]{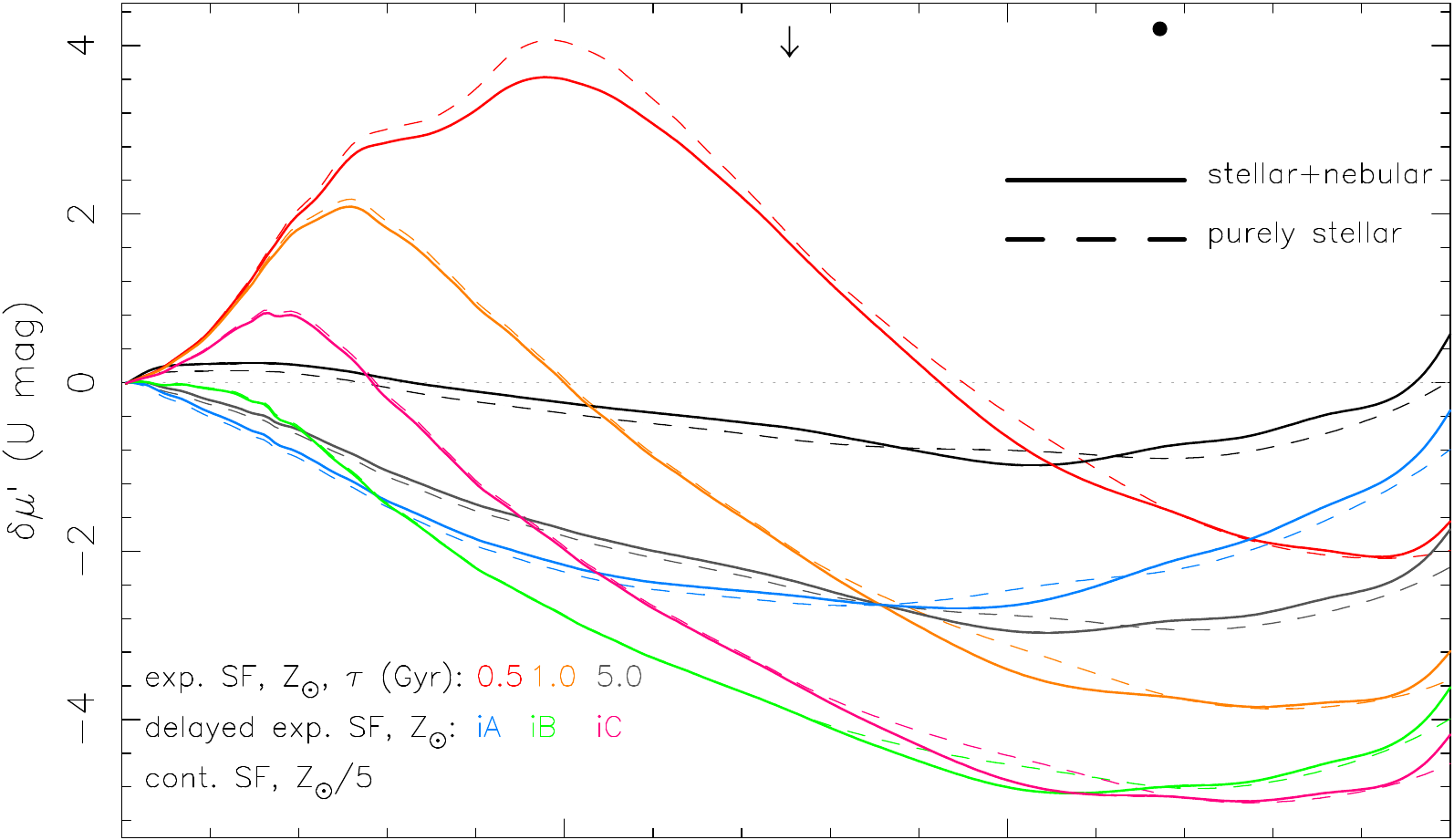}}
\put(0,111){\includegraphics[clip, width=8.6cm]{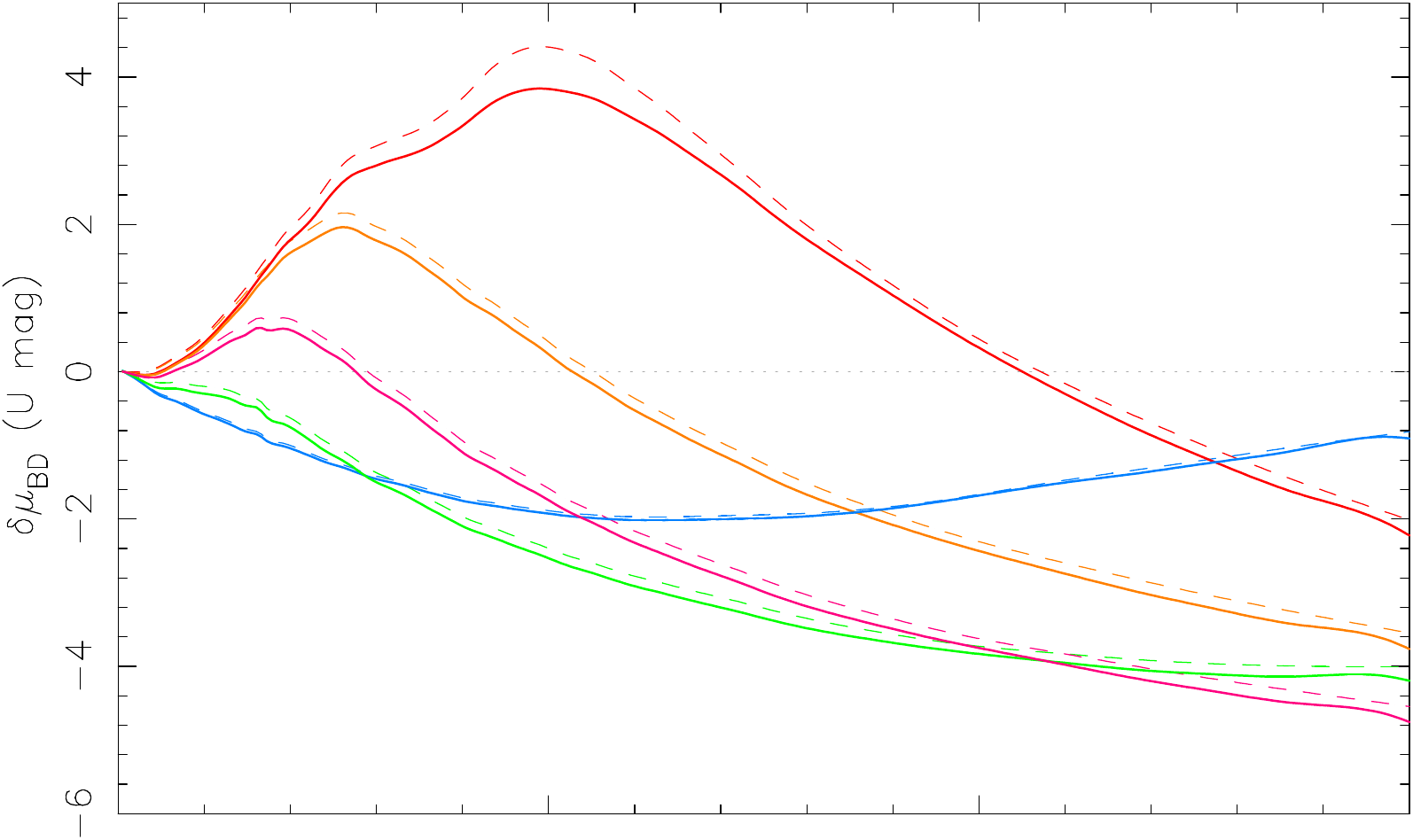}}
\put(0,59){\includegraphics[clip, width=8.6cm]{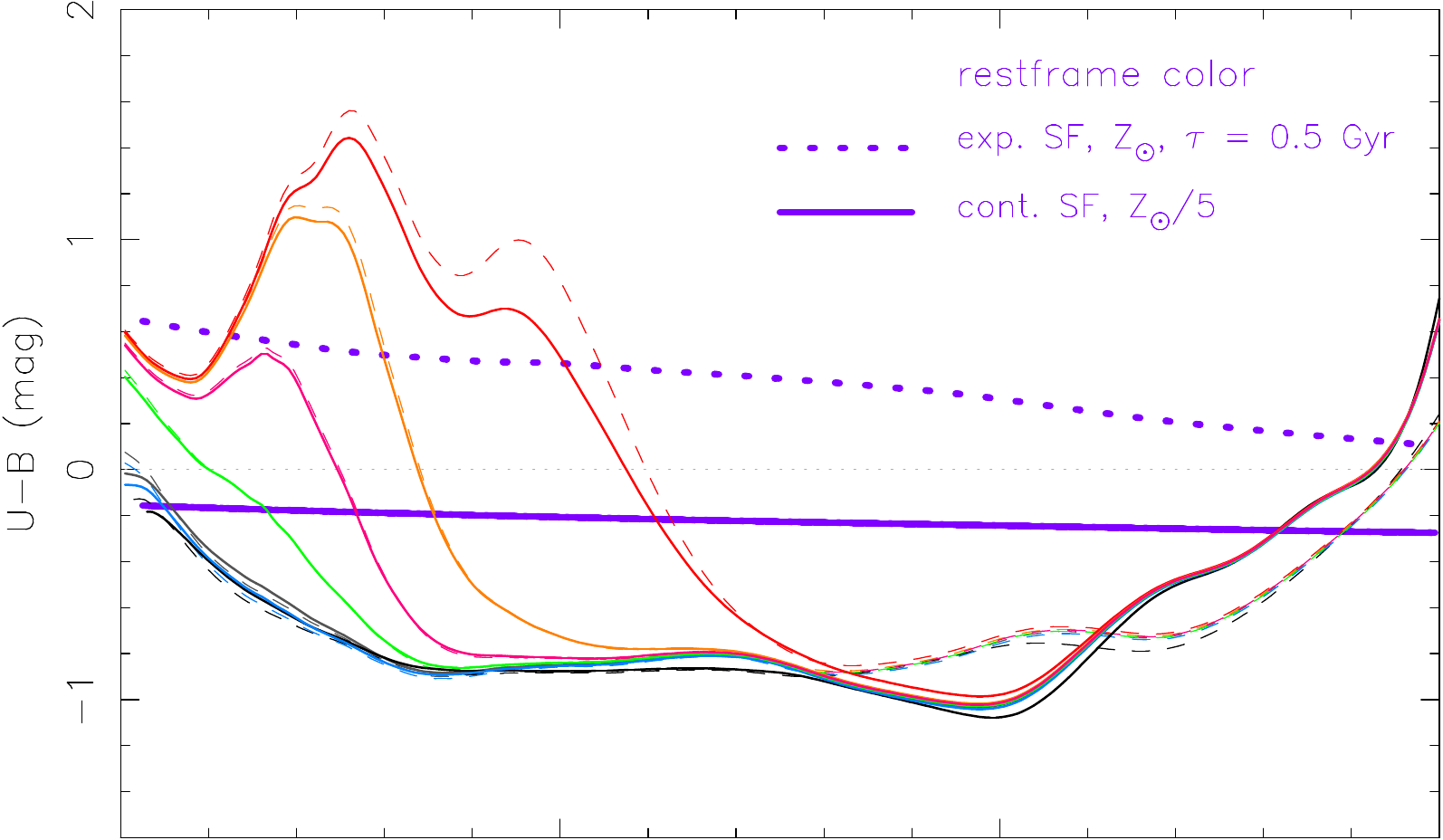}}
\put(0,0){\includegraphics[clip, width=8.63cm]{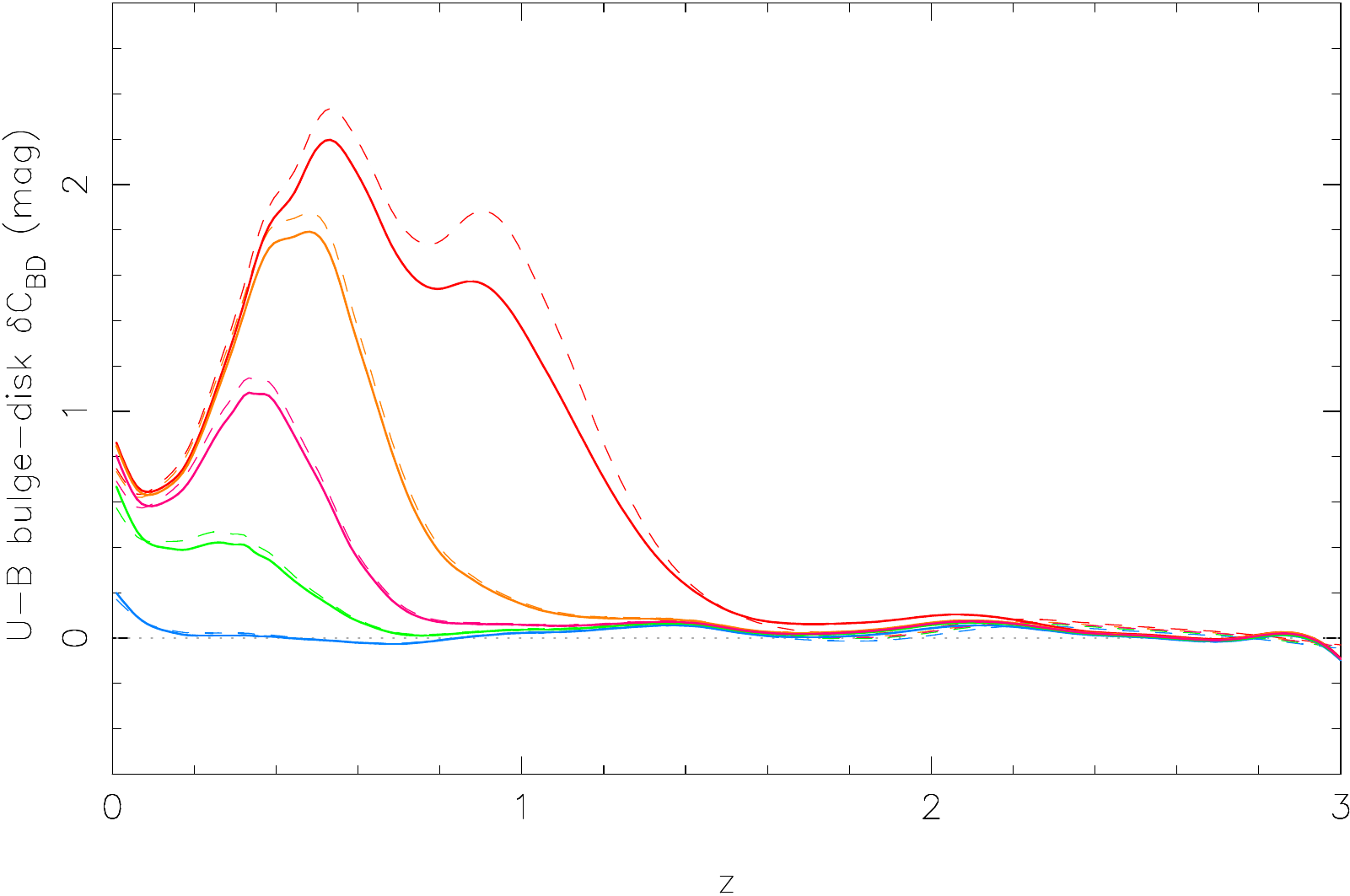}}
\put(98,165){\includegraphics[clip, width=8.6cm]{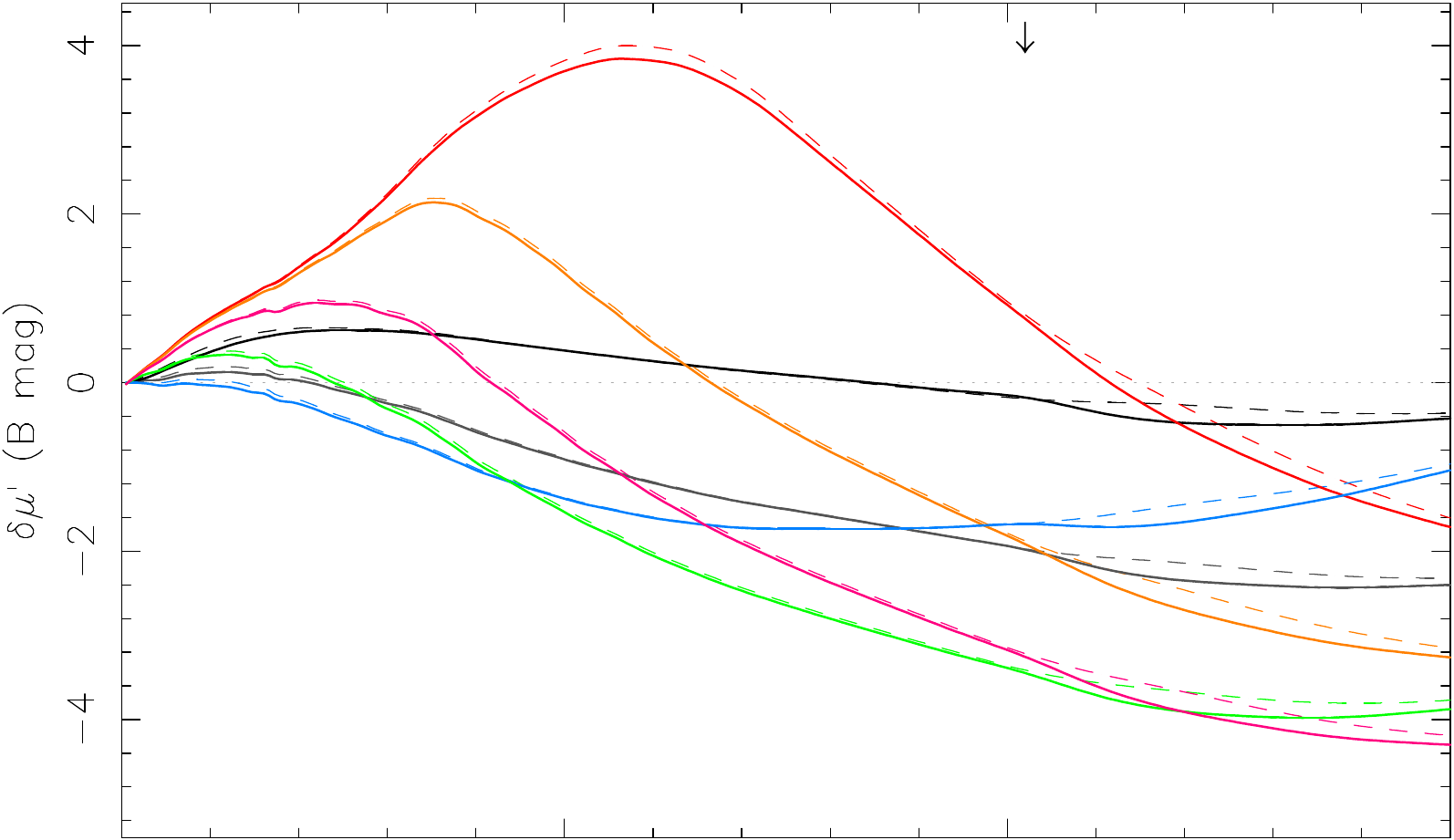}}
\put(98,111){\includegraphics[clip, width=8.6cm]{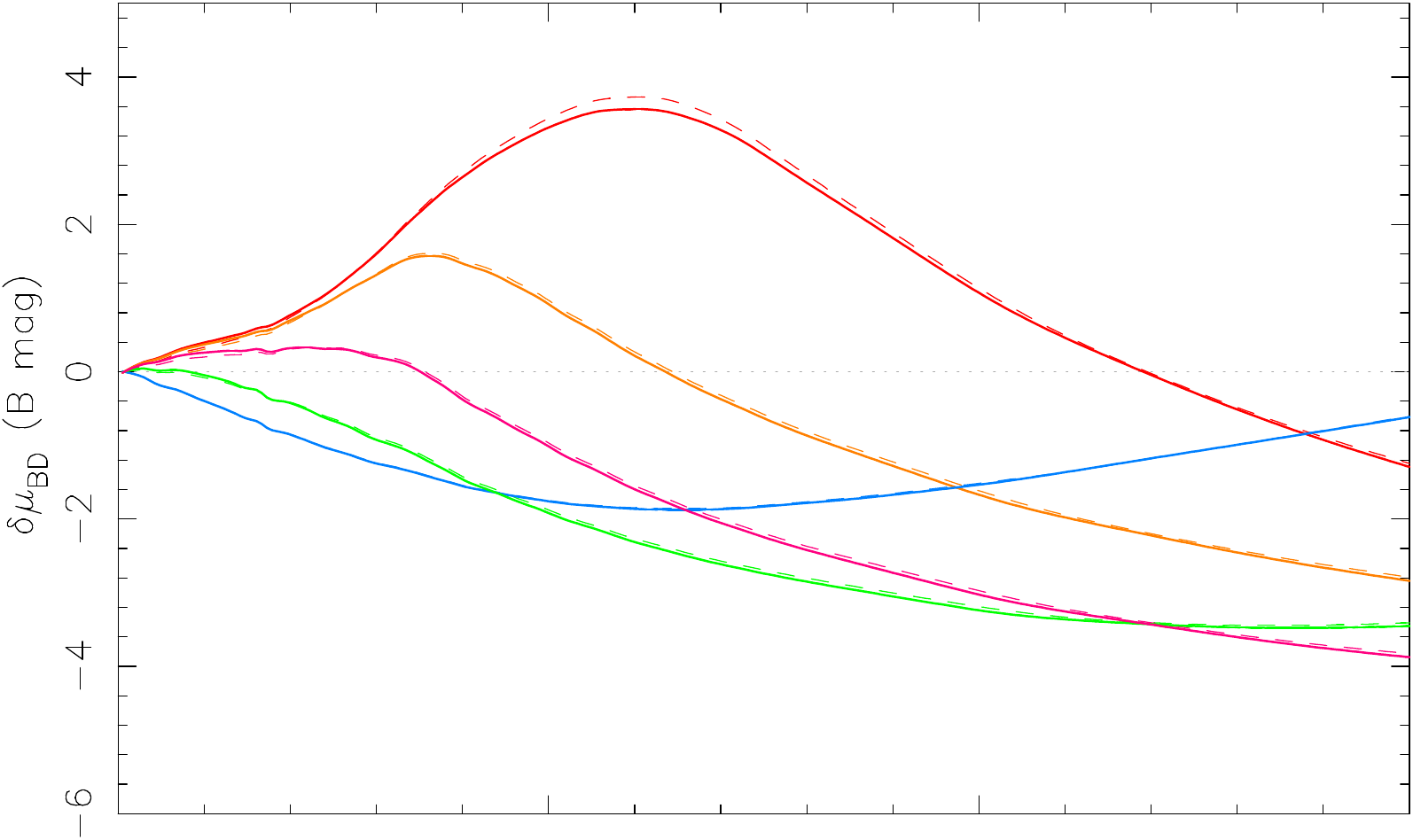}}
\put(98,59){\includegraphics[clip, width=8.6cm]{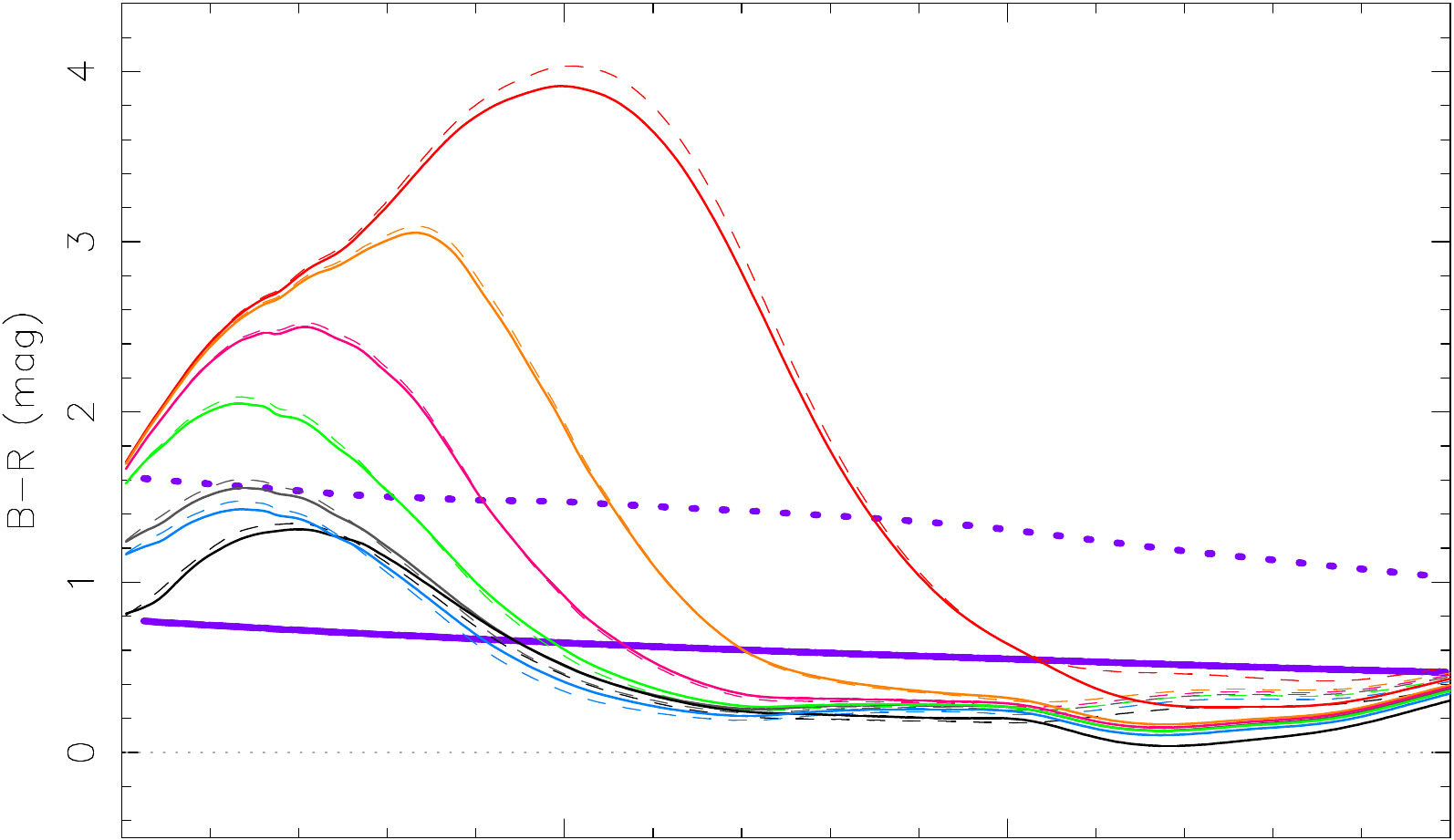}}
\put(98,0){\includegraphics[clip, width=8.63cm]{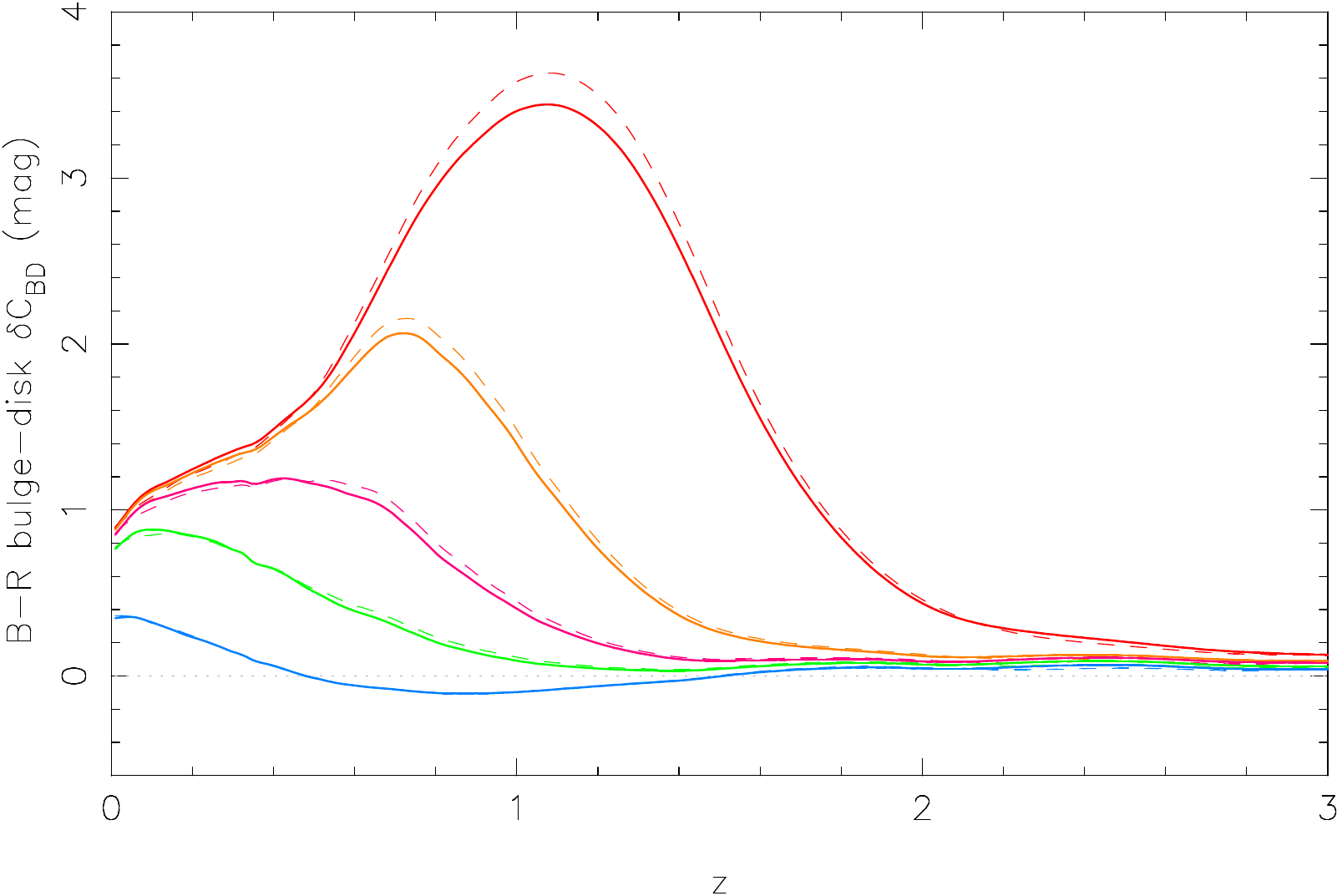}}
\PutLabel{87}{212}{\hvss \textcolor{black}{a}}
\PutLabel{87}{158}{\hvss \textcolor{black}{b}}
\PutLabel{87}{106}{\hvss \textcolor{black}{c}}
\PutLabel{87}{52}{\hvss \textcolor{black}{d}}
\PutLabel{185}{212}{\hvss \textcolor{black}{e}}
\PutLabel{185}{158}{\hvss \textcolor{black}{f}}
\PutLabel{185}{106}{\hvss \textcolor{black}{g}}
\PutLabel{185}{52}{\hvss \textcolor{black}{h}}
\end{picture}
\caption{Predictions from EvCon models for SEDs including nebular emission (solid curves) and purely stellar SEDs (dashed curves)
for the parametric SFHs in Fig.~\ref{fig:SFHs}. Zero intrinsic extinction and attenuation by the intergalactic medium have been assumed throughout.
\brem{a:} Variation \dmu\ ($U$ mag) of the reduced surface brightness $\mu$\arcmin. The arrow and bullet mark the \zet\ at which the Ly$\alpha$ (1216 \AA) line and the LyC edge (912 \AA) enter the blue edge of the $U$ filter (1.508 and 2.34, respectively).
\brem{b:} bulge-to-disk surface brightness contrast \dmBD\ ($U$ mag) when assuming the contSF SFH scenario for the disk. \brem{c:} ObsF $U$-$B$ color.
The evolution of the rest-frame color for the contSF and $\tau$0.5 SFH scenario is shown by the purple solid and dotted curve, respectively). \brem{d:} ObsF bulge-to-disk color contrast in $U$-$B$. The meaning of panels \brem{e}-\brem{h} is analogous.}
\label{ub:zcons}
\end{figure*}
\end{center}

\begin{center}
\begin{figure*}[h]
\begin{picture}(200,224)
\put(0,165){\includegraphics[clip, width=8.6cm]{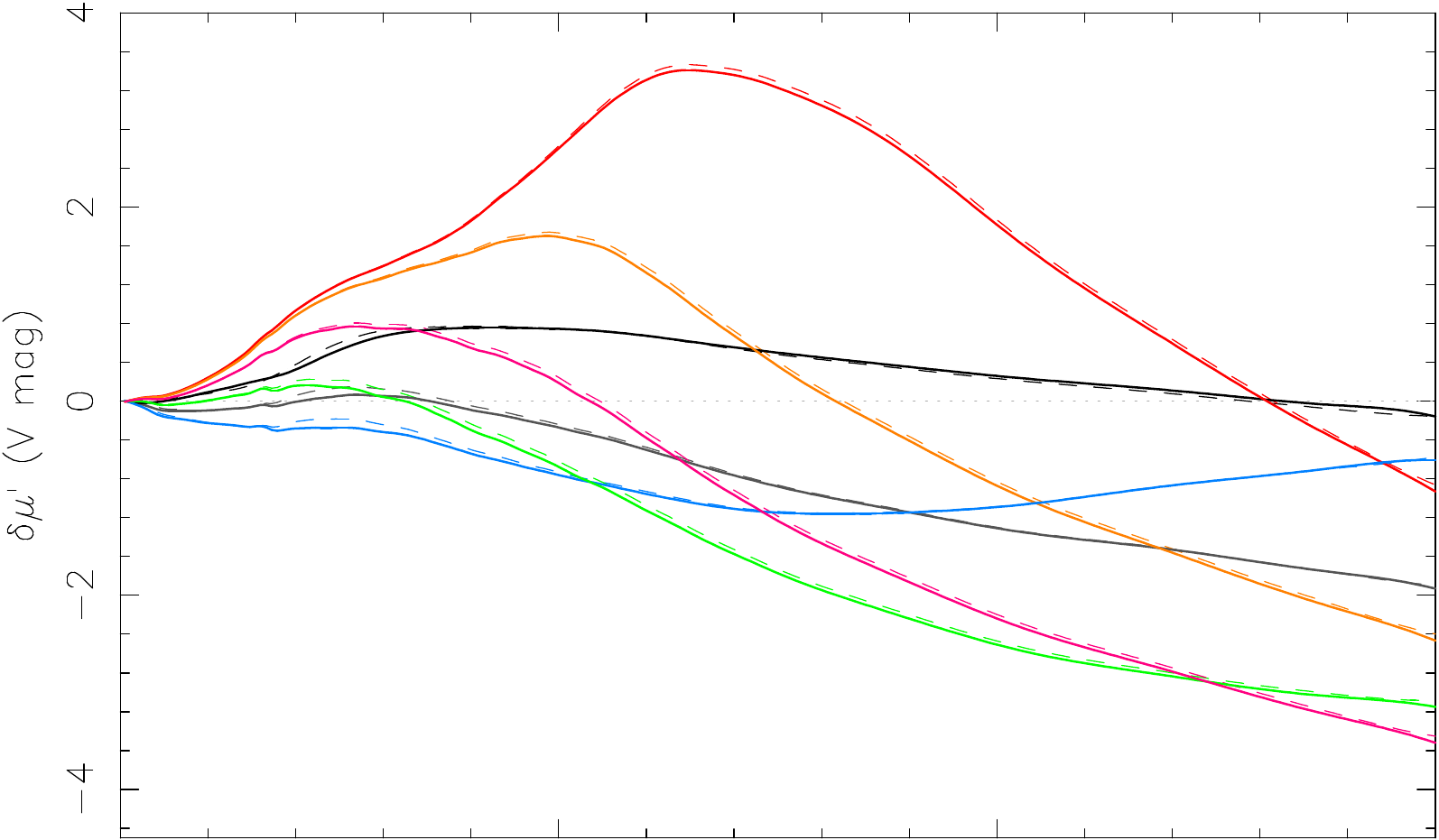}}
\put(0,111){\includegraphics[clip, width=8.6cm]{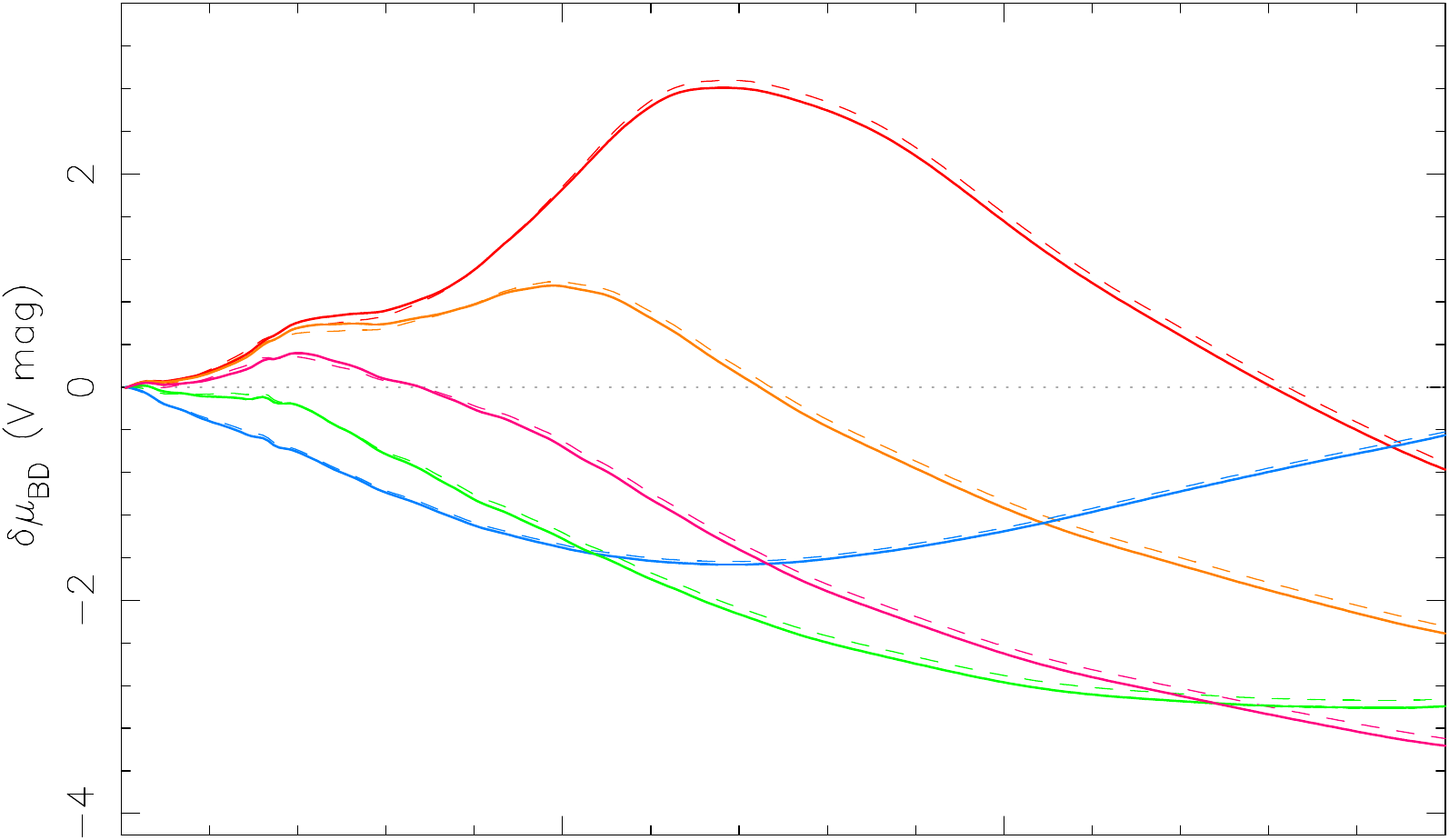}}
\put(0,59){\includegraphics[clip, width=8.6cm]{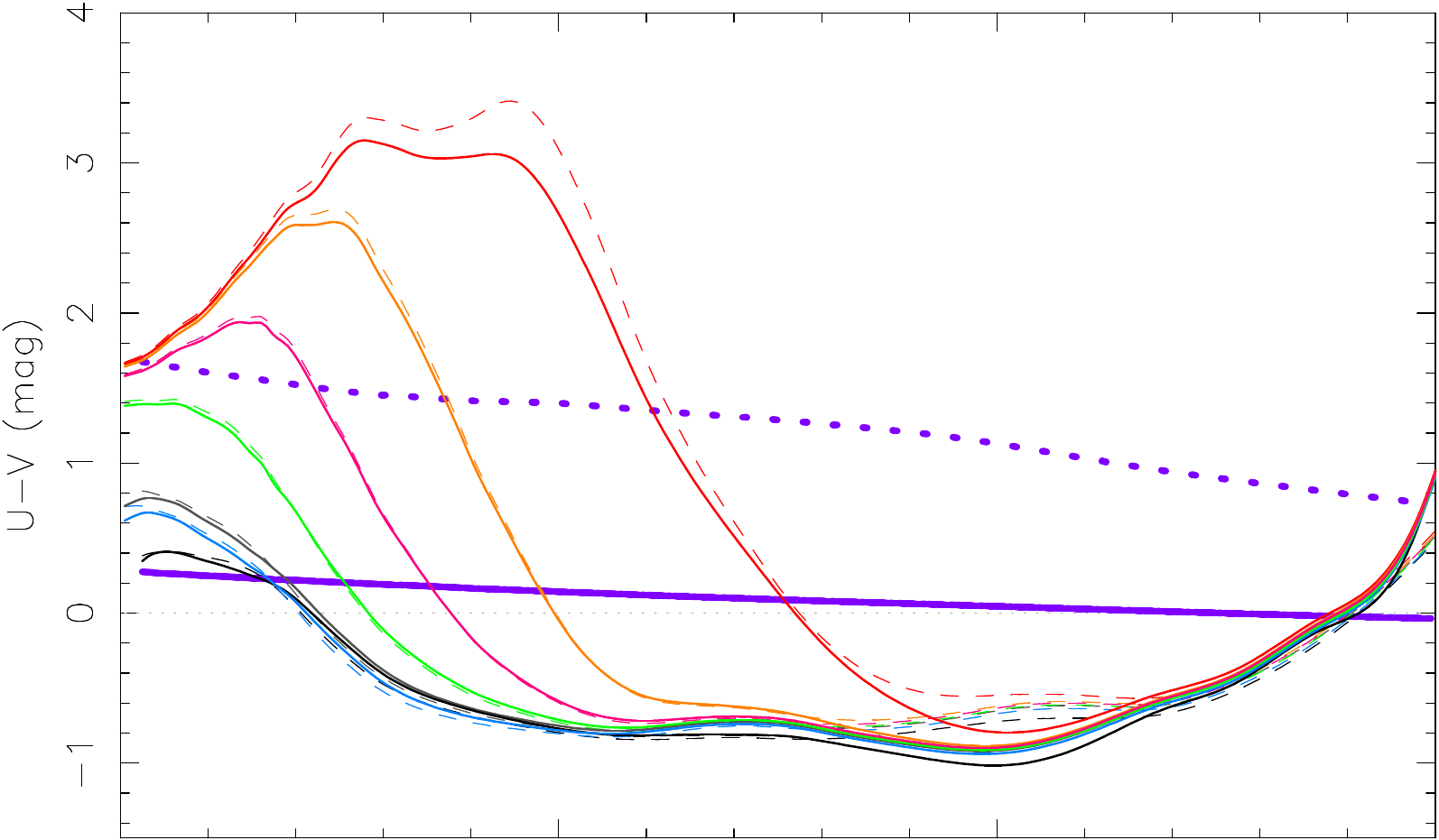}}
\put(0,0){\includegraphics[clip, width=8.63cm]{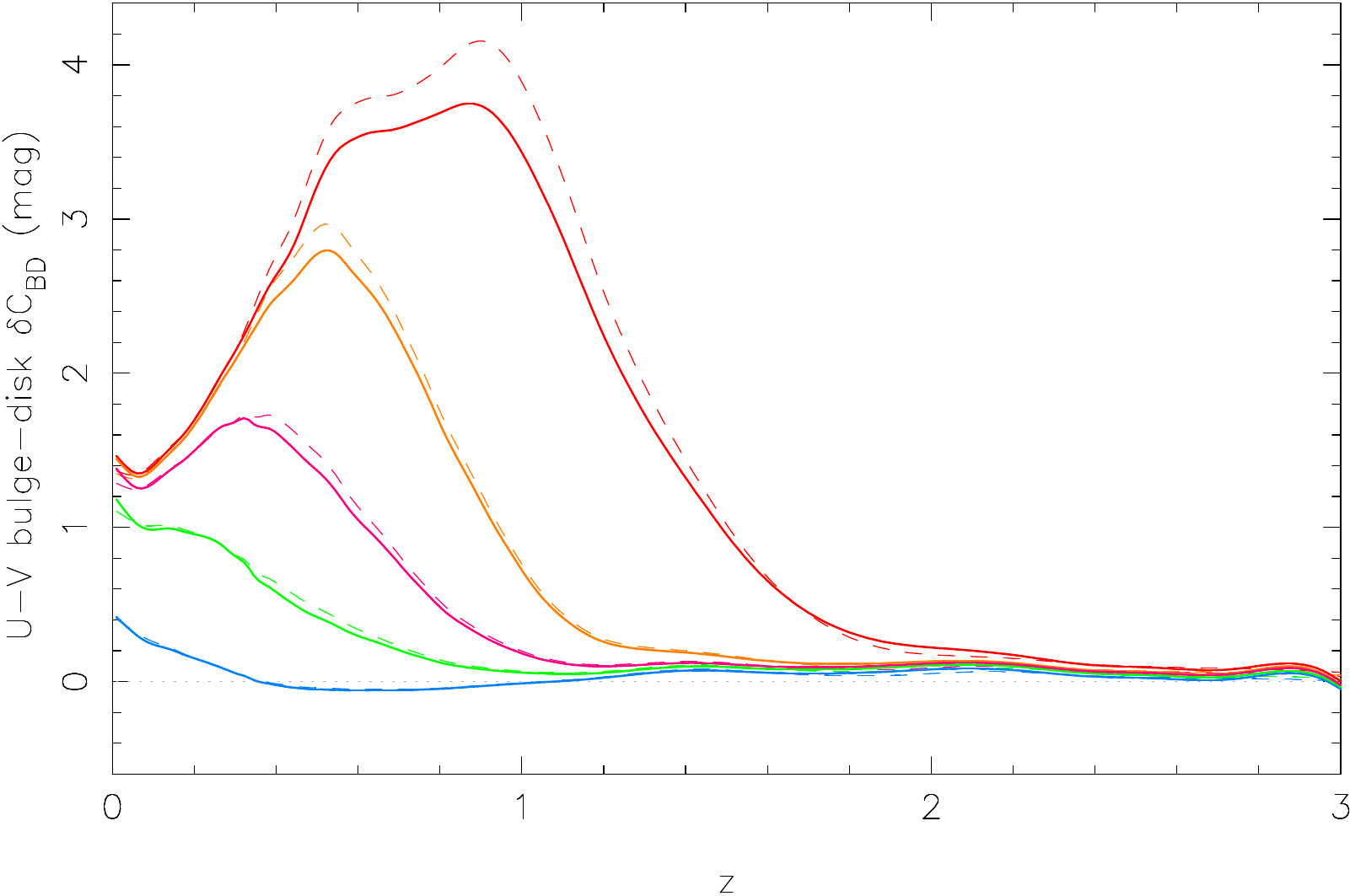}}
\put(98,165){\includegraphics[clip, width=8.6cm]{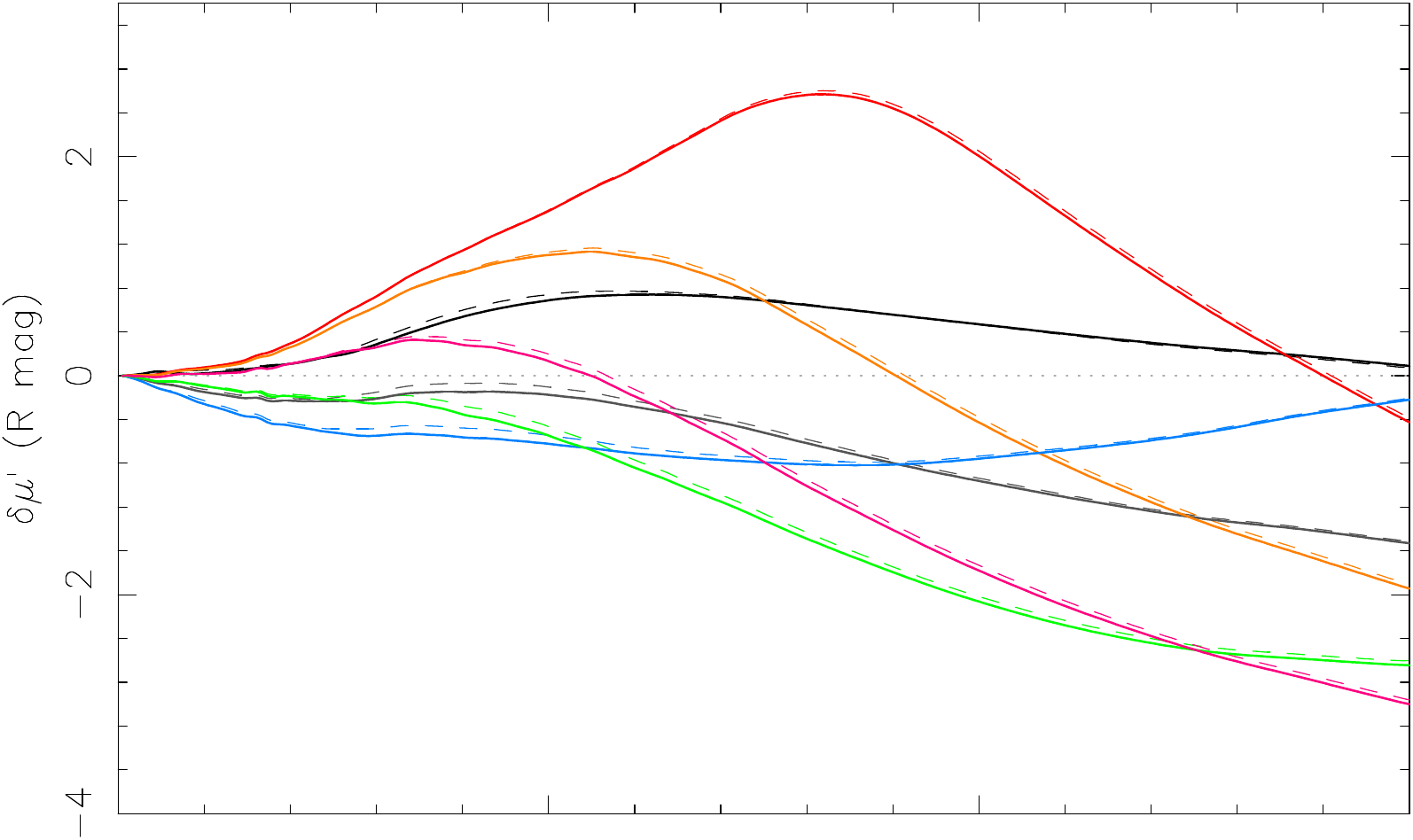}}
\put(98,111){\includegraphics[clip, width=8.6cm]{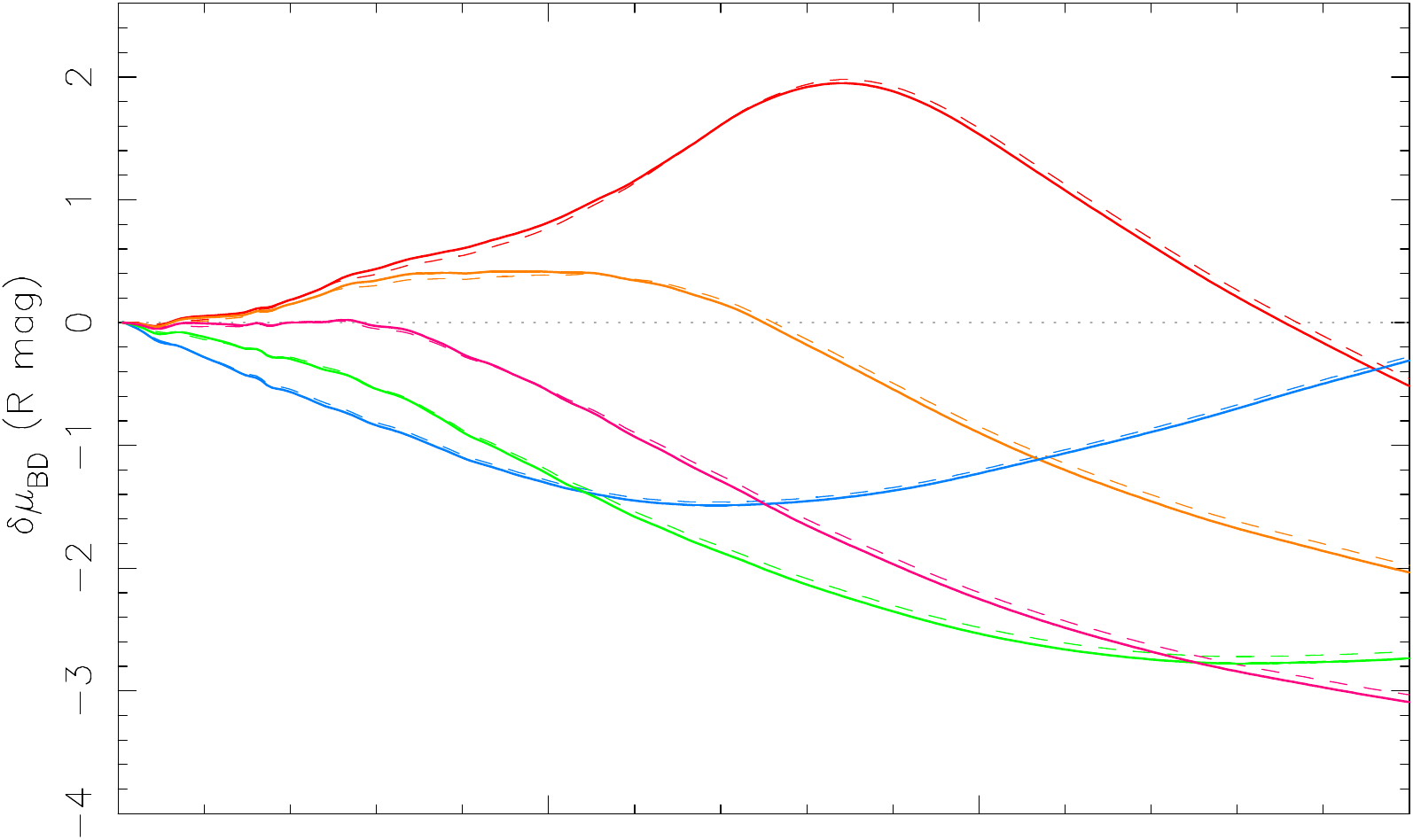}}
\put(98,59){\includegraphics[clip, width=8.6cm]{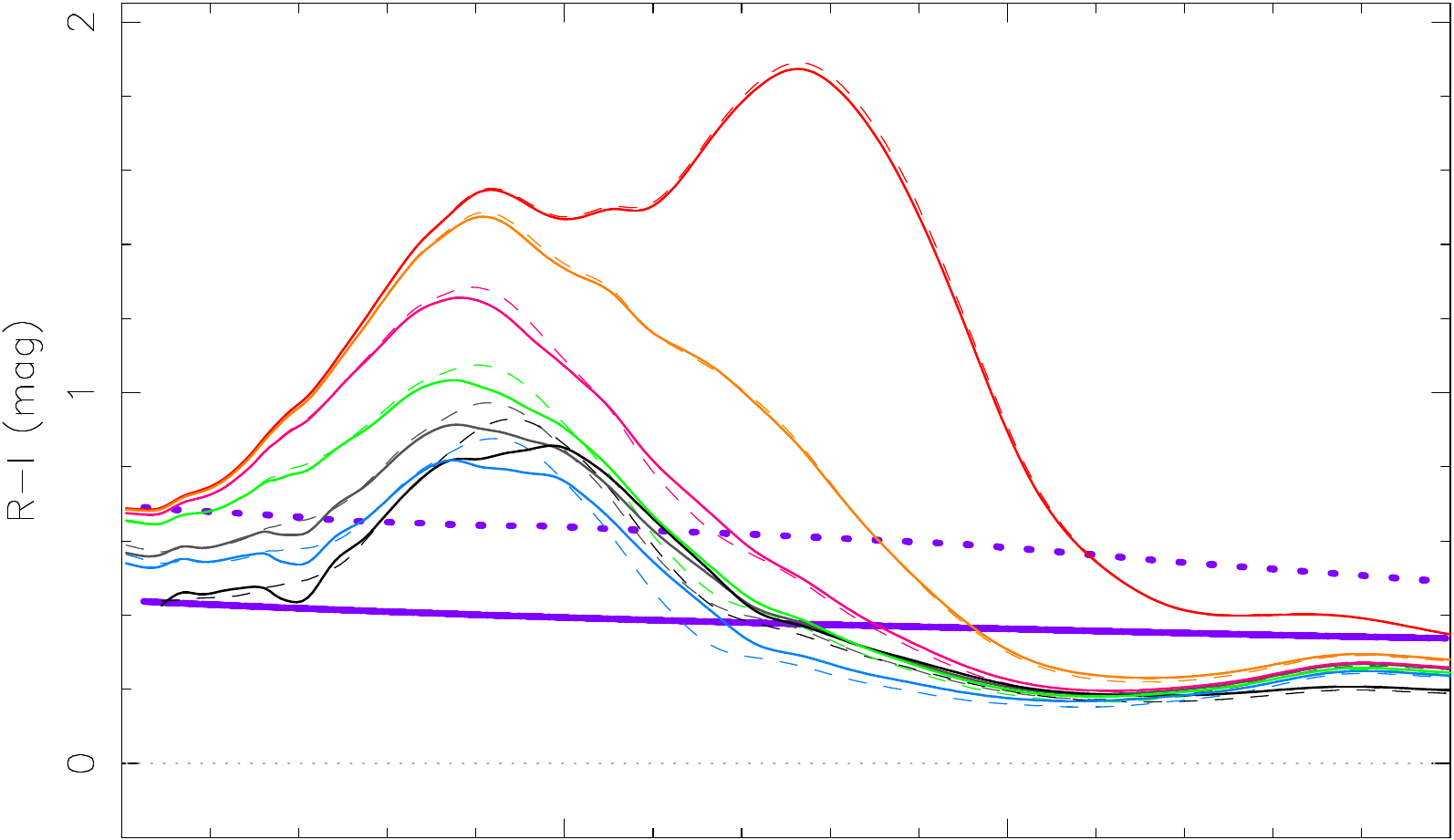}}
\put(98,0){\includegraphics[clip, width=8.63cm]{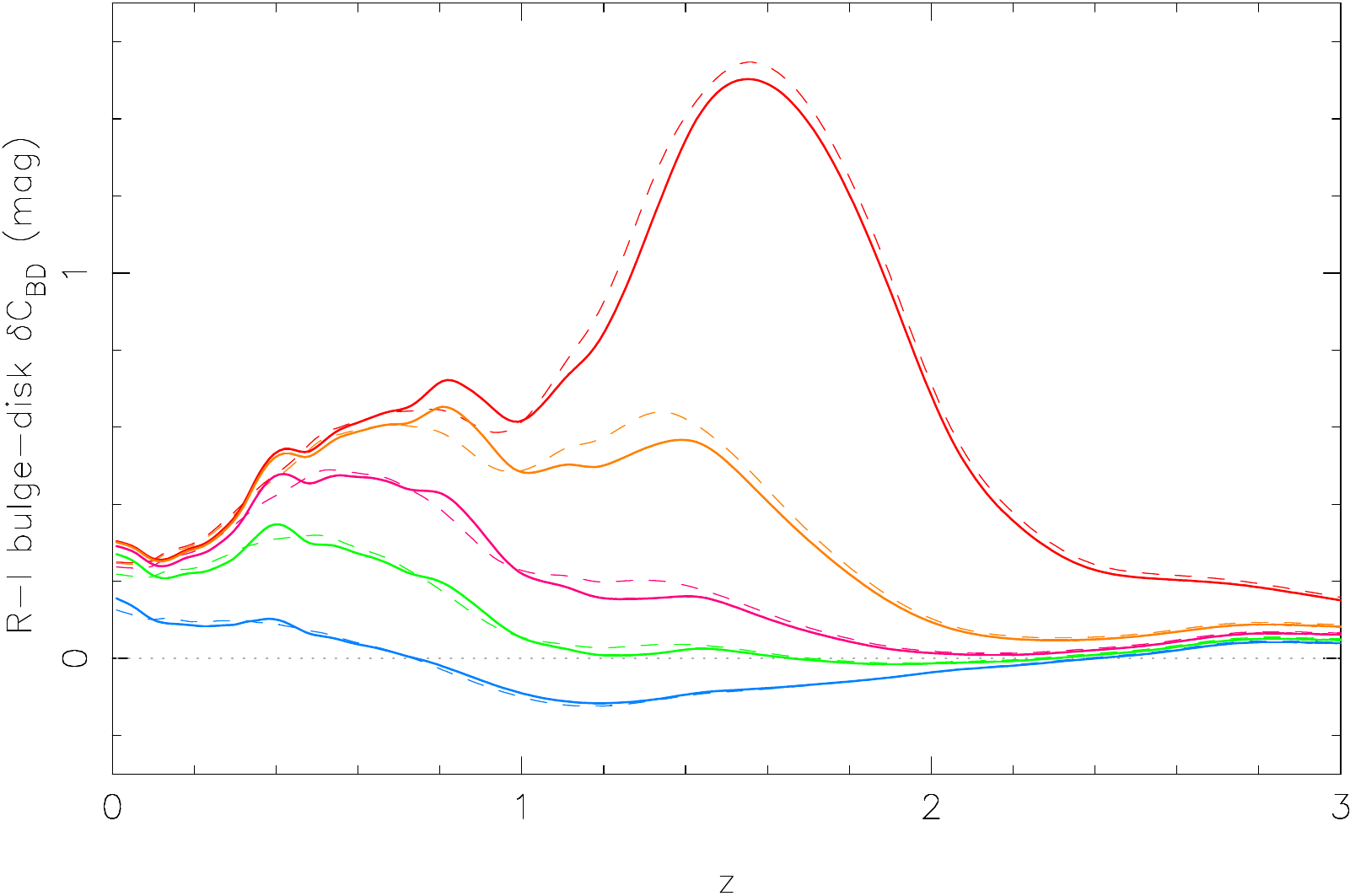}}
\end{picture}
\noindent{{\bf Fig. A.5.} Continued.}
\label{vr:zcons} 
\end{figure*}
\end{center}

\begin{center}
\begin{figure*}[h]
\begin{picture}(200,224)
\put(0,165){\includegraphics[clip, width=8.6cm]{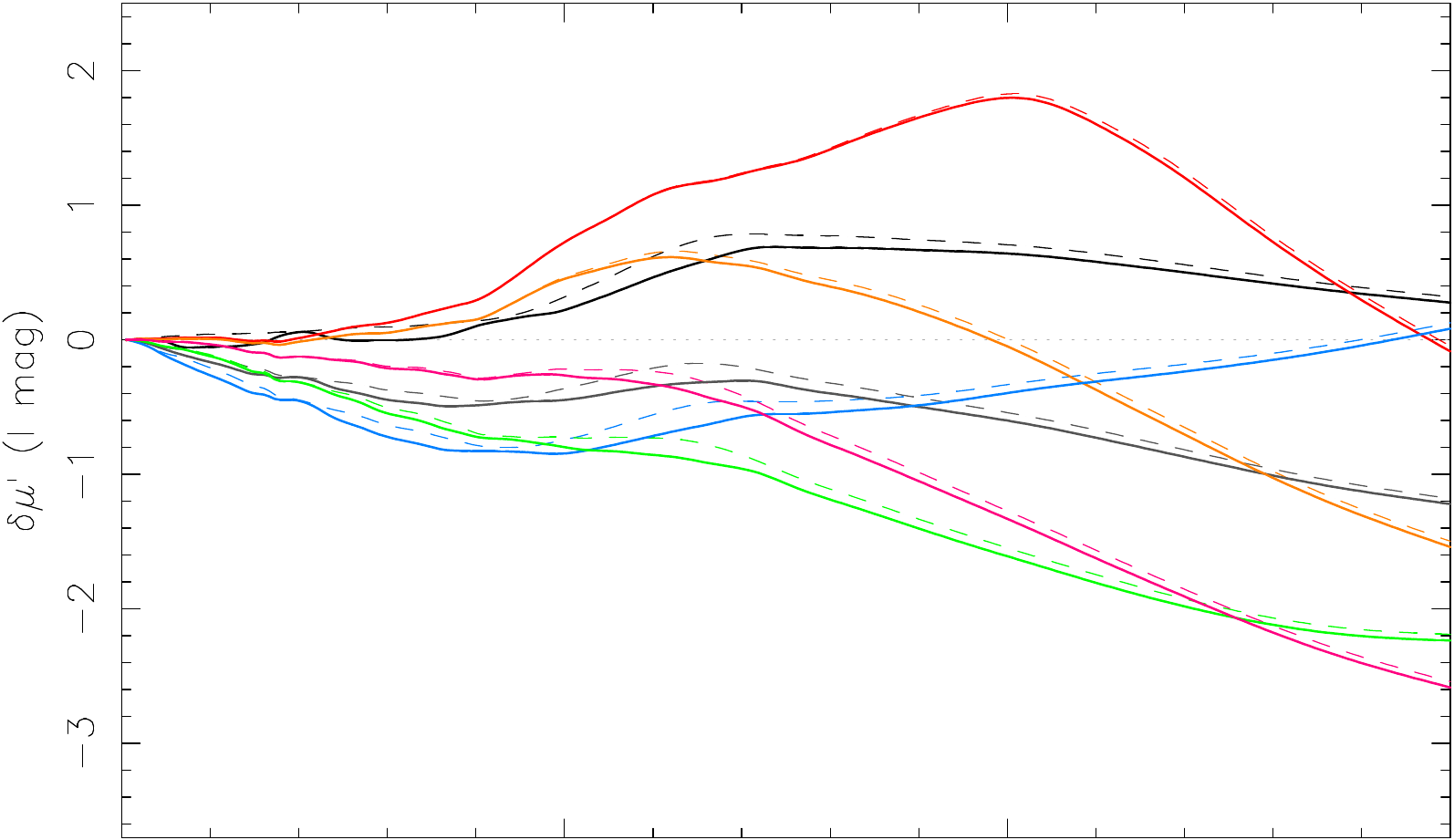}}
\put(0,111){\includegraphics[clip, width=8.6cm]{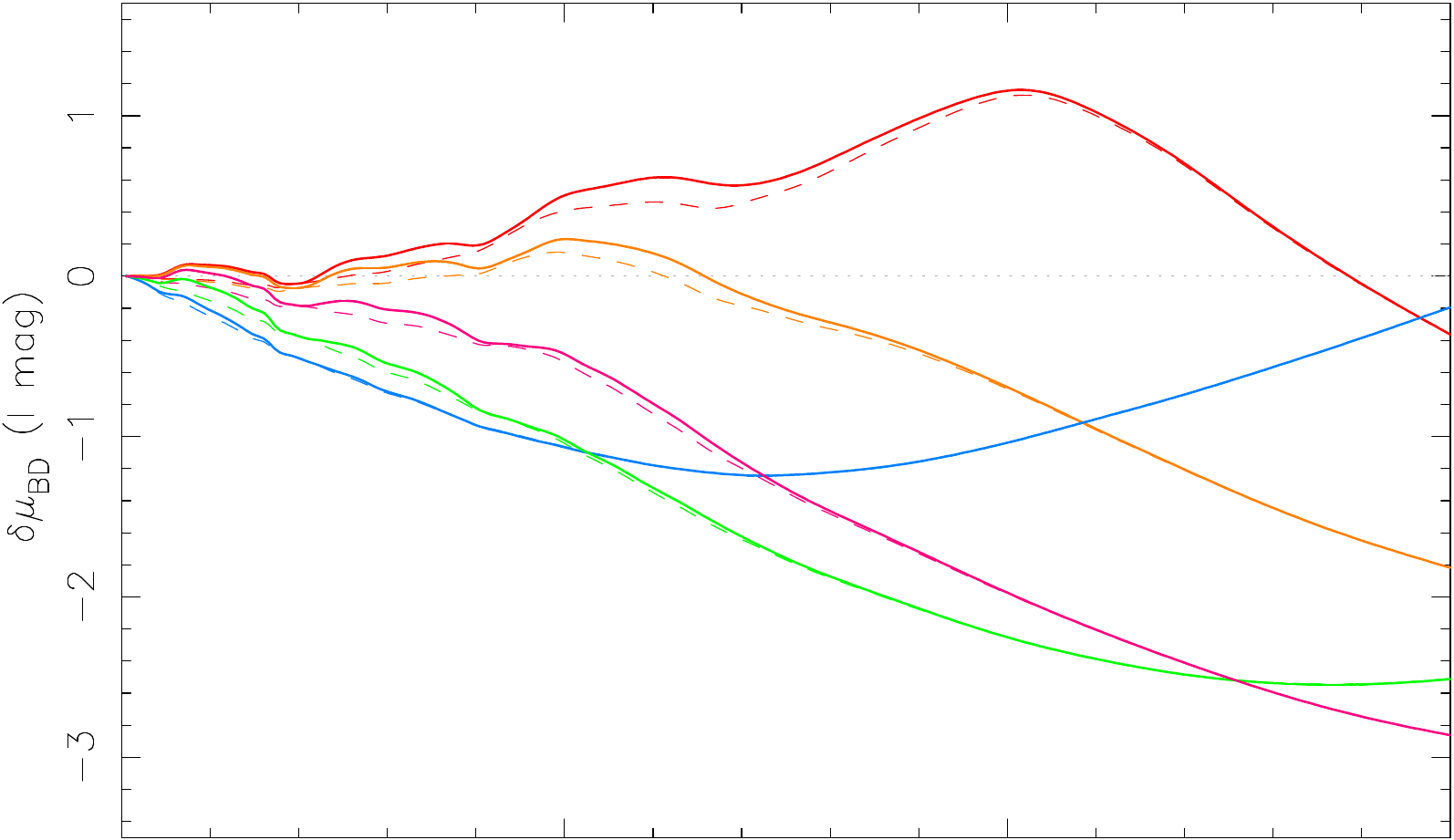}}
\put(0,59){\includegraphics[clip, width=8.6cm]{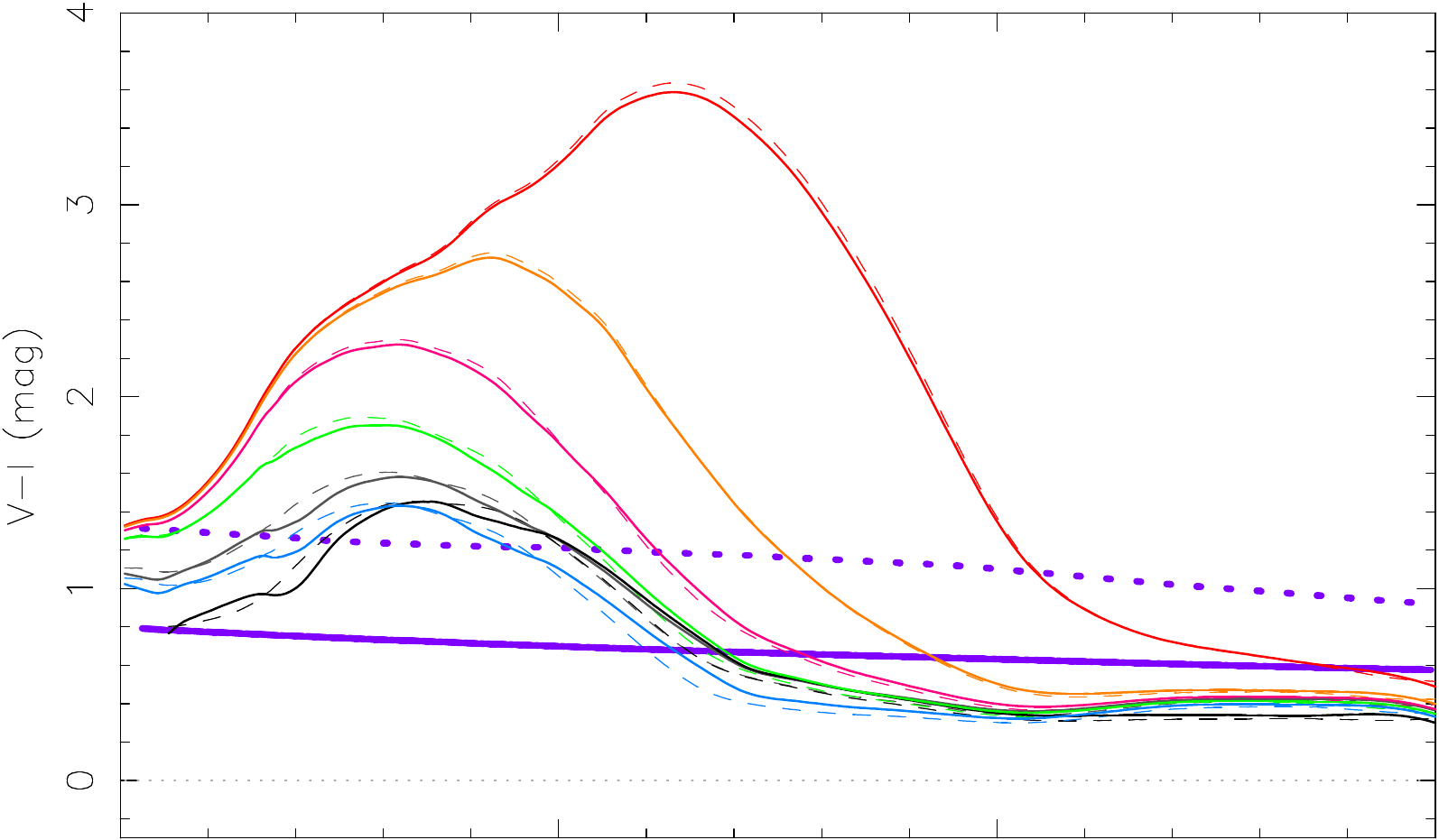}}
\put(0,0){\includegraphics[clip, width=8.63cm]{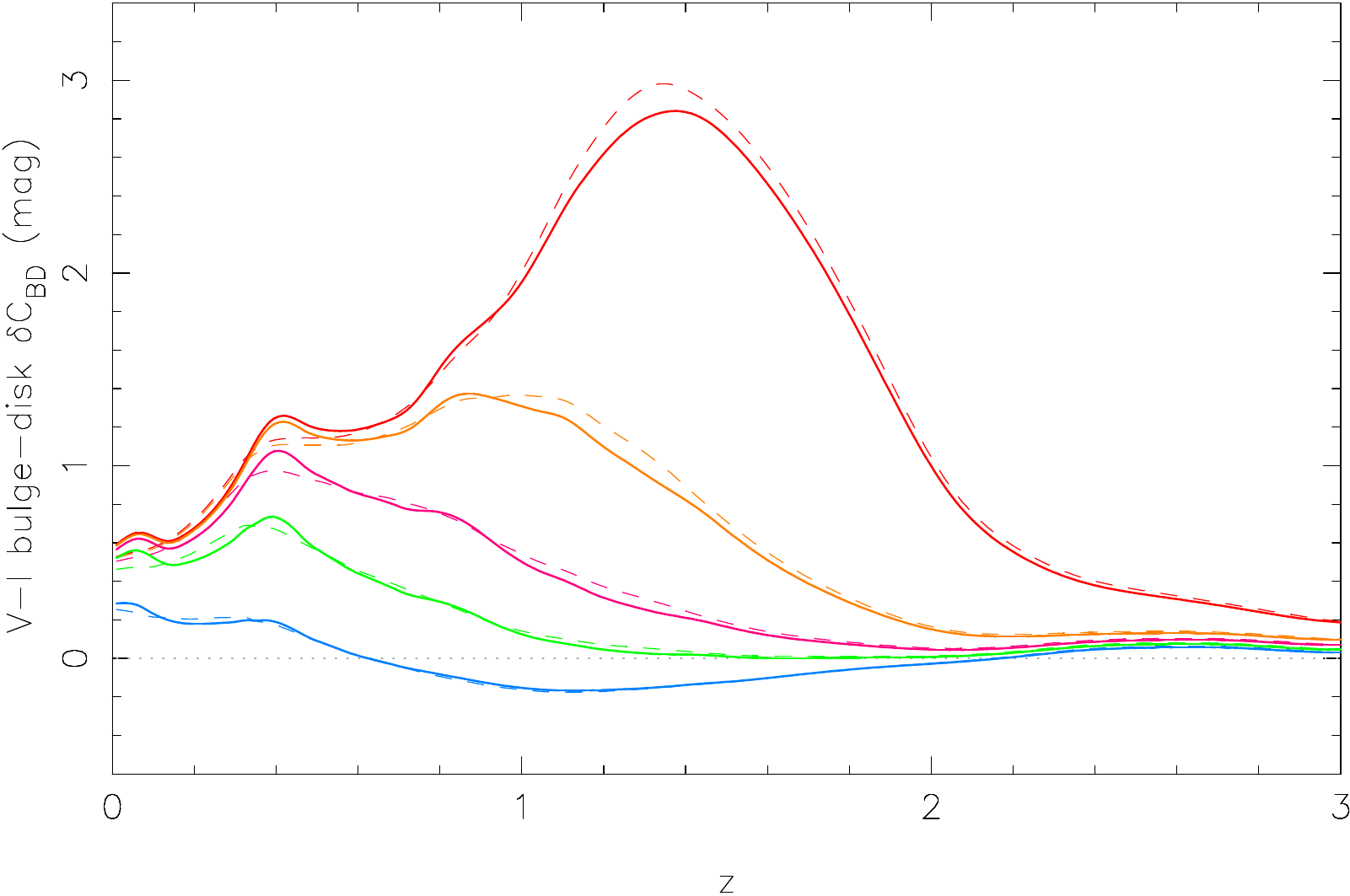}}
\put(98,165){\includegraphics[clip, width=8.6cm]{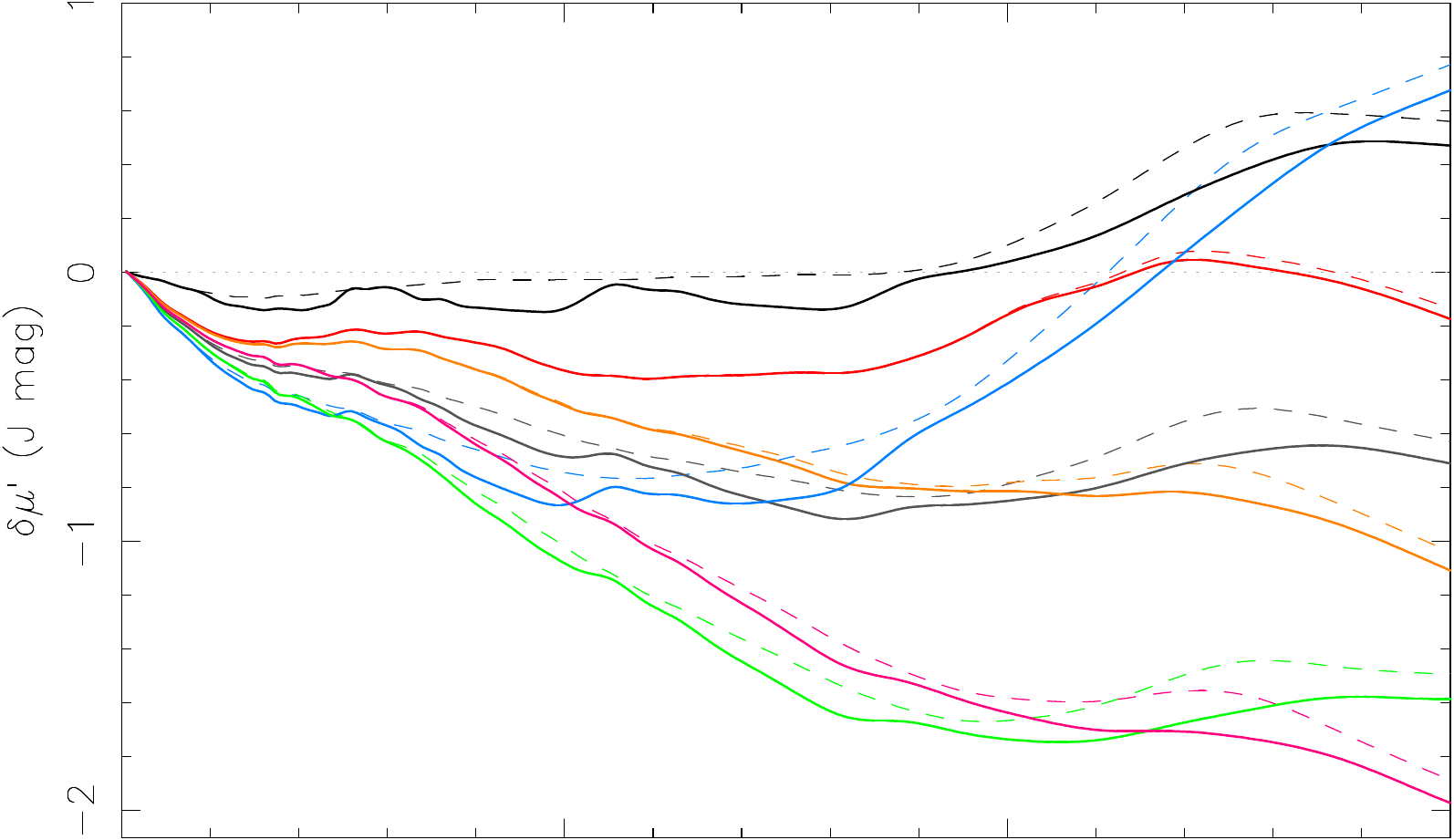}}
\put(98,111){\includegraphics[clip, width=8.6cm]{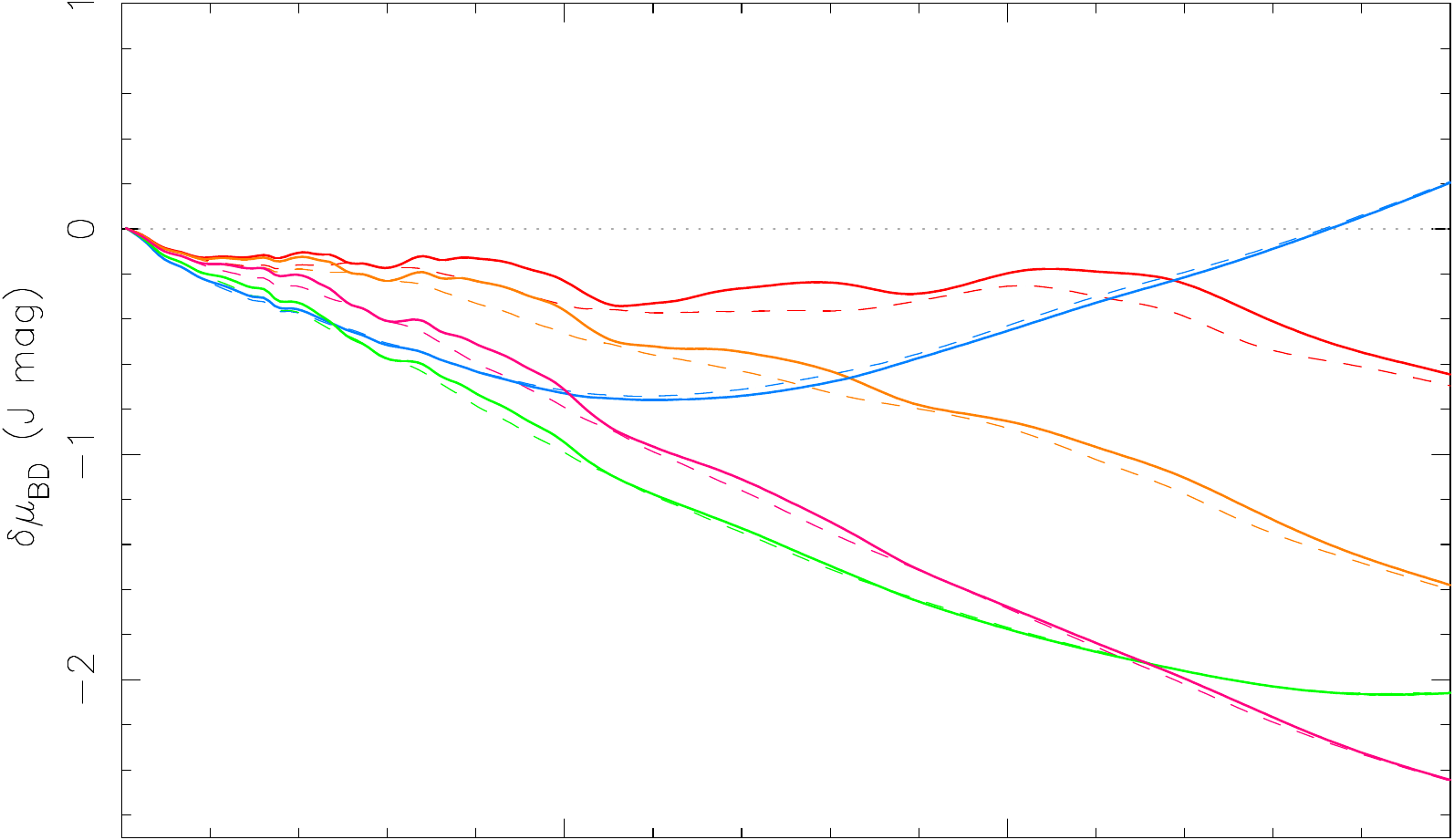}}
\put(98,59){\includegraphics[clip, width=8.6cm]{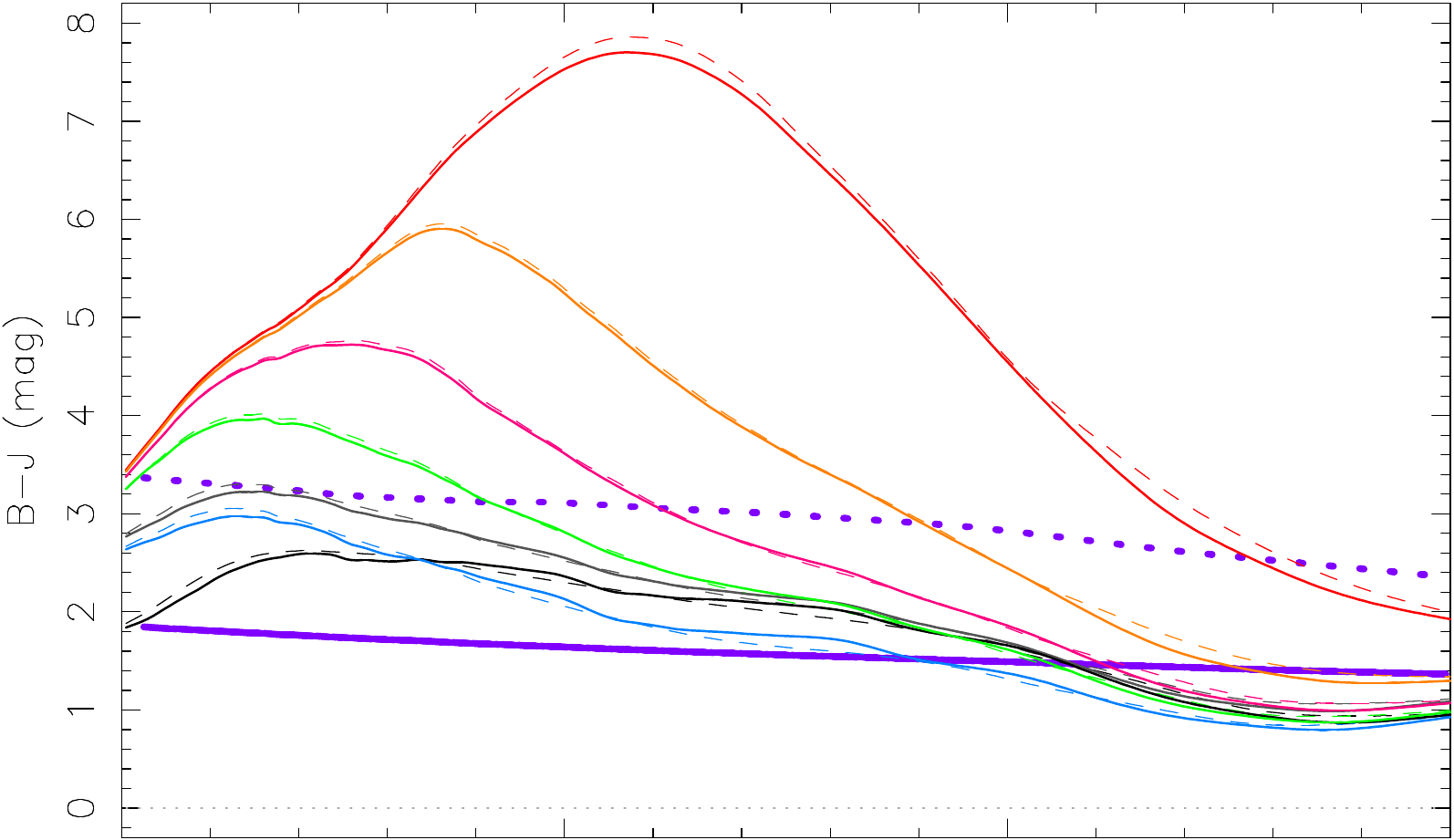}}
\put(98,0){\includegraphics[clip, width=8.63cm]{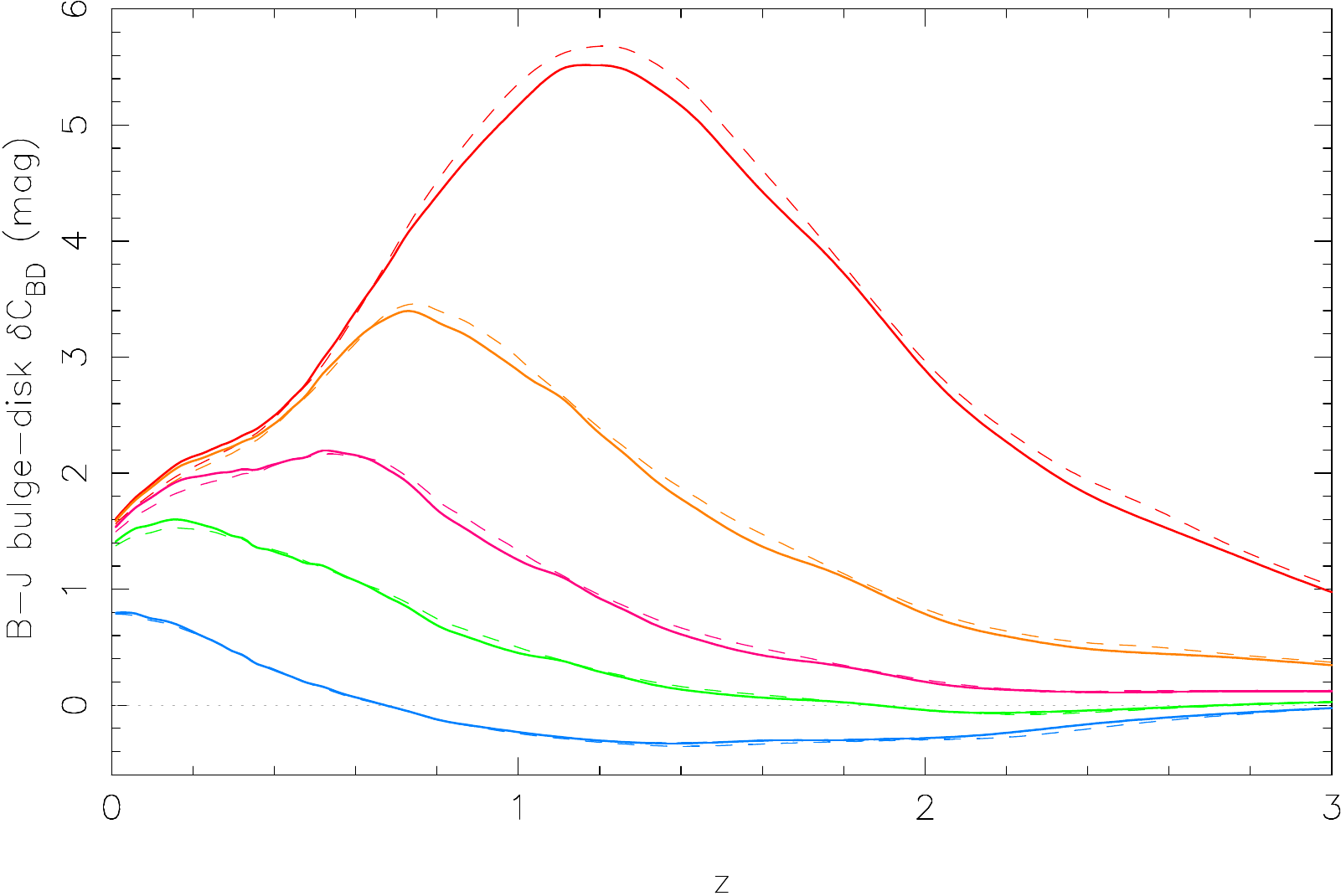}}
\end{picture}
\noindent{{\bf Fig. A.5.} Continued.}
\label{ij:zcons} 
\end{figure*}
\end{center}

\begin{center}
\begin{figure*}[h]
\begin{picture}(200,217)
\put(0,165){\includegraphics[clip, width=8.6cm]{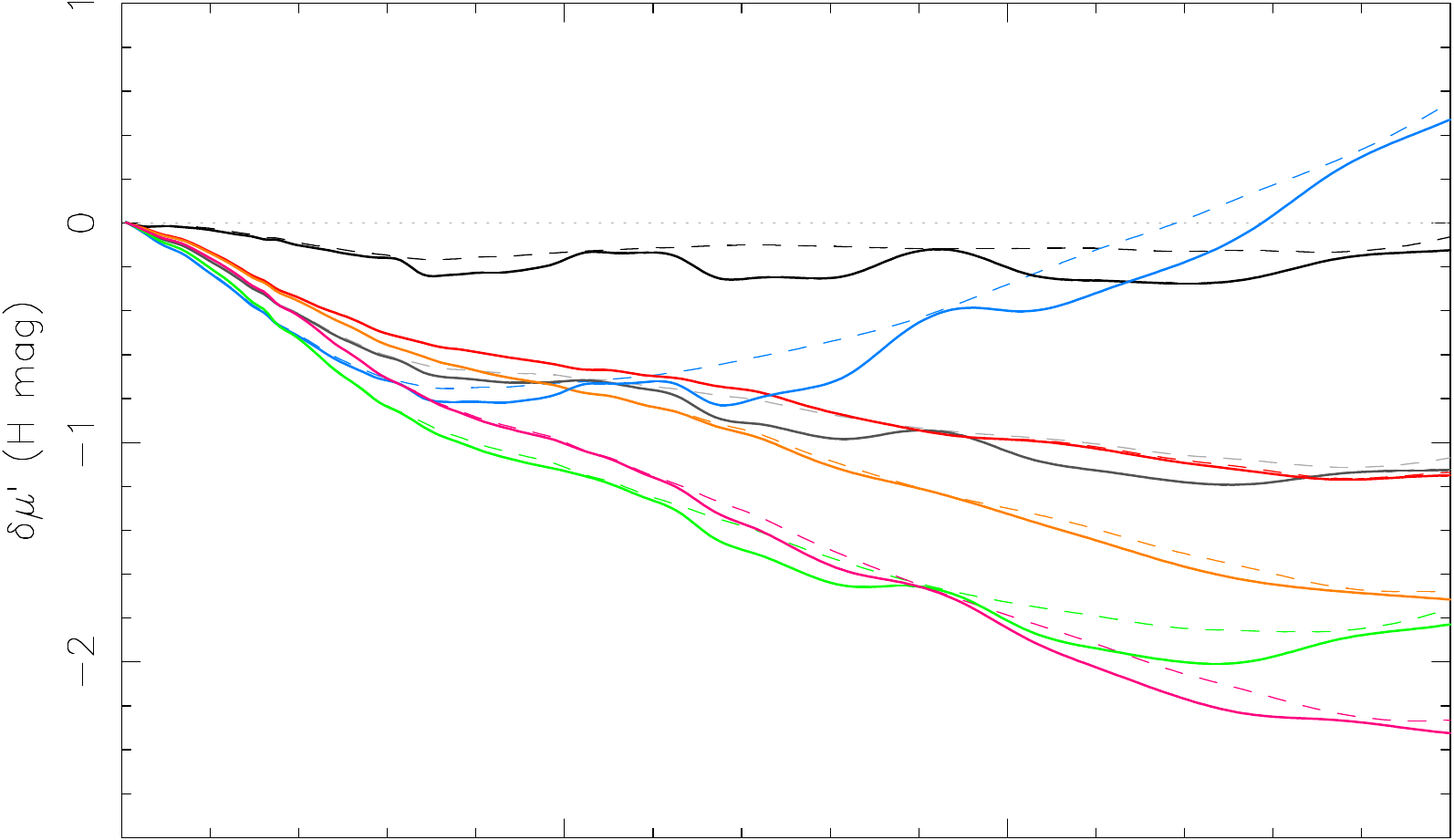}}
\put(0,111){\includegraphics[clip, width=8.6cm]{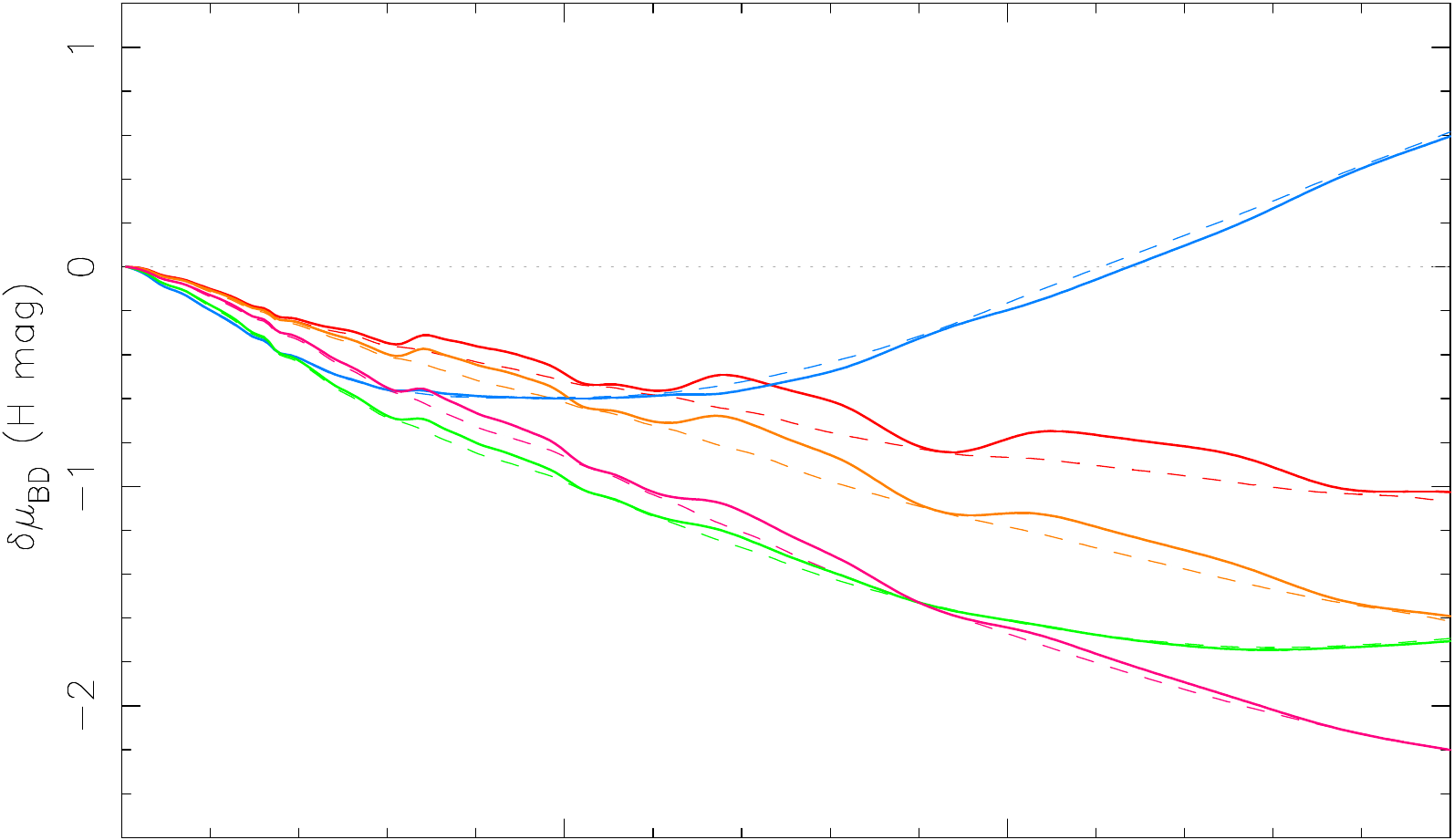}}
\put(0,59){\includegraphics[clip, width=8.6cm]{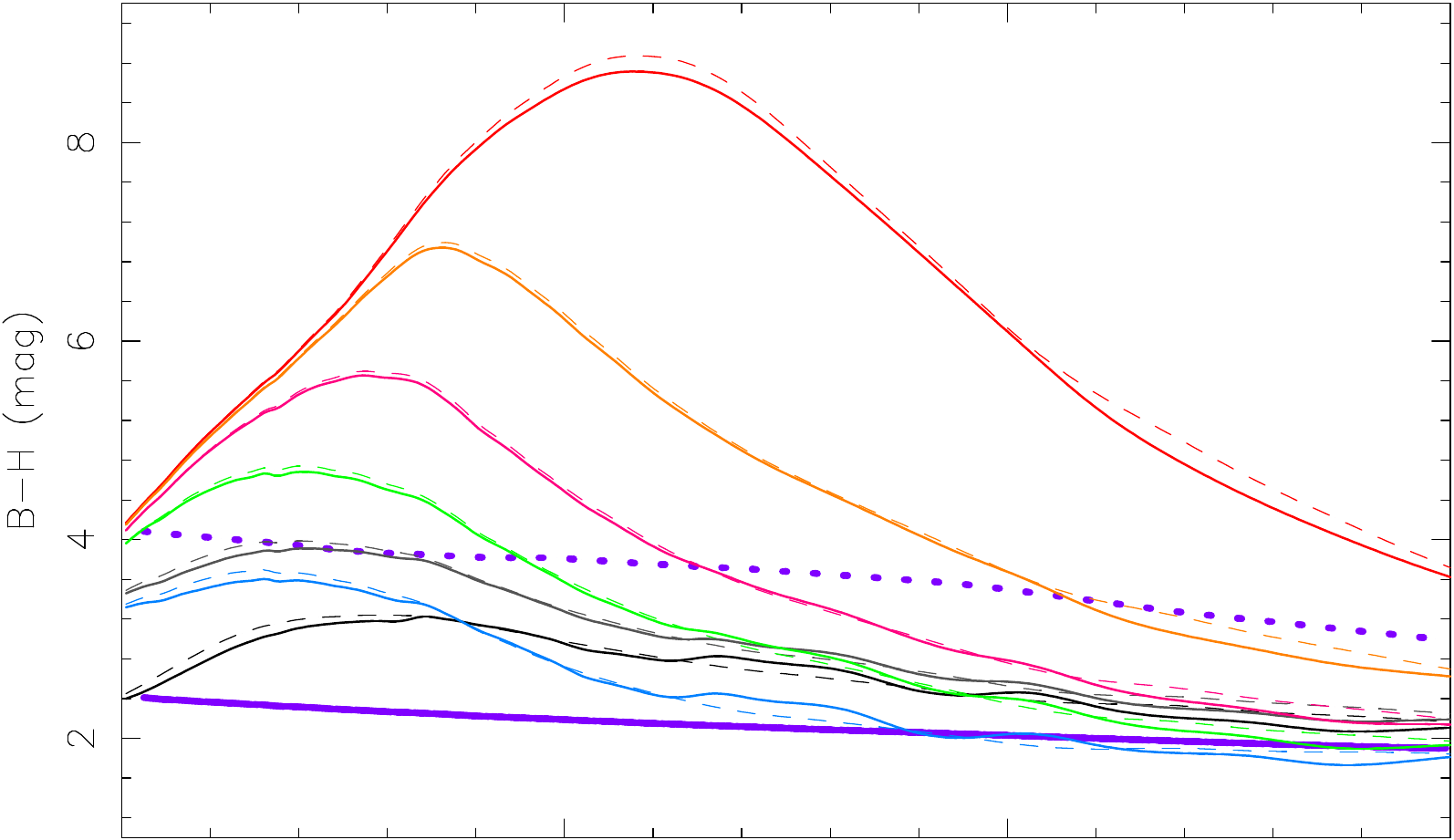}}
\put(0,0){\includegraphics[clip, width=8.63cm]{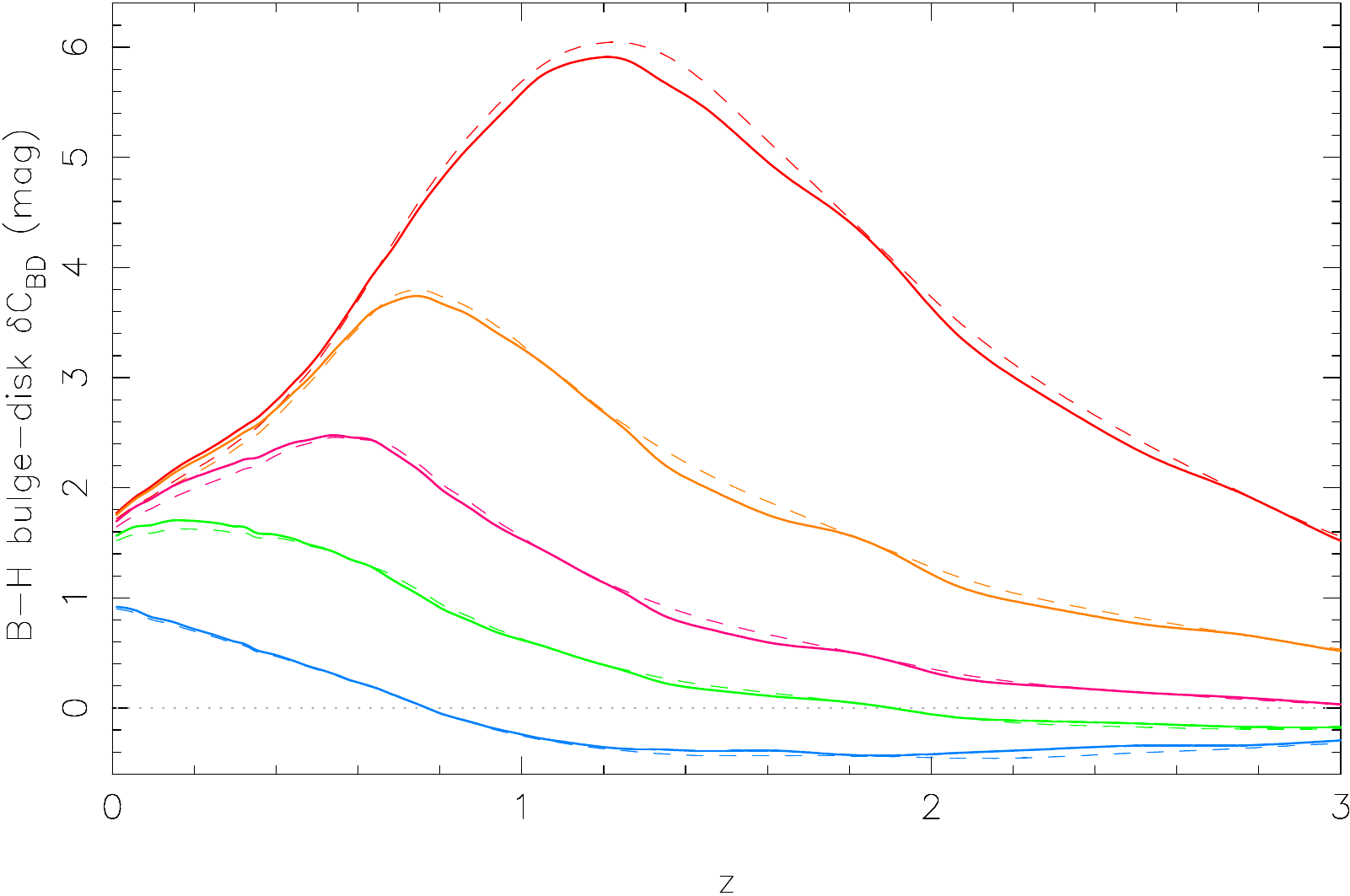}}
\put(98,165){\includegraphics[clip, width=8.6cm]{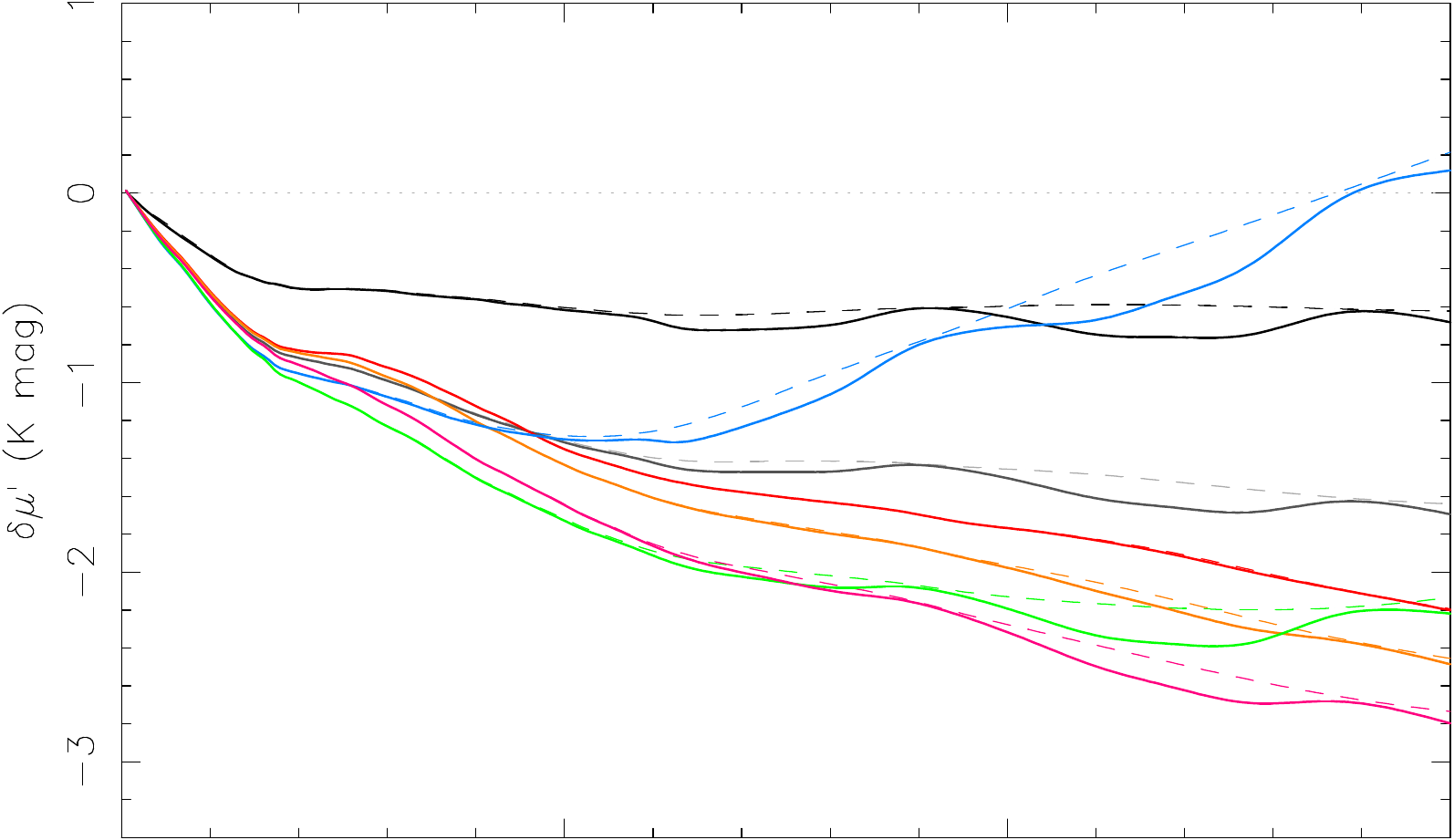}}
\put(98,111){\includegraphics[clip, width=8.6cm]{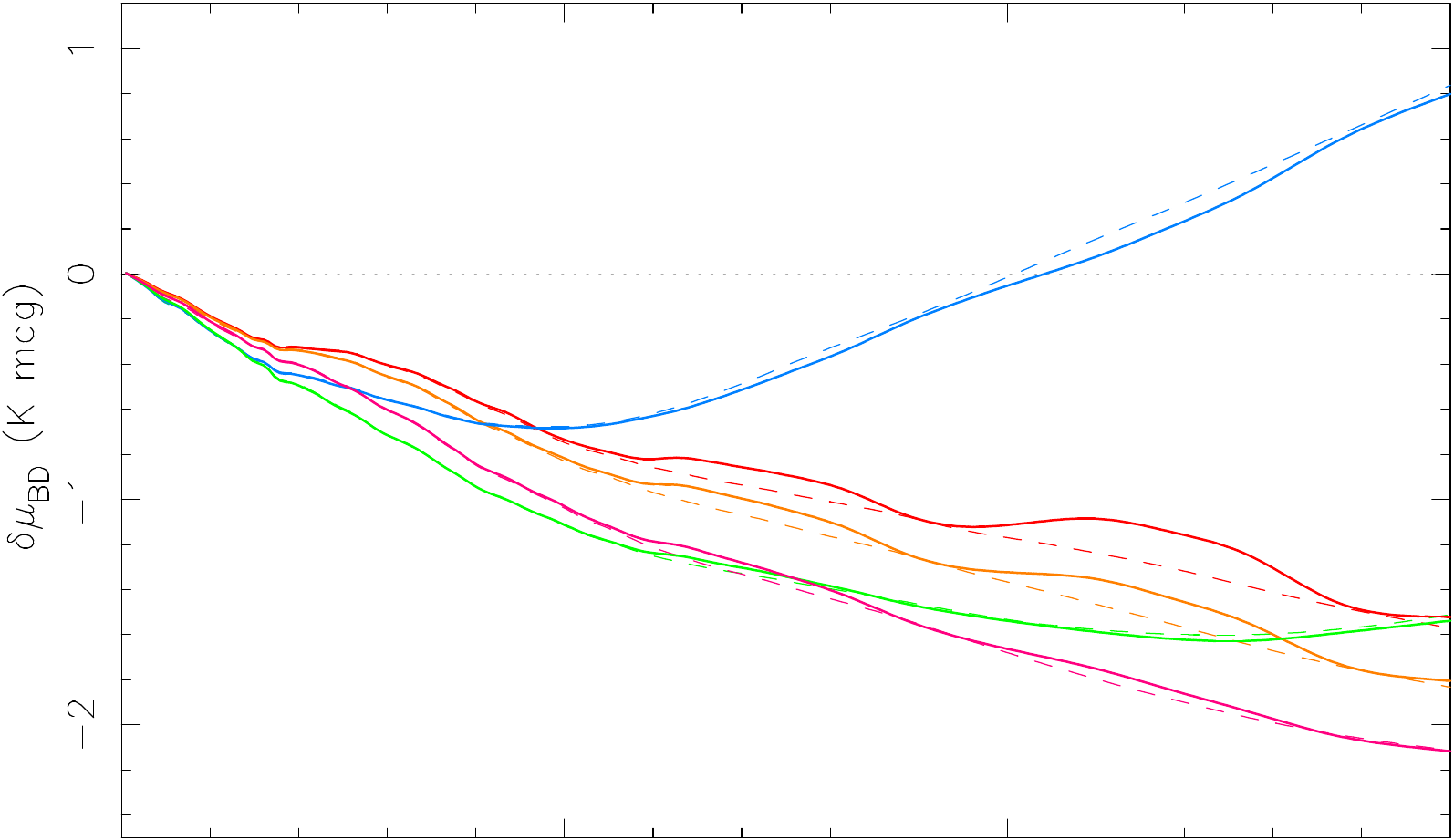}}
\put(98,59){\includegraphics[clip, width=8.6cm]{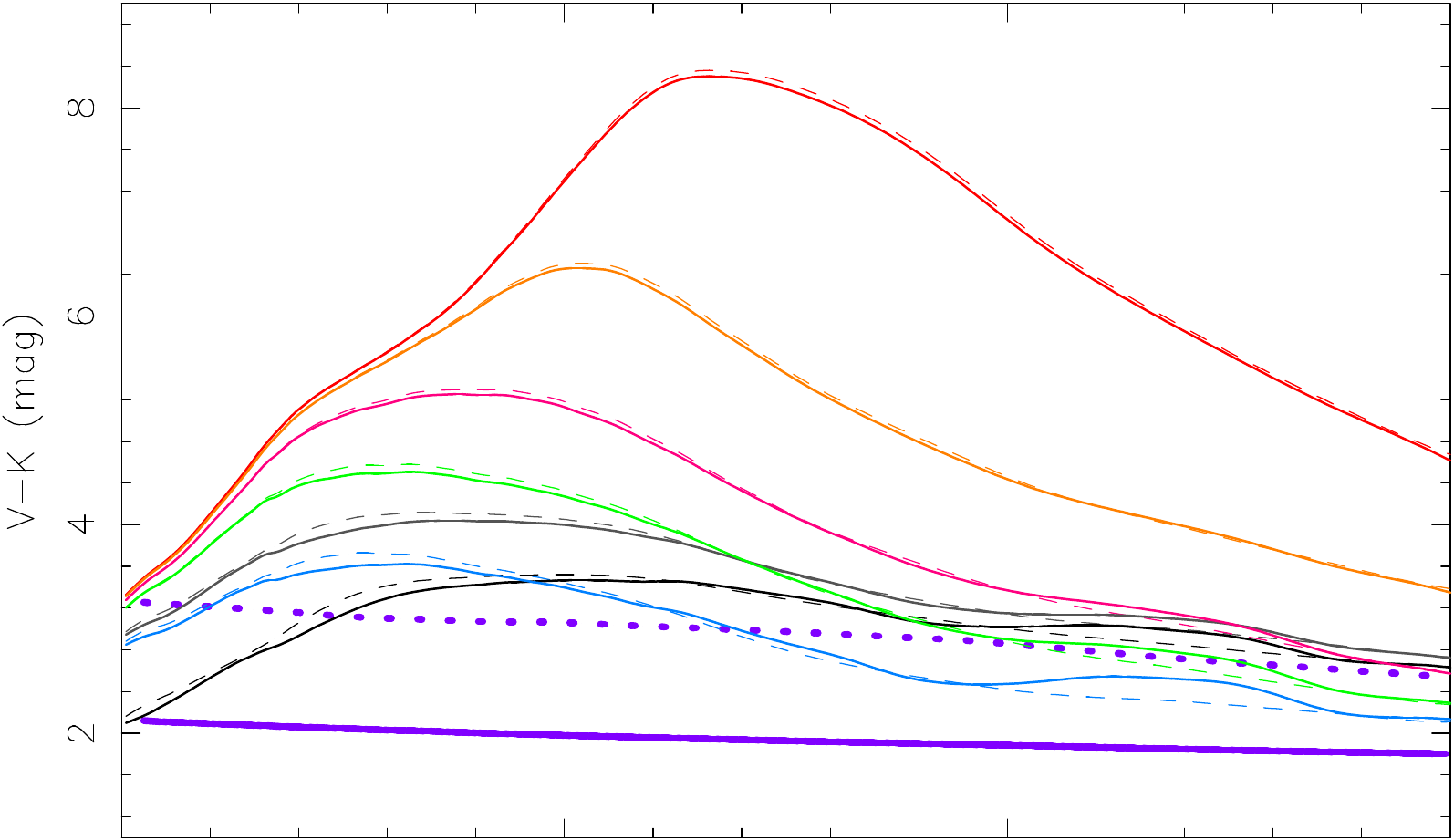}}
\put(98,0){\includegraphics[clip, width=8.63cm]{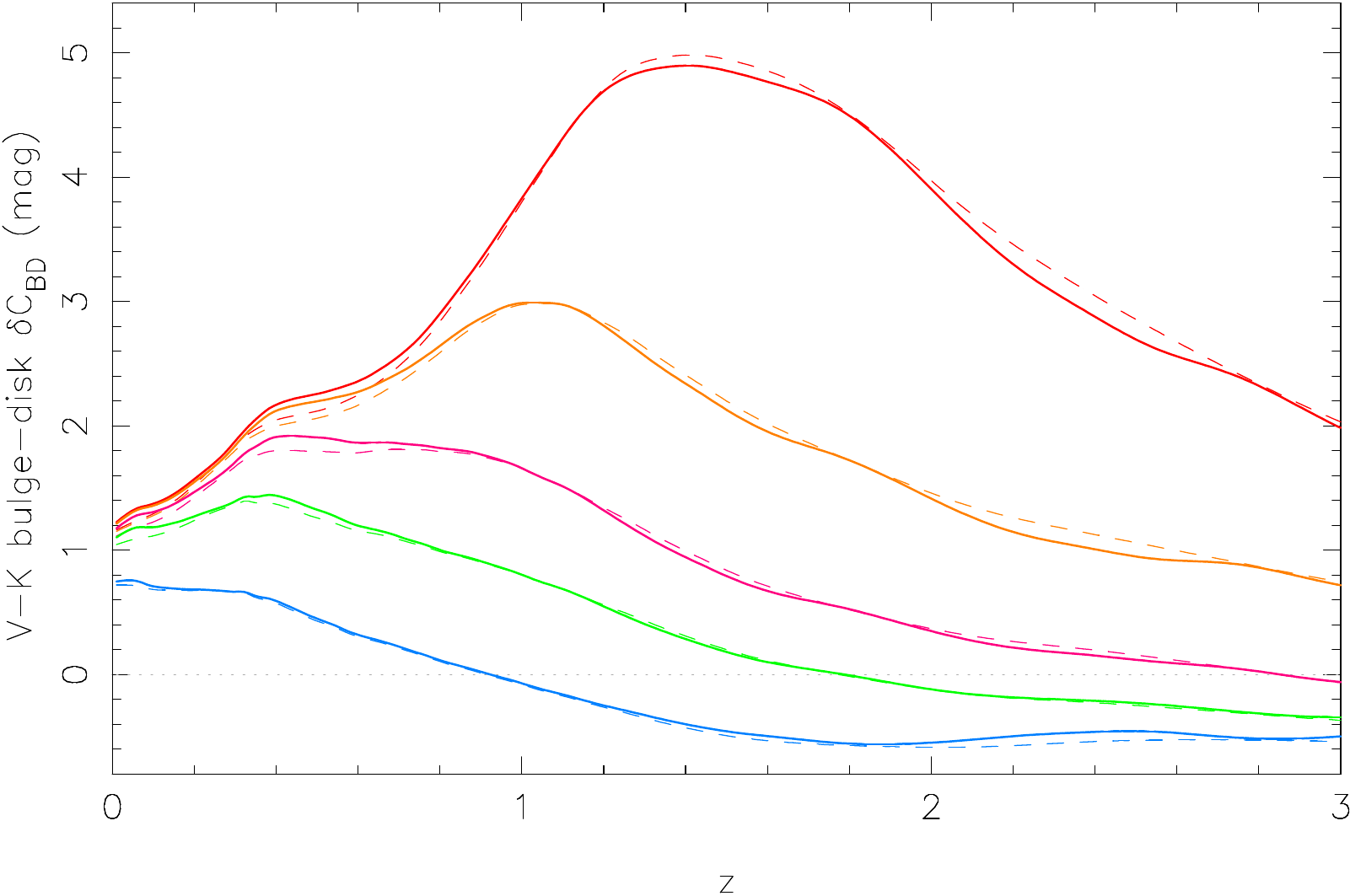}}
\end{picture}
\noindent{{\bf Fig. A.5.} Continued.}
\label{hk:zcons} 
\end{figure*}
\end{center}

Supplementing the discussion in Sect. \ref{sub:eSED}, Figs.~\ref{ub:zcons}-\ref{hk:zcons} show EvCon predictions for the seven SFHs in Fig.~\ref{fig:SFHs} for eight filters ($U$, $B$, $V$, Cousins $R$ and $I$, $J$, $H$, $K$).
The meaning of the diagrams in each column (from top to bottom) is as follows: \brem{a)} \dmu\ versus \zet\ for purely stellar models and those including
nebular emission (dashed and solid curve, respectively). All figures follow the color coding in Fig.~\ref{ub:zcons}\irem{a}. Whenever applicable, the arrow and the
bullet mark the redshift at which the Lyman-$\alpha$ and Lyman limit (912 \AA) approximately enter the $U$ filter (\zet=1.508 and 2.34, respectively).
\brem{b)} Surface brightness contrast \dmBD\! between bulge and disk (\dmu(bulge)-\dmu(disk), whereby \dmBD\! at \zet=0 is taken to be 0 mag) when assuming that the disk forms according to the contSF model. \brem{c)} ObsF color versus \zet. Following color indices are considered: $U$-$B$, $B$-$R$, $U$-$V$, $R$-$I$, $V$-$I$, $B$-$J$, $B$-$H$ and $V$-$K$. The purple solid and dashed curves show the true (rest-frame) color at a given \zet\ for, respectively, the contSF and $\tau$0.5 SFH scenario.
\brem{d)} Bulge-to-disk color contrast \dcol\ (mag).

\begin{center}
\begin{figure*}[h]
\begin{picture}(200,217)
\put(0,165){\includegraphics[clip, width=8.6cm]{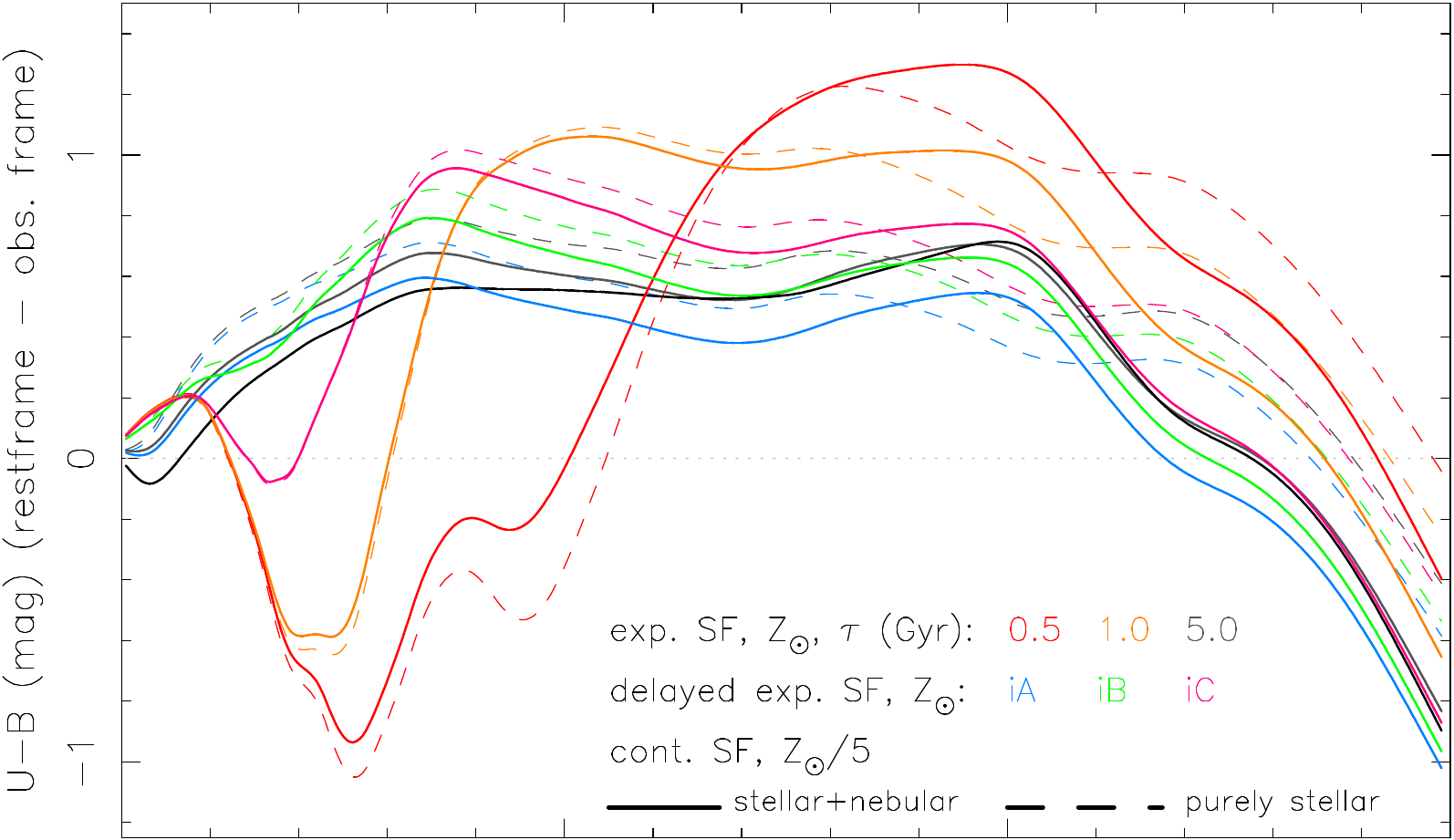}}
\put(0,111){\includegraphics[clip, width=8.6cm]{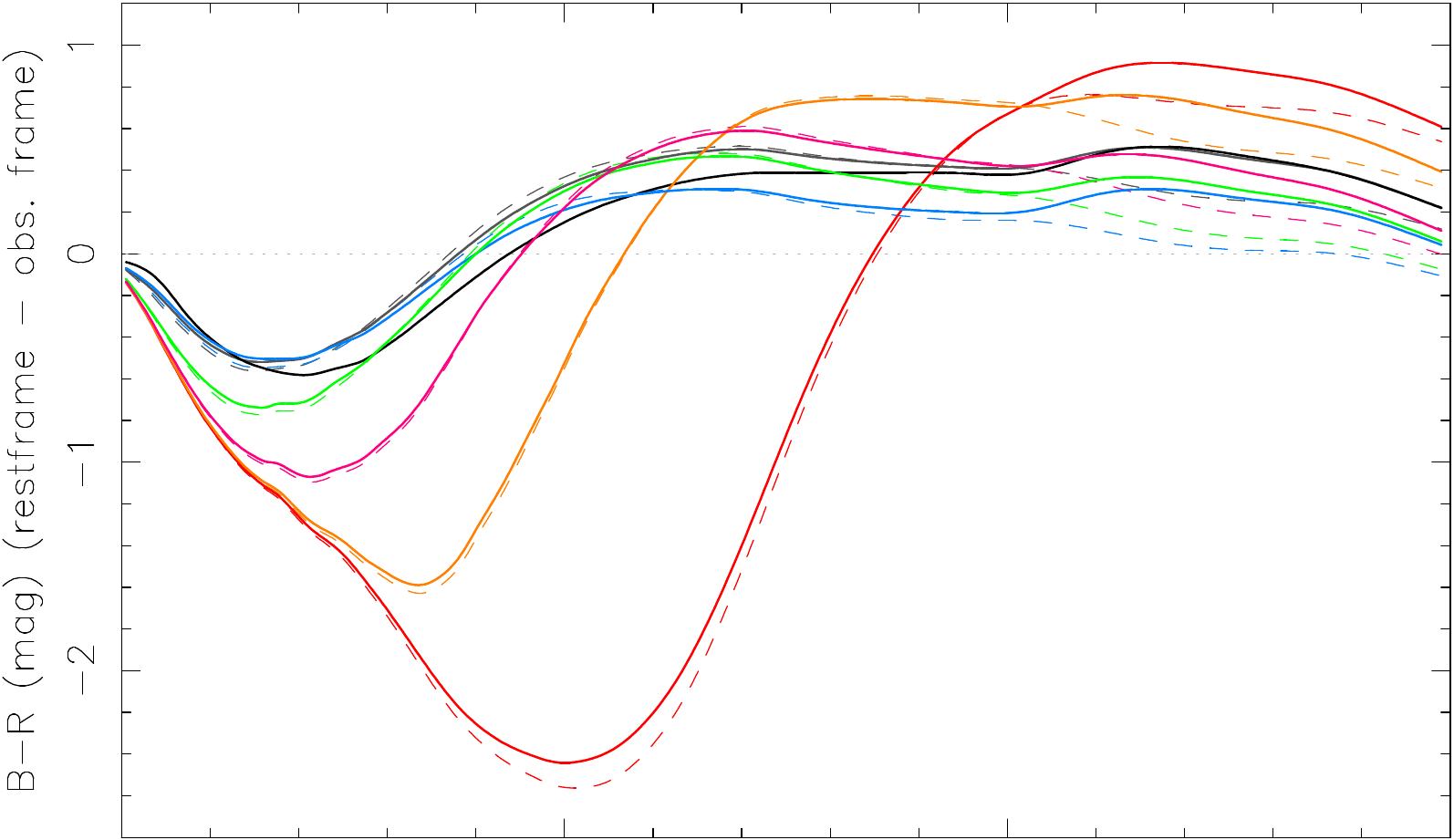}}
\put(0,59){\includegraphics[clip, width=8.6cm]{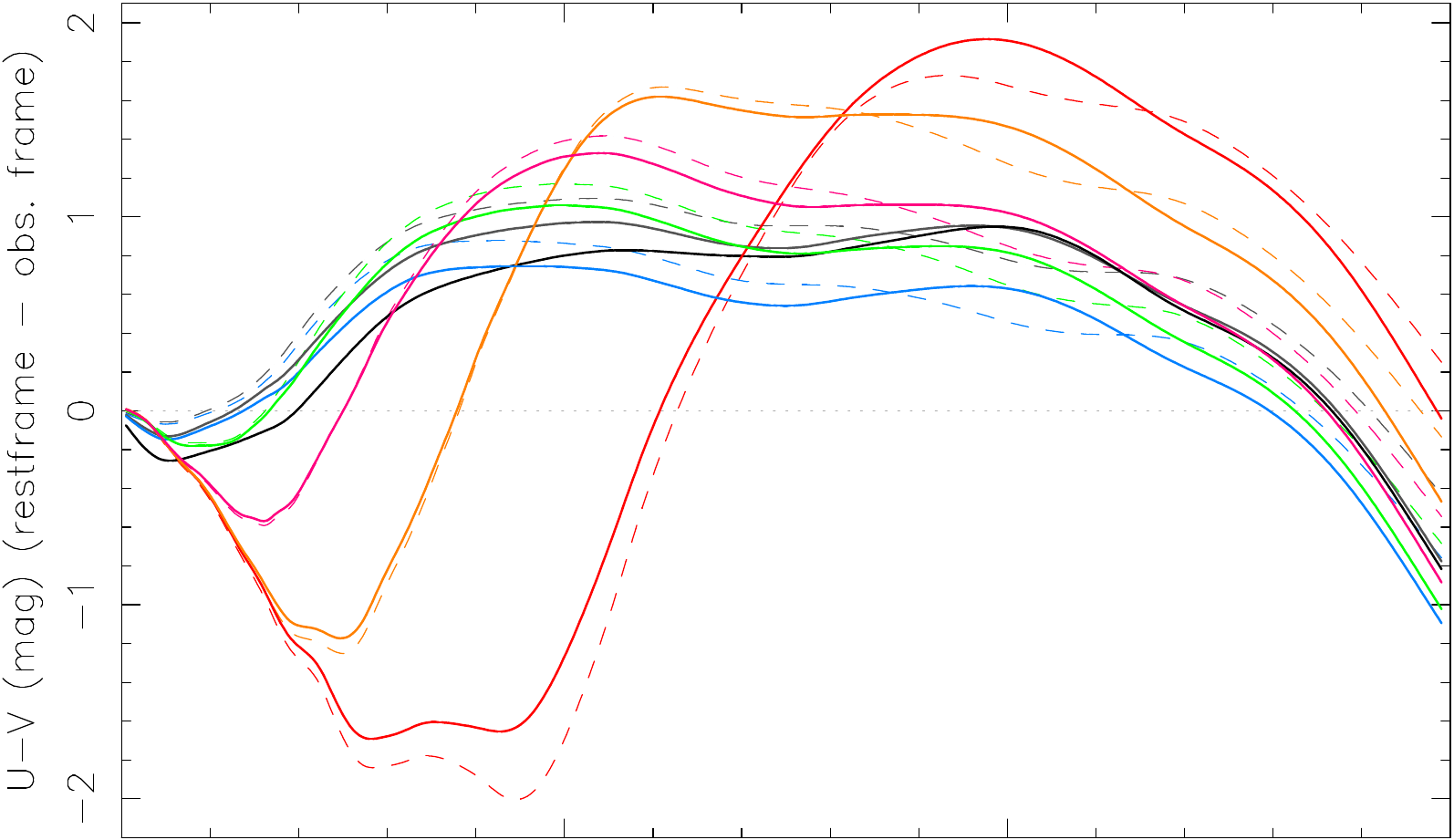}}
\put(0,0){\includegraphics[clip, width=8.63cm]{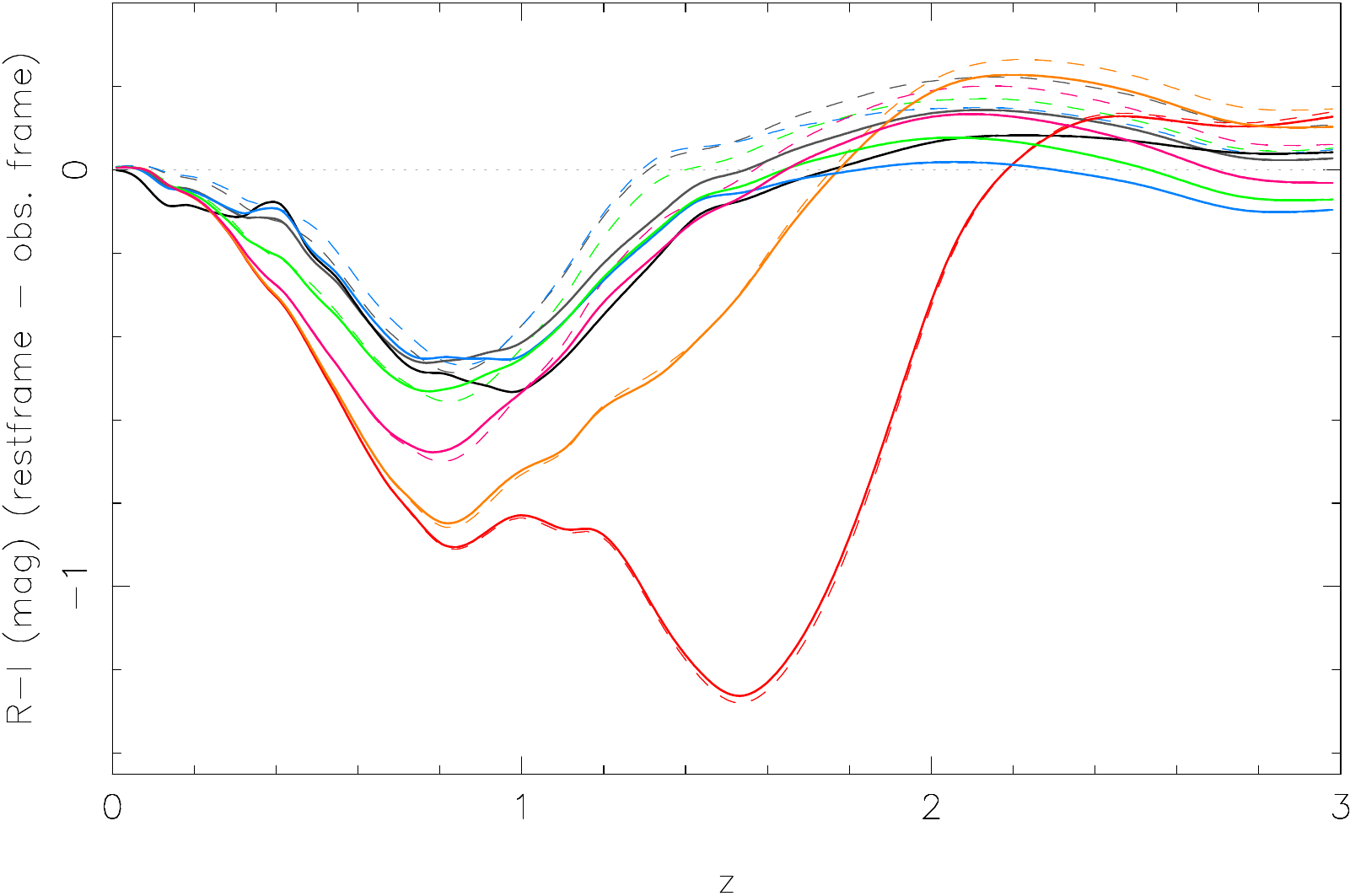}}
\put(98,165){\includegraphics[clip, width=8.6cm]{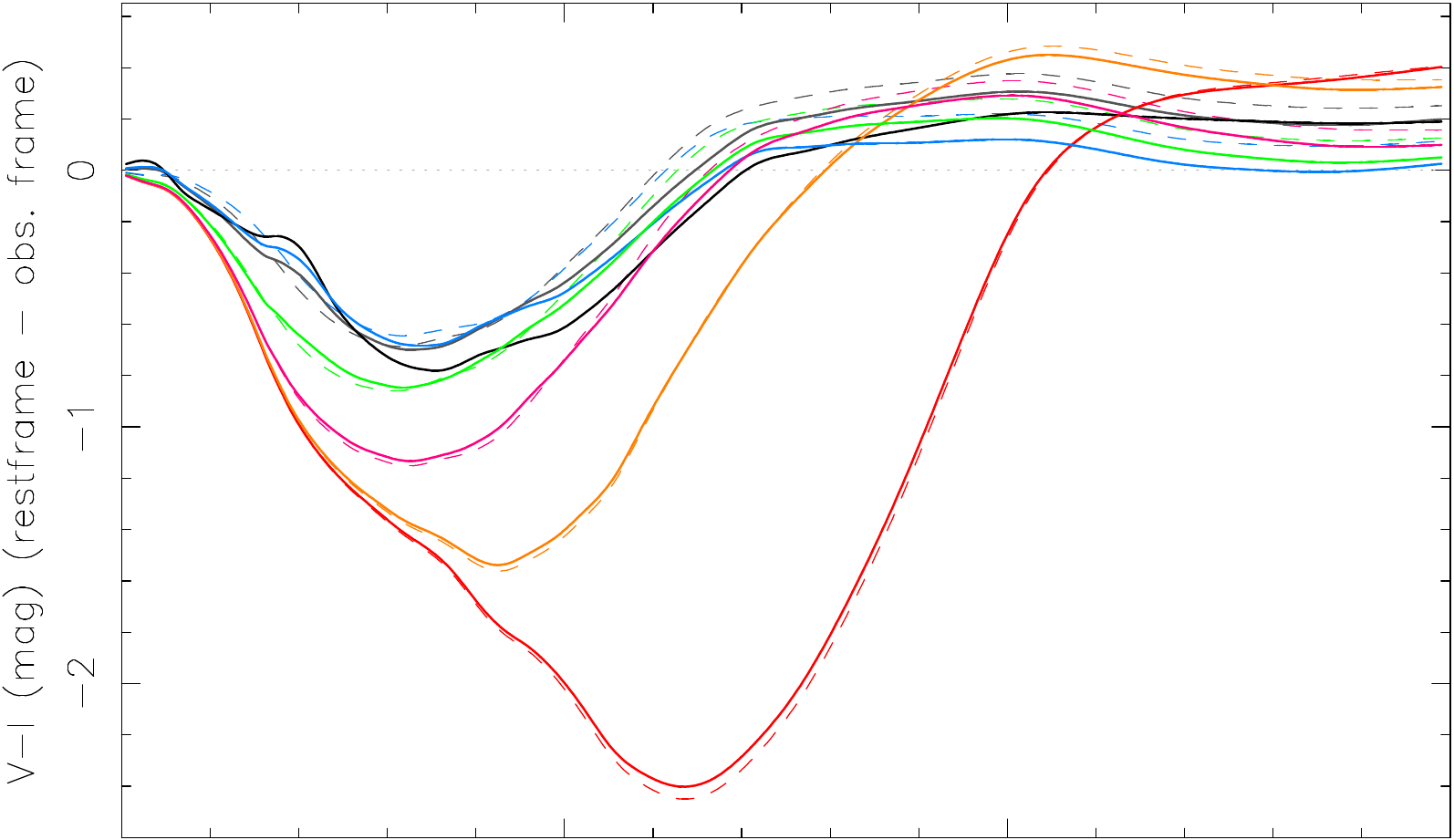}}
\put(98,111){\includegraphics[clip, width=8.6cm]{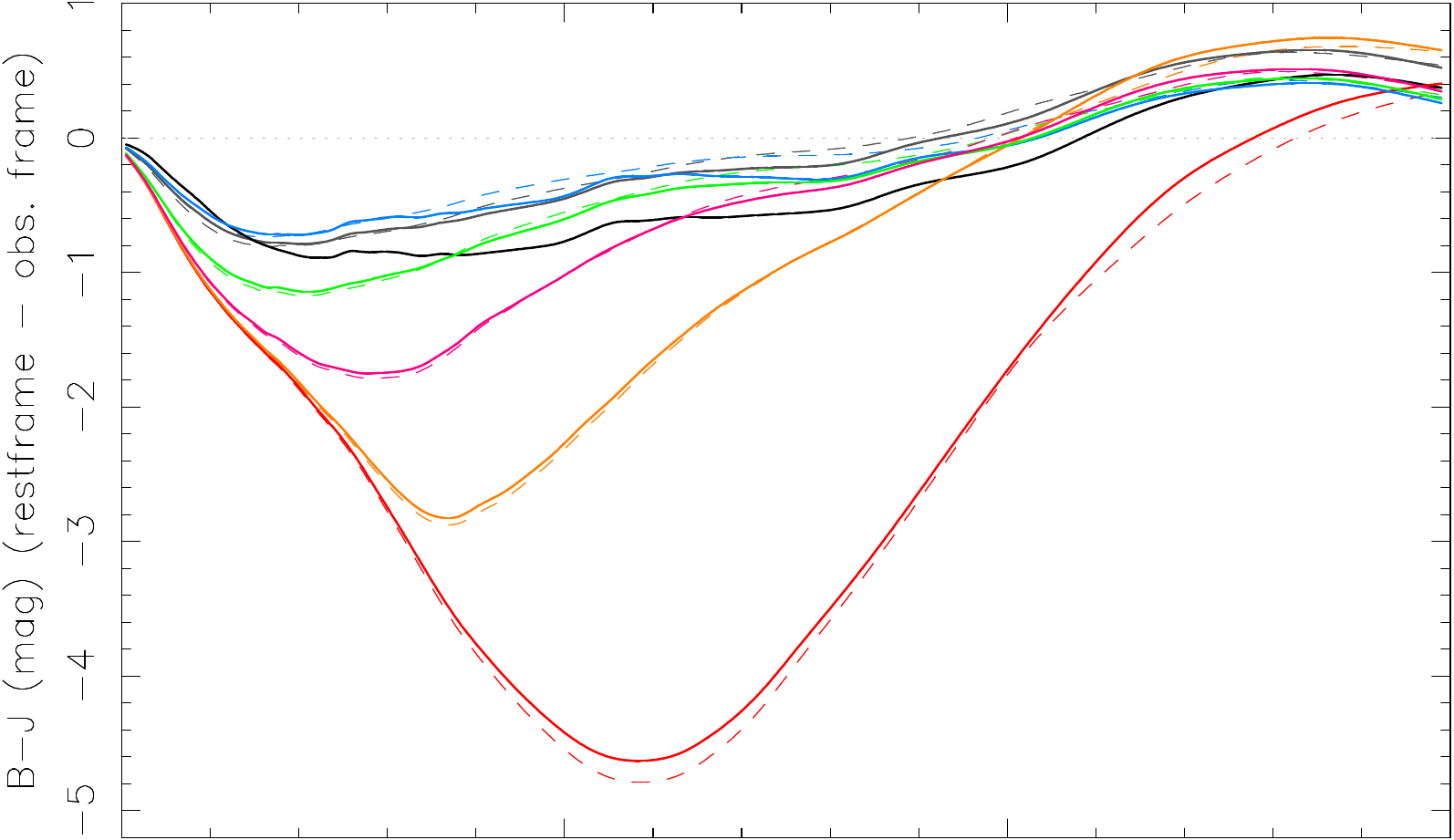}}
\put(98,59){\includegraphics[clip, width=8.6cm]{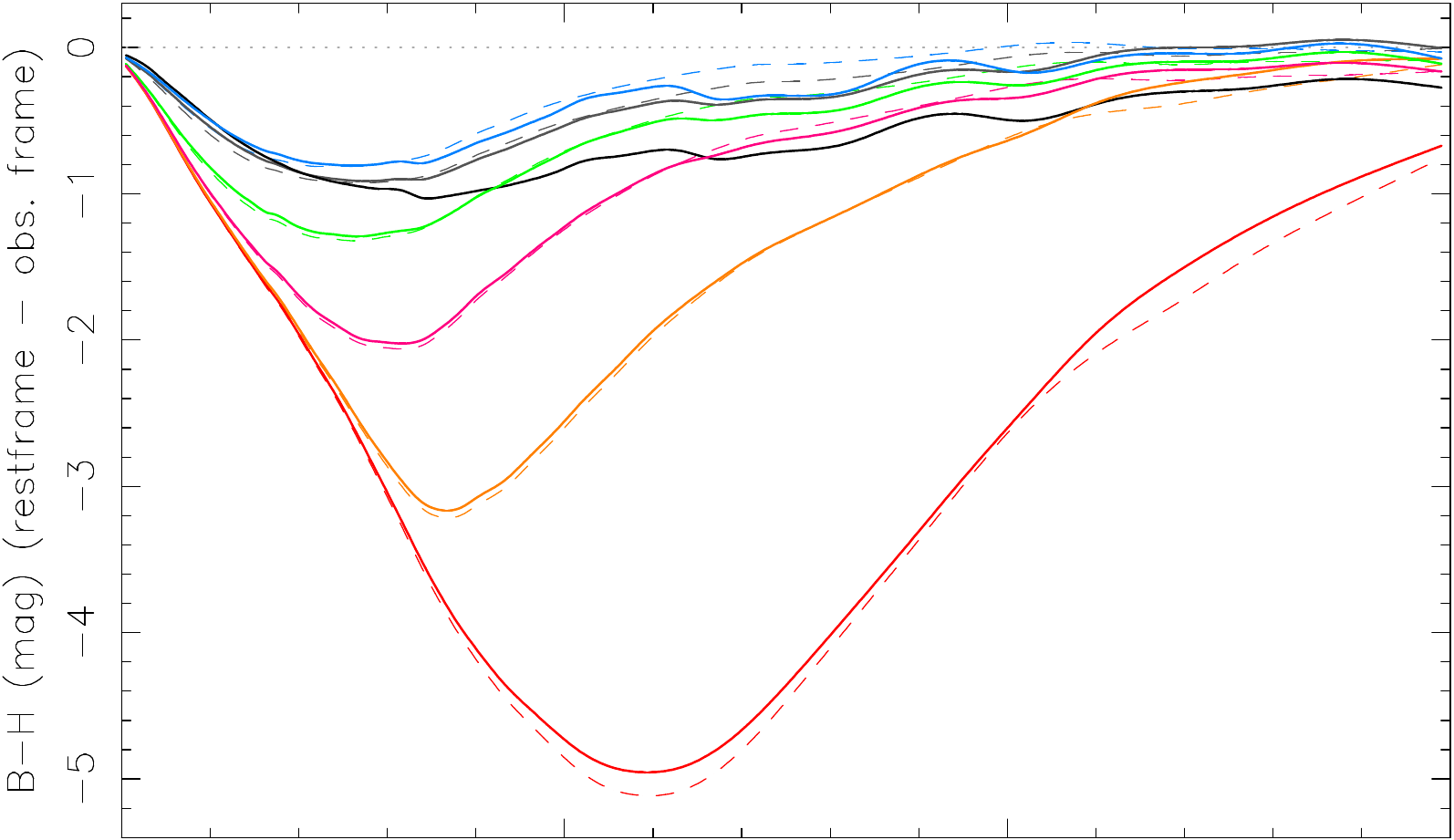}}
\put(98,0){\includegraphics[clip, width=8.63cm]{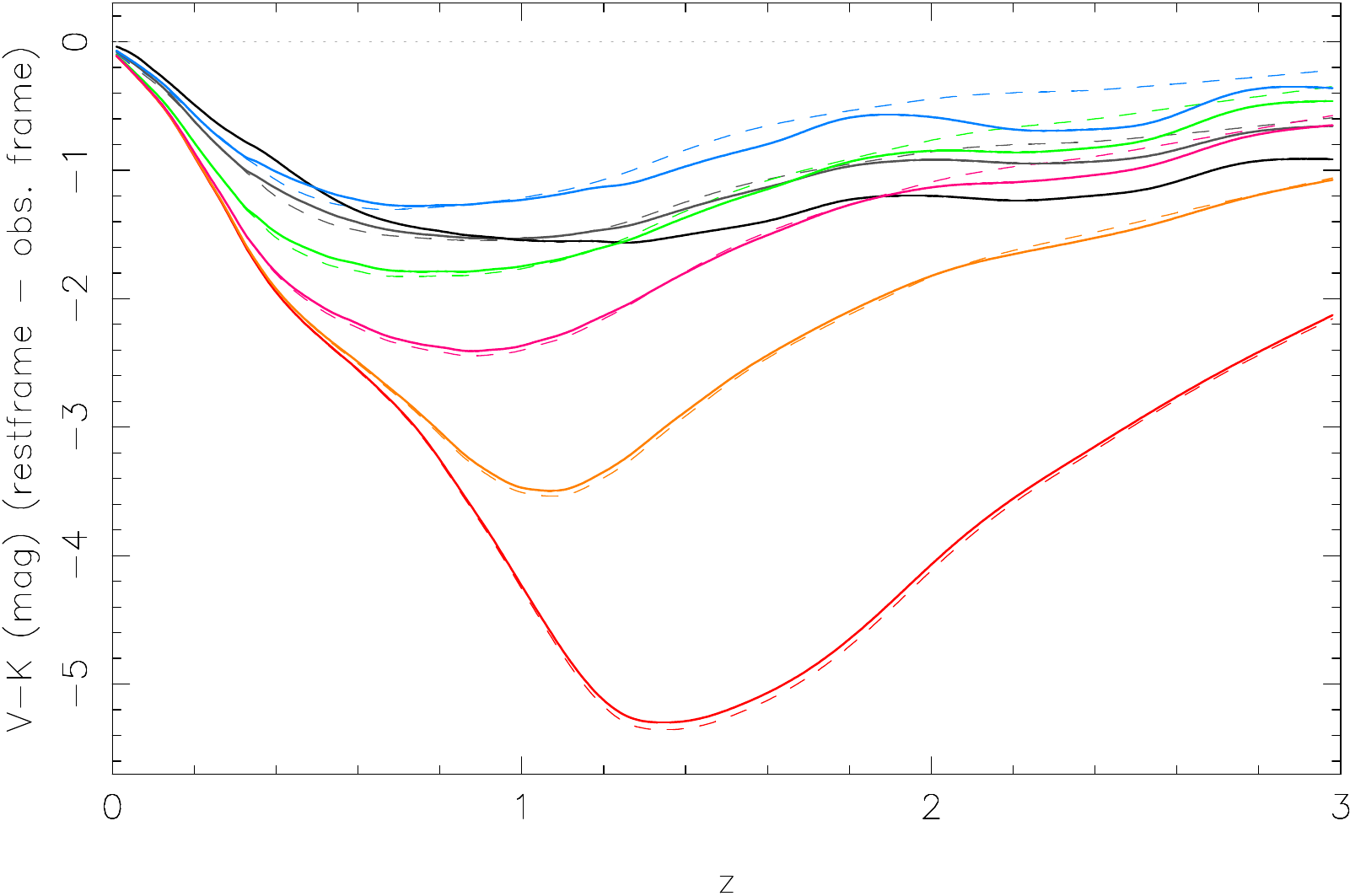}}
\end{picture}
\caption{Difference between the rest frame and the ObsF color for the SFH parameterizations in Fig.~\ref{fig:SFHs},
as inferred from EvCon simulations. Similar to Fig.~\ref{ub:zcons}, predictions including nebular emission are shown with solid curves and those based on purely stellar SEDs with dashed curves.}
\label{dif:zcons} 
\end{figure*}
\end{center}

Finally, Fig.~\ref{dif:zcons} displays the difference between the true (rest-frame) and ObsF color, that is, it facilitates the conversion of ObsF into rest-frame colors. We recall that colors are computed on the assumption of zero intrinsic extinction and attenuation by the intergalactic medium and refer to models that assume constant metallicity and the semi-empirical line flux ratios relative to \hb\ adopted in \pegase~2.

%\newpage

%\clearpage
%\newpage
\FloatBarrier
% :::::::::::::::::::::::::::::::::::::::::::::::::::::::::::::::::::::::::::::::::::::::::::::::::
\subsection{Notes on the enhancement of broadband magnitudes by nebular emission \label{ap:nebSED}}
% :::::::::::::::::::::::::::::::::::::::::::::::::::::::::::::::::::::::::::::::::::::::::::::::::
In this section we provide empirical estimates on the enhancement $\delta\mu$ (mag) of broadband magnitudes by nebular emission as a function of
the rest-frame \ewha.
The latter offers a handy metrics for the fractional contribution of nebular emission since it scales
\footnote{This only applies to the idealized picture of a point source that forms stars at a constant SFR and is optically thick to LyC radiation.
The situation is different in triaxial stellar systems hosting nuclear SF activity where, because of dilution by the stellar background
along the line-of-sight, emission-line EWs are centrally strongly reduced and show a positive radial gradient \citep{P02,P13}.}
with the specific SFR. An additional advantage of it is that it can be determined even from non-flux-calibrated spectra (or narrowband imaging) and, for a foreground absorbing screen geometry, is insensitive to intrinsic extinction.

The $\delta\mu$ expected from an emission line with a rest-frame EW$_0$ is
\begin{equation}
\delta\mu \simeq -2.5\,\log \left[ 1 + \frac{{\rm EW}_0(1+z)}{\Delta\lambda} \right]
\end{equation}
where $\Delta\lambda$ denotes the effective width of a broadband filter \citep[e.g.,][]{Papovich01}.
In reality, because multiple strong lines may fall within a filter transmission curve, with the EW of only one of those (typically, the \ha\ or [O{\sc iii}]${5007}$) approximately known from narrowband photometry, an estimate of $\delta\mu$ is more complex and requires assumptions on relative line fluxes and the contribution of the nebular continuum.

Evolutionary synthesis models (\pegase~2 in the case of this study) offer a possibility to examine the connection between \ewha\ and $\delta\mu$ in different optical and NIR filters. We note that \pegase\ assumes that 70\% of the LyC radiation produced by stars is reprocessed into nebular emission in a warm gas phase with standard conditions (the rest is absorbed by dust) and allow for the computation of the nebular SED assuming semi-empirical and theoretical line flux ratios relative to H$\beta$ \citep[see][for details]{FRV97,GRV87,Stasinska84}. Clearly, the modeling of nebular emission in high-\zet\ galaxies could require significantly different prescriptions \citep[e.g.,][]{ChaLon01,Brinchmann22}.

In the following, we study the variation in $\delta\mu$ as a function of age and \ewha\ based on a comparison of broadband magnitudes computed for purely stellar synthetic SEDs with those that also comprise nebular (line and continuum) emission. Both sets of SEDs were computed for three SFHs and assume a Salpeter IMF between 0.1 and 100 \msun\ for a metallicity of \zsun/20, \zsun/5 and \zsun.

\begin{center}
\begin{figure}
\begin{picture}(86,132)
\put(2,74){\includegraphics[clip, width=8.4cm]{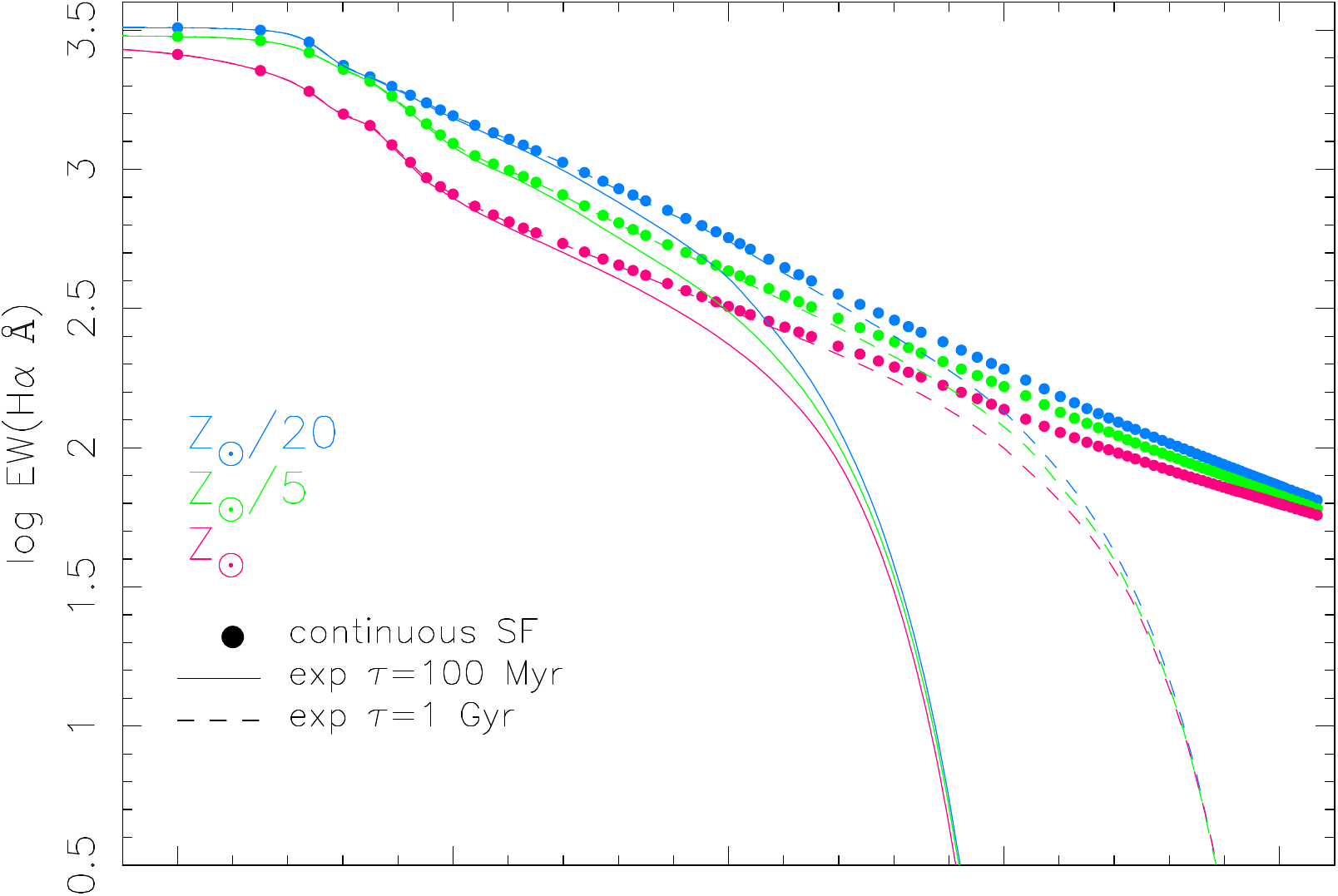}}
\put(2,0){\includegraphics[clip, width=8.42cm]{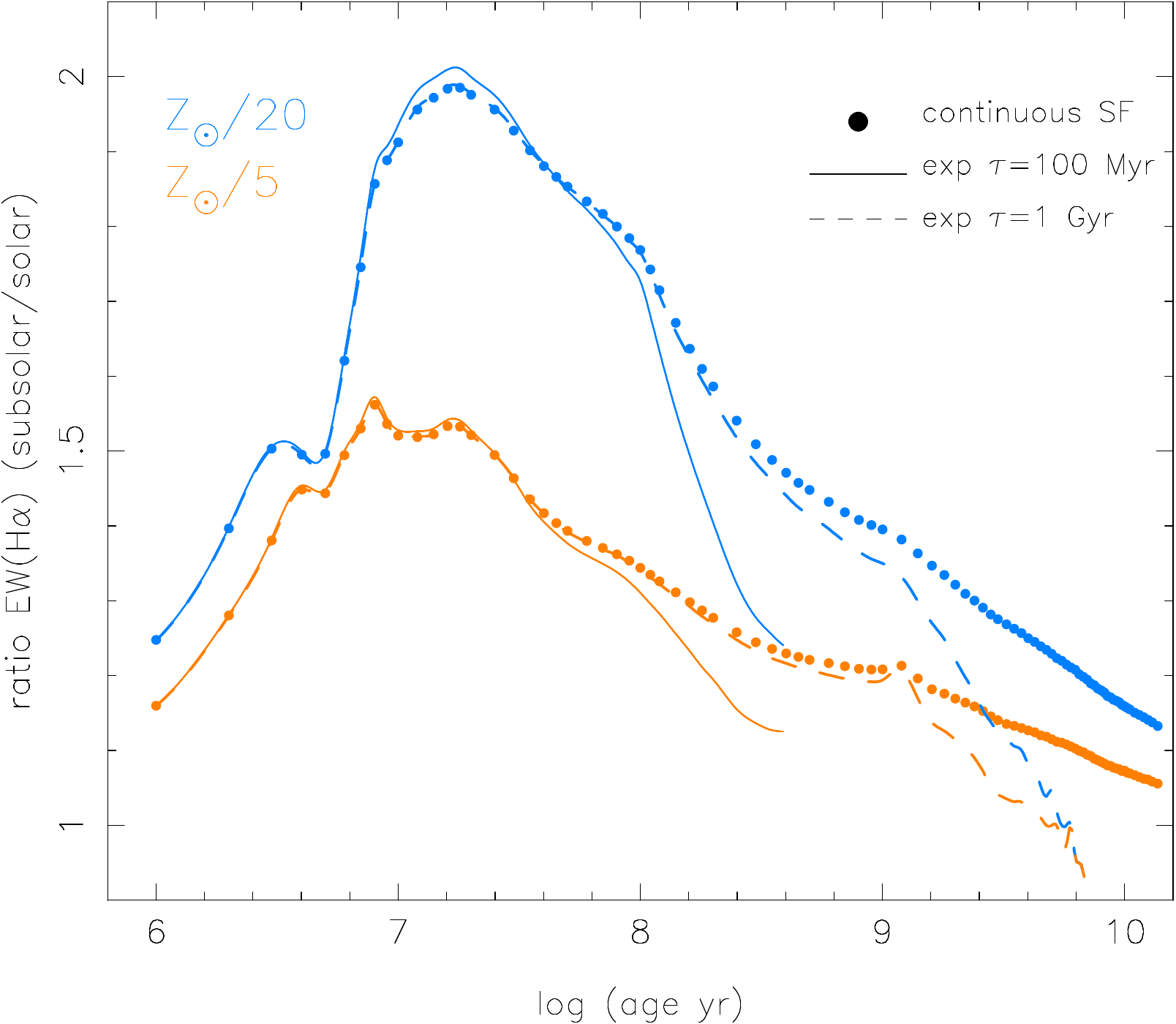}}
\end{picture}
\caption{Illustration of the effect of stellar metallicity on \ewha. \brem{upper panel:} Evolution of \ewha\ for SFH models involving continuous SF at a constant SFR (dots), and exponentially decreasing SFR with an e-folding time of 0.1 Gyr and 1 Gyr (solid and dashed curve, respectively). Predictions from \pegase~2 \citep{FRV97} refer to a fixed stellar metallicity of \zsun, \zsun/5 and \zsun/20 (red, green and blue, respectively) and a Salpeter IMF between 0.1 and 100 \msun. \brem{lower panel:} Ratio between the  \ewha\ for subsolar- and solar-metallicity models as a function of age. At an age of $\sim$16 Myr subsolar metallicity models imply a by a factor of up to $\sim$2 larger \ewha\ than solar metallicity models.}
\label{ew1}
\end{figure}
\end{center}

Figure~\ref{ew1} shows the evolution of \ewha\ in the case of continuous SF at constant SFR (dots) and that predicted for an exponentially decreasing SFR with an e-folding time of 0.1 and 1 Gyr (solid and dashed curve, respectively). As shown in several previous studies \citep[e.g.,][]{Lei99,Cid11,GP17}, the \ewha\ is tightly linked to the SFH, dropping steeply with time from initially $\sim$3000 \AA\ to less than 3 \AA\ after $\sim$0.7 (6) Gyr for the second (third) scenario, whereas maintaining a significant level ($\sim$60 \AA) even at 13.7 Gyr in the case of constant SFR. Additionally, because of its dependence on the LyC photon production rate it is at a given sSFR inversely related to stellar metallicity\footnote{An implication of this is that conversion of \ha\ luminosity into SFR using the standard calibration (which refers to solar-metallicity) entails an overestimation of the SFR and sSFR for galaxies with subsolar-metallicity \citep[see, e.g.,][for a discussion]{BFvA05}. A related effect is that a negative radial stellar metallicity gradient in a star-forming galaxy leads to or amplifies existing positive \ewha\ gradients \citep{Breda20a}.}.
For instance, \citet{WeilbacherUF01} show that the ratio between \ha\ luminosity and SFR increases by a factor of $\sim$4 from 2\,\zsun\ to \zsun/20.
The critical importance of metallicity is also illustrated in the lower panel of Fig.~\ref{ew1} where we compare the evolution of \ewha\ for subsolar- and solar-metallicity models.

\begin{center}
\begin{figure*}
\begin{picture}(86,62)
\put(8,0){\includegraphics[clip, height=6cm]{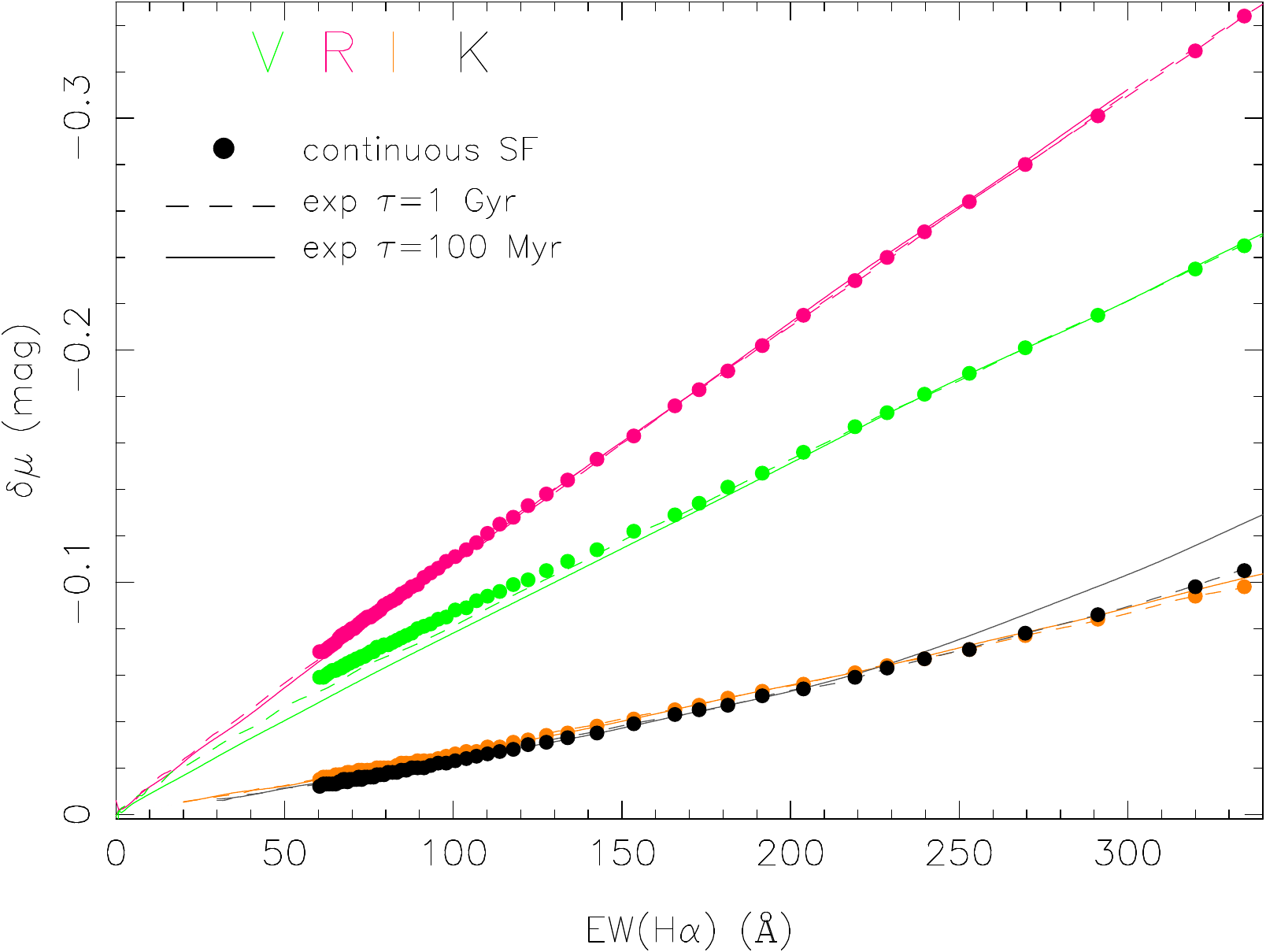}}
\put(96,0){\includegraphics[clip, height=6.24cm]{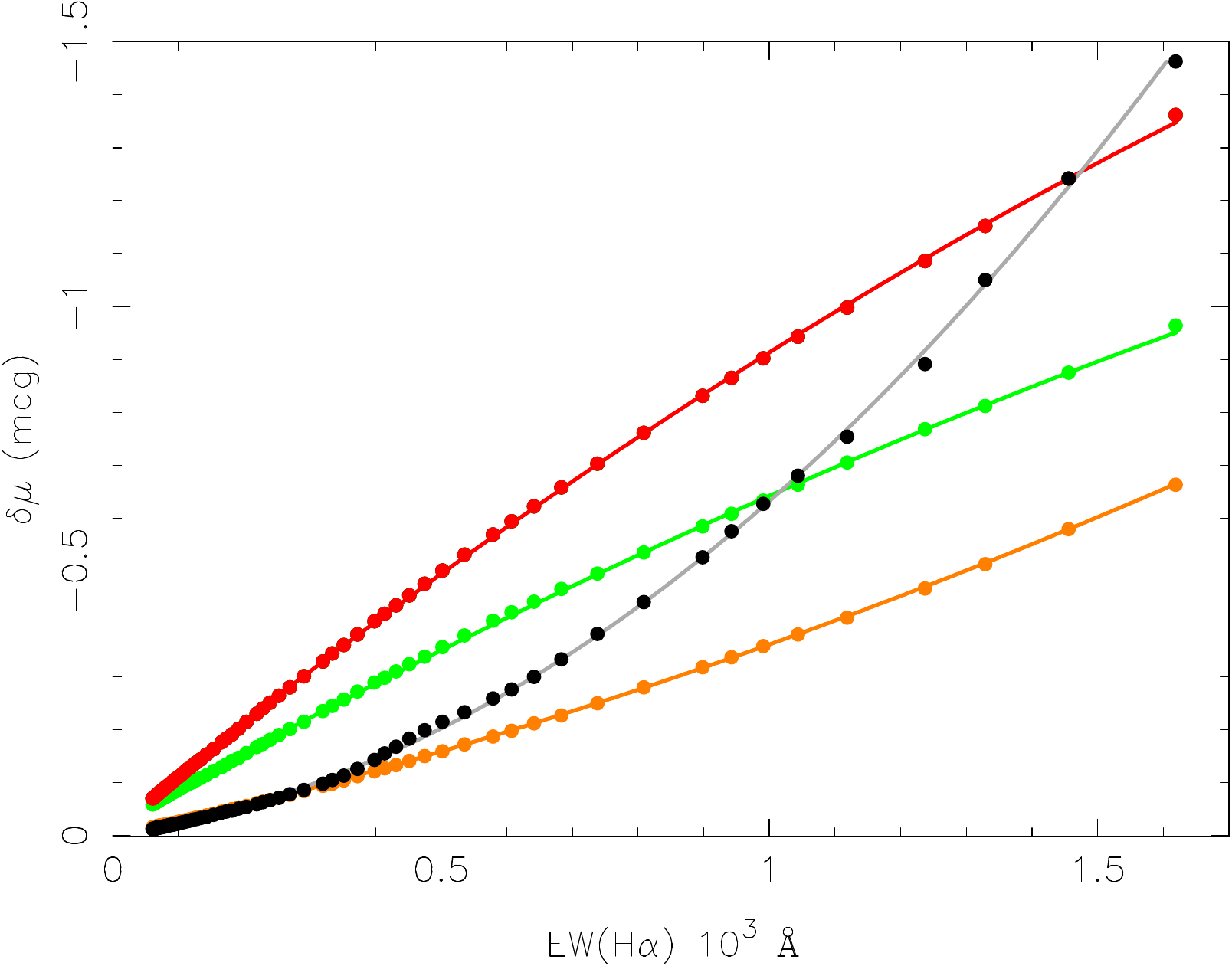}}
\end{picture}
\caption{Enhancement $\delta\mu$ (mag) of broadband $V$, $R$, $I$ and $K$ magnitudes due to nebular (line and continuum) emission
vs. \ewha\ (\AA). The left-hand panel shows a close-up view within 60$\leq$\ewha\,(\AA)$\leq$340 whereas determinations for an \ewha\ of up to 1600 \AA\ are
shown in the right-hand panel. Symbols and lines in the left-hand panel have the same meaning as in Fig.~\ref{ew1}.
Curves in the right-hand panel show polynomial fits in the range 60$\leq$\ewha\,(\AA)$\leq$1600 for predictions referring to continuous SF and a metallicity \zsun/5 (cf. Table~\ref{tab:dmu}).}
\label{ew2}
\end{figure*}
\end{center}

The enhancement of $V$, Cousins $R$ and $I$, and $K$ magnitudes as a function of rest-frame \ewha\ in the case of \zsun/5 is shown in Fig.~\ref{ew2}.
The left--hand panel displays determinations in the range 60 $\leq$ \ewha\ (\AA) $\leq$ 340, which is representative for galactic disks  \citep{Sanchez13Metals3D,BP18} and normal star-forming galaxies, whereas the right-hand panel covers an \ewha\ of up to $\sim$1600 \AA, in the range of determinations for extremely metal-poor BCDs and green peas \citep[e.g.,][]{Salzer1989b,Terlevich91,GdP03,Izotov04,Izotov06,Amorin2012}.
% ::::::::::::::::::::::::::::::::::::::::::: Fig. ew2 :::::::::::::::::::::::::::::::::::::::::::::
% ::::::::::::::::::::::::::::::::::::::::::: Fig. ew2 :::::::::::::::::::::::::::::::::::::::::::::
From the left-hand panel it can be appreciated that the assumed SFH has little influence on the relation between \ewha\ and $\delta\mu$.
The substantial photometric effect of nebular emission in high-sSFR systems with a rest-frame \ewha$\sim 10^3$ \AA\
(e.g., starburst galaxies or massive SF clumps in higher-\zet\ spiral galaxies; cf. Sect. \ref{sub:nebSED}) is reflected in a $\delta\mu$ of --0.37 mag in the $I$ band,
--0.63 mag in the $V$ and $K$ band and --0.9 mag in the $R$ band. The increase in ObsF EWs by a factor of 1+\zet\ due to the stretching of the
continuum is further enhancing the effect of nebular emission (cf. Fig.~\ref{zmag}).

The relationship between $\delta\mu$ and \ewha\ can be approximated with second-order polynomials of the form
$\delta\mu$ (mag) = $a_0$ + $a_1$\,k\ewha\ + $a_2$\,k\ewha$^2$, where k\ewha\ is in k\AA. Table~\ref{tab:dmu} lists the coefficients applying to the range 60$\leq$\ewha\,(\AA)$\leq$1600 for continuous SF and the three stellar metallicities considered (\zsun/20, \zsun/5 and \zsun).

\begin{table*}[h]\label{tab:dmu}{\parbox{14cm}{\caption{Coefficients of second-order  polynomial fits to $\delta\mu$ (mag) as a function EW (k\AA)}}
%\rule[-0.5ex]{0cm}{0.5ex}
}\\[-1ex]
\begin{tabular*}{13.2cm}{lccccccccc} %\label{label}
\hline
\pano &   \multicolumn{3}{c}{\zsun/20} & \multicolumn{3}{c}{\zsun/5} & \multicolumn{3}{c}{\zsun}\\
     & $a_0$ & $a_1$ & $a_2$& $a_0$ & $a_1$ & $a_2$& $a_0$ & $a_1$ & $a_2$ \\
%     \kato
\hline
\pano
$V$  &--0.024 & --0.661 & --0.050 & --0.019 & --0.700 & \ 0.077 & --0.023 & --0.714 & \ 0.101 \\
$R$  &--0.014 & --1.006 &  \ 0.111 & --0.015 & --1.019 & \ 0.121 & --0.016 & --1.028 & \ 0.125 \\
$I$  & \ 0.010 & --0.311 & --0.061 &  \ 0.005 & --0.289 & --0.077 & \ 0.007 & --0.274 & --0.107 \\
$K$  & \ 0.031 & --0.377 & --0.300 & --0.005 & --0.162 & --0.465 & --0.014 & \ 0.119 & --0.822 \\
\hline
\label{tab:dmu}
\end{tabular*}
\end{table*}

\FloatBarrier
% :::::::::::::::::::::::::::::::::::::::::::::::::::::::::::::::::::
\section{Bulge-disk decomposition of synthetic galaxies\label{ap:BD}}
% :::::::::::::::::::::::::::::::::::::::::::::::::::::::::::::::::::
The synthetic galaxies photometrically analyzed in Sect.~\ref{BD} (Fig.~\ref{fig:BDdec1}) were constructed assuming an exponentially decreasing SFR with an e-folding time of 1 Gyr for a bulge of solar metallicity ($\tau$1 model) and continuous SF at a constant SFR for a disk of \zsun/5 (contSF model). As is apparent from Fig.~\ref{fig:BDdec2}, the trends documented in Sect.~\ref{BD} do not qualitatively change when alternative combinations for the SFH in the bulge and the disk are adopted.

The SFHs considered Fig.~\ref{fig:BDdec2} involve a shorter formation timescale for the bulge ($\tau$0.5) in the left panels, the $\tau$1 model for the bulge and a smooth decline in the SFR in the disk according to $\tau$5 (middle panels), and a $\tau$1 model for the bulge combined with a delayed-exponential iB model for the disk (right panels).
In all cases, the stellar metallicity of the bulge and the disk were kept constant at \zsun\ and \zsun/5, respectively.
Results in the left and middle column are based on purely stellar synthetic galaxy models, whereas those in the right-hand column additionally include nebular emission.

In addition to the quantities shown in Fig.~\ref{fig:BDdec1}, namely the concentration index after Eq.~\ref{eq:CI}, reduced effective radius
R\arcmin$_{\rm eff}$, S\'ersic index $\eta$ and the logarithm of the \BD\ and \BT\ ratios, Fig.~\ref{fig:BDdec2} includes
the reduced Petrosian radius R\arcmin$_{\rm Petrosian}$,
the concentration index log(R$_{80}$/R$_{20}$) where  R$_{80}$ and R$_{20}$ denote the radii enclosing 80\% and 20\% of the total luminosity,
and the \citet{Trujillo01} concentration parameter.
The layout follows that of Fig.~\ref{fig:BDdec1}, with quantities obtained from simulations for single-age SEDs shown with dotted curves, and those based on EvCon simulations with solid curves. Similar to Fig.~\ref{fig:BDdec1}, thick green and gray curves show the true (rest-frame) properties of the galaxy in the $V$ and $H$ band, as obtained from EvCon simulations.

\begin{figure*}
\begin{picture}(200,78)
\put(0,35){\includegraphics[clip, width=3.0cm]{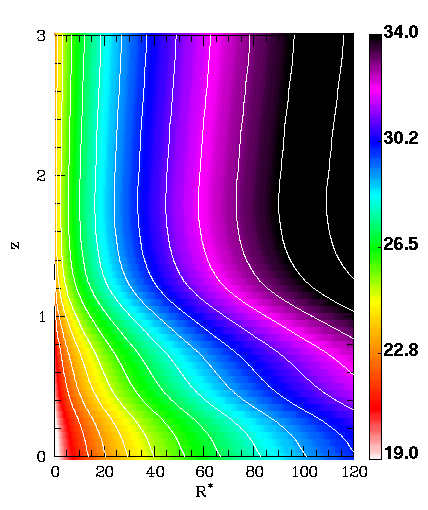}}
\put(30,35){\includegraphics[clip, width=3.0cm]{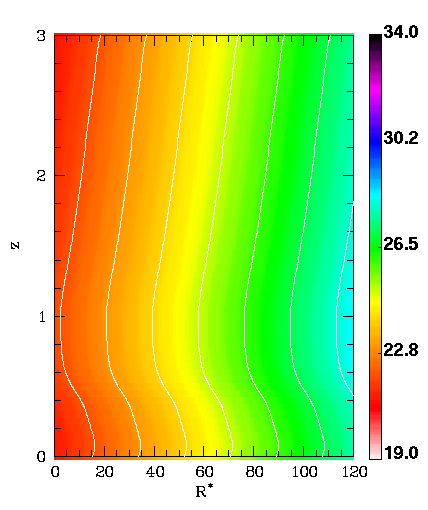}}
\put(60,35){\includegraphics[clip, width=3.0cm]{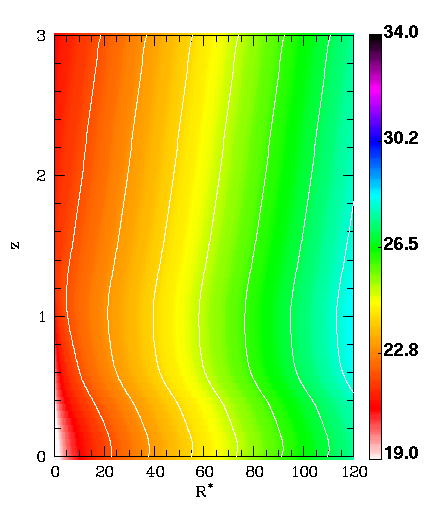}}
\put(0,0){\includegraphics[clip, width=3.0cm]{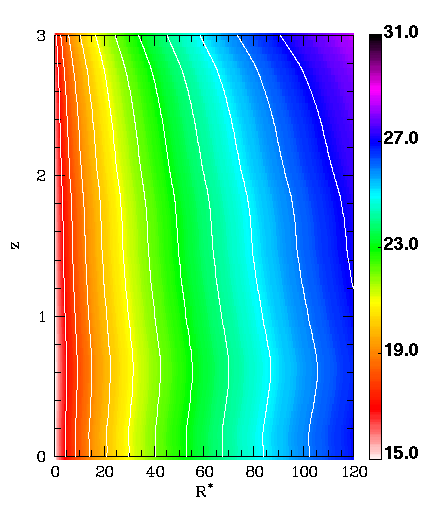}}
\put(30,0){\includegraphics[clip, width=3.0cm]{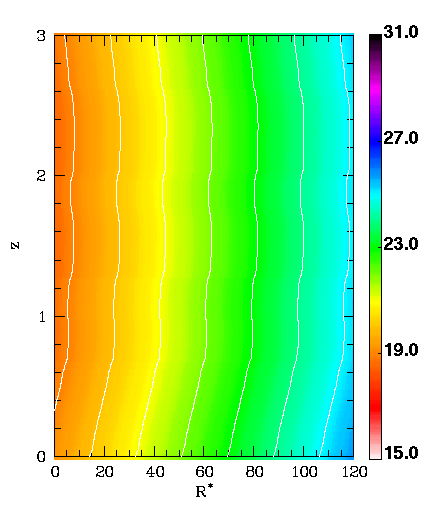}}
\put(60,0){\includegraphics[clip, width=3.0cm]{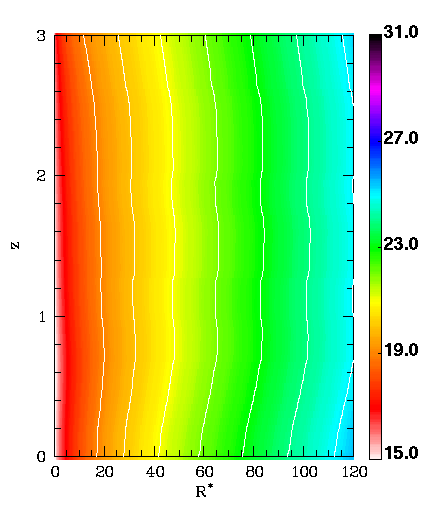}}

\put(95,35){\includegraphics[clip, width=3.0cm]{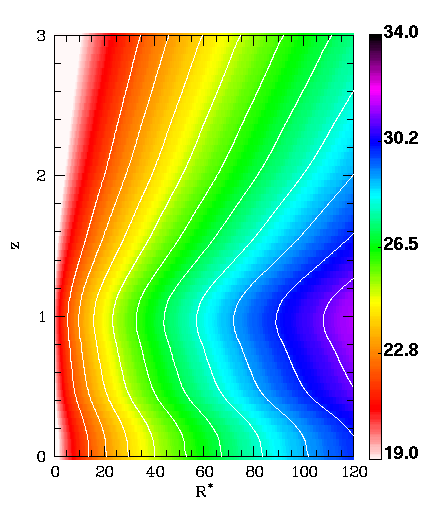}}
\put(125,35){\includegraphics[clip, width=3.0cm]{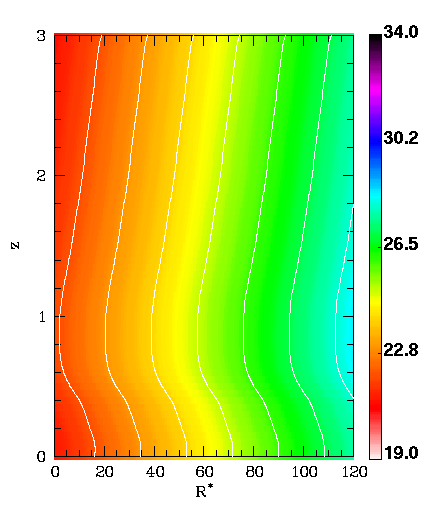}}
\put(155,35){\includegraphics[clip, width=3.0cm]{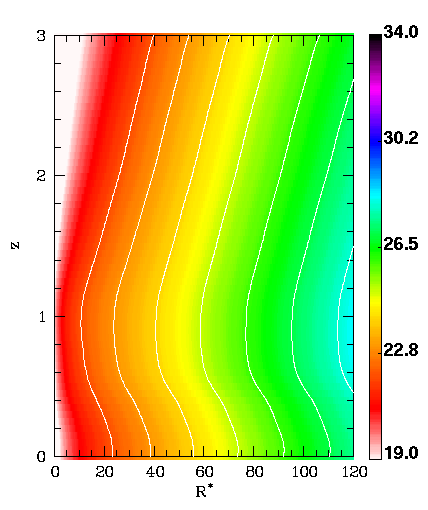}}
\put(95,0){\includegraphics[clip, width=3.0cm]{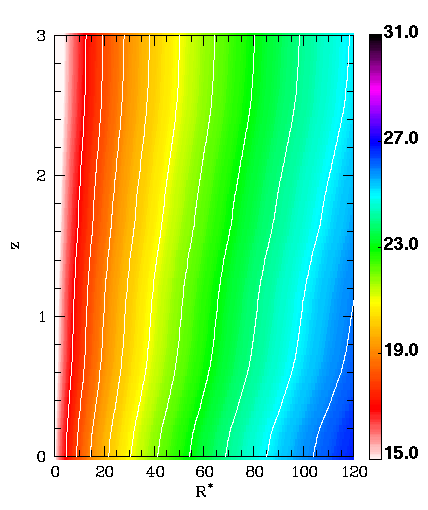}}
\put(125,0){\includegraphics[clip, width=3.0cm]{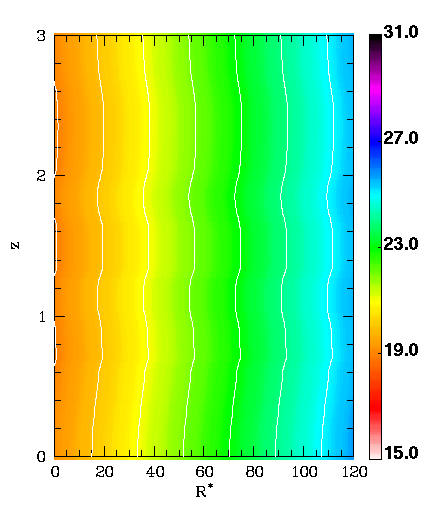}}
\put(155,0){\includegraphics[clip, width=3.0cm]{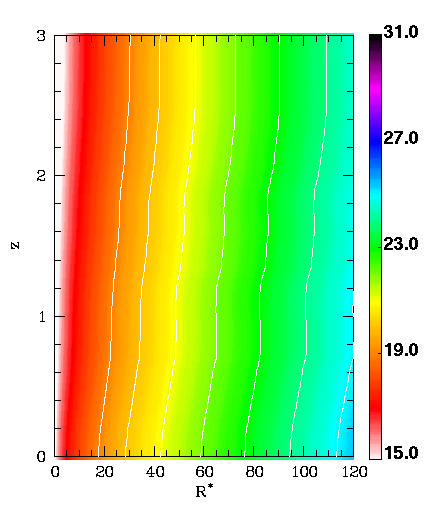}}

\PutLabel{20}{74}{\nlx \textcolor{black}{simulation with single-age (13.7 Gyr) SEDs}}
\PutLabel{120}{74}{\nlx \textcolor{black}{EvCon simulation}}
\PutLabel{-2}{65}{\nlx \textcolor{black}{V}}
\PutLabel{-2}{30}{\nlx \textcolor{black}{H}}
\PutLabel{12}{70}{\mlx \textcolor{black}{bulge}}
\PutLabel{42}{70}{\mlx \textcolor{black}{disk}}
\PutLabel{69}{70}{\mlx \textcolor{black}{bulge+disk}}
\PutLabel{107}{70}{\mlx \textcolor{black}{bulge}}
\PutLabel{137}{70}{\mlx \textcolor{black}{disk}}
\PutLabel{164}{70}{\mlx \textcolor{black}{bulge+disk}}
\end{picture}
\caption{Variation in the reduced $V$ and $H$ surface brightness (\sbb) of the bulge, disk and their sum as a function of galactocentric radius \rr\ (\arcsec) and redshift
for simulations based on SEDs corresponding to a local (13.7 Gyr old) galaxy (left) and those in an EvCon manner (right). Nebular emission has been taken into account. Contours referring to the bulge go from 22 (17) to 35 (27) \sbb\ in $V$ ($H$), and those overlaid with the disk and the total emission from 22 (19) to 28 (25) \sbb\ in $V$ ($H$) in increments of 1 mag.}
\label{ap:BDimages}
\end{figure*}

\begin{figure}
\begin{picture}(86,80)
\put(0,37){\includegraphics[clip, width=4.0cm]{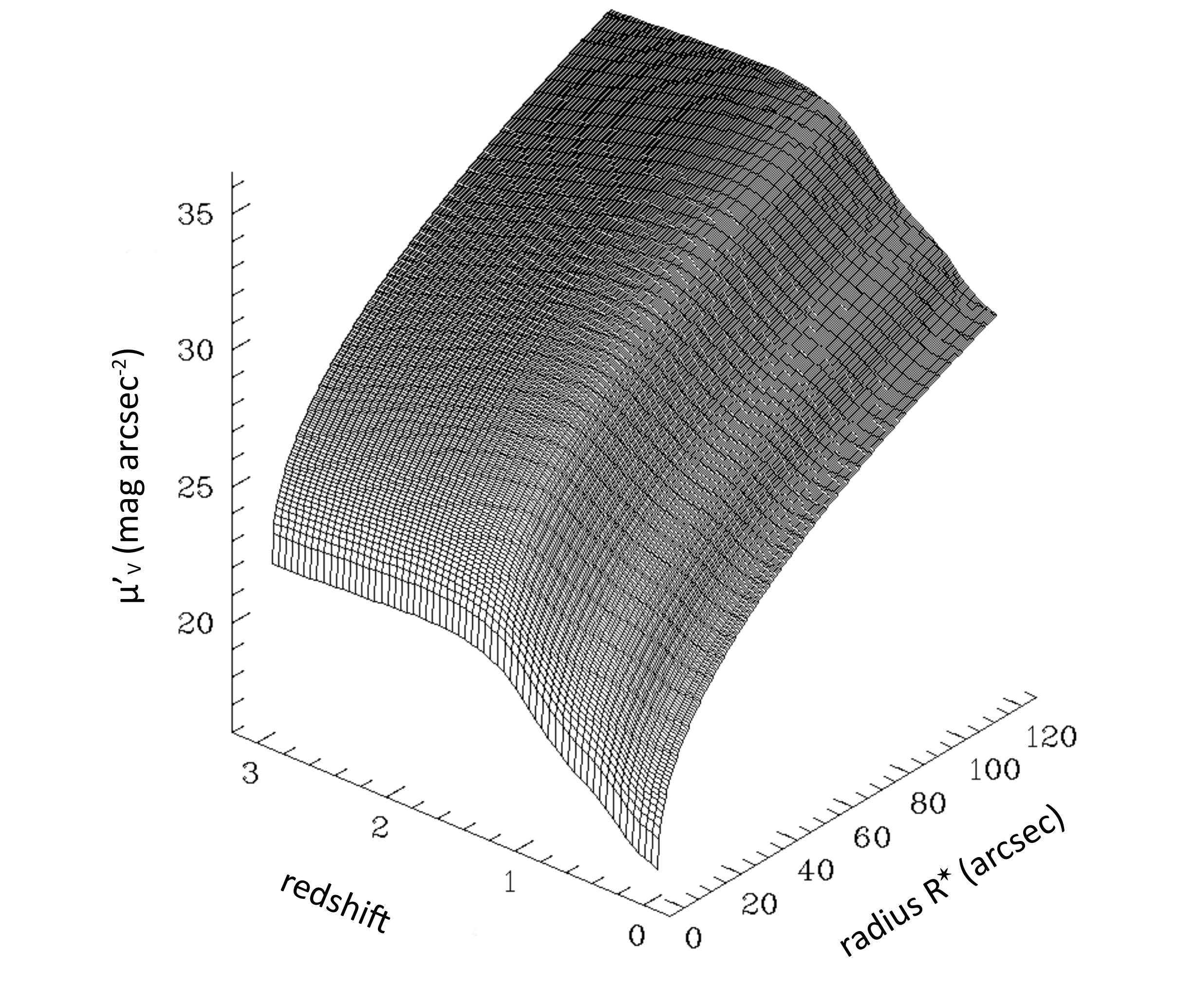}}
\put(45,37){\includegraphics[clip, width=4.0cm]{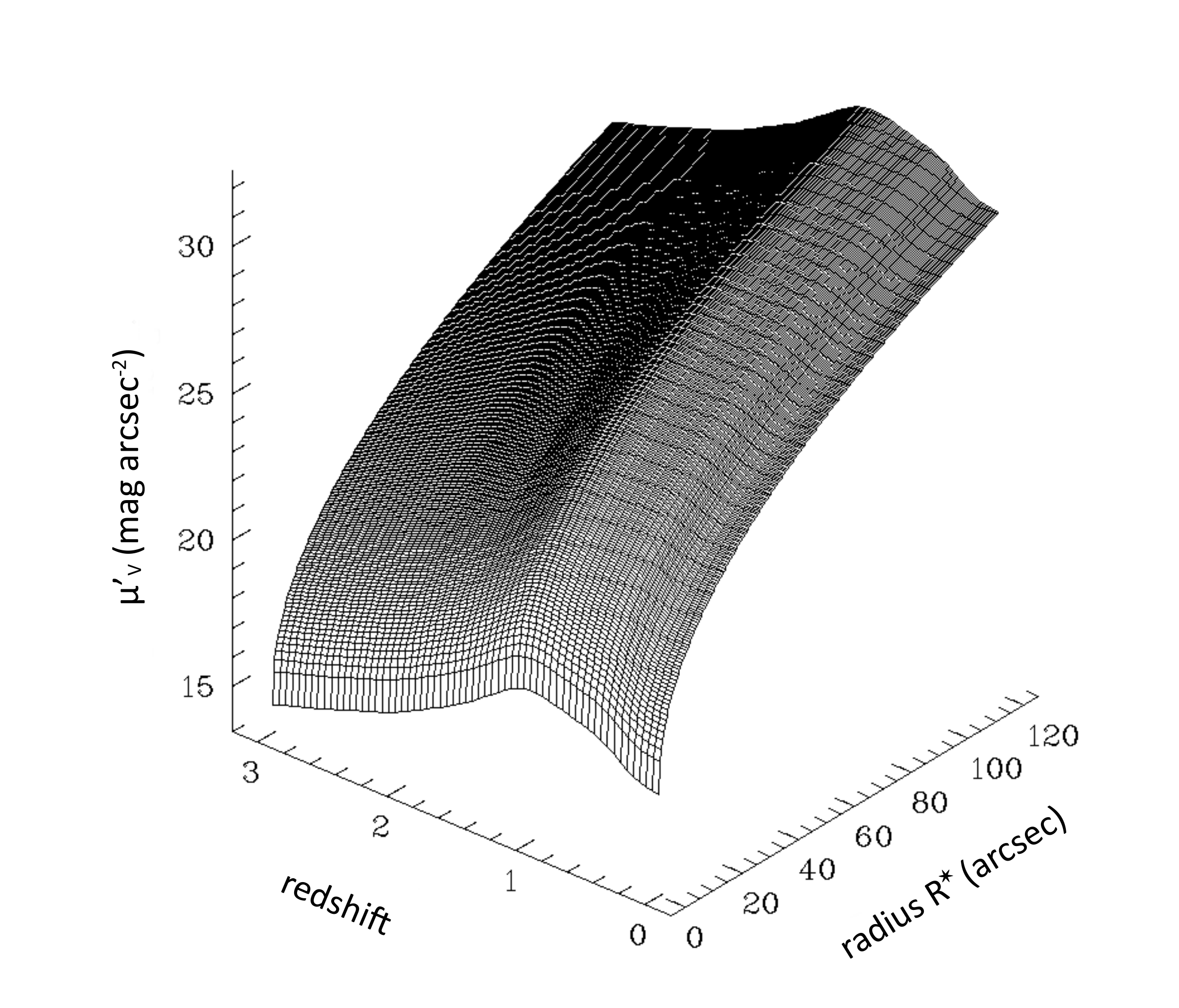}}
\put(0,0){\includegraphics[clip, width=4.0cm]{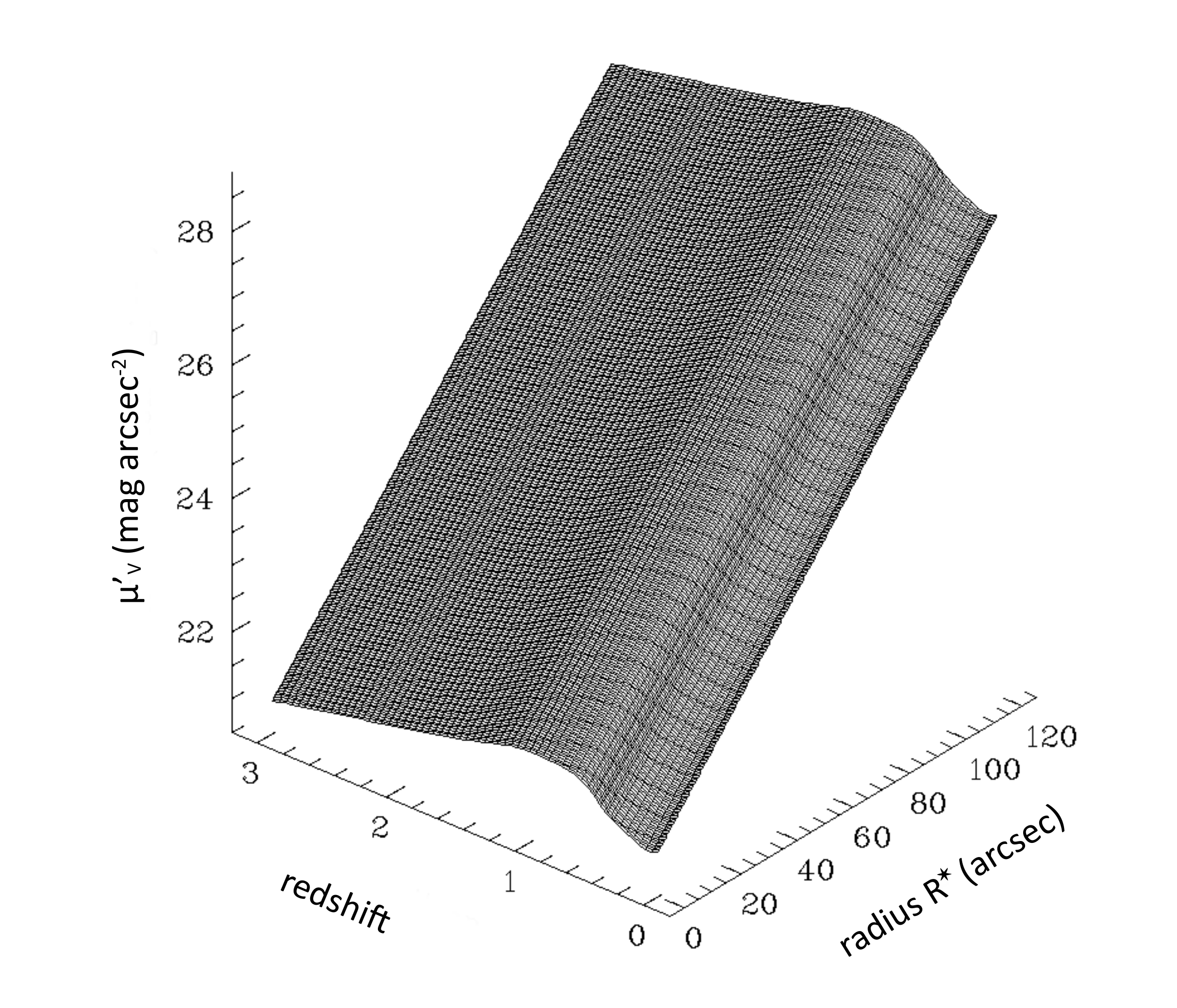}}
\put(45,0){\includegraphics[clip, width=4.0cm]{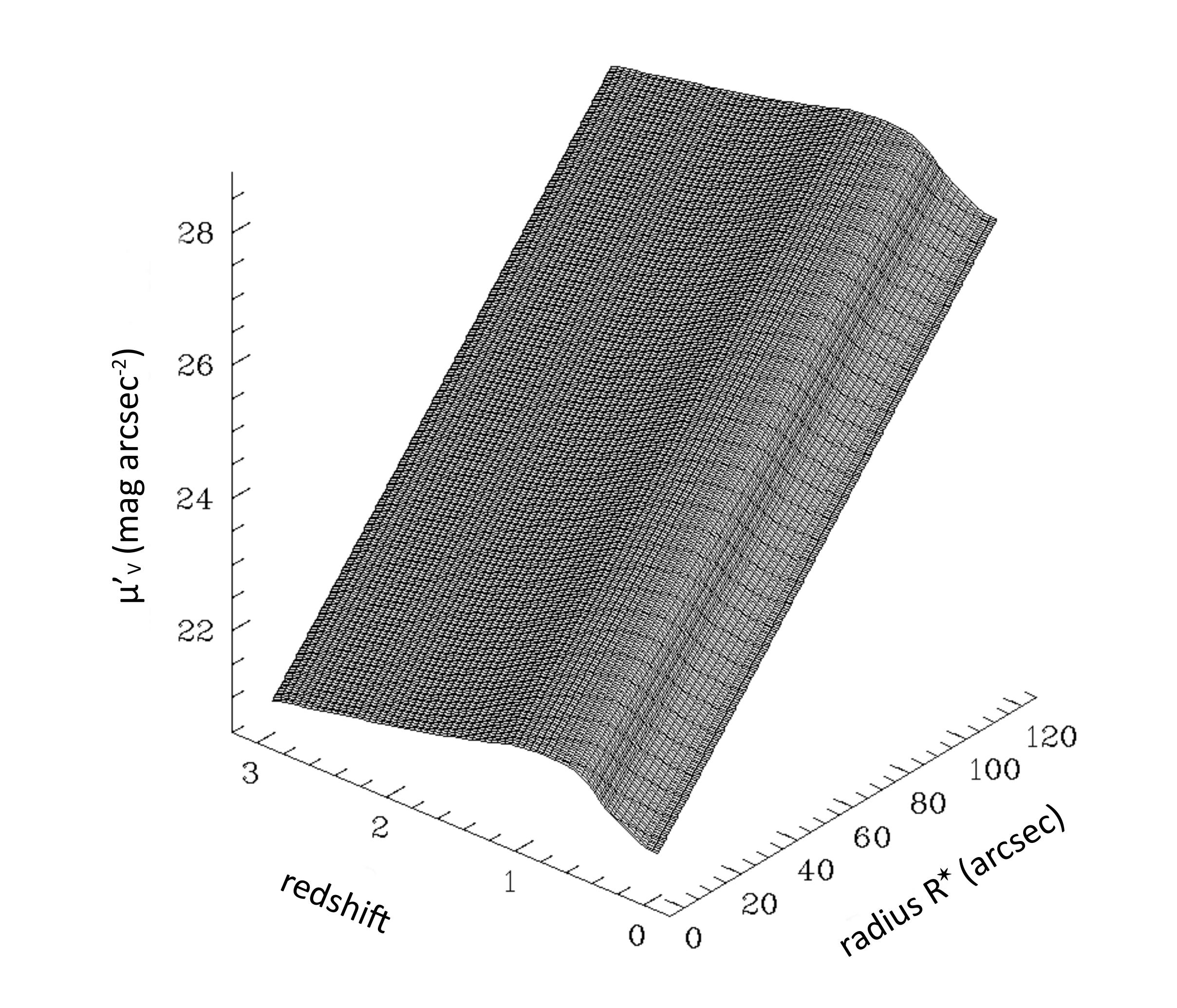}}

\PutLabel{2}{76}{\mlx \textcolor{black}{simulation with single-age SEDs}}
\PutLabel{50}{76}{\mlx \textcolor{black}{EvCon simulation}}
\PutLabel{-2}{65}{\mlx \textcolor{red}{bulge}}
\PutLabel{-2}{30}{\mlx \textcolor{blue}{disk}}
\end{picture}
\caption{Illustration of the reduced $V$-band surface brightness of the bulge and disk (upper and lower row, respectively) as a function of galactocentric radius \rr\ and redshift
in the case of simulations based on 13.7 Gyr old SEDs (left) and those employing an EvCon approach (right). The disk follows a similar evolution across \zet\ for both types of simulations, showing a minor dimming from \zet=3 to \zet$\sim$1, followed by a brightening by $\sim$0.7 mag at a lower redshift.
The situation is different for the bulge, which in the case of single-age simulations experiences between \zet=0 and \zet$\sim$1 a dimming by $>$5 mag (cf. Fig.~\ref{sSED}).
EvCon simulations (cf. Fig.~\ref{eSED}) indicate a dimming by $\sim$1.5 $V$ mag from \zet=0 and \zet$\sim$1 that is followed by a brightening by up to 2 mag at a larger redshift.}
\label{ap:per}
\end{figure}

\begin{figure*}
\begin{picture}(200,200)
\put(0,183){\includegraphics[clip, width=6.1cm]{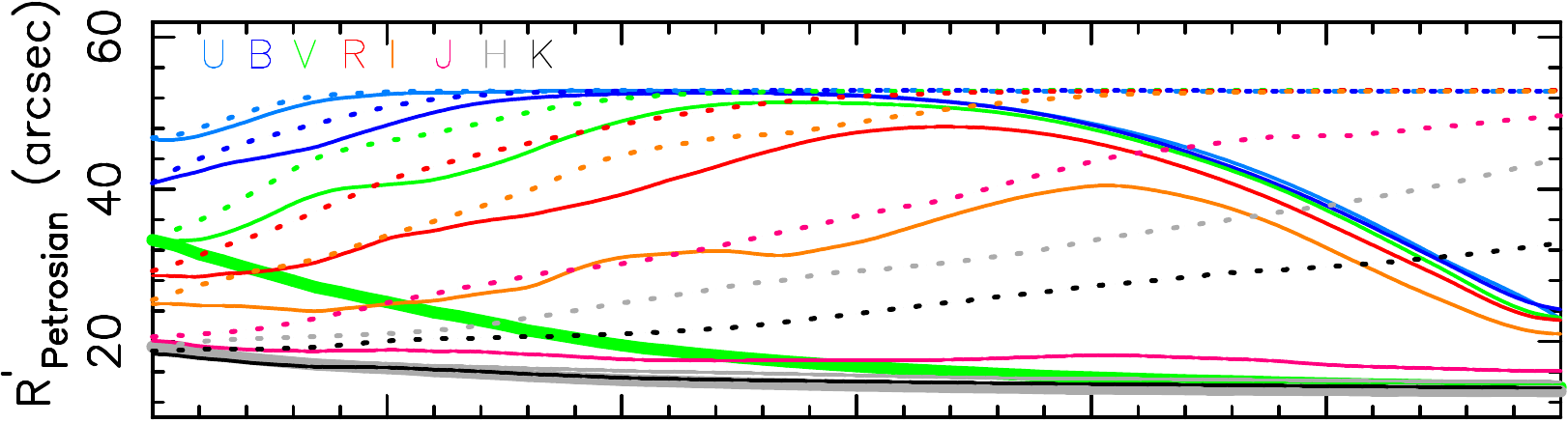}}
\put(0,165){\includegraphics[clip, width=6.1cm]{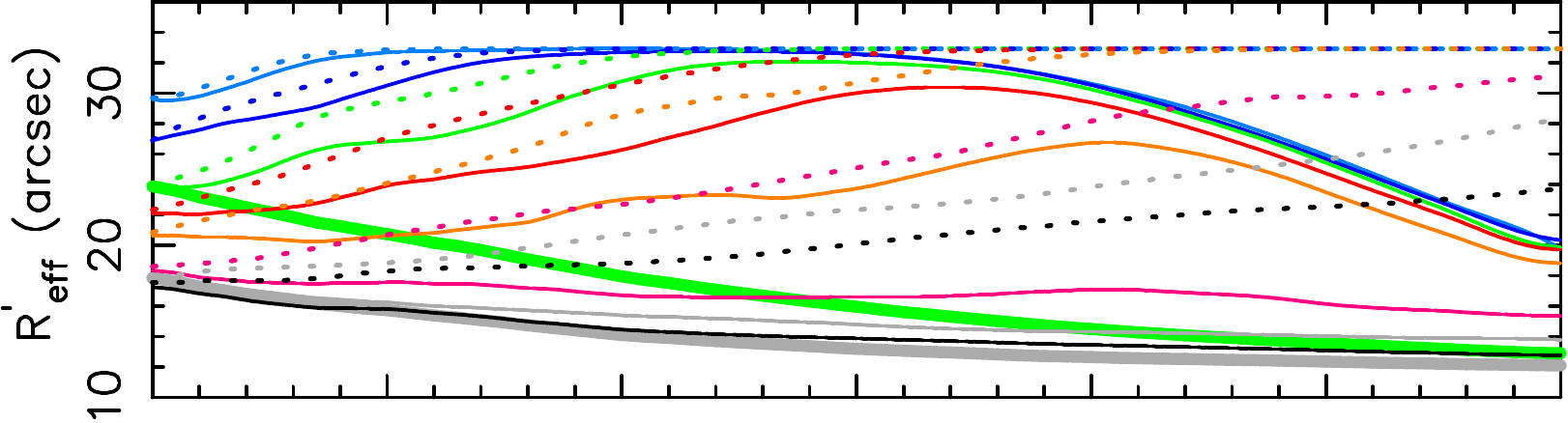}}
\put(0,147){\includegraphics[clip, width=6.1cm]{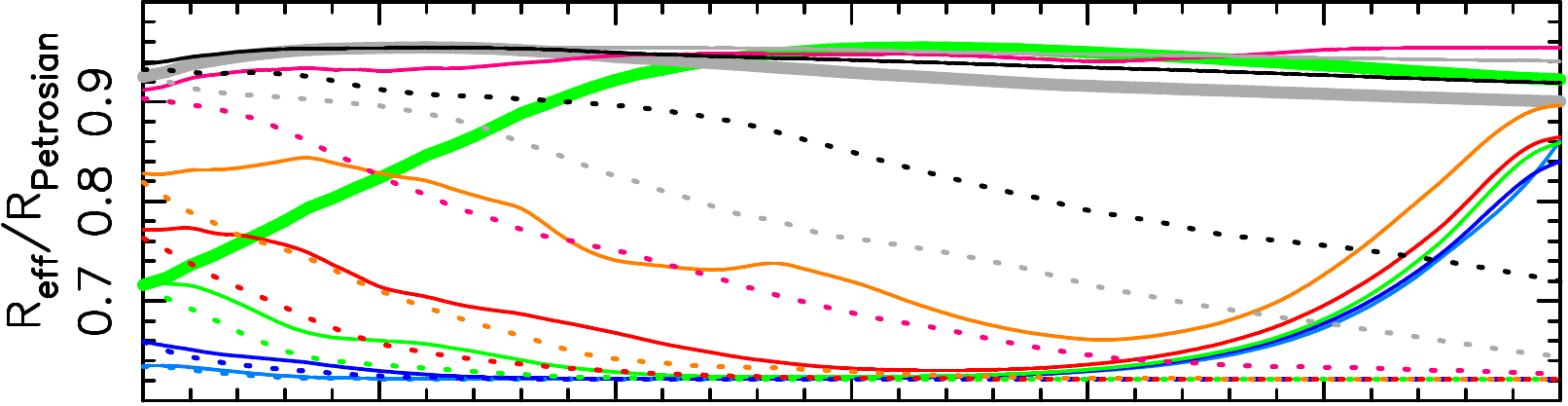}}
\put(0,128){\includegraphics[clip, width=6.1cm]{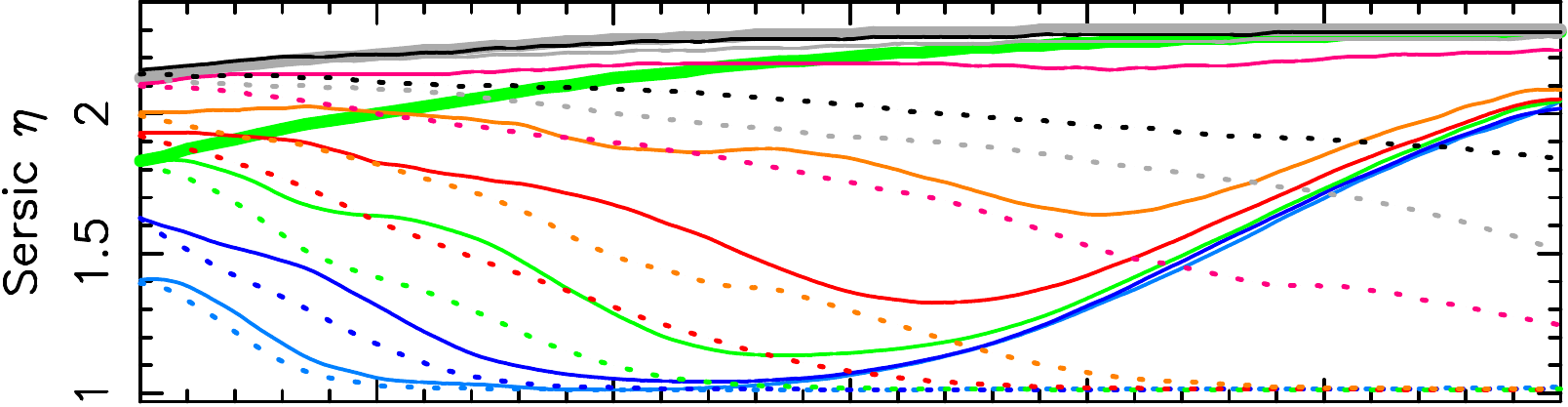}}
\put(0,110){\includegraphics[clip, width=6.1cm]{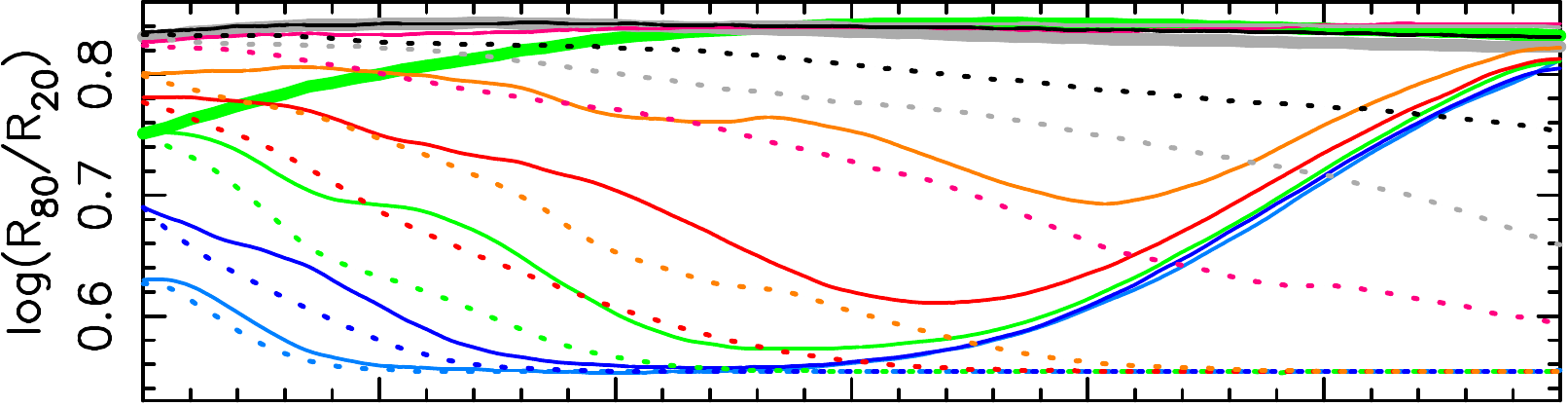}}
\put(0,91){\includegraphics[clip, width=6.1cm]{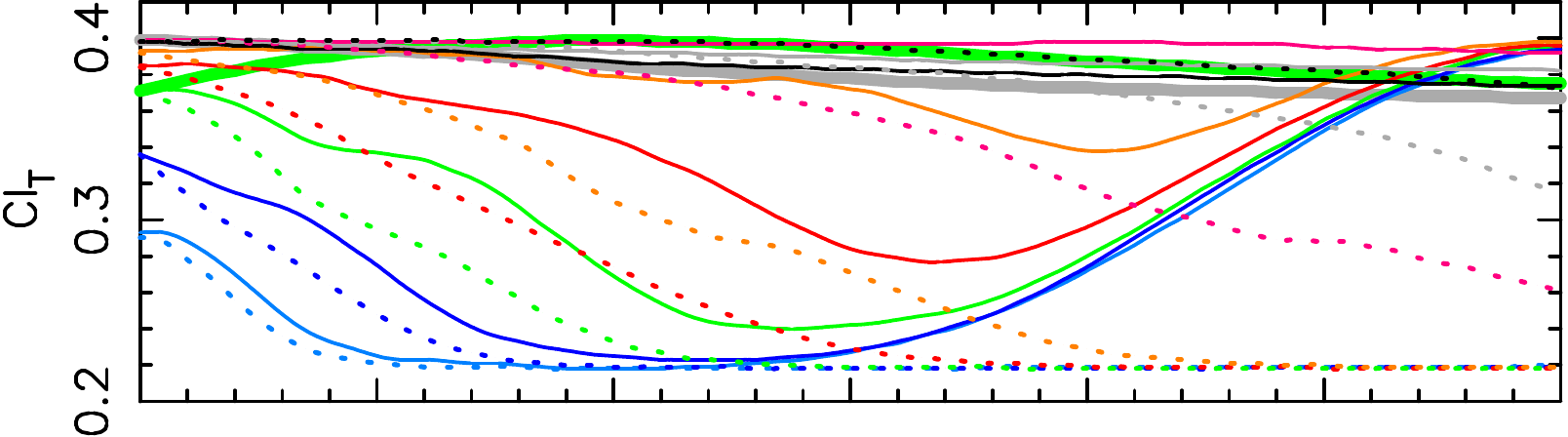}}
\put(0,73){\includegraphics[clip, width=6.1cm]{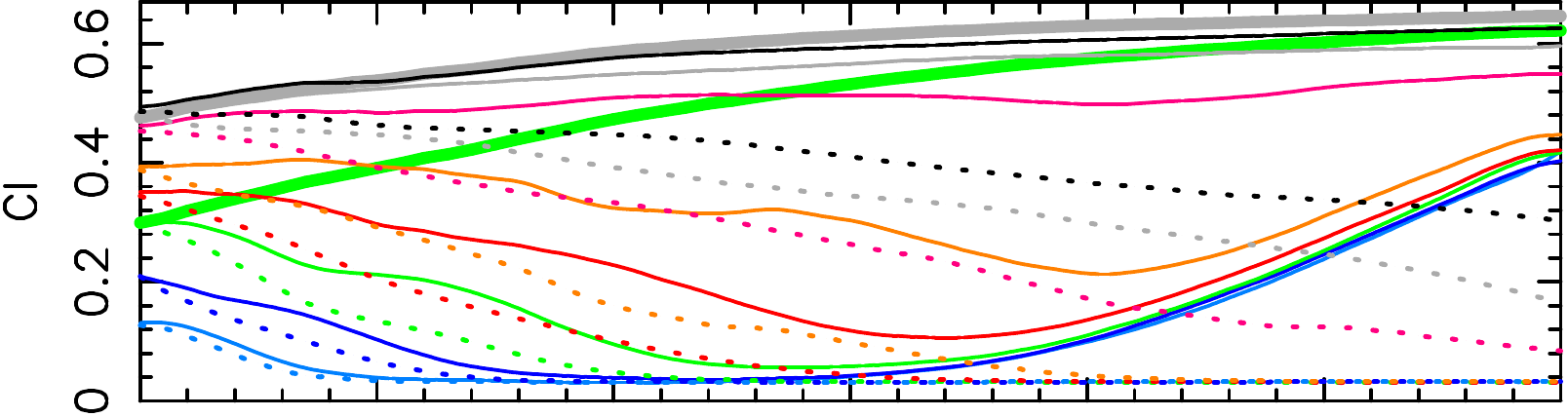}}
\put(0,38){\includegraphics[clip, width=6.1cm]{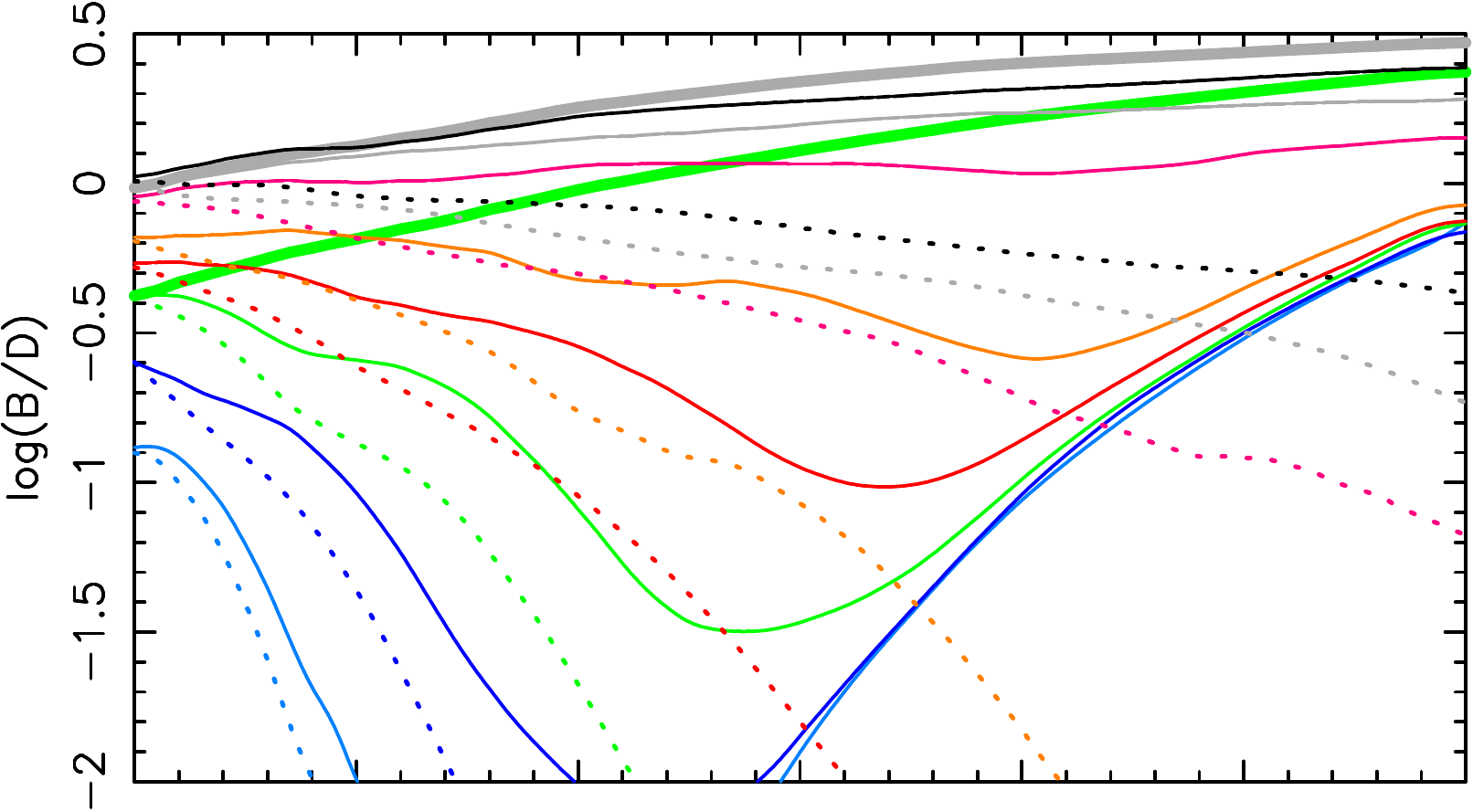}}
\put(0,0){\includegraphics[clip, width=6.12cm]{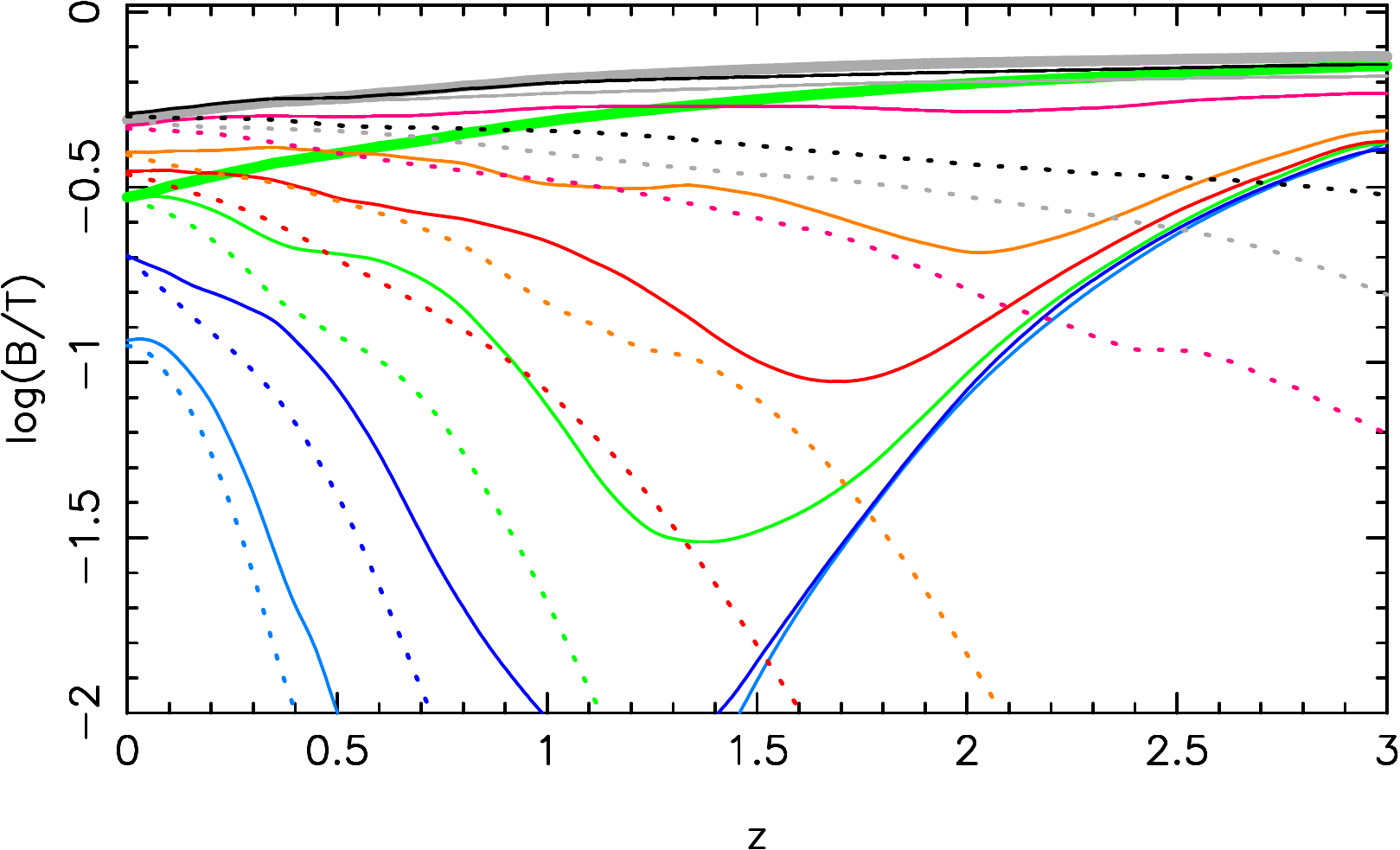}}
\put(62,183){\includegraphics[clip, width=6.1cm]{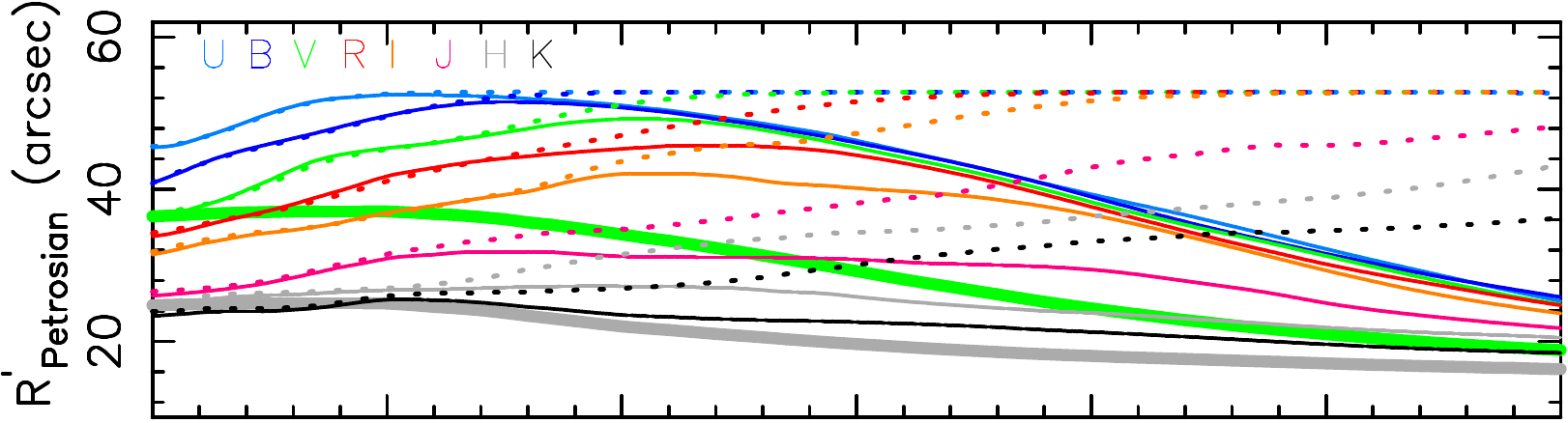}}
\put(62,165){\includegraphics[clip, width=6.1cm]{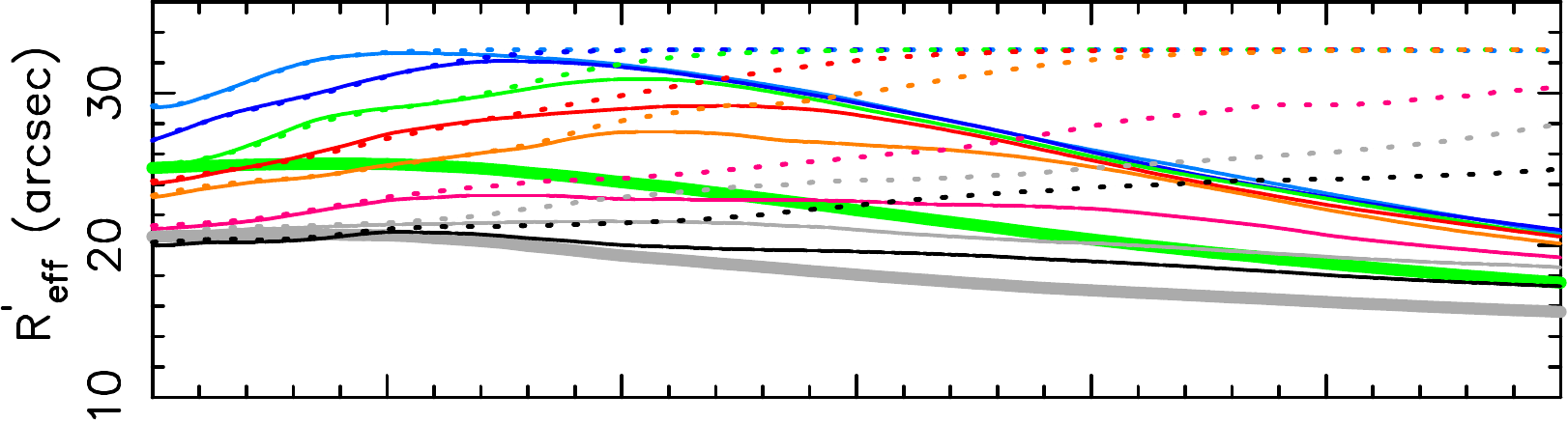}}
\put(62,147){\includegraphics[clip, width=6.1cm]{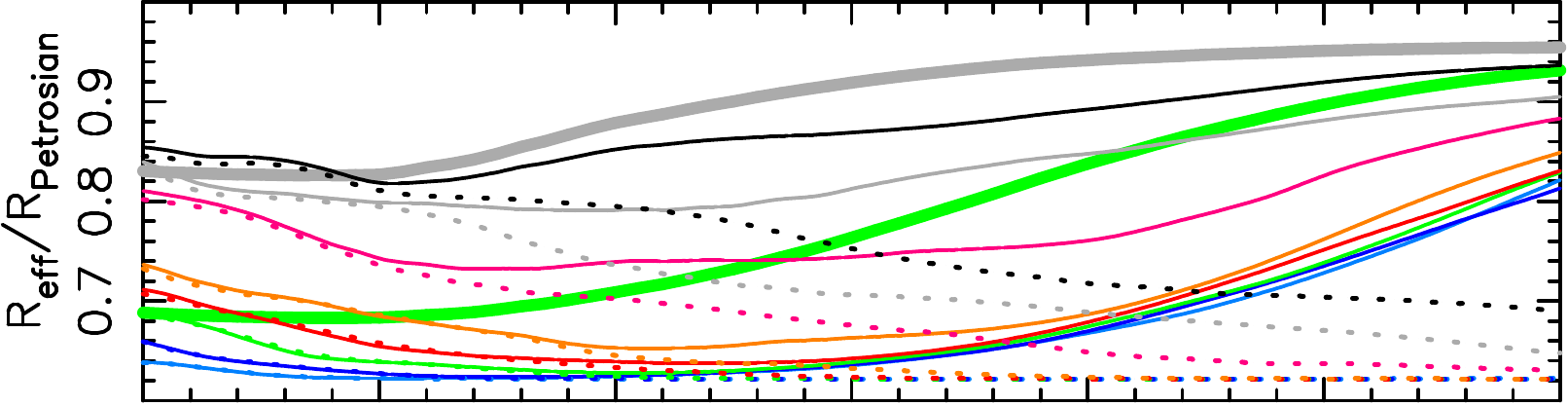}}
\put(62,128){\includegraphics[clip, width=6.1cm]{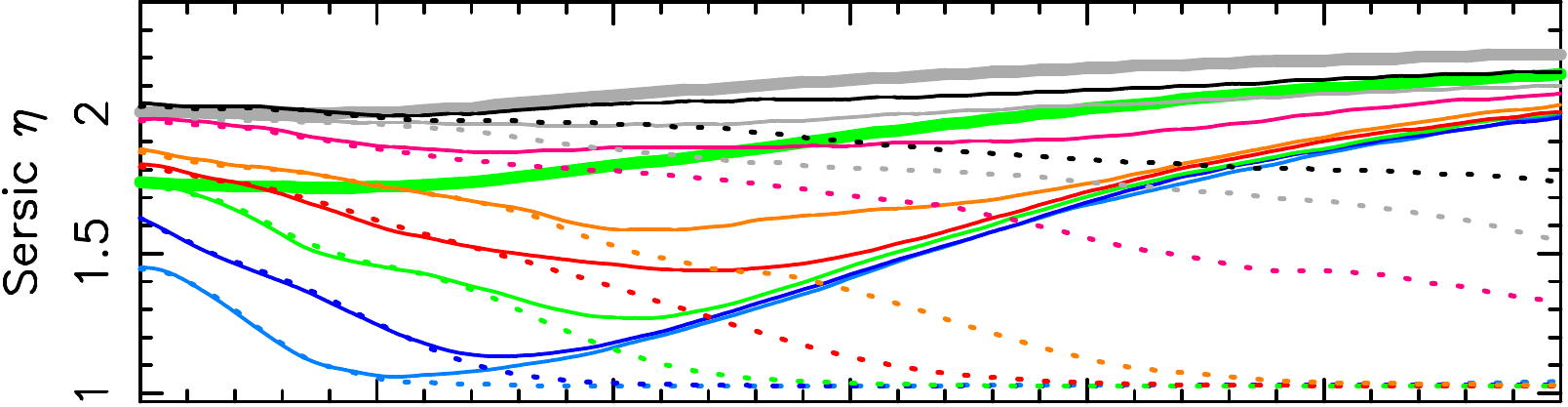}}
\put(62,110){\includegraphics[clip, width=6.1cm]{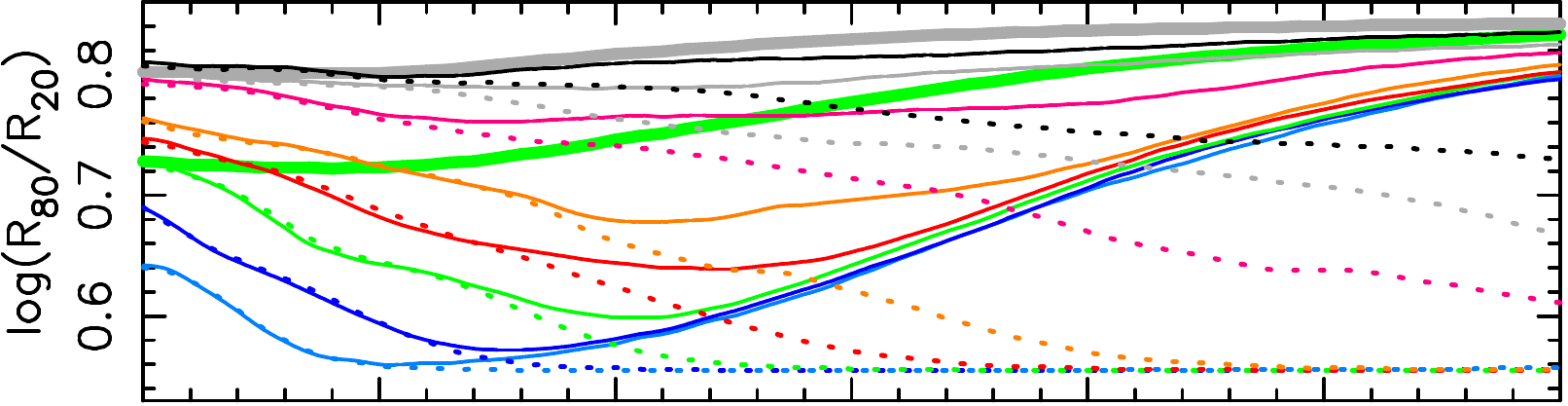}}
\put(62,91){\includegraphics[clip, width=6.1cm]{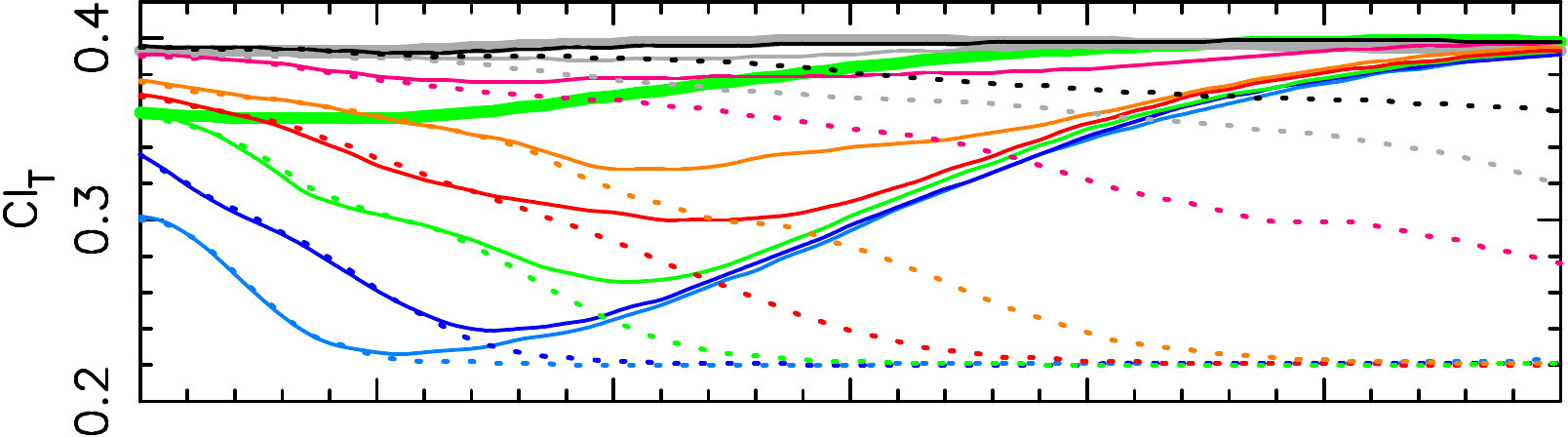}}
\put(62,73){\includegraphics[clip, width=6.1cm]{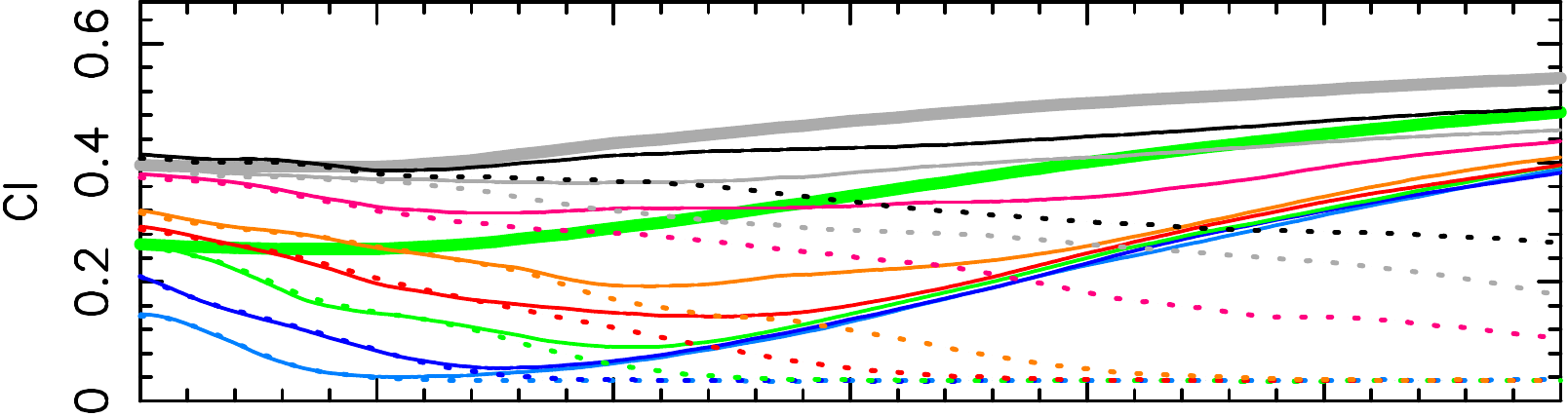}}
\put(62,38){\includegraphics[clip, width=6.1cm]{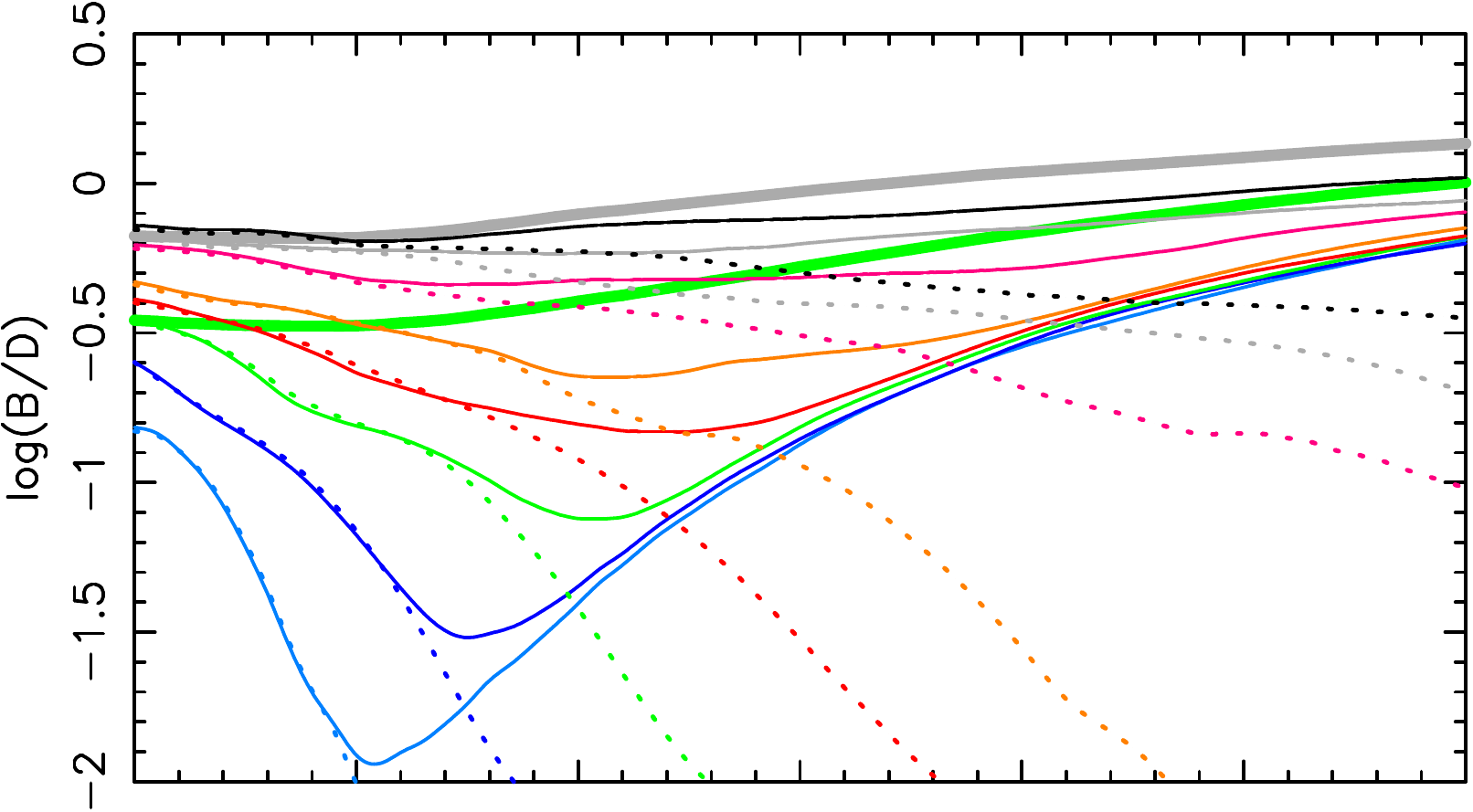}}
\put(62,0){\includegraphics[clip, width=6.12cm]{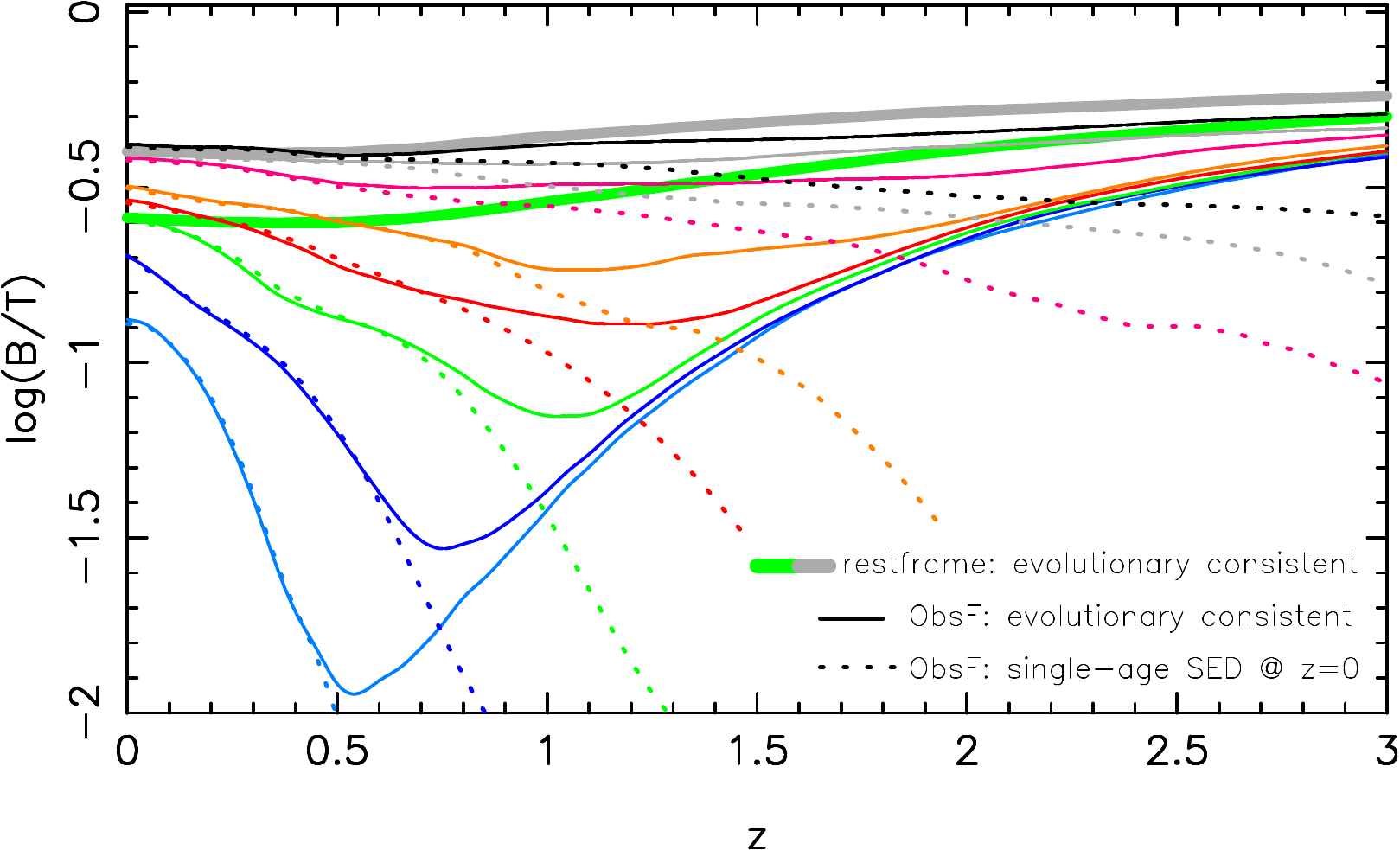}}
\put(124,183){\includegraphics[clip, width=6.1cm]{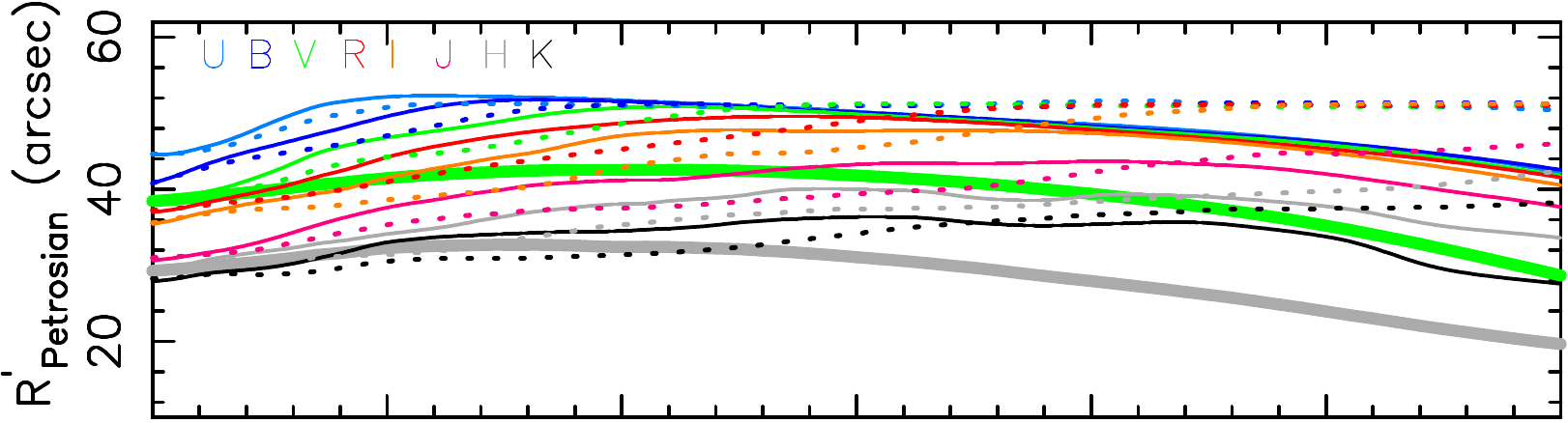}}
\put(124,165){\includegraphics[clip, width=6.1cm]{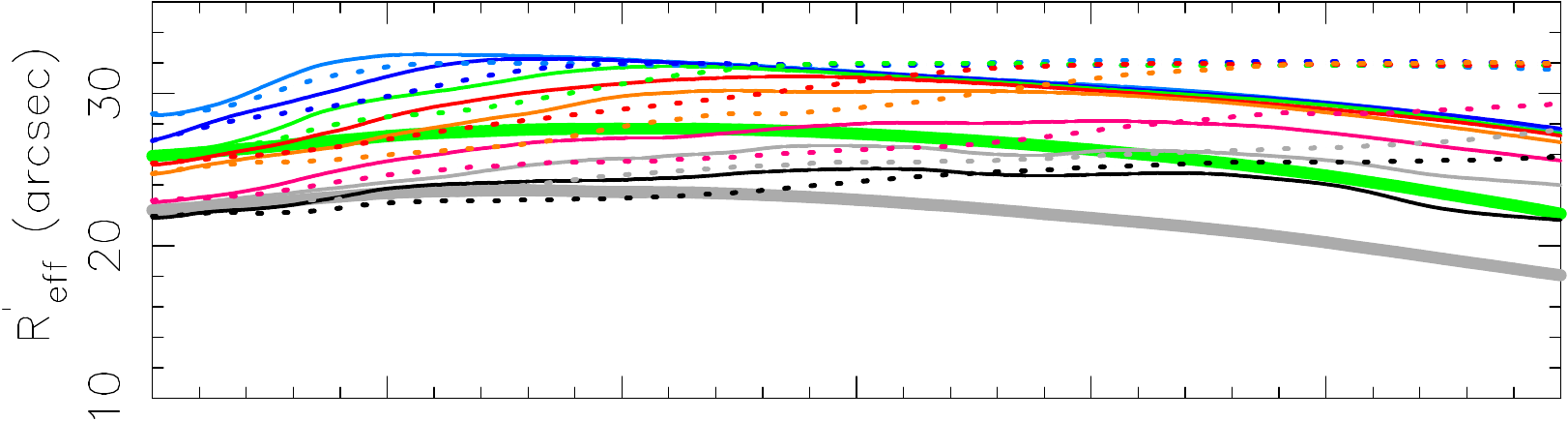}}
\put(124,147){\includegraphics[clip, width=6.1cm]{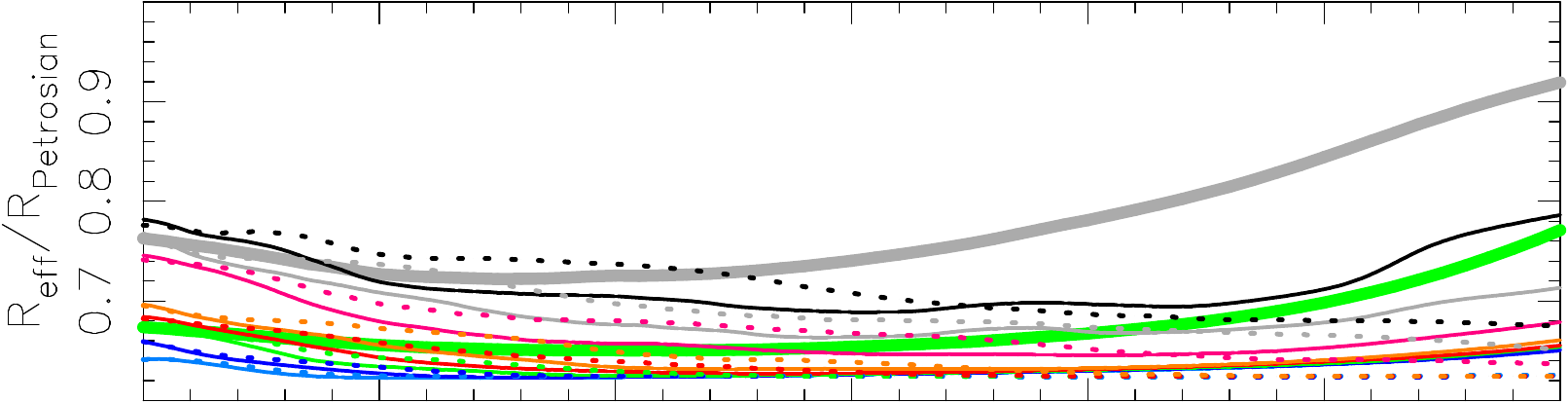}}
\put(124,128){\includegraphics[clip, width=6.1cm]{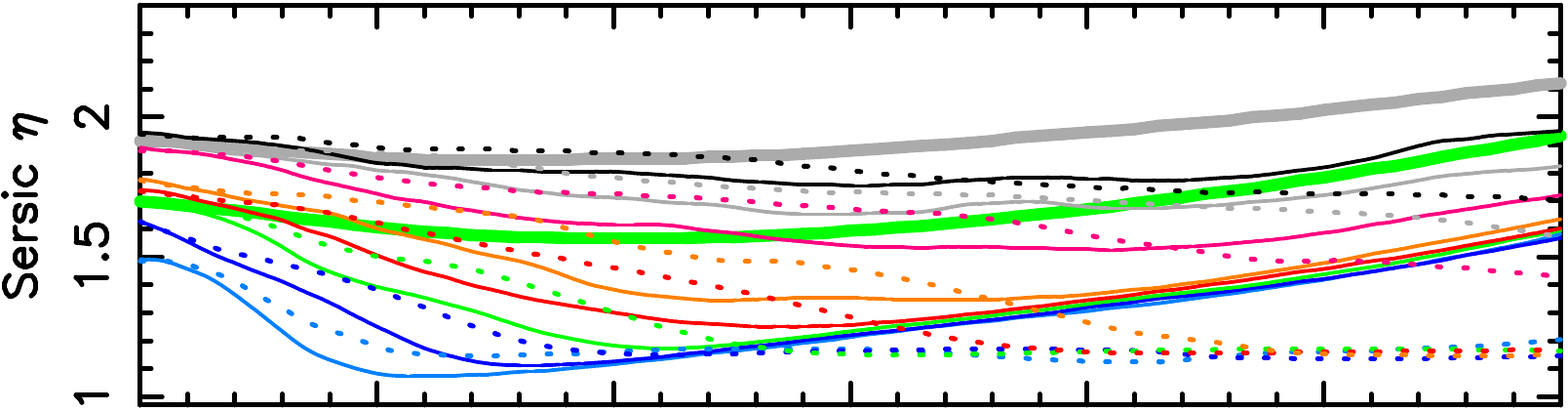}}
\put(124,110){\includegraphics[clip, width=6.1cm]{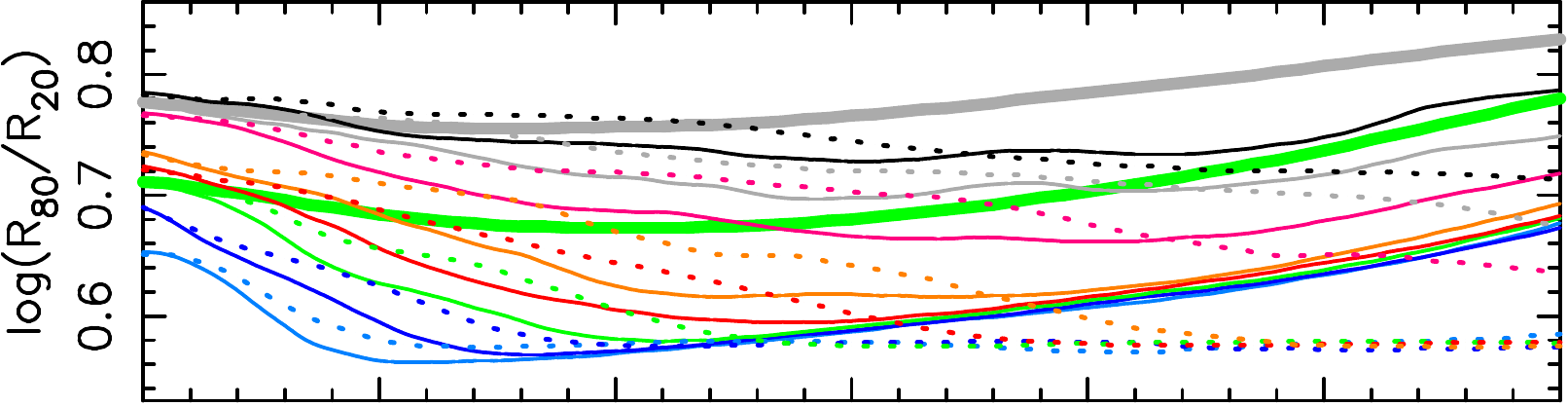}}
\put(124,91){\includegraphics[clip, width=6.1cm]{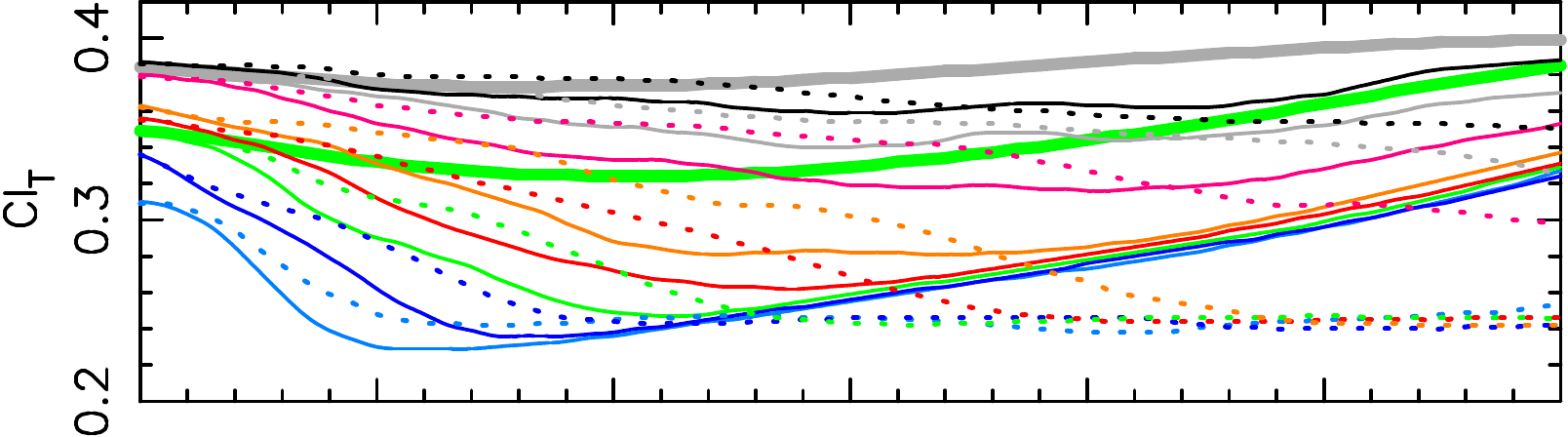}}
\put(124,73){\includegraphics[clip, width=6.1cm]{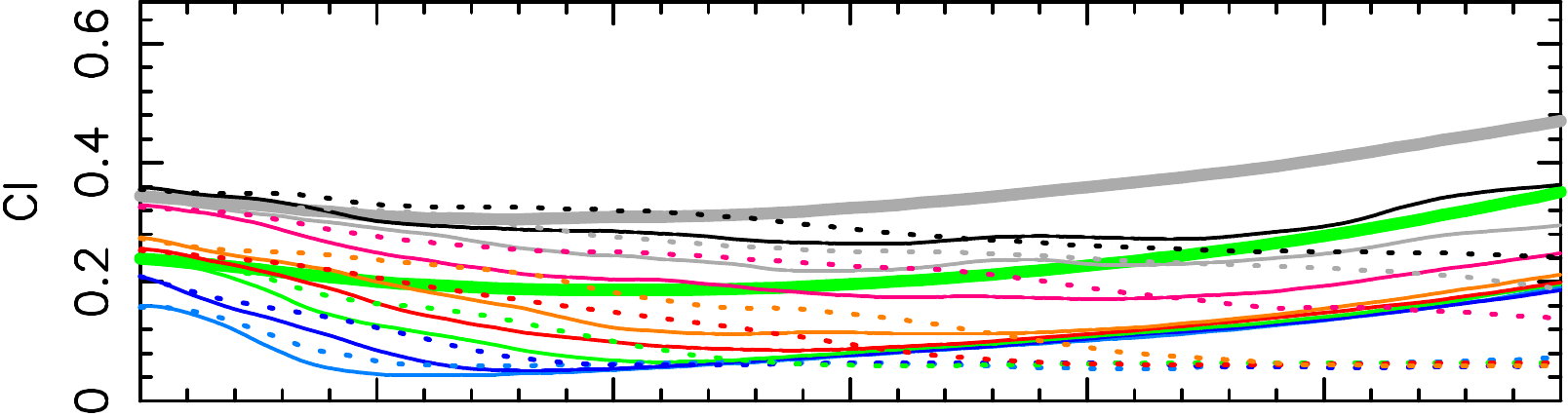}}
\put(124,38){\includegraphics[clip, width=6.1cm]{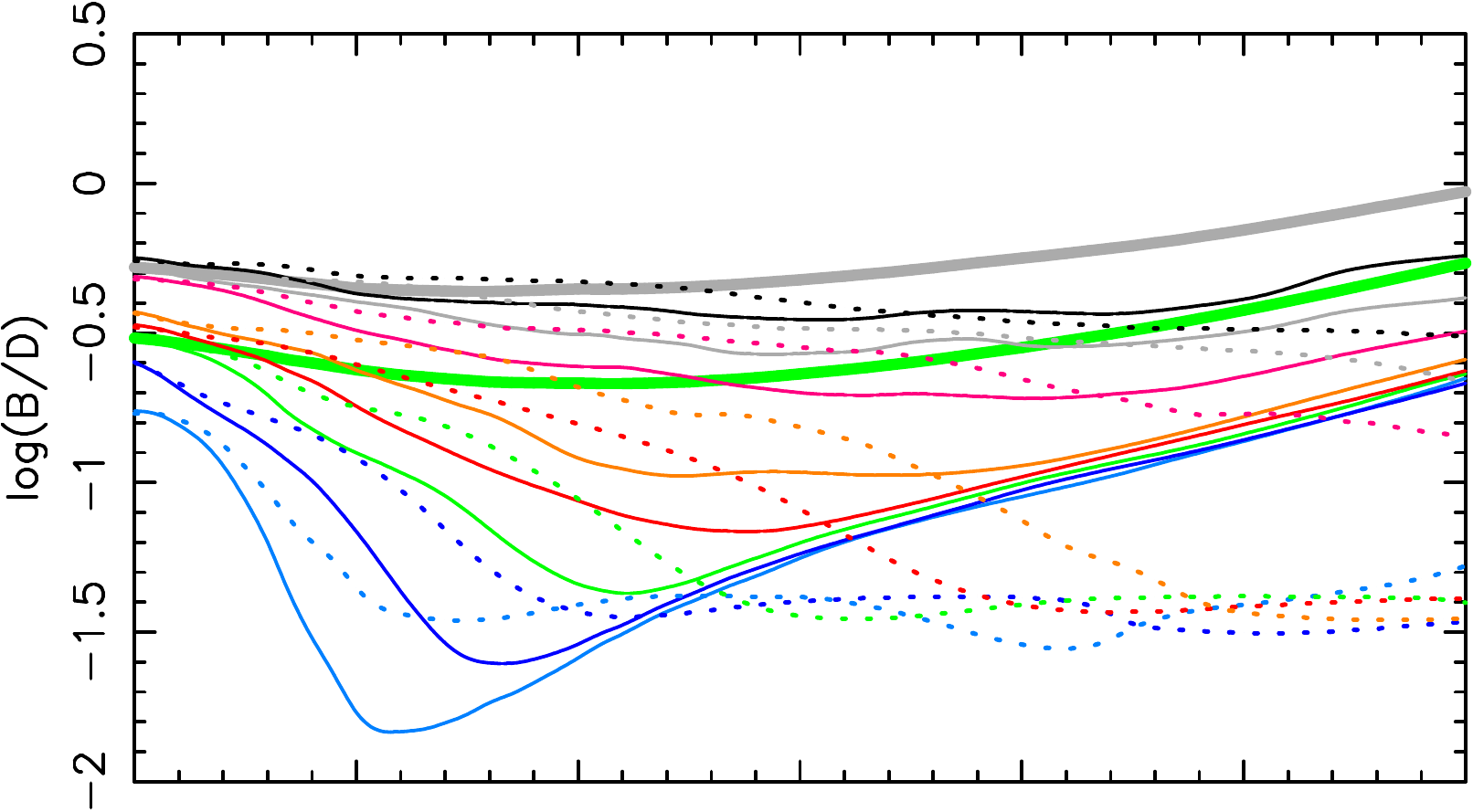}}
\put(124,0){\includegraphics[clip, width=6.12cm]{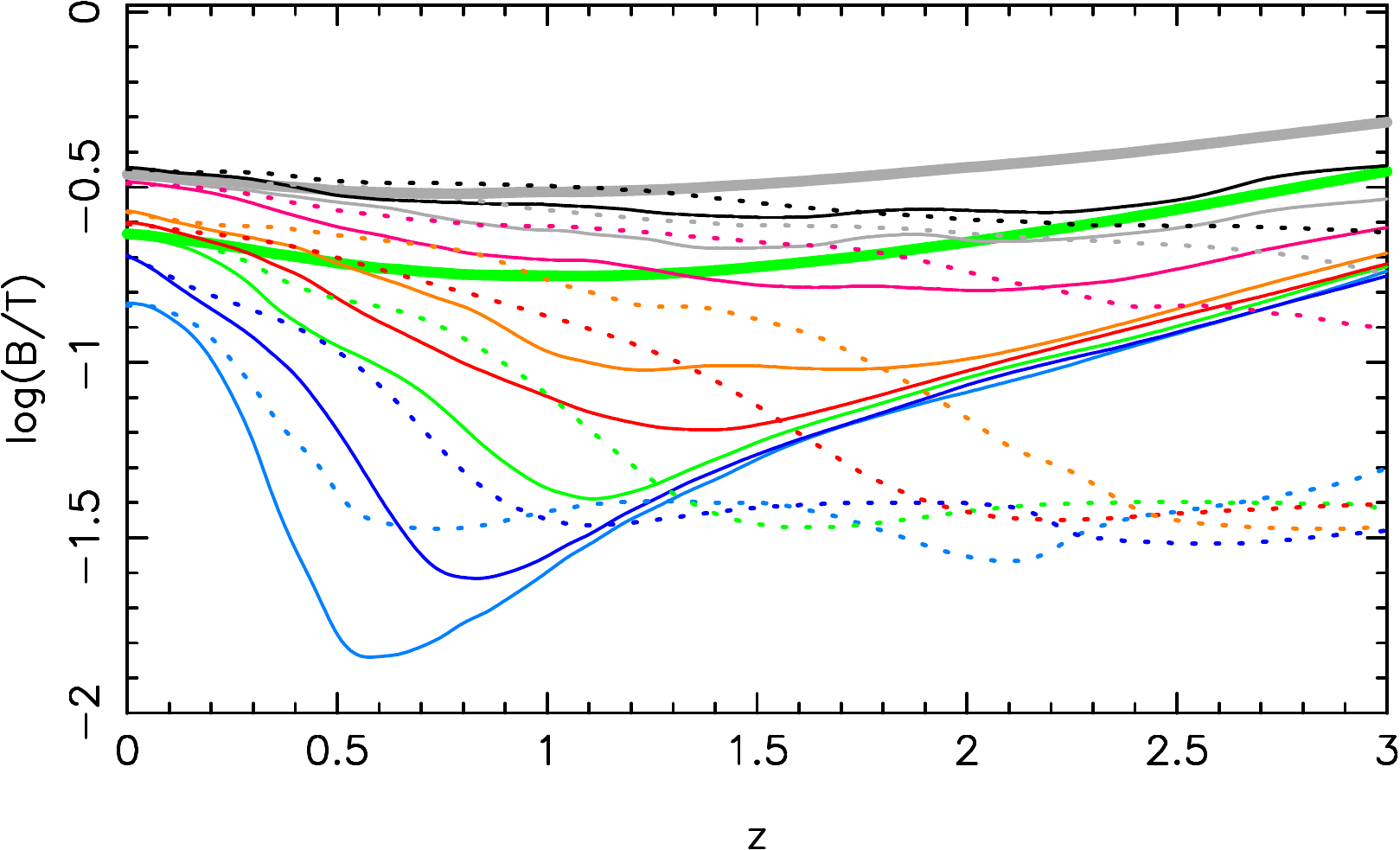}}
\PutLabel{186}{184}{\mlx \textcolor{black}{(a)}}
\PutLabel{186}{167}{\mlx \textcolor{black}{(b)}}
\PutLabel{186}{148}{\mlx \textcolor{black}{(c)}}
\PutLabel{186}{129}{\mlx \textcolor{black}{(d)}}
\PutLabel{186}{111}{\mlx \textcolor{black}{(e)}}
\PutLabel{186}{93}{\mlx \textcolor{black}{(f)}}
\PutLabel{186}{75}{\mlx \textcolor{black}{(g)}}
\PutLabel{186}{40}{\mlx \textcolor{black}{(h)}}
\PutLabel{186}{7}{\mlx \textcolor{black}{(i)}}
\end{picture}
%\PutWin{130}{125}{5.4cm}{
\caption{Photometric quantities for synthetic galaxies that consist of a bulge and a disk forming according to
the $\tau$0.5+contSF (left column), $\tau$1+$\tau$5 (middle column) and $\tau$1+iB (right column) model (cf. Fig.~\ref{fig:SFHs}).
Results from EvCon and single-age simulations are shown, respectively, with thin solid and dotted curves.
Thick curves show rest-frame values in $V$ (green) and $H$ (gray), as obtained from EvCon models.
Simulated galaxies in the first two columns comprise only stellar emission, whereas those in the right-hand column additionally include nebular emission.}
\label{fig:BDdec2} 
\end{figure*}
\FloatBarrier
% ::::::::::::::::::::::::::::::::::::::::::::::::::::::::::::::::::::::::::::::::::::::::::::::::::::::::::::::::::::::::::::::::::::::::::
\section{Supplementary notes on IFS-based simulations \label{ap:simIFS}}  % \brem{ap:simIFS} 
% ::::::::::::::::::::::::::::::::::::::::::::::::::::::::::::::::::::::::::::::::::::::::::::::::::::::::::::::::::::::::::::::::::::::::::
This section provides a concise description of the procedure followed for computing spatially resolved panchromatic (91 \AA\ -- 160 $\mu$m) SED cubes from IFS data.
In the case of \object{NGC 309} (Sect.~\ref{simIFS}) the IFS data come from the CALIFA Survey \citep{Sanchez12-DR1,GB15CALIFA} and cover a spectral range
between 0.34 $\mu$m and 0.75 $\mu$m (COMB setup) at a mean spectral resolution of $R\sim 850$. 

For \object{Mrk 1172} (Sect.~\ref{dis:morph}), \object{Haro 11} (Sect.~\ref{dis:CG}) and
\object{Arp 220} (Sect.~\ref{dis:DGs}) we used archival data from MUSE \citep{Bacon14} that were retrieved from the
European Southern Observatory (ESO) data archive in reduced form and covering a spectral range between 0.465 $\mu$m and 0.94 $\mu$m at $R\sim 2000$ in the visual.
These data were acquired in the framework of the ESO observing programs
099.B-0411 (PI: Johnston), 094.B-0944 and 096.B-0923 (PI: \"Ostlin) and 0103.B-0391 (PI: Arribas).

The IFS data cubes were processed with the pipeline {\sc Porto3D} \citep{P13,GP16-ETGs}, which invokes the population synthesis code \starlight\ \citep{Cid05} for fitting the stellar component, and additional routines described in \citet{BP18}. Spectral modeling was carried out using libraries of 236 simple stellar population templates from \citet{BruCha03}
for a Chabrier IMF that cover an age between 1 Myr and 13.5 Gyr and a metallicity between \zsun/50 and 2.5\,\zsun, and by assuming the \citet[][]{Cardelli89} extinction curve (hereafter CCM). Subsequently, the tool {\sc RemoveYoung} \citep{GP16-RY} was used to compute for each population vector (the best-fitting combination of simple stellar populations, i.e., a discretized approximation to the SFH) the UV-through-IR SED of the stellar continuum. This synthetic SED takes into account the attenuation of the stellar component in the UV, based on the intrinsic extinction $A_V$ inferred by \starlight.

At the next stage, emission line fluxes, as determined with {\sc Porto3D} after correction for underlying stellar absorption, were added to the stellar SED. 
Fluxes for emission lines outside the observed spectral range were computed assuming their theoretical ratio relative to H$\alpha$ flux, based on prescriptions
encoded in \pegase~2, and attenuated assuming CCM for the nebular component. The nebular extinction was estimated from the observed H$\alpha$/H$\beta$ ratio for regions
with an \ewha$>$6 \AA\ whereas set to the stellar extinction $A_V$ otherwise. The 6~\AA\ cutoff is because lower EWs can arise from the diffuse ionized gas
and photoionization by post-asymptotic giant branch sources \citep[e.g.,][]{bin94,Cid11}.
It is important to bear in mind that the choice of the intrinsic extinction model both for the nebular and stellar component can significantly 
influence the predicted rest-frame SED of a galaxy in the UV. For example, because of its different slope and the weakness of the 2175 \AA\ absorption feature
\citep[see, e.g.,][for a recent review]{SN20}, the \citet{Calzetti00} attenuation curve yields a lower dimming of the intrinsic NUV SED than the CCM model and
therefore also has a stronger effect on the \cmod\ for star-forming galaxies.

A slightly different approach was taken for the starburst galaxy \object{Haro 11}: because of the intense nebular emission in this BCD, in which the \ewha\ locally exceeds $10^3$ \AA\ (Fig.~\ref{cp-Haro11}), the nebular continuum is no longer negligible, as it can affect spectral modeling in the optical \citep{P98,Izotov11,PO12,GP17}
and contributes a significant fraction of the rest-frame UV luminosity.
In this case, \starlight\ was applied to the pure stellar component, after bidimensional subtraction of nebular (line+continuum) emission from the MUSE IFS data cube.
Subsequently, the synthetic stellar and nebular SED were co-added and used for simulating the $V$--$I$ color map of this system at different redshifts.
\end{appendix}
\end{document}